\documentclass{aastex631}

\usepackage{soul}
\usepackage{longtable}
\usepackage{longfigure}
\usepackage{rotating}
\usepackage[toc,page]{appendix}
\received{January 26, 2024}
\revised{April 12, 2024}
\accepted{April 26, 2024}

\submitjournal{PSJ}

\graphicspath{{./}{figures/}}

\begin{document}

\title{The population of small near-Earth objects: composition, source regions and rotational properties}

\correspondingauthor{Juan A. Sanchez}
\email{jsanchez@psi.edu}

\author{Juan A. Sanchez}
\altaffiliation{Visiting Astronomer at the Infrared Telescope Facility, which is operated by the University of Hawaii under Cooperative Agreement no. NNX-08AE38A with 
the National Aeronautics and Space Administration, Science Mission Directorate, Planetary Astronomy Program.}
\affiliation{Planetary Science Institute, 1700 East Fort Lowell Road, Tucson, AZ 85719, USA}

\author{Vishnu Reddy}
\altaffiliation{Visiting Astronomer at the Infrared Telescope Facility, which is operated by the University of Hawaii under Cooperative Agreement no. NNX-08AE38A with 
the National Aeronautics and Space Administration, Science Mission Directorate, Planetary Astronomy Program.}
\affiliation{Lunar and Planetary Laboratory, University of Arizona, 1629 E University Blvd, Tucson, AZ 85721-0092}

\author{Audrey Thirouin}
\affiliation{Lowell Observatory, 1400 West Mars Hill Road, Flagstaff, AZ 86004, USA}

\author{William F. Bottke}
\affiliation{Department of Space Studies, Southwest Research Institute, 1050 Walnut Street, Suite 300 Boulder, CO 80302, USA}

\author{Theodore Kareta}
\altaffiliation{Visiting Astronomer at the Infrared Telescope Facility, which is operated by the University of Hawaii under Cooperative Agreement no. NNX-08AE38A with 
the National Aeronautics and Space Administration, Science Mission Directorate, Planetary Astronomy Program.}
\affiliation{Lowell Observatory, 1400 West Mars Hill Road, Flagstaff, AZ 86004, USA}

\author{Mario De Florio}
\affiliation{Division of Applied Mathematics, Brown University, 170 Hope St, Providence, RI 02906, USA}

\author{Benjamin N. L. Sharkey}
\altaffiliation{Visiting Astronomer at the Infrared Telescope Facility, which is operated by the University of Hawaii under Cooperative Agreement no. NNX-08AE38A with 
the National Aeronautics and Space Administration, Science Mission Directorate, Planetary Astronomy Program.}
\affiliation{Department of Astronomy, University of Maryland 4296 Stadium Dr. PSC (Bldg 415) Rm 1113 College Park, MD 20742-2421, USA}

\author{Adam Battle}
\altaffiliation{Visiting Astronomer at the Infrared Telescope Facility, which is operated by the University of Hawaii under Cooperative Agreement no. NNX-08AE38A with 
the National Aeronautics and Space Administration, Science Mission Directorate, Planetary Astronomy Program.}
\affiliation{Lunar and Planetary Laboratory, University of Arizona, 1629 E University Blvd, Tucson, AZ 85721-0092}

\author{David C. Cantillo}
\altaffiliation{Visiting Astronomer at the Infrared Telescope Facility, which is operated by the University of Hawaii under Cooperative Agreement no. NNX-08AE38A with 
the National Aeronautics and Space Administration, Science Mission Directorate, Planetary Astronomy Program.}
\affiliation{Lunar and Planetary Laboratory, University of Arizona, 1629 E University Blvd, Tucson, AZ 85721-0092}

\author{Neil Pearson}
\affiliation{Planetary Science Institute, 1700 East Fort Lowell Road, Tucson, AZ 85719, USA}

\begin{abstract}

The study of small ($<$300 m) near-Earth objects (NEOs) is important because they are more closely related than larger objects to the precursors of meteorites that fall on Earth.  Collisions of these bodies with Earth are also more frequent. Although such collisions cannot produce massive extinction events, they can still produce significant local damage. 
Here we present the results of a photometric and spectroscopic survey of small NEOs, which include near-infrared (NIR) spectra of 84 objects with a mean diameter of 126 m and photometric data of 59 objects with a mean diameter of 87 m. We found that S-complex asteroids are the most abundant among the NEOs, comprising $\sim$66\% of the sample. Most 
asteroids in the S-complex were found to have compositions consistent with LL-chondrites. Our study revealed the existence of NEOs with spectral characteristics similar to those in 
the S-complex, but that could be hidden within the C- or X-complex due to their weak absorption bands. We suggest that the presence of metal or shock-darkening could be 
responsible for the attenuation of the absorption bands. These objects have been grouped into a new subclass within the S-complex called Sx-types. The dynamical modeling showed 
that 83\% of the NEOs escaped from the $\nu_{6}$ resonance, 16\% from the 3:1 and just 1\% from the 5:2 resonance. Lightcurves and rotational periods were derived from the photometric data. No clear trend between the axis ratio and the absolute magnitude or rotational period of the NEOs was found.

\end{abstract}


\section{Introduction} \label{sec:intro}

The population of small near-Earth objects (NEOs) constitutes the main reservoir of meteorites that fall on Earth. Because of this, their study is important to understand 
how and where they formed, as well as identifying the regions in the solar system that contribute most to their delivery to the near-Earth space. It is widely accepted that a collision with a 10-km object over 66 million years ago was responsible for the extinction of $\sim$75\% of animal and plant species on Earth \citep[e.g.,][]{1980Sci...208.1095A, 1991Geo....19..867H, 1994lmip.book.....D, 2002Geo....30..999S}. However, smaller objects, of tens of meters, can also pose a risk for civilization and result in significant local damage. The Tunguska event occurred in 1908 that 
devastated an area of over 2,200 km$^{2}$ in Central Siberia is thought to have been caused by an $\sim$60 m diameter object that exploded in the atmosphere \citep[e.g.,][]{1993Natur.361...40C, 2019Icar..327....4J}. More recently, the explosion of an $\sim$20 m meteor in Chelyabinsk, Russia, injured 
hundreds of people and caused extensive damage to the city. Events like these have demonstrated the importance of finding and characterizing small NEOs, since collisions with these objects are more frequent than with larger bodies. 

 NEO surveys have been very successful at finding new objects with over 34,600 NEOs discovered as of April 2024 (IAU Minor Planet Center Page). However, the 
physical characterization of these objects has lagged behind with only a few studies dedicated to determine their rotation periods, taxonomy and composition
 \citep[e.g.,][]{2014Icar..228..217T, 2016AJ....152..163T, 2018ApJS..239....4T,  2019AJ....158..196D,  2019Icar..324...41B, 2023MNRAS.520.3143H}. Currently, taxonomic classifications are available for $\sim$15\% of the known NEO population, but actual compositional analysis and/or established meteorite affinities are only available for a smaller fraction of NEOs. 

Most near-infrared (NIR) spectroscopic surveys of NEOs carried out so far studied objects with a very broad size range, with samples often dominated by km-sized asteroids, as 
these are the most accesible for ground-based telescopes \citep[e.g.,][]{2008Natur.454..858V, 2010A&A...517A..23D, 2013Icar..222..273D, 2014Icar..228..217T,  2019Icar..324...41B}. The present work seeks to complement previous efforts by focusing on the study of small NEOs (absolute magnitudes H$>$20), while also 
addressing the U.S. Congress mandate, which in 2005 directed NASA to find and characterize at least 90\% of potentially hazardous NEOs sized 140 meters or larger. At 
the same time, we also investigate if the trends observed among larger NEOs in terms of taxonomic distribution, composition and source regions remain the same for 
smaller objects.

Our study combines both spectroscopic and photometric data of small NEOs obtained over the course of $\sim$7 years. In the spectroscopic study (section 2) we analyze the NEO spectra employing different techniques, 
including thermal modeling, spectral band parameters, curve matching and machine learning. We carry out the compositional analysis of the NEOs and look for possible 
meteorite analogs. Laboratory spectra of meteorite samples prepared by us are also used to help in the interpretation of the telescopic data. In addition, dynamical modeling 
is preformed to identify the source regions of the NEOs. In the photometric study (section 3) we derive the lightcurves of a fraction of the NEOs and obtain rotational periods and 
lightcurve amplitudes. Moreover, we investigate possible relationships between axis ratios and absolute magnitudes and rotational periods. A summary of our main 
results is presented in section 4.

\begin{figure*}[!ht]
\begin{center}
\includegraphics[height=9cm]{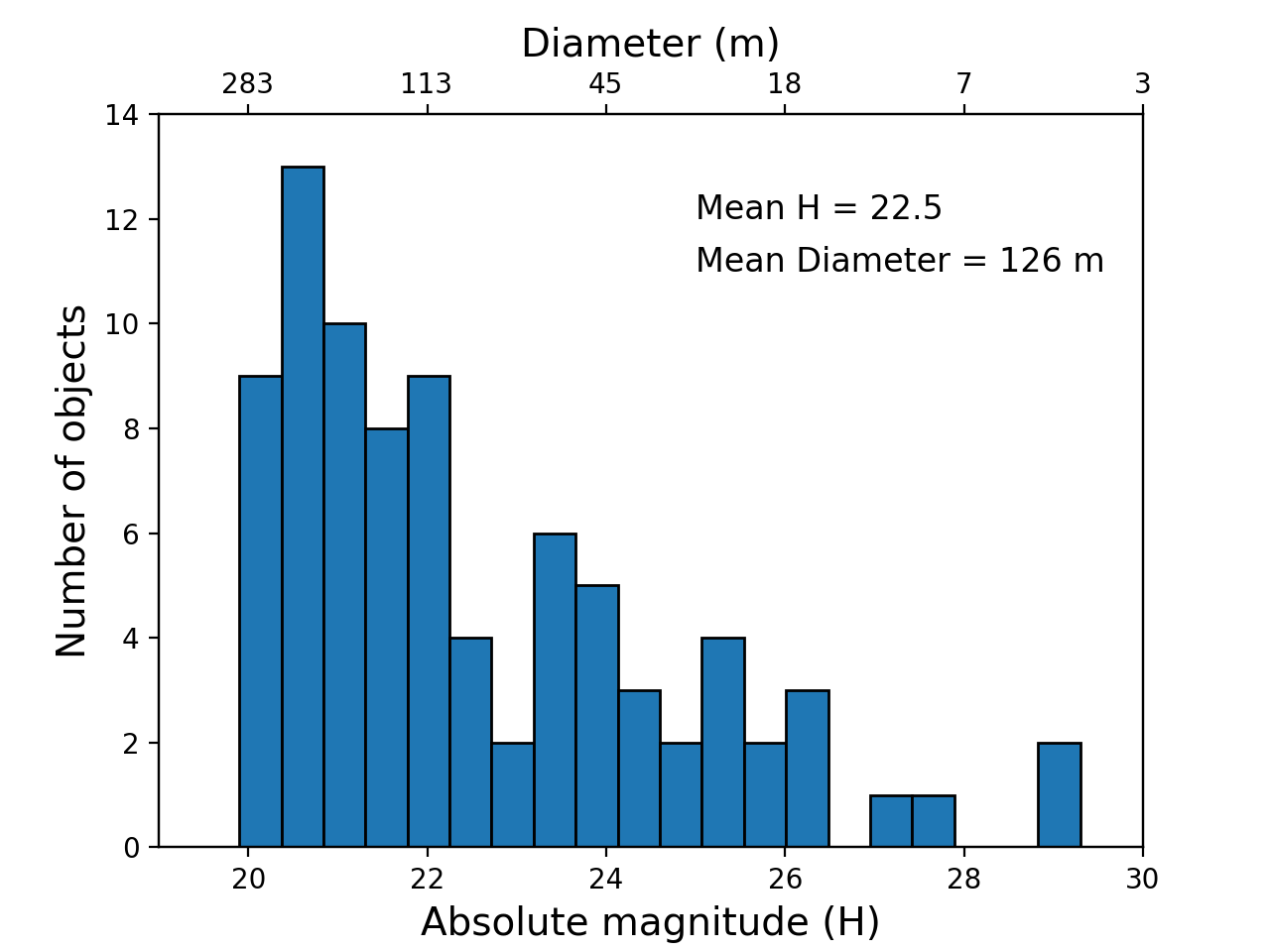}

\caption{\label{f:H_D_Histogram_spec}{\small Distribution of absolute magnitudes (H) and diameters for the NEOs included in the spectroscopic study. Diameters 
were calculated from the absolute magnitudes and the mean geometric albedo of the sample ($P_{V}$=0.22).}}

\end{center}
\end{figure*}

\vspace{-0.5cm}

\section{Spectroscopic Study} \label{sec:specstudy}

\subsection{Observations and data reduction} \label{sec:obs}

All NEOs presented in this study were observed with the SpeX instrument \citep{2003PASP..115..362R} on NASA Infrared Telescope Facility (IRTF) between 
October 2013 and January 2021. In 2014, SpeX was upgraded and the Raytheon Aladdin 3 1024x1024 InSb array in the spectrograph was replaced by a Teledyne 2048x2048 Hawaii-2RG array. NIR spectra (0.7-2.5 $\mu$m) were obtained in low-resolution (R$\sim$150) prism 
mode with a 0.8” slit width. During the observations, the slit was oriented along the parallactic angle in order to minimize the effects of differential atmospheric refraction. Spectra were obtained in 
two different slit positions (A-B) following the sequence ABBA. Depending on the magnitude of the asteroid the integration time varied between 120 and 200 seconds. In 
order to correct the telluric bands from the asteroid spectra, a G-type local extinction star was observed before and after the asteroid. NIR spectra of a solar analog 
were also obtained to correct for possible spectral slope variations that could be introduced by the use of a non-solar local extinction star. For asteroids fainter 
than V.mag $\sim$16.5, guiding was done using the MIT Optical Rapid Imaging System (MORIS) instrument, a high-speed visible-wavelength camera mounted 
on SpeX. Using non-sidereal tracking rates, MORIS can image fast-moving fainter targets ($>$10”/sec) with no trailing. This instrument has a pixel scale of 
0.11$\arcsec$/pixel for a field of view of 1$\arcmin$$\times$1$\arcmin$. MORIS images were saved and used to obtain the lightcurves of the asteroids (Section 3). For observation carried out prior of 2017, a 0.8 $\mu$m cut-on dichroic was used with MORIS, this dichroic was later replaced with a 0.7 $\mu$m dichroic during 
semester 2017A. For each night, calibration images including flat fields and argon arc-lamp spectra were also acquired. Observational circumstances for the 84 NEOs 
included in the spectroscopic study are presented in Table 1. The distribution of absolute magnitudes for these NEOs are shown in Figure 
\ref{f:H_D_Histogram_spec}. 

Spectroscopic data were reduced using the IDL-based software Spextool \citep{2004PASP..116..362C} and several python scripts following the same 
procedure described in \cite{2013Icar..225..131S}. The data reduction procedure includes the following steps: (1) sky background removal by subtracting the A-B image 
pairs, (2) flat-fielding, (3) cosmic ray and spurious hit removals, (4) wavelength calibration, (5) division of asteroid spectra by the spectra of the local extinction star and 
solar analog star, and (6) co-adding of individual spectra. NIR spectra of the observed NEOs are shown in Appendix A.

\vspace{-0.3cm}

\startlongtable
\begin{deluxetable*}{ccccccccc}

\tablecaption{\label{t:Table1} {\small Observational circumstances. The columns in this table are: object number and designation, date, absolute magnitude (H), 
V-magnitude (V), phase angle ($\alpha$), heliocentric distance (r), airmass and solar analog used.}}

\tablewidth{0pt}

\tablehead{Number&Designation&Date (UT)&H (mag)&V (mag)&$\alpha$ $(^{\circ})$&r (au)&Airmass&Solar Analog \\}

\startdata
85990&1999 JV6&11-Jan-2015&20.2&16.7&44.5&1.04&1.14&SAO 120107 \\
163348&2002 NN4&01-Jul-2020&20.1&18.0&32.5&1.16&1.21&SAO 83469 \\
363599&2004 FG11&09-Apr-2016&21.0&17.1&62.1&1.03&1.29&SAO 120107 \\
412995&1999 LP28&12-Dec-2018&20.1&17.1&15.3&1.14&1.27&SAO 93936 \\
436724&2011 UW158&08-Aug-2015&19.9&16.7&71.4&1.03&1.11&SAO 31899 \\
437844&1999 MN&21-Jun-2015&21.2&17.9&20.9&1.12&1.39&SAO 120107 \\
438908&2009 XO&12-May-2020&20.7&14.9&19.0&1.05&1.29&SAO 120107 \\
459872&2014 EK24$^{a}$&20-Feb-2015&23.4&17.9&35.9&1.02&1.11&SAO 120107 \\
467336&2002 LT38&10-Jun-2016&20.5&15.5&7.1&1.43&1.11&SAO 120107\\
469737&2005 NW44$^{a}$&13-Jul-2016&20.4&18.1&29.2&1.16&1.25&SAO 120107\\
471240&2011 BT15&08-Jan-2014&21.7&17.3&60.2&1.01&1.04&SAO 120107\\
496816&1989 UP&14-Oct-2017&20.6&17.0&21.7&1.10&1.13&SAO 93936 \\
501647&2014 SD224&17-Dec-2020&22.4&17.4&32.1&1.03&1.26&SAO 93936 \\
515742&2015 CU$^{b}$&20-Feb-2015&20.8&18.0&14.5&1.15&1.55&SAO 120107 \\
515767&2015 JA2&01-Feb-2019&21.2&17.3&13.6&1.09&1.27&SAO 93936 \\
528159&2008 HS3&09-May-2019&21.6&15.2&12.2&1.05&1.10&SAO 120107 \\
&2000 TU28&19-Oct-2020&21.1&16.7&27.1&1.06&1.22&SAO 93936 \\
&2001 YV3&17-Dec-2020&20.6&17.2&48.7&1.04&1.10&SAO 93936 \\
&2002 LY1$^{a}$&10-Jun-2016&22.3&17.8&55.3&1.04&1.08&SAO 120107 \\
&2005 NE21&11-Jul-2019&21.3&17.8&11.2&1.15&1.38&SAO 120107 \\
&2005 TF&03-Nov-2016&20.3&16.5&7.9&1.11&1.05&SAO 93936 \\ 
&2006 XY&16-Dec-2017&24.2&15.8&47.2&0.99&1.45&SAO 93936 \\ 
&2007 EC&19-Jan-2015&22.2&16.6&55.0&1.00&1.28&SAO 120107 \\
&2012 ER14&14-Oct-2013&20.5&17.1&40.4&1.07&1.05&SAO 147208 \\
&2013 CW32&01-Feb-2019&22.1&15.9&10.0&1.03&1.06&SAO 93936 \\
&2013 RS43$^{b}$&13-Sept-2013&27.0&18.2&49.6&1.01&1.28&SAO 147208 \\
&2013 XA22&31-May-2020&22.9&17.1&31.4&1.05&1.21&SAO 120107 \\
&2014 PL51&25-Aug-2014&20.4&16.9&35.3&1.09&1.29&SAO 93936 \\
&2014 PR62&22-Aug-2014&20.6&17.4&13.6&1.15&1.26&SAO 93936 \\
&2014 QH33$^{b}$&25-Aug-2014&24.3&18.3&8.7&1.05&1.22&SAO 93936 \\
&2014 QL32$^{b}$&26-Aug-2014&22.8&18.7&26.5&1.08&1.14&SAO 93936 \\
&2014 QZ265$^{b}$&26-Aug-2014&25.3&19.2&35.5&1.04&1.12&SAO 93936 \\
&2014 SF304$^{b}$&03-Oct-2014&27.3&17.7&17.5&1.01&1.03&SAO 93936 \\
&2014 SS1&24-Sept-2014&21.7&17.2&37.0&1.05&1.09&SAO 93936 \\
&2014 UV210$^{b}$&16-Dec-2014&26.9&18.7&5.4&1.00&1.06&SAO 120107 \\
&2014 VH2$^{a}$&25-Nov-2014&21.6&18.3&6.6&1.14&1.01&SAO 93936 \\
&2014 VQ&18-Nov-2014&20.4&16.2&14.6&1.08&1.18&SAO 93936 \\
&2014 WC201&01-Dec-2014&26.2&16.7&38.7&0.99&1.26&SAO 93936 \\
&2014 WN4$^{a}$&18-Nov-2014&23.3&18.0&17.0&1.04&1.15&SAO 93936 \\
&2014 WO4$^{b}$&18-Nov-2014&24.4&17.9&17.0&1.02&1.14&SAO 93936 \\
&2014 WO7$^{b}$&25-Nov-2014&20.5&18.9&10.7&1.26&1.27&SAO 93936 \\ 
&2014 WP4$^{b}$&18-Nov-2014&24.3&18.8&22.1&1.03&1.27&SAO 93936 \\ 
&2014 WY119$^{a}$&25-Nov-2014&26.3&18.1&37.0&0.99&1.22&SAO 93936 \\
&2014 WZ120&25-Nov-2014&20.5&16.7&26.3&1.07&1.08&SAO 93936 \\
&2014 XB6$^{b}$&16-Dec-2014&26.5&20.3&63.8&0.99&1.31&SAO 120107 \\
&2015 AK45$^{a}$&26-Jan-2015&26.4&17.9&22.0&0.99&1.07&SAO 120107 \\
&2015 AP43&09-Jun-2019&20.0&17.2&33.0&1.12&1.42&SAO 120107 \\
&2015 BC&19-Jan-2015&24.0&15.3&40.0&0.99&1.13&SAO 120107 \\
&2015 BK509&24-Feb-2020&22.4&18.0&44.0&1.03&1.03&SAO 93936 \\
&2015 CA40$^{b}$&20-Feb-2015&24.6&18.2&47.0&1.00&1.66&SAO 120107 \\
&2015 CN13$^{b}$&20-Feb-2015&23.3&18.4&23.7&1.04&1.35&SAO 120107 \\
&2015 FL&10-Apr-2015&20.8&16.7&64.0&1.02&1.31&SAO 120107 \\
&2015 GY$^{b}$&25-Apr-2015&21.7&19.1&45.6&1.09&1.08&SAO 120107 \\
&2015 HA1&21-May-2015&21.2&17.3&36.2&1.07&1.17&SAO 120107 \\
&2015 HE10$^{b}$&25-Apr-2015&26.1&19.1&39.2&1.02&1.77&SAO 120107 \\
&2015 HW11$^{b}$&26-Apr-2015&23.3&18.7&37.6&1.05&1.32&SAO 120107 \\
&2015 JW$^{b}$&21-May-2015&25.8&19.5&2.7&1.06&1.37&SAO 120107 \\
&2015 KA$^{b}$&21-May-2015&26.2&19.4&27.1&1.03&1.10&SAO 120107 \\
&2015 LK24&22-Jun-2015&21.6&16.9&35.6&1.06&1.59&SAO 120107 \\
&2015 MC$^{b}$&22-Jun-2015&24.1&18.8&19.2&1.07&1.39&SAO 120107 \\
&2015 NA14&21-Jul-2015&22.0&17.7&31.2&1.08&1.27&SAO 120107 \\
&2015 SA$^{b}$&19-Sept-2015&25.3&18.7&42.7&1.02&1.08&SAO 31899 \\
&2015 SE$^{b}$&19-Sept-2015&26.4&18.8&44.7&1.01&1.88&SAO 31899 \\
&2015 SZ$^{b}$&12-Oct-2015&23.5&18.6&43.0&1.03&1.09&SAO 31899 \\
&2015 TB25$^{a}$&12-Oct-2015&24.5&17.8&28.5&1.02&1.02&SAO 31899 \\
&2015 TC25&12-Oct-2015&29.3&17.5&32.0&1.00&1.58&SAO 31899 \\
&2015 TE$^{b}$&12-Oct-2015&22.5&18.6&37.5&1.06&1.54&SAO 31899 \\
&2015 TF$^{a}$&12-Oct-2015&22.2&18.7&3.6&1.15&1.50&SAO 31899 \\
&2015 VE66&22-Nov-2015&24.1&16.3&13.8&1.01&1.22&SAO 93936 \\
&2015 WF13$^{a}$&08-Dec-2015&23.2&17.1&42.6&1.01&1.54&SAO 93936 \\ 
&2015 XC$^{a}$&08-Dec-2015&25.2&17.4&74.1&0.99&1.54&SAO 93936 \\ 
&2016 BC14$^{a}$&07-Mar-2016&20.9&17.8&64.6&1.03&1.10&SAO 120107 \\ 
&2016 CM194&13-Feb-2016&27.7&13.7&59.6&0.99&1.13&SAO 93936 \\
&2016 CO247&15-Jan-2021&20.6&16.6&15.3&1.08&1.07&SAO 93936 \\
&2016 EB1$^{a}$&07-Mar-2016&25.3&18.3&10.4&1.02&1.19&SAO 93936 \\
&2016 EF28$^{a}$&07-Mar-2016&21.7&17.9&21.9&1.08&1.02&SAO 120107 \\
&2016 EV27$^{a}$&03-Apr-2016&23.4&17.8&21.0&1.04&1.05&SAO 120107 \\
&2016 FV13$^{a}$&09-Apr-2016&25.8&17.9&47.3&1.01&1.24&SAO 120107 \\
&2016 GU&09-Apr-2016&25.7&17.1&9.0&1.02&1.20&SAO 120107 \\
&2016 JD18$^{b}$&13-May-2016&24.6&18.4&32.0&1.04&1.20&SAO 120107 \\
&2016 LG$^{a}$&10-Jun-2016&25.3&18.5&57.8&1.02&1.05&SAO 120107 \\
&2017 BS5&23-Jul-2017&24.2&15.2&33.7&1.02&1.50&SAO 53622 \\
&2017 BW&06-Feb-2017&23.5&17.0&18.5&1.02&1.13&SAO 93936 \\
&2017 BY93&22-Feb-2017&23.1&15.6&31.0&1.00&1.21&SAO 120107 \\
&2017 CR32&20-Mar-2017&22.2&17.4&10.3&1.07&1.06&SAO 120107 \\ 
&2017 DR34$^{a}$&24-Feb-2017&29.2&18.2&8.3&0.99&1.08&SAO 120107 \\
&2017 FU64$^{a}$&03-Apr-2017&23.8&18.0&26.3&1.04&1.21&SAO 93936 \\
&2017 OL1&01-Aug-2017&21.7&17.0&11.2&1.09&1.31&SAO 53622 \\
&2017 OP68&13-Sept-2017&21.0&15.8&16.0&1.06&1.28&SAO 93936 \\
&2017 RR15&14-Oct-2017&20.7&17.5&19.5&1.12&1.03&SAO 93936 \\
&2017 WX12&16-Dec-2017&22.1&16.3&21.0&1.02&1.14&SAO 93936 \\
&2018 XG5&09-May-2019&20.3&16.8&49.0&1.06&1.10&SAO 120107 \\
&2018 XS4&17-Dec-2018&25.2&16.6&42.4&0.99&1.03&SAO 93936 \\
&2019 AN5&17-Aug-2020&21.1&17.1&57.1&1.04&1.37&SAO 109542 \\
&2019 GT3&10-Sept-2019&21.0&16.5&42.0&1.05&1.31&SAO 93936 \\
&2019 JL3&18-May-2019&25.0&17.2&50.0&1.02&1.05&SAO 120107 \\
&2019 JX7&09-Jun-2019&21.5&17.9&41.3&1.08&1.02&SAO 120107 \\
&2019 RC&10-Sept-2019&21.8&18.0&49.2&1.05&1.28&SAO 93936 \\
&2019 SH6&04-Jan-2020&20.0&17.3&53.0&1.05&1.13&SAO 93936 \\
&2019 UO13&03-Nov-2019&23.8&16.7&24.8&1.01&1.52&SAO 93936 \\
&2019 YM3&02-Jan-2020&23.3&17.9&17.6&1.03&1.02&SAO 93936 \\
&2020 DZ1&24-Feb-2020&24.0&17.6&14.5&1.02&1.35&SAO 93936 \\
&2020 HS6&12-May-2020&22.2&17.6&44.8&1.05&1.12&SAO 120107 \\
&2020 KC5&29-May-2020&27.4&17.6&67.1&1.02&1.23&SAO 120107 \\
&2020 RO6&15-Jan-2021&22.4&17.6&35.8&1.03&1.12&SAO 93936 \\
&2020 SN&22-Sept-2020&24.8&17.2&5.8&1.03&1.09&SAO 93936 \\
&2020 ST1&17-Nov-2020&22.0&16.4&12.8&1.04&1.22&SAO 93936 \\
&2020 YQ3&15-Jan-2021&20.0&17.9&21.1&1.17&1.3&SAO 93936 \\
\enddata
\tablenotetext{a}{Asteroid observed with the 0.8 $\mu$m dichroic.} \tablenotetext{b}{The S/N of the spectrum was deemed too low to be published. Only photometric data could be used.}
\end{deluxetable*}

\begin{figure*}[h]
\begin{center}
\includegraphics[height=17cm]{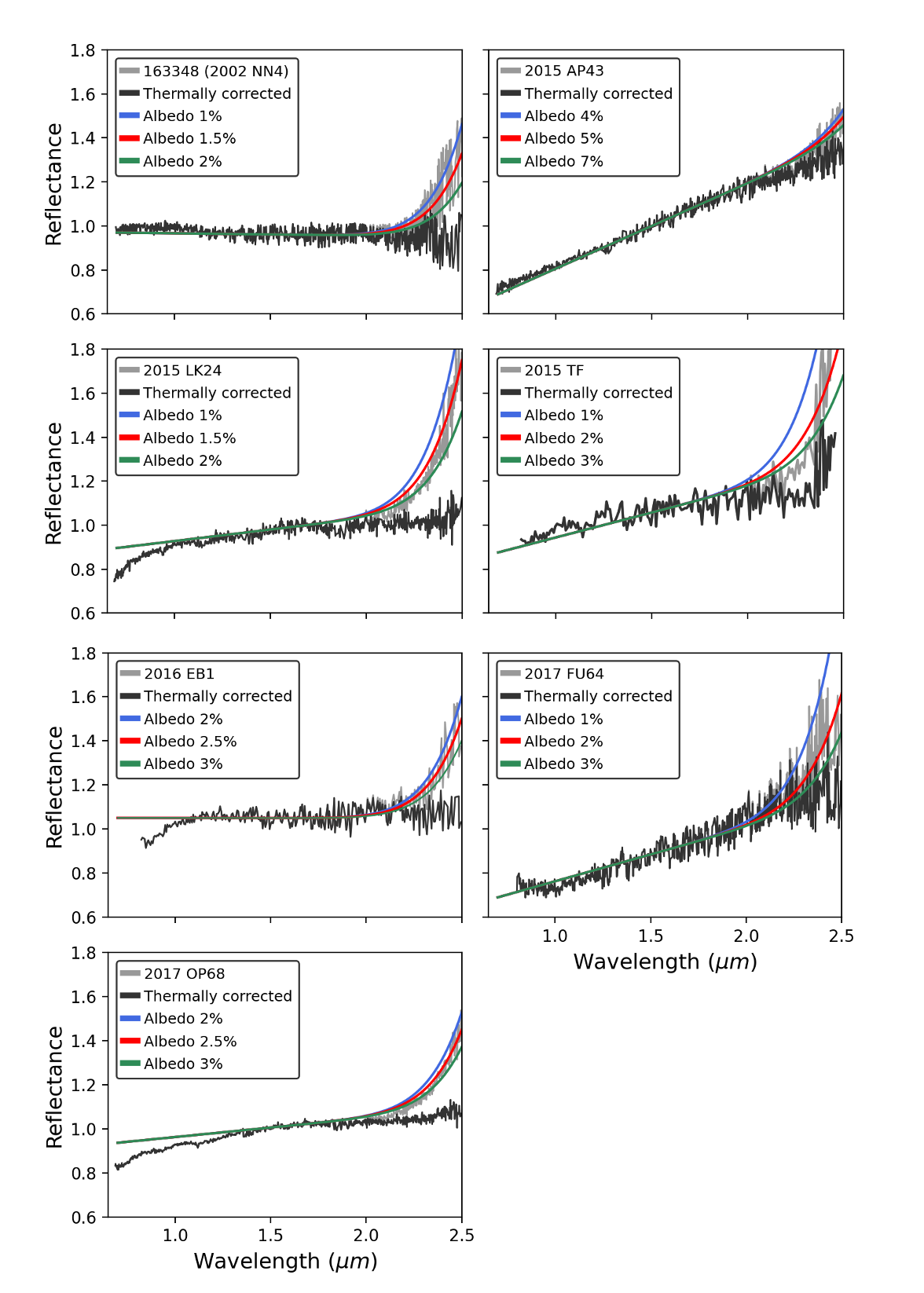}

\caption{\label{f:Thermal_model}{\small NIR spectra of asteroids 163348 (2002 NN4), 2015 AP43, 2015 LK24, 2015 TF, 2016 EB1, 2017 FU64 and 2017 OP68. Modeled albedo curves that encompass the observed thermal fluxes and thermally corrected spectra are shown.}}

\end{center}
\end{figure*}

\vspace{-0.7cm}

\subsection{Thermal modeling} \label{sec:tm}

Seven objects in our sample show a thermal excess at wavelengths $>$2 $\mu$m. The location and magnitude of the thermal contribution in the near-infrared is an indicator of the surface temperature of the body. Thus, the amount of thermal flux emitted by an asteroid can be used to constrain its albedo 
\citep[e.g.,][]{2009M&PS...44.1917R, 2012Icar..219..382R,  2022PSJ.....3..105K, 2023PSJ.....4...91L}. The thermal flux of the asteroids was modeled 
following the same method described in \cite{2009M&PS...44.1917R, 2012Icar..219..382R}, which is a modified version of the Standard Thermal Model 
(STM) \citep{1989aste.conf..128L}. Thermal models were generated for a given solar distance and phase angle for albedos ranging from 1\% to 10\%. An 
emissivity $\epsilon$=0.90 and a beaming parameter $\eta$=0.75 was used for all the models. The spectra along with the modeled albedo curves that can 
be fit to encompass the observed thermal fluxes are shown in Figure \ref{f:Thermal_model}. The quality of the fit is largely dependent on the signal-to-noise ratio 
(S/N) of the spectra at long wavelengths. Spectra with a high point-to-point scatter at wavelengths $>$2$\mu$m generally produce a poor fit (e.g., 2015 TF). 
This was one of the reasons why we choose to use a fixed value for $\eta$ instead of leaving it as a free parameter, because we noticed that it produced a better fit for 
data with a low S/N. Geometric albedos obtained from the thermal modeling were used along with the absolute magnitudes of the NEOs to estimate their diameter using the 
following equation \citep{1992imps.rept.....T, 2007Icar..190..250P}:

\begin{equation}
D = \left[\frac{1329}{\sqrt{P_{V}}}\right] \times 10^{-H/5}
\end{equation}

where $P_{V}$ is the geometric albedo and $H$ the absolute magnitude of the asteroid. Derived albedos and diameters are presented in Table 2. The final step was to use the models to remove the thermal excess from each 
spectrum, so the taxonomic classification can be applied.

\begin{table}[h!]
\begin{center}
\caption{\label{t:Table2} {\small Albedos and diameters resulting from the thermal modeling.}}
\hspace*{-1.7cm}
\begin{tabular}{ccccc}
\tableline
Number&Designation&Albedo&H (mag)&Diameter (m) \\  \hline

163348&2002 NN4&0.015$\pm$0.005&20.1$\pm$0.5&1036$\pm$295 \\

&2015 AP43&0.050$\pm$0.020&20.0$\pm$0.5&594$\pm$181 \\

&2015 LK24&0.015$\pm$0.005&21.6$\pm$0.5&519$\pm$148 \\

&2015 TF&0.020$\pm$0.010&22.2$\pm$0.5&341$\pm$116 \\

&2016 EB1&0.025$\pm$0.005&25.3$\pm$0.5&73$\pm$18 \\

&2017 FU64&0.020$\pm$0.010&23.8$\pm$0.5&163$\pm$56 \\

&2017 OP68&0.025$\pm$0.005&21.0$\pm$0.5&530$\pm$133 \\

\tableline
\end{tabular}
\end{center}
\end{table}

\begin{figure*}[h]
\begin{center}
\includegraphics[height=10cm]{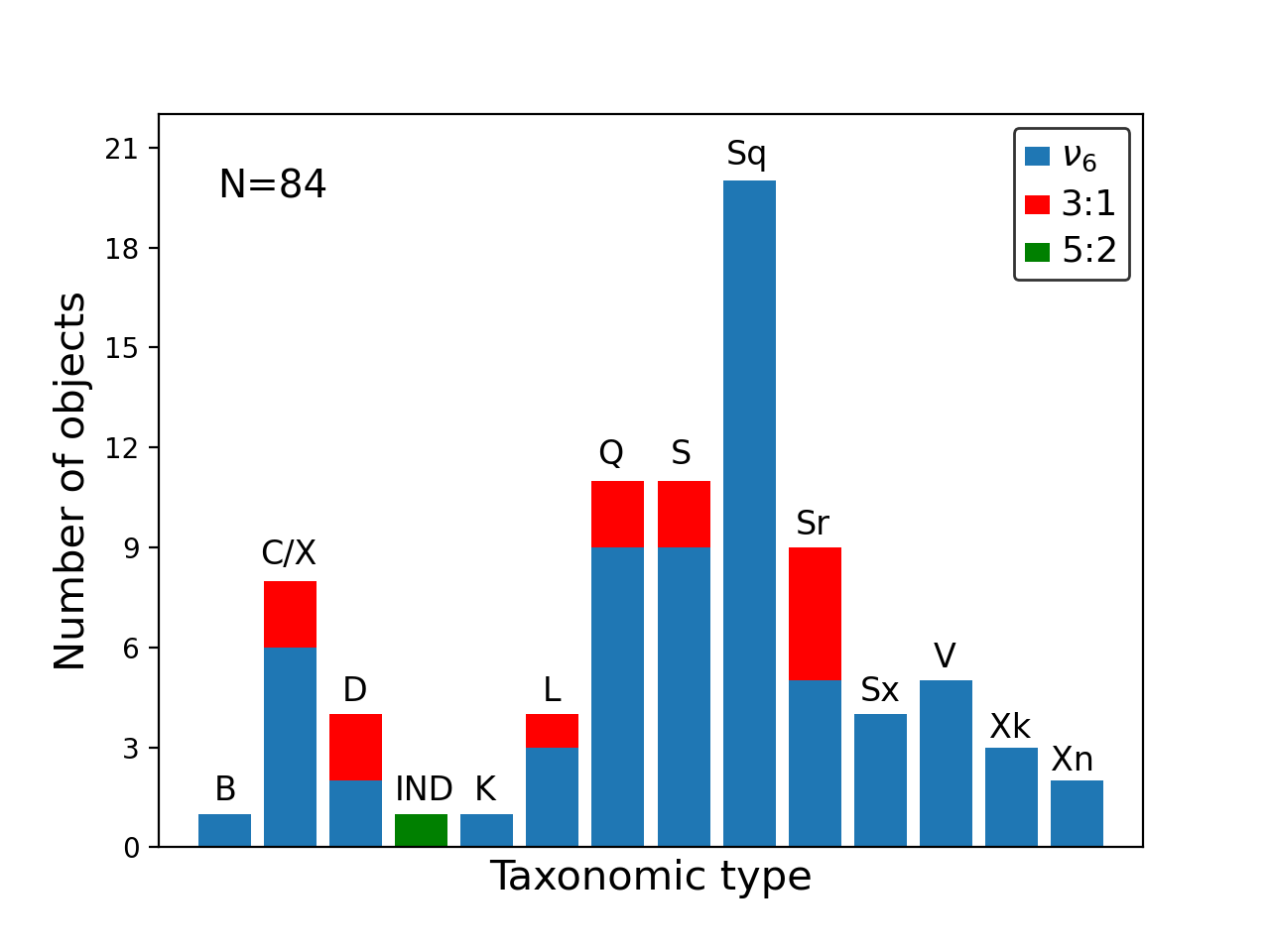}

\caption{\label{f:Tax_Hist}{\small Distribution of taxonomic types found in the present study. The label C/X corresponds to objects with ambiguous C- or X-complex 
classification. The label IND is assigned to objects whose taxonomic classification is indeterminate. The Sx-types are a new subclass of objects introduced in 
the present study. The source regions of the NEOs are indicated with different colors.}}

\end{center}
\end{figure*}

\subsection{Taxonomic classification} \label{sec:tc}

The taxonomic classification of the NEOs was done using the Bus-DeMeo Taxonomy Classification Web tool 
$\footnote{http://smass.mit.edu/busdemeoclass.html}$. This online tool is an implementation of the Bus-DeMeo taxonomy \citep{2009Icar..202..160D}, 
which uses principal components analysis (PCA) to classify asteroids among 25 different classes. The Bus-DeMeo taxonomy requires visible and NIR spectra (0.45-2.45 $\mu$m) for the classification; however, most of our spectra cover the wavelength range of $\sim$0.7-2.45 $\mu$m. For objects whose NIR spectra show an absorption band at $\sim$1 $\mu$m, such as those in the S-complex, V-types, etc., and do not show features in the visible, we extrapolated the data to 0.45 $\mu$m. For objects whose spectra are featureless in the NIR, like most taxonomic types in the C- and X-complexes, we cannot use this procedure because in the visible the spectra of these objects can have very different shapes that cannot be reproduced with a simple extrapolation. For this reason, for those objects the taxonomic classification was done using only data from 0.8-2.45 $\mu$m. The main drawback of not having visible data is not being able to distinguish between most of the types that comprise the C- and X-complexes, with the exception of the Xe-, Xk- and Xn-types that exhibit a weak absorption band at $\sim$0.9 $\mu$m. Due to this limitation, in the present work we have combined the C- and X-complexes into the same group (C/X).

\begin{table}[h!]

\caption{\label{t:Table4}{\small Taxonomic classification and meteorite analogs for the studied NEOs. If the classification result is indeterminate the letters IND are 
used. Meteorite abbreviations are: H-, L- and LL-ordinary chondrites (H/L/LL-OC), howardite (How), non-cumulate eucrite (NC-Euc), cumulate eucrite (C-Euc), 
carbonaceous chondrite (CC), aubrite (A), enstatite chondrite (EC), primitive achondrite (PA), metal-rich chondrite (MR-OC), stony-iron meteorite (SI), iron 
meteorite (I). The possible presence of shock darkening is indicated as SD. If no meteorite analog was identified the letters IND are used.}}

\hspace*{-1cm}
\begin{tabular}{cccc|ccc}
\tableline

Number&Designation&Taxonomy&Meteorite Analog&Designation&Taxonomy&Meteorite Analog \\  \hline

85990&1999 JV6&Xk&NC-Euc, SI, I&2015 TB25&L&CC \\
163348&2002 NN4&B&CC&2015 TC25&Xn&A \\
363599&2004 FG11&Q&LL-OC&2015 TF&C/X&CC \\
412995&1999 LP28&Sqw&LL-OC&2015 VE66&Sr&L-OC \\
436724&2011 UW158&S&H-OC&2015 WF13&V&NC-Euc \\
437844&1999 MN&Q&LL-OC&2015 XC&V&C-Euc \\
438908&2009 XO&Q&LL-OC&2016 BC14&S&L-OC \\
459872&2014 EK24&Sq&LL-OC&2016 CM194&Sx&L-OC, SD, PA \\
467336&2002 LT38&Sq&LL-OC&2016 CO247&C/X&A, EC \\
469737&2005 NW44&Sq&LL-OC&2016 EB1&C/X&CC \\
471240&2011 BT15&Sr&L-OC&2016 EF28&V&NC-Euc \\
496816&1989 UP&Sq&LL-OC&2016 EV27&Q&LL-OC \\
501647&2014 SD224&Sq&LL-OC&2016 FV13&Q&L-OC \\
515767&2015 JA2&Sq&LL-OC&2016 GU&Xn&A, EC \\
528159&2008 HS3&Sq&LL-OC&2016 LG&Q&LL-OC \\
&2000 TU28&Sq&LL-OC&2017 BS5&S&LL-OC  \\
&2001 YV3&Q&LL-OC&2017 BW&L&CC\\
&2002 LY1&C/X&A, EC, CC&2017 BY93&Sq&L-OC \\
&2005 NE21&S&LL-OC&2017 CR32&L&CC  \\
&2005 TF&Q&LL-OC&2017 DR34&S&H-OC \\
&2006 XY&C/X&A, EC, CC&2017 FU64&D&CC \\
&2007 EC&Sq&LL-OC&2017 OL1&S&L-OC \\
&2012 ER14&D&CC&2017 OP68&C/X&CC \\
&2013 CW32&Sx&L-OC, MR-OC&2017 RR15&V&C-Euc \\
&2013 XA22&Sq&L-OC&2017 WX12&Sx&LL-OC, SD, MR-OC \\
&2014 PL51&Q&LL-OC&2018 XG5&Sr&L-OC \\
&2014 PR62&K&LL-OC&2018 XS4&Sr&H-OC \\
&2014 SS1&S&L-OC&2019 AN5&Sqw&LL-OC \\
&2014 VH2&L&CC&2019 GT3&Srw&L-OC \\
&2014 VQ&Sqw&LL-OC&2019 JL3&Sqw&LL-OC \\
&2014 WC201&Sq&L-OC&2019 JX7&Q&LL-OC \\
&2014 WN4&C/X&A, EC, CC&2019 RC&D&CC \\
&2014 WY119&Sq&LL-OC&2019 SH6&Sx&LL-OC, SD, MR-OC \\
&2014 WZ120&Sr&H-OC, PA&2019 UO13&Sr&L-OC\\
&2015 AK45&Sr&L-OC&2019 YM3&Qw&LL-OC \\
&2015 AP43&D&CC&2020 DZ1&S&LL-OC \\
&2015 BC&Xk&How, SI, I&2020 HS6&Sq&LL-OC \\
&2015 BK509&Srw&LL-OC&2020 KC5&S&L-OC \\
&2015 FL&IND&IND&2020 RO6&V&NC-Euc \\
&2015 HA1&Sq&LL-OC&2020 SN&Xk&A, EC \\
&2015 LK24&C/X&CC&2020 ST1&Sw&L-OC \\
&2015 NA14&S&L-OC&2020 YQ3&Sq&LL-OC \\

\tableline
\end{tabular}
\end{table}

\begin{figure*}[!ht]
\begin{center}
\includegraphics[height=10cm]{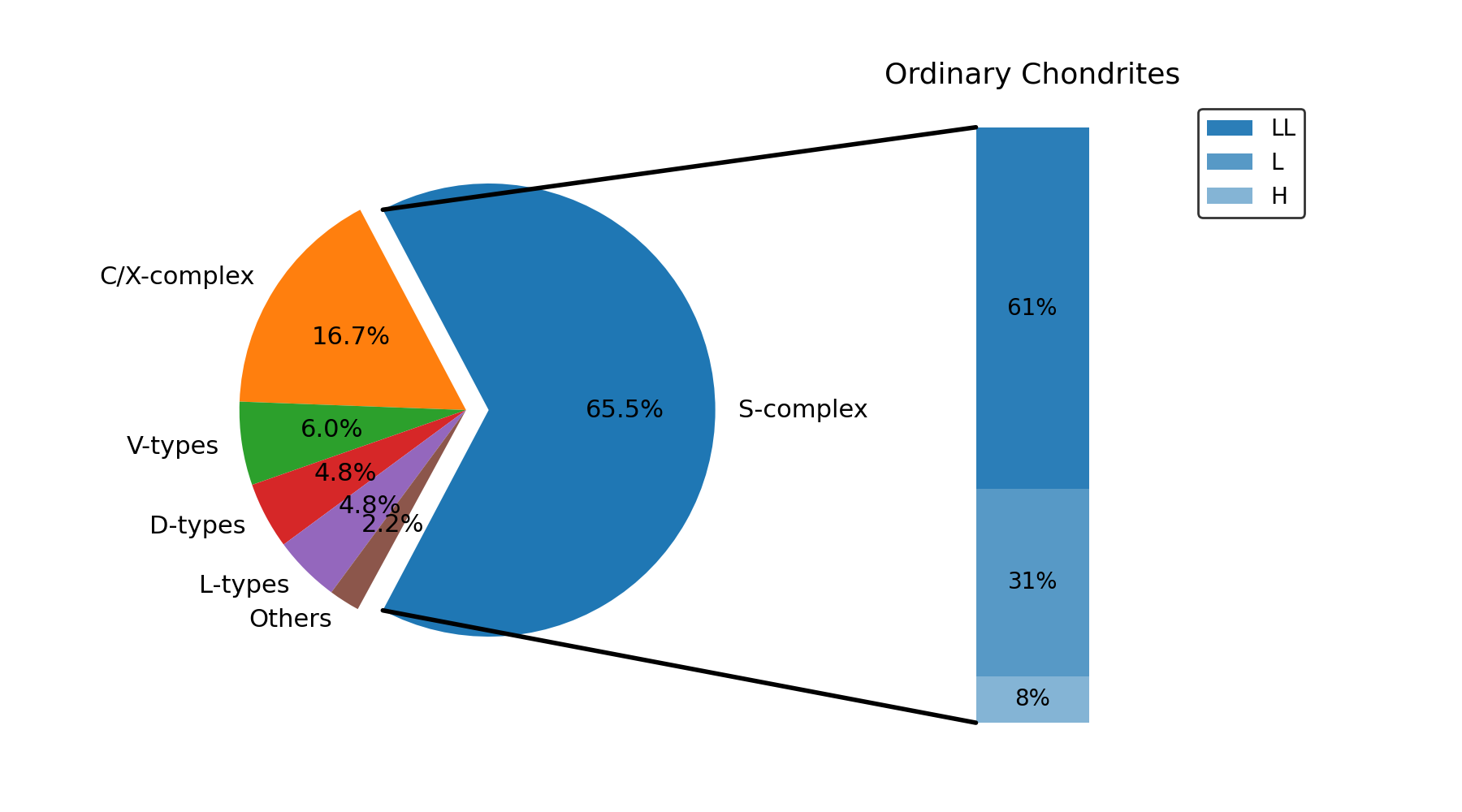}

\caption{\label{f:Pie_chart}{\small Approximate fractions of main taxonomic types and ordinary chondrites. In this figure, asteroids in the S-complex include 
Q-types and a new subclass of objects introduced in the present study called Sx-types. Asteroids in the C/X-complex include those with ambiguous C- or X-complex 
classification plus asteroids classified as Xk-, Xn- and B-types. The fraction corresponding to "others" includes K-types and objects whose 
taxonomic classification is indeterminate.}}

\end{center}
\end{figure*}

The other instance where only data from 0.8-2.45 $\mu$m was used for taxonomic classification was when the spectra were obtained using the 0.8 $\mu$m dichroic. In those cases, more data is missing and it is not possible to extrapolate the data. This, however, does not represent a major problem for the classification of asteroids whose spectra show prominent absorption bands such as those belonging to the S-complex or V-types.

The taxonomic type assigned to each NEO is presented in Table 3. If during the classification process more than one taxonomic type was assigned to an asteroid, a visual 
inspection was performed to determine the class that best represents the spectral characteristics of the asteroid. We also compared the average absolute residuals between the NEO and the reference spectra 
obtained from the Bus-DeMeo Taxonomy Classification Web tool. There was one case where the classification was indeterminate and it has been indicated in Table 3.

The classification of NEOs 2013 CW32, 2016 CM194, 2017 WX12, 2019 SH6 yielded very ambiguous results, with all of them being assigned multiple taxonomies including C-, X- and S-complex, as well as K- and L-types. This happened because, even though the spectra of these objects are similar to those in the S-complex, their absorption 
bands are much weaker. For this reason, and in order to highlight these unusual characteristics, in the present study we refer to these objects as Sx-types. Thus, we define 
Sx-types as objects whose NIR spectra are similar to those in the S-complex, but that due to their weak absorption bands could fall in the region of the PC2’ vs PC1’  diagram 
where  the C- and X-complex normally fall. The composition of these objects and the possible causes for their weak absorption bands are discussed in sections 2.4.2 and 2.6, respectively.  

Figure \ref{f:Tax_Hist} shows the distribution of taxonomic types, we found that the most common taxonomic type in our sample is the Sq-type, followed by the S-, Q-, and Sr-types. The fraction of the main taxonomic types is shown in Figure \ref{f:Pie_chart}, NEOs classified as S-complex represent $\sim$66\% of the entire sample, objects in the C/X-complex represent $\sim$17\% and the other $\sim$17\% less common taxonomic types. Our results are consistent with the work done by \cite{2019Icar..324...41B} who found a fractional distribution of 60\% S, 20\% C and 20\% of other types for NEOs in the size range of 100 m to 10 km. It is worth 
noting that in the original Bus-DeMeo classification \citep{2009Icar..202..160D} Q-types are not considered part of the S-complex, but they are in
\cite{2022Icar..38014971D}. \cite{2019Icar..324...41B} followed the same definition as \cite{2009Icar..202..160D} whereas in the present work we use that of 
\cite{2022Icar..38014971D}. The new subclass of Sx-types is also included within the S-complex in Figure \ref{f:Pie_chart}.

For the majority of the NEOs in this study the albedo is unknown, however, having information about their taxonomic type and the mean albedo for those 
types, it is possible to do an estimation of their size using equation (1). Thus, diameters shown in Figure \ref{f:H_D_Histogram_spec} were calculated from the absolute magnitudes of the NEOs and the mean albedos of their taxonomic types found by \cite{2022AJ....163..165M}. The mean geometric albedo of the spectroscopic sample was found to be $P_{V}$=0.22.

\subsection{Compositional analysis and meteorite analogs} \label{sec:comp}

\subsubsection{S-complex asteroids}

Asteroids in the S-complex include those that were classified as S-, Sq-, Sr-, Sx- and Q-types. The spectra of all these objects show absorption bands at $\sim$1 and 2
$\mu$m due to the minerals olivine and pyroxene. Olivine has three overlapping bands centered near 1.04-1.1 $\mu$m and pyroxene has two absorption bands centered 
near 0.9-1 $\mu$m and 1.9-2 $\mu$m \citep{1974JGR....79.4829A, 1993macf.book.....B}. These absorption bands are caused by the presence of Fe$^{2+}$ cations. Diagnostic spectral band parameters including Band I and II
centers and Band Area Ratio (BAR) were measured using a Python code following the procedure described in \cite{2020AJ....159..146S}. Band centers were 
measured after dividing out the linear continuum by fitting 3$^{rd}$ and 4$^{th}$ order polynomials over the bottom of the absorption bands. Band areas are 
defined as the area between the linear continuum and the data curve and are calculated using trapezoidal numerical integration. The BAR was calculated 
as the ratio of the area of Band II to that of Band I. A temperature correction derived by \cite{2012Icar..220...36S} was applied to the BAR in order to account 
for differences between the surface temperature of the NEOs and the room temperature at which spectral calibrations used for compositional analysis were 
obtained. The uncertainties associated with the band parameters are given by the standard deviation of the mean calculated from multiple measurements of 
each band parameter. Spectral band parameters are shown in Table 4.

\begin{figure*}[!ht]
\begin{center}
\includegraphics[height=9cm]{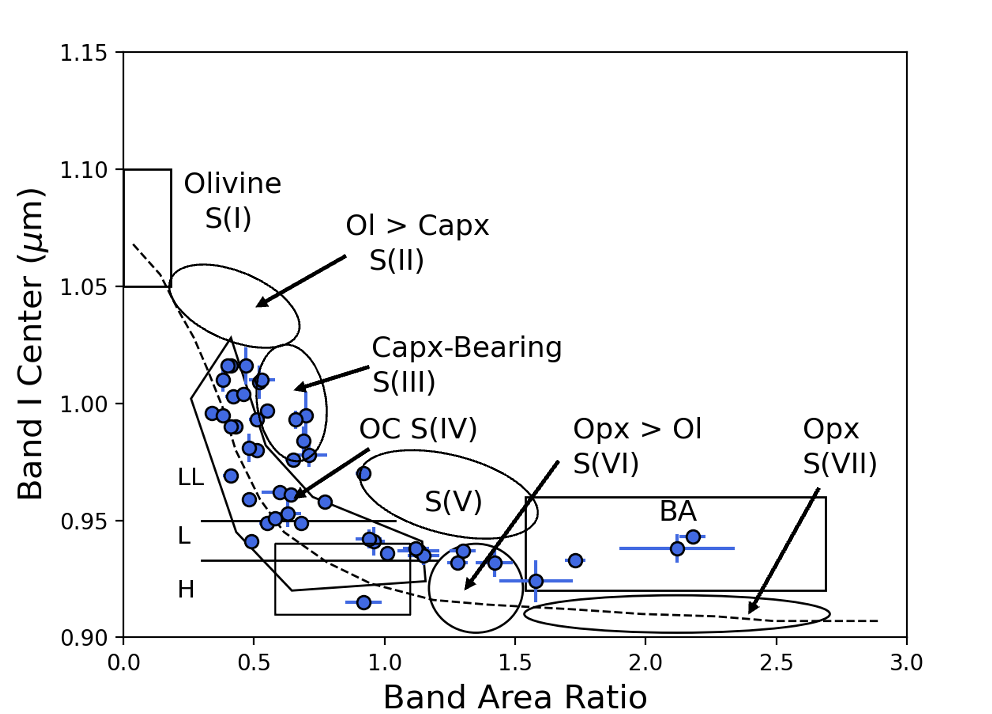}

\caption{\label{f:BIC_BAR}{\small Band I center versus BAR from \cite{1993Icar..106..573G} for NEOs whose spectra exhibit absorption bands centered at 
$\sim$1- and 2-$\mu$m due to the presence of olivine and pyroxene. The polygonal region corresponds to the S(IV) subtype associated with ordinary chondrites 
(OC). The horizontal lines represent the approximate boundaries for ordinary chondrites found by \cite{2020AJ....159..146S}. The rectangular zone overlapping the 
S(IV) subtype represents the spectral zone for acapulcoite-lodranite clan meteorites found by \cite{2019M&PS...54..157L}. The rectangular zone (BA) includes the 
pyroxene-dominated basaltic achondrite assemblages \citep{1993Icar..106..573G}. The dashed curve indicates the location of the olivine-orthopyroxene mixing line \citep{1986JGR....9111641C}.}}

\end{center}
\end{figure*}

Prior the compositional analysis it is useful to plot the spectral band parameters in the Band I center versus BAR diagram from \cite{1993Icar..106..573G} to 
determine the S-asteroid subtype. The results shown in Figure \ref{f:BIC_BAR} indicate that most of the NEOs in the S-complex belong to the S(IV) subtype, which 
also represents the mafic silicate components of ordinary chondrites \citep{1993Icar..106..573G}. These meteorites are the most common type to fall to Earth representing 
about 86\% of all meteorites. They are divided into three subgroups (H, L, LL) based on the abundance of Fe and the ratio of metallic Fe (Fe$^{0}$) to oxidized Fe (FeO) 
\citep{2006mess.book...19W}. 

NEOs 363599, 2005 NE21, 2005 TF, 2015 HA1, 2019 JL3 and 2020 YQ3 were found to fall in the S(III) subtype. Objects in this 
subtype can contain a calcic pyroxene component, some ordinary chondrites can fall in this region as well \citep[e.g.,][]{2010Icar..208..789D, 2020AJ....159..146S}. In 
the particular case of 2005 NE21 it is worth mentioning that the spectrum shows a low S/N and large scattering in the 2-$\mu$m band, which can lead to an overestimation of 
the BAR moving the object from the S(IV) to the S(III) subtype. 

One NEO, 2019 UO13, was found to be an S(V) subtype. This class of asteroids are thought to have 
experienced intrusion of early partial melts into their crust and upper mantle \citep{1993Icar..106..573G}. A highly metamorphosed H-chondrite or a calcic pyroxene-bearing 
lodranite are possible meteorite analogs for these objects. The NEO 2014 WZ120 falls slightly bellow the S(IV) subtype, which is possible for some ordinary chondrites, although 
we cannot rule out an affinity with primitives achondrites (acapulcoites and lodranites) that are also located in this region \citep{2019M&PS...54..157L}. NEOs 436724 and 2018 
XS4 are both classified as S(VI) subtype. Objects in this subtype contain low-Ca pyroxene and less olivine than olivine-poor ordinary chondrites \citep{1993Icar..106..573G}. 
Winonaites and the inclusions present in the IAB iron meteorites have been proposed as possible meteorite analogs for this subtype \citep{1993Icar..106..573G}. Like in the 
case of 2005 NE21, 436724 also shows a large scattering in the 2-$\mu$m band, which could be causing an overestimation of the BAR.

\startlongtable
\begin{deluxetable*}{cccccccccc}
\tablecaption{\label{t:Table4}{\small Spectral band parameters and composition for the NEOs. The columns in this table are: object number and designation, Band I center (BIC), 
Band II center (BIIC), Band Area Ratio (BAR), molar contents of fayalite (Fa), ferrosilite (Fs) and wollastonite (Wo), and ol/(ol +px) ratio. The uncertainties for Fa, Fs, Wo and 
the ol/(ol+px) ratio are from \cite{2020AJ....159..146S} and \cite{2009MandPS...44.1331B}.}}
\tablewidth{0pt}
\tablehead{Number&Designation&BIC ($\mu m$)&BIIC ($\mu m$)&BAR&Fa (mol$\%)$&Fs (mol$\%$)&Wo (mol$\%$)&ol/(ol+px) \\ }
\startdata
85990&1999 JV6&0.938$\pm$0.006&2.041$\pm$0.019&2.12$\pm$0.22&-&46$\pm$3&11$\pm$1&- \\
163348&2002 NN4&-&-&-&-&-&-&- \\
363599&2004 FG11&0.997$\pm$0.003&2.063$\pm$0.011&0.55$\pm$0.02&30.2$\pm$2.0&25.1$\pm$1.4&-&0.59$\pm$0.04 \\
412995&1999 LP28&1.023$\pm$0.005&-&-&30.6$\pm$2.0&25.4$\pm$1.4&-&- \\
436724&2011 UW158&0.932$\pm$0.006&2.024$\pm$0.023&1.42$\pm$0.07&21.5$\pm$2.0&19.1$\pm$1.4&-&0.38$\pm$0.04 \\
437844&1999 MN&0.980$\pm$0.003&2.036$\pm$0.014&0.51$\pm$0.02&28.9$\pm$2.0&24.3$\pm$1.4&-&0.60$\pm$0.04 \\
438908&2009 XO&1.016$\pm$0.002&2.022$\pm$0.012&0.41$\pm$0.01&30.7$\pm$2.0&25.4$\pm$1.4&-&0.62$\pm$0.04 \\
459872&2014 EK24&0.974$\pm$0.004&1.931$\pm$0.025&0.92$\pm$0.06&27.5$\pm$2.0&23.3$\pm$1.4&-&0.56$\pm$0.04 \\
467336&2002 LT38&0.990$\pm$0.003&2.004$\pm$0.007&0.43$\pm$0.02&29.7$\pm$2.0&24.8$\pm$1.4&-&0.61$\pm$0.04 \\
469737&2005 NW44&0.982$\pm$0.005&1.960$\pm$0.010&0.85$\pm$0.04&28.4$\pm$2.0&23.9$\pm$1.4&-&0.57$\pm$0.04 \\
471240&2011 BT15&0.941$\pm$0.003&1.942$\pm$0.023&0.49$\pm$0.01&23.3$\pm$2.0&20.4$\pm$1.4&-&0.60$\pm$0.04 \\
496816&1989 UP&1.010$\pm$0.005&1.991$\pm$0.012&0.38$\pm$0.01&30.6$\pm$2.0&25.4$\pm$1.4&-&0.63$\pm$0.04 \\
501647&2014 SD224&0.981$\pm$0.006&2.018$\pm$0.014&0.48$\pm$0.02&29.0$\pm$2.0&24.3$\pm$1.4&-&0.60$\pm$0.04 \\
515767&2015 JA2&0.992$\pm$0.005&-&-&29.9$\pm$2.0&24.9$\pm$1.4&-&- \\
528159&2008 HS3&0.996$\pm$0.003&1.956$\pm$0.012&0.34$\pm$0.01&30.1$\pm$2.0&25.1$\pm$1.4&-&0.64$\pm$0.04 \\
&2000 TU28&0.990$\pm$0.002&2.018$\pm$0.006&0.41$\pm$0.02&29.7$\pm$2.0&24.8$\pm$1.4&-&0.62$\pm$0.04 \\
&2001 YV3&0.995$\pm$0.001&2.036$\pm$0.011&0.38$\pm$0.01&30.1$\pm$2.0&25.0$\pm$1.4&-&0.63$\pm$0.04 \\
&2002 LY1&-&-&-&-&-&-&- \\
&2005 NE21&0.976$\pm$0.002&1.903$\pm$0.024&0.65$\pm$0.03&28.5$\pm$2.0&24.0$\pm$1.4&-&0.56$\pm$0.04 \\
&2005 TF&1.009$\pm$0.007&2.003$\pm$0.010&0.52$\pm$0.02&30.6$\pm$2.0&25.4$\pm$1.4&-&0.59$\pm$0.04 \\ 
&2006 XY&-&-&-&-&-&-&- \\
&2007 EC&1.003$\pm$0.002&2.020$\pm$0.011&0.42$\pm$0.03&30.4$\pm$2.0&25.3$\pm$1.4&-&0.62$\pm$0.04 \\ 
&2012 ER14&-&-&-&-&-&-&- \\
&2013 CW32&0.941$\pm$0.006&1.988$\pm$0.013&0.98$\pm$0.04&23.3$\pm$2.0&20.4$\pm$1.4&-&0.49$\pm$0.04 \\ 
&2013 XA22&0.958$\pm$0.002&1.992$\pm$0.012&0.77$\pm$0.01&26.2$\pm$2.0&22.4$\pm$1.4&-&0.53$\pm$0.04 \\ 
&2014 PL51&1.004$\pm$0.003&1.939$\pm$0.015&0.46$\pm$0.02&30.5$\pm$2.0&25.3$\pm$1.4&-&0.61$\pm$0.04 \\ 
&2014 PR62&1.001$\pm$0.009&-&-&30.3$\pm$2.0&25.2$\pm$1.4&-&- \\
&2014 SS1&0.949$\pm$0.001&1.992$\pm$0.015&0.55$\pm$0.03&24.8$\pm$2.0&21.4$\pm$1.4&-&0.59$\pm$0.04 \\
&2014 VH2&-&-&-&-&-&-&- \\
&2014 VQ&1.016$\pm$0.001&2.022$\pm$0.009&0.40$\pm$0.02&30.7$\pm$2.0&25.4$\pm$1.4&-&0.62$\pm$0.04 \\
&2014 WC201&0.936$\pm$0.003&1.940$\pm$0.014&1.01$\pm$0.02&22.3$\pm$2.0&19.7$\pm$1.4&-&0.48$\pm$0.04 \\
&2014 WN4&-&-&-&-&-&-&- \\
&2014 WY119&0.971$\pm$0.010&-&-&27.2$\pm$2.0&23.1$\pm$1.4&-&- \\
&2014 WZ120&0.915$\pm$0.001&1.952$\pm$0.024&0.92$\pm$0.07&17.4$\pm$2.0&16.2$\pm$1.4&-&0.50$\pm$0.04 \\
&2015 AK45&0.974$\pm$0.002&1.989$\pm$0.016&1.17$\pm$0.13&27.5$\pm$2.0&23.3$\pm$1.4&-&0.52$\pm$0.04 \\
&2015 AP43&-&-&-&-&-&-&- \\
&2015 BC&0.925$\pm$0.009&2.001$\pm$0.026&1.58$\pm$0.14&-&32$\pm$3&5$\pm$1&- \\
&2015 BK509&0.962$\pm$0.003&1.947$\pm$0.036&0.60$\pm$0.07&26.8$\pm$2.0&22.8$\pm$1.4&-&0.58$\pm$0.04 \\
&2015 FL&-&-&-&-&-&-&- \\
&2015 HA1&1.010$\pm$0.001&1.994$\pm$0.010&0.53$\pm$0.05&30.6$\pm$2.0&25.4$\pm$1.4&-&0.59$\pm$0.04 \\
&2015 LK24&-&-&-&-&-&-&- \\
&2015 NA14&0.942$\pm$0.004&1.994$\pm$0.018&0.94$\pm$0.05&23.5$\pm$2.0&20.5$\pm$1.4&-&0.49$\pm$0.04 \\
&2015 TB25&-&-&-&-&-&-&- \\
&2015 TC25&0.905$\pm$0.003&-&-&-&-&-&- \\
&2015 TF&-&-&-&-&-&-&- \\
&2015 VE66&0.937$\pm$0.002&1.927$\pm$0.014&1.13$\pm$0.08&22.5$\pm$2.0&19.8$\pm$1.4&-&0.45$\pm$0.04 \\
&2015 WF13&0.943$\pm$0.001&1.984$\pm$0.008&3.64$\pm$0.13&-&44$\pm$3&10$\pm$1&- \\ 
&2015 XC&0.947$\pm$0.006&1.947$\pm$0.007&3.29$\pm$0.10&-&36$\pm$3&7$\pm$1&- \\ 
&2016 BC14&0.942$\pm$0.002&1.926$\pm$0.026&1.40$\pm$0.08&22.9$\pm$2.0&20.1$\pm$1.4&-&0.49$\pm$0.04 \\
&2016 CM194&0.937$\pm$0.001&1.992$\pm$0.008&1.30$\pm$0.05&22.5$\pm$2.0&19.8$\pm$1.4&-&0.41$\pm$0.04 \\
&2016 CO247&-&-&-&-&-&-&- \\
&2016 EB1&-&-&-&-&-&-&- \\
&2016 EF28&0.949$\pm$0.004&1.989$\pm$0.008&3.54$\pm$0.10&-&45$\pm$3&10$\pm$1&- \\ 
&2016 EV27&1.011$\pm$0.003&1.940$\pm$0.015&0.74$\pm$0.05&30.4$\pm$2.0&25.3$\pm$1.4&-&0.58$\pm$0.04 \\
&2016 FV13&0.936$\pm$0.005&1.970$\pm$0.010&0.94$\pm$0.07&21.8$\pm$2.0&19.3$\pm$1.4&-&0.55$\pm$0.04 \\
&2016 GU&0.910$\pm$0.004&-&-&-&-&-&- \\
&2016 LG&0.989$\pm$0.004&2.057$\pm$0.011&0.41$\pm$0.02&29.0$\pm$2.0&24.4$\pm$1.4&-&0.63$\pm$0.04 \\
&2017 BS5&0.992$\pm$0.002&-&-&29.9$\pm$2.0&24.9$\pm$1.4&-&- \\
&2017 BW&-&-&-&-&-&-&- \\
&2017 BY93&0.951$\pm$0.002&1.967$\pm$0.011&0.58$\pm$0.02&25.1$\pm$2.0&21.6$\pm$1.4&-&0.58$\pm$0.04 \\
&2017 CR32&-&-&-&-&-&-&- \\ 
&2017 DR34&0.930$\pm$0.002&1.920$\pm$0.018&0.81$\pm$0.06&20.6$\pm$2.0&18.5$\pm$1.4&-&0.57$\pm$0.04 \\
&2017 FU64&-&-&-&-&-&-&- \\
&2017 OL1&0.953$\pm$0.006&1.987$\pm$0.007&0.63$\pm$0.05&25.5$\pm$2.0&21.9$\pm$1.4&-&0.57$\pm$0.04 \\
&2017 OP68&0.932$\pm$0.005&1.138$\pm$0.010&-&-&-&-&- \\
&2017 RR15&0.933$\pm$0.003&1.943$\pm$0.006&1.73$\pm$0.04&-&41$\pm$3&9$\pm$1&- \\ 
&2017 WX12&0.995$\pm$0.010&2.044$\pm$0.013&0.70$\pm$0.03&30.1$\pm$2.0&25.0$\pm$1.4&-&0.55$\pm$0.04 \\
&2018 XG5&0.935$\pm$0.004&1.956$\pm$0.005&1.15$\pm$0.06&22.1$\pm$2.0&19.5$\pm$1.4&-&0.44$\pm$0.04 \\
&2018 XS4&0.932$\pm$0.001&1.987$\pm$0.009&1.28$\pm$0.04&21.5$\pm$2.0&19.1$\pm$1.4&-&0.41$\pm$0.04 \\
&2019 AN5&1.016$\pm$0.008&2.050$\pm$0.017&0.47$\pm$0.03&30.7$\pm$2.0&25.4$\pm$1.4&-&0.61$\pm$0.04 \\
&2019 GT3&0.949$\pm$0.002&1.958$\pm$0.010&0.68$\pm$0.02&24.8$\pm$2.0&21.4$\pm$1.4&-&0.56$\pm$0.04 \\
&2019 JL3&0.984$\pm$0.006&1.998$\pm$0.011&0.69$\pm$0.02&29.3$\pm$2.0&24.5$\pm$1.4&-&0.55$\pm$0.04 \\
&2019 JX7&0.993$\pm$0.002&2.047$\pm$0.012&0.51$\pm$0.03&29.9$\pm$2.0&25.0$\pm$1.4&-&0.60$\pm$0.04 \\
&2019 RC&-&-&-&-&-&-&- \\
&2019 SH6&0.978$\pm$0.005&1.979$\pm$0.024&0.71$\pm$0.07&28.7$\pm$2.0&24.1$\pm$1.4&-&0.55$\pm$0.04 \\
&2019 UO13&0.970$\pm$0.001&1.991$\pm$0.005&0.92$\pm$0.03&27.9$\pm$2.0&23.5$\pm$1.4&-&0.50$\pm$0.04 \\
&2019 YM3&0.969$\pm$0.001&2.008$\pm$0.009&0.41$\pm$0.03&27.7$\pm$2.0&23.4$\pm$1.4&-&0.62$\pm$0.04 \\
&2020 DZ1&0.984$\pm$0.004&-&-&29.3$\pm$2.0&24.5$\pm$1.4&-&- \\
&2020 HS6&0.959$\pm$0.002&1.955$\pm$0.006&0.48$\pm$0.02&26.4$\pm$2.0&22.5$\pm$1.4&-&0.60$\pm$0.04 \\
&2020 KC5&0.961$\pm$0.003&1.981$\pm$0.007&0.64$\pm$0.04&26.7$\pm$2.0&22.7$\pm$1.4&-&0.57$\pm$0.04 \\
&2020 RO6&0.943$\pm$0.003&1.974$\pm$0.008&2.18$\pm$0.05&-&50$\pm$3&12$\pm$1&- \\
&2020 SN&0.891$\pm$0.003&-&-&-&-&-&- \\
&2020 ST1&0.938$\pm$0.001&1.998$\pm$0.004&1.12$\pm$0.05&22.7$\pm$2.0&20.0$\pm$1.4&-&0.45$\pm$0.04 \\
&2020 YQ3&0.993$\pm$0.004&1.935$\pm$0.012&0.66$\pm$0.02&29.9$\pm$2.0&25.0$\pm$1.4&-&0.56$\pm$0.04 \\
\enddata
\end{deluxetable*}

\begin{figure*}[h]
\begin{center}
\includegraphics[height=11cm]{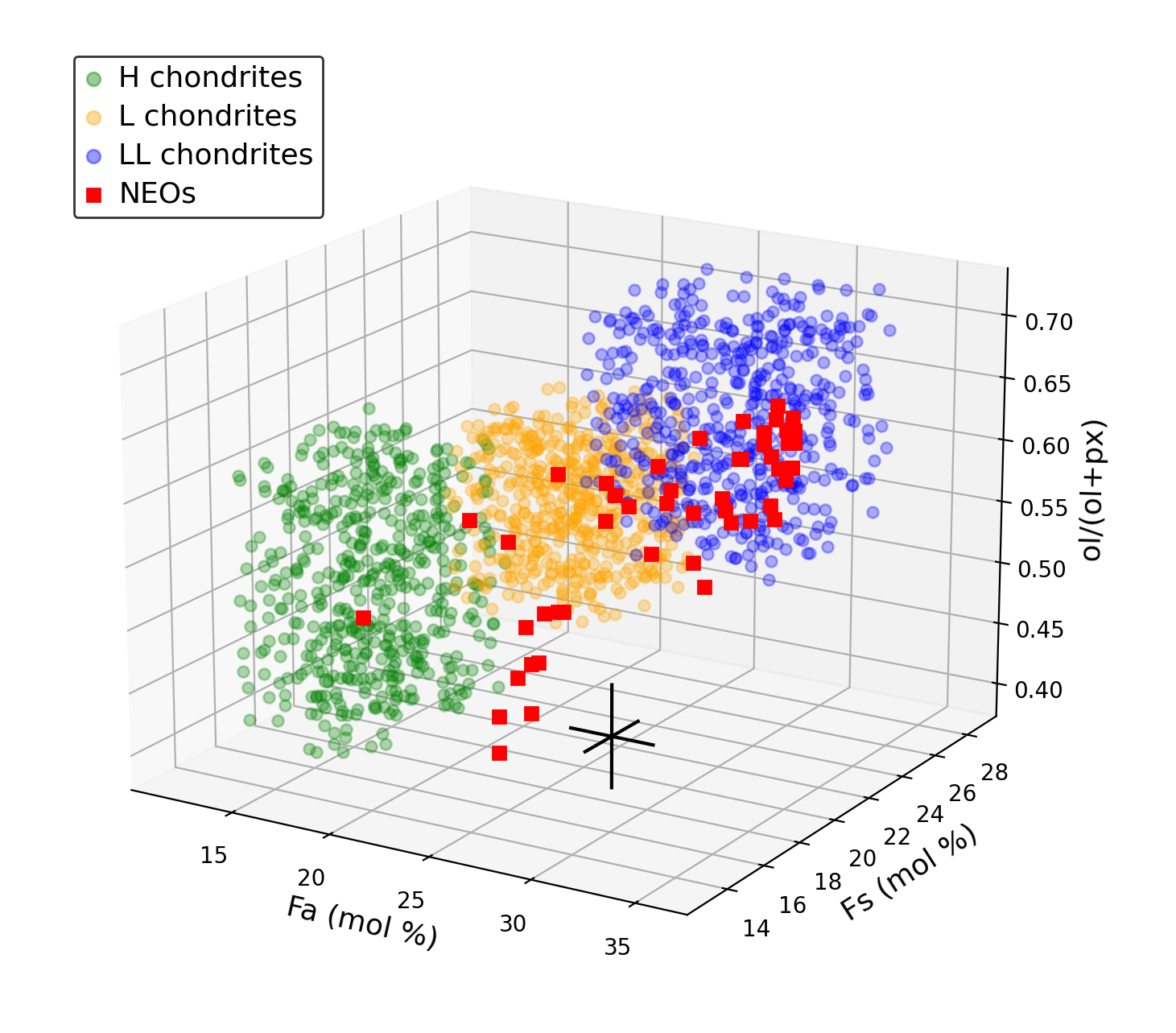}

\caption{\label{f:Fa_Fs_olopx}{\small Molar contents of fayalite (Fa) and ferrosilite (Fs) and ol/(ol+px) ratio for NEOs with an ordinary chondrite-like composition 
(red squares). The synthetic dataset corresponding to the three ordinary chondrite subtypes (H, L, LL) created to train the machine learning classifier is 
also shown. The error bars correspond to the uncertainties derived by \cite{2020AJ....159..146S}, 2.0 mol\% for Fa, 1.4 mol\% for Fs and 0.04 for the ol/(ol+px) ratio.}}

\end{center}
\end{figure*}

Since the majority of 
the S-complex NEOs in our sample fall in the S(IV) subtype (or the vicinity), as a first approximation, the composition of these objects can be determined with spectral calibrations 
developed from ordinary chondrites. For this, we used the Band I center and BAR along with the equations 
derived by \cite{2020AJ....159..146S}. These equations are a modified version of the original spectral calibrations obtained by \cite{2010Icar..208..789D} from the 
analysis of ordinary chondrites. The equations obtained by \cite{2020AJ....159..146S} are more appropriated for low S/N spectra and can be used with incomplete 
spectra like the ones obtained with the 0.8 $\mu$m dichroic. The Band I center was used to determine the olivine and pyroxene chemistries, which are given by the 
molar \% of fayalite (Fa) and ferrosilite (Fs), respectively. For those objects where the BAR was measured, the olivine to pyroxene ratio (ol/(ol+px)) was also calculated.  
Figure \ref{f:Fa_Fs_olopx} shows the olivine and pyroxene chemistries versus the ol/(ol+px) ratio for the NEOs. These values are also presented in Table 4. As 
mentioned before, there are at least a couple of cases where the BAR values could have been overestimated. Becasue this would result in an underestimation of the 
ol/(ol+px) ratios, these values must be taken as a lower limit.

\begin{figure*}[!ht]
\begin{center}
\includegraphics[height=9cm]{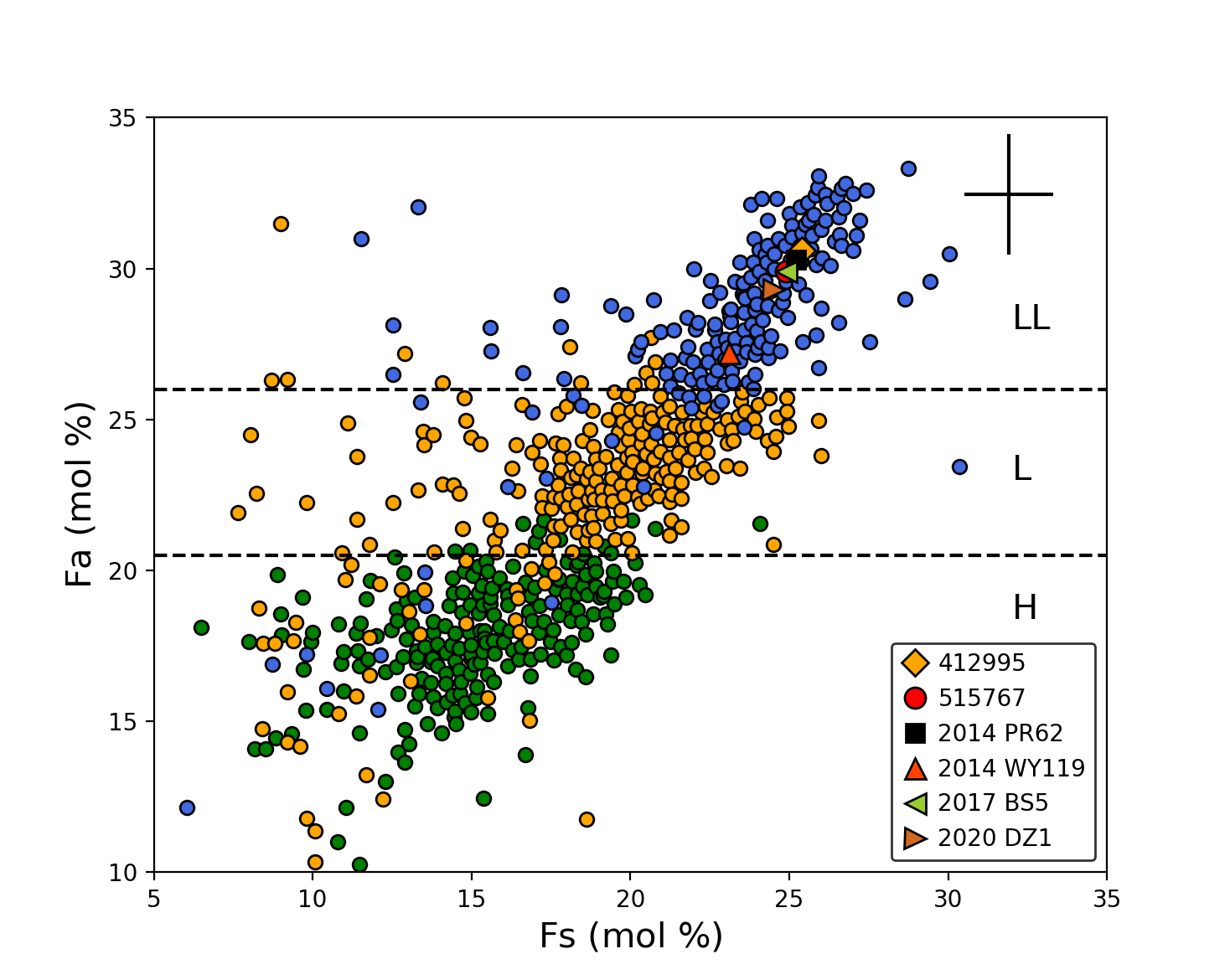}

\caption{\label{f:Nakamura}{\small Molar content of fayalite (Fa) vs. ferrosilite (Fs) for NEOs with an ordinary chondrite-like composition whose spectra do not 
show the 2-$\mu$m band. Measured values for H- (green), L- (orange), and LL- (blue) ordinary chondrites from 
\cite{2011ScienceNakamura} are also included. The error bars correspond to the uncertainties derived by \cite{2020AJ....159..146S}, 2.0 mol\% for Fa, and 
1.4 mol\% for Fs. Figure adapted from \cite{2011ScienceNakamura}.}}

\end{center}
\end{figure*}

There are a few NEOs in our sample whose spectra do not show the 2-$\mu$m absorption band, in those cases only the Band I centers were used for the 
compositional analysis. The molar contents of Fa and Fs of these asteroids are compared with data from \cite{2011ScienceNakamura} in Figure \ref{f:Nakamura}.

\begin{figure*}[!h]
\begin{center}
\includegraphics[height=9.5cm]{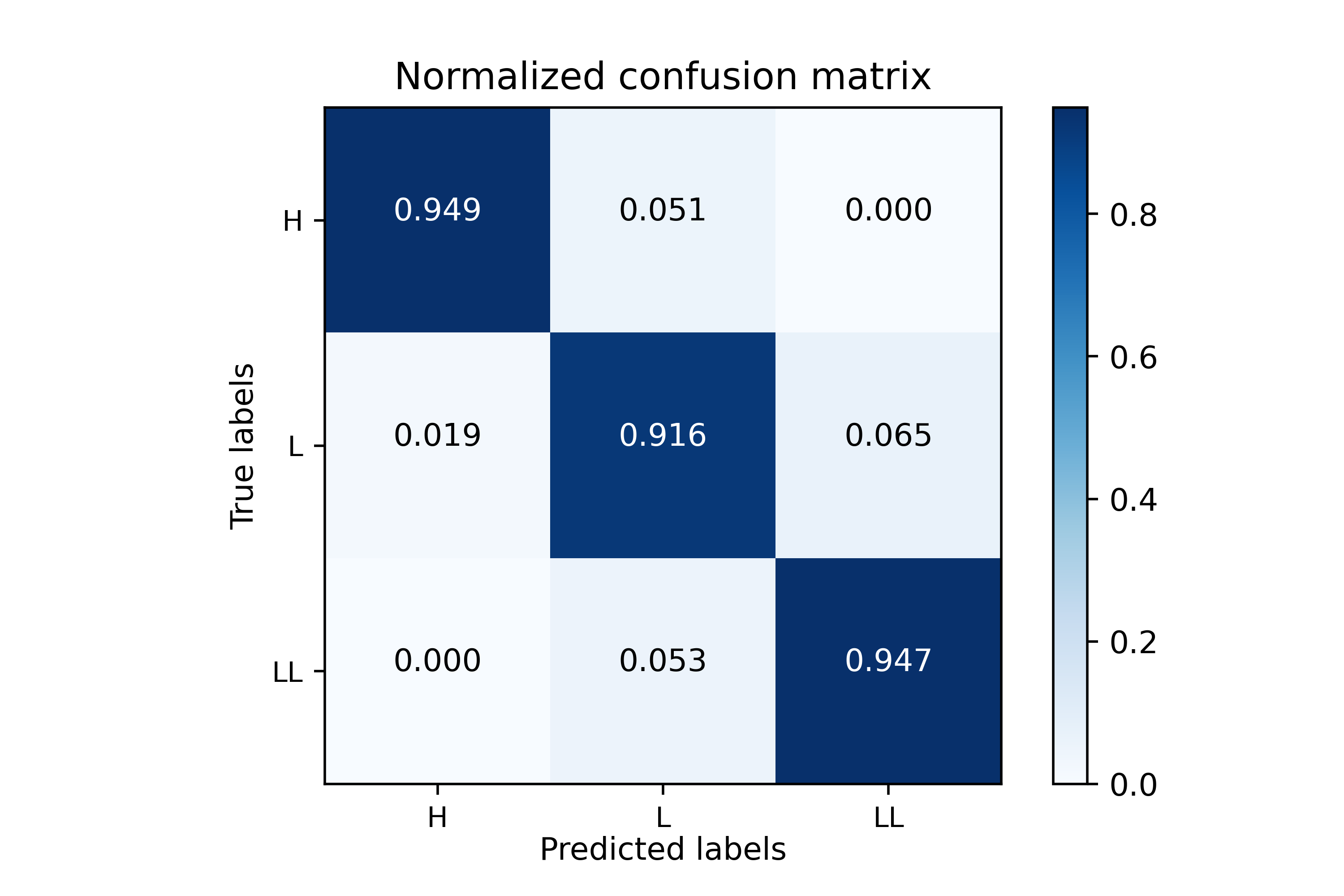}

\caption{\label{f:conf_matrix}{\small Normalized confusion matrix resulting from the multinomial logistic regression model. Labels correspond to the three 
ordinary chondrite subtypes (H, L, LL).}}

\end{center}
\end{figure*}

\begin{figure*}[!ht]
\begin{center}
\hspace{-20mm}
\includegraphics[height=13cm]{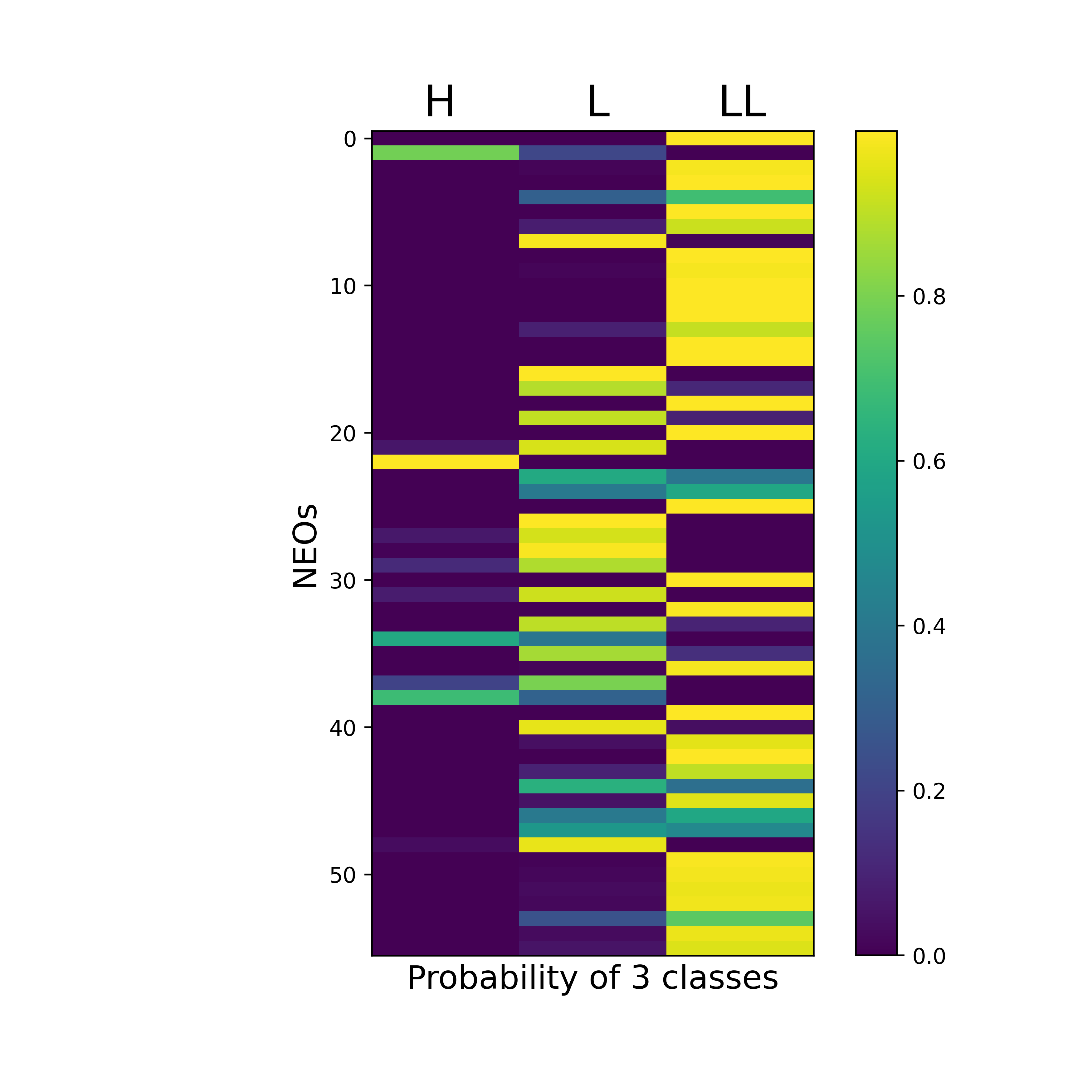}

\caption{\label{f:Proba_NEO}{\small Probability of NEOs belonging to the three ordinary chondrite subtypes (H, L, LL). Each row represents 
a NEO and each column an ordinary chondrite subtype.}}

\end{center}
\end{figure*}

In order to classify these objects into the three ordinary chondrite subtypes, we built a supervised machine learning algorithm based on multinomial logistic 
regression. This method is a generalization of logistic regression and it is used for multi-class classification \citep[e.g.,][]{2006ISS....B}. Multinomial logistic regression uses the softmax 
function to convert a vector of K real numbers into a probability distribution of K possible outcomes. This allow us not only to predict the 
class of each asteroid (H, L or LL), but also the probability of the asteroid belonging to that specific class.  

Given an input feature $x^{(i)}$ and a class label $y^{(i)}$ $\in$ \{1, 2, ..., K\}, where K is the number of classes, we want to estimate the probability that $y^{(i)}$ equals k for each value of k=1, 2, …, K. The predicted probability for the $kth$ class is given by:

\begin{equation}
P(y^{(i)}=k\vert x^{(i)}; \theta)= \frac{e^{{ \theta^{(k)}}^T  x^{(i)}}}{\sum_{j=1}^{K} e^{ {\theta^{(j)}}^T  x^{(i)}}}
\end{equation}

where ${\theta^Tx}$ is the regression result, i.e., the sum of the variables weighted by the coefficients. The difference between the actual and the predicted 
values give us the error of our model, which is calculated by the cost function:

\begin{equation}
J(\theta) = -\left[\sum_{i=1}^{m}\sum_{k=1}^{K}1\{y^{(i)}=k\} {\rm log}\frac{e^{{ \theta^{(k)}}^T  x^{(i)}}}{\sum_{j=1}^{K}e^{ {\theta^{(j)}}^T  x^{(i)}} }  \right]
\end{equation}

Since the cost shows how poorly the model is estimating the labels, the idea is to train our model to minimize this cost. 

We started by creating a synthetic multi-class dataset to train and test the model. For each ordinary chondrite subtype, 
500 data points with three input variables (ol/(ol+px), Fa, Fs) were generated using the boundaries found by \cite{2020AJ....159..146S}. The synthetic data is shown in 
Figure \ref{f:Fa_Fs_olopx}. This dataset was then split into a training and a testing set. For the testing set 20\% of the whole sample was randomly selected. This provides a more 
accurate evaluation on out-of-sample accuracy because the testing dataset is not part of the dataset that has been used to train the data. In order to prevent overfitting we 
employed regularization, i.e, we used the L2 (Ridge) penalty, which is a type of regularization that adds a penalty term equal to the sum of the squares of the weights to the 
cost function. The strength of the penalty is controlled by the hyper-parameter C (the inverse of the regularization strength). The model was fit to the training set using the 
Limited-memory Broyden-Fletcher-Goldfarb-Shanno (lbfgs) optimization algorithm with a 10-fold cross validation. A grid search method was used to find the optimal value of C. Finally, we evaluated the performance of the model on the testing set and the F1 score and logarithmic loss (log loss) were 
calculated. The F1 score is the harmonic mean of the precision and recall and it ranges from 0-1, where an F1 score reaches its best value at 1 (perfect precision and recall). 
The log loss is used to evaluate the performance of the classifier where the predicted output is a probability value between 0 and 1. The resulting log loss value quantifies 
how close the predicted probabilities are to the true class labels; a perfect model would yield a value of 0. 

Figure \ref{f:conf_matrix} shows the normalized confusion matrix resulting from evaluating the model on the testing set. Approximately 95\% of the H-chondrites were 
correctly classified by the model and $\sim$5\% of them were classified as L-chondrites. A slightly lower number of L-chondrites ($\sim$92\%) were correctly classified by the 
model and the remaining $\sim$8\% were classified as H- and LL-chondrites. This comes as no surprise since L-chondrites overlap with both H- and LL-chondrites. 
Approximately 95\% of the LL-chondrites were correctly classified and $\sim$5\% of them were classified as L-chondrites. The F1 score and log loss yielded values of 0.94 and 
0.14, respectively. 

The model was applied to all the NEOs with an ordinary chondrite-like composition (56 in total). For the few objects whose spectra do not show the 2-$\mu$m absorption band and only the Fa and Fs were calculated, we modified our model to perform the classification using 
only these two input variables. The probabilities of the NEOs belonging to the three subtypes (H, L, LL) are shown in Figure \ref{f:Proba_NEO}. The best meteorite analogs for 
each NEO are included in Table 3. We found that 8\% of the objects in the S-complex were classified as H-, 31\% as L- and 61\% as LL-chondrites (Figure 
\ref{f:Pie_chart}). These NEOs have diameters ranging from $\sim$8-284 m with a mean value of $\sim$138 m.

\cite{2013Icar..222..273D} analyzed 47 NEOs with ordinary chondrite-like compositions and found a proportion of 15\% H-, 10\% L- and 60\% LL-chondrites 
among these bodies. The 
remaining 15\% could not be distinguished between L- and LL-chondrites. \cite{2014Icar..228..217T} used a larger sample consisting of 109 NEOs and found 
a proportion of 22.9\% H-, 10.1\% L- and 40.4\% LL-ordinary chondrites. The rest of the objects were found to have overlapping (8.3\% H/L and 12.8\% L/LL) 
or potentially inconsistent compositions (3.7\% H and 1.8\% LL). \cite{2019Icar..324...41B}, on the other hand, spectrally modeled 194 NEOs as ordinary 
chondrites and found that $\sim$29\% were consistent with H-, $\sim$20\% with L- and  $\sim$51\% with LL-chondrites. Our results agree with previous 
studies that showed that LL-chondrites are dominant among NEOs with ordinary chondrite-like compositions. The different proportions of each ordinary 
chondrite subtype are probably the result of several factors, such as the use of a sample with smaller objects, the intrinsic difficulty of classifying ordinary chondrites that 
can have overlapping compositions and the use of different procedures for the compositional analysis.

\cite{2008Natur.454..858V} first noticed that LL-chondrites were the most common type among S-complex NEOs. This result was unexpected because 
LL-chondrites only represent 10\% of all ordinary chondrites falls. This discrepancy has been attributed to the size of the NEOs studied (typically tens of 
meters to kilometers), which might be too large to be the immediate parent bodies of the meteorites that fall on Earth. The Yarkovsky effect is more efficient at 
delivering meter-sized objects from the asteroid belt to the near-Earth space. Therefore, it is possible that smaller NEOs could have 
compositions more similar to what we see among ordinary chondrites meteorites \citep[e.g.,][]{2008Natur.454..858V, 2019Icar..324...41B}. If this is true, then this means that most of the NEOs in our sample with ordinary chondrite-like compositions are too large to be the parent bodies of the 
ordinary chondrites meteorites.

\subsubsection{Sx-type asteroids}

In section 2.3 we introduced this new subclass of the S-complex to make the distinction that their NIR spectra exhibit absorption bands that are much weaker than 
the typical objects in the S-complex. The four NEOs that were assigned this class are 2013 CW32, 2016 CM194, 2017 WX12 and 2019 SH6. The composition of these NEOs was determined in the same way as the other objects in the S-complex.

In the Band I center vs. BAR plot, 2013 CW32 is located in the region corresponding to the S(IV) subtypes (Figure \ref{f:BIC_BAR}). For this object we found that L-chondrites are the closest in composition. 2016 CM194 was classified as an S(VI) subtype in the Band I center vs. BAR plot. The compositional analysis of this NEO yielded an olivine and pyroxene chemistry similar to L-chondrites. However, given its classification as an S(VI) subtype, it could also have an affinity with primitive achondrites. Both 2017 WX12 and 2019 SH6 fall in the S(III) subtype of the Band I center vs. BAR plot. For these two objects the olivine and pyroxene 
chemistry as well as the ol/(ol+px) ratio are consistent with LL-chondrites. The possible reasons behind the particular spectral characteristics of these asteroids are 
investigated in more detail in section 2.6.

\subsubsection{V-type asteroids}

NEOs 2015 WF13, 2015 XC, 2016 EF28, 2017 RR15 and 2020 RO6 were classified as V-types. In the Band I center versus BAR plot, 2017 RR15 and 2020 RO6, which are
the only ones that were observed without the 0.8 $\mu$m dichroic, are located in the basaltic achondrites region (Figure \ref{f:BIC_BAR}). V-type asteroids are 
linked to asteroid 4 Vesta and to howardite, eucrite, and diogenite (HEDs) meteorites based on their spectral similarities 
\citep[e.g.,][]{1970Sci...168.1445M, 1977GeCoA..41.1271C, 2010Icar..208..773M}. Eucrites are basaltic rocks composed mainly of calcium-rich plagioclase feldspar, augite 
and pigeonite. Diogenites are orthopyroxene-rich rocks that formed deeper than the eucrites (lower crust/upper mantle) and cooled slowly. Howardites are physical mixtures of eucrites 
and diogenites \citep[e.g.,][]{1998LPI....29.1220M, 2015ChEG...75..155M}.

\begin{figure*}[h]
\begin{center}
\includegraphics[height=9cm]{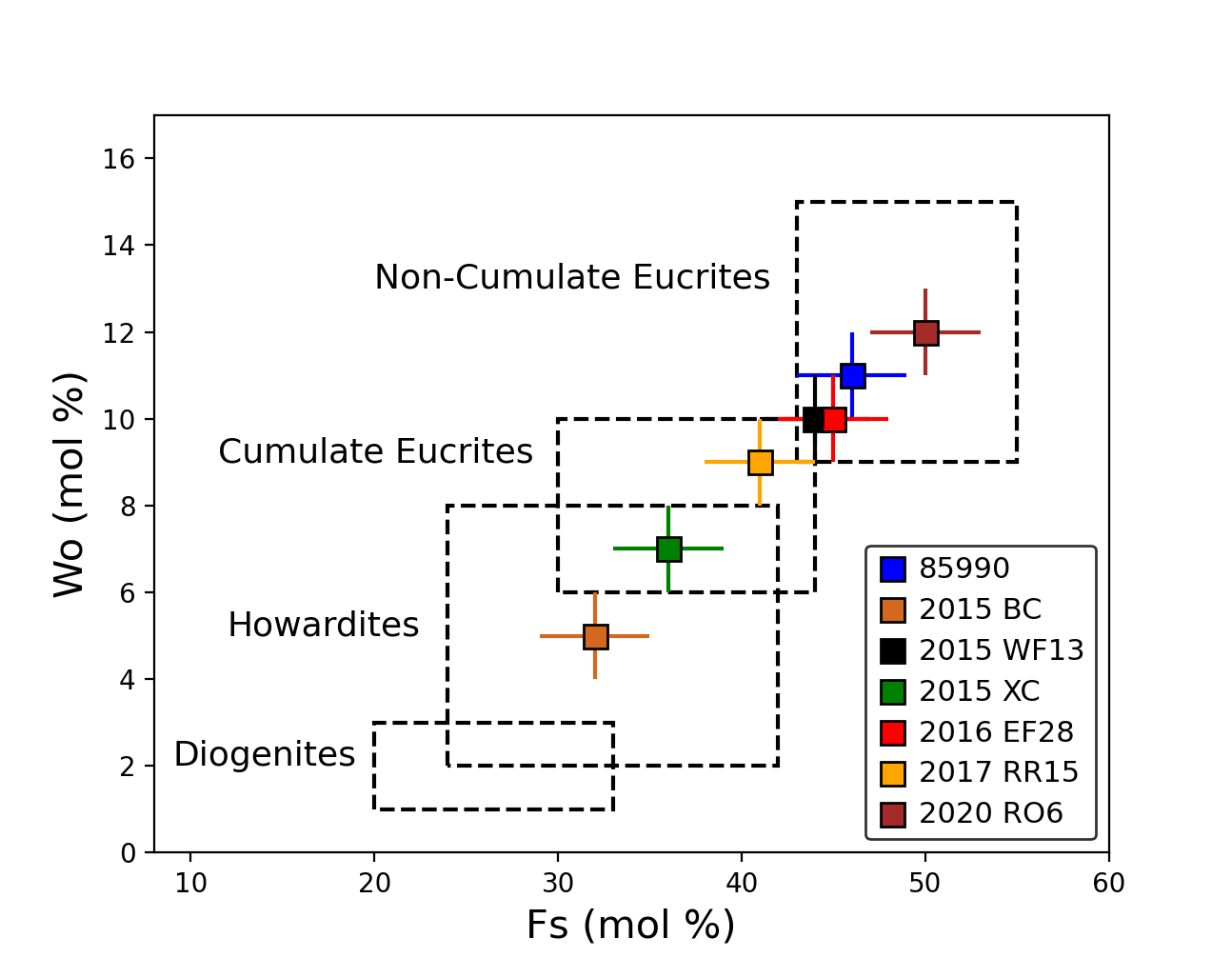}

\caption{\label{f:Wo_Fs}{\small Molar contents of wollastonite (Wo) vs. ferrosilite (Fs) for NEOs 85990, 2015 BC, 2015 WF13, 2015 XC, 2016 EF28, 2017 RR15 
and 2020 RO6. The error bars correspond to the values determined by \cite{2009MandPS...44.1331B}, 3 mol\% for Fs and 1 mol \% for Wo. The approximated range of 
pyroxene chemistries for howardites, non-cumulate eucrites, cumulate eucrites, and diogenites from \cite{2013Icar..225..131S} are indicated as dashed-line boxes.}}

\end{center}
\end{figure*}

The compositional analysis of the V-type NEOs was done using their band centers. Since these parameters can be affected by temperature variations, we first  
applied the temperature corrections derived by \cite{2012Icar..217..153R} from the analysis of HED meteorites. For those asteroids whose spectra were obtained without 
the 0.8 $\mu$m dichroic, both band centers were used along with the equations of \cite{2009MandPS...44.1331B}. These equations are used to determine the pyroxene chemistry, which is 
given by the molar contents of ferrosilite (Fs) and wollastonite (Wo). For asteroids whose spectra were obtained with the 0.8 $\mu$m dichroic, only the Band II centers were used to 
determine their composition (Table 4). 

Figure \ref{f:Wo_Fs} shows the molar contents of wollastonite (Wo) vs. ferrosilite (Fs) for the V-type NEOs and the regions corresponding to the howardites, non-cumulate eucrites, 
cumulate eucrites, and diogenites. In order to determine the probability of each V-type belonging to a specific class, we built a machine learning model like the one used with the 
S-complex asteroids. In this case, the synthetic multi-class dataset was generated with two input variables (Wo and Fs) and the boundaries shown in Figure \ref{f:Wo_Fs}. 

NEOs 2015 WF13, 2016 EF28 and 2020 RO6 were classified as non-cumulate eucrites. These are the most common type of eucrites and represent surface or near-surface basalts \citep{2015ChEG...75..155M}. NEOs 2015 XC and 2017 RR15, on the other hand, were classified as cumulate eucrites, which are thought represent subsurface cumulate layers formed by fractional crystallization \citep{2015ChEG...75..155M}. For 
2015 XC, Figure \ref{f:Wo_Fs} shows some overlap between the cumulate eucrite and the howardite regions, however, our model favors cumulate eucrites as the best meteorite 
analogs with a 66\% probability.

\subsubsection{C/X-complex and B-type asteroids}

\begin{figure*}[!ht]
\begin{center}
\includegraphics[height=13cm]{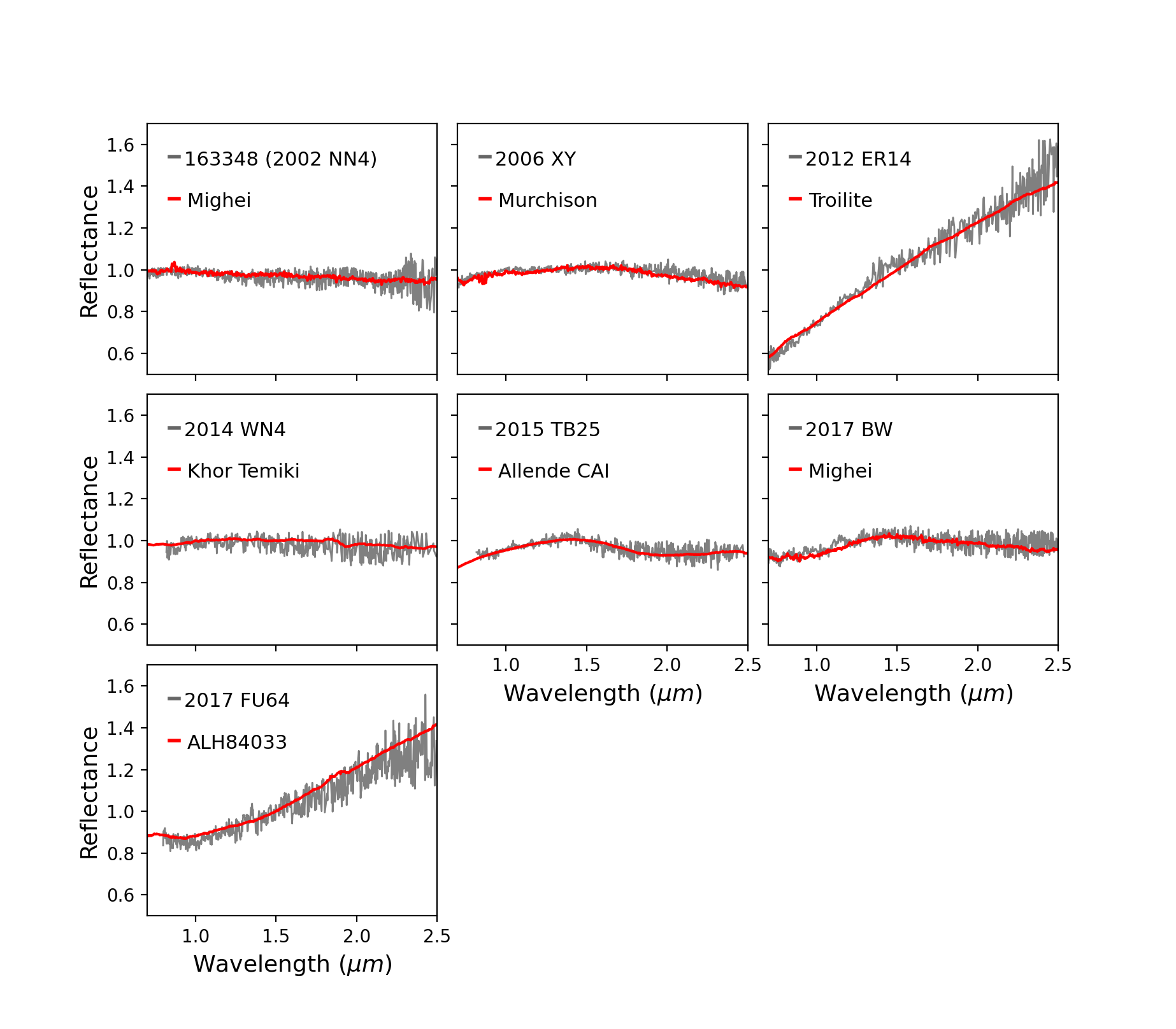}

\caption{\label{f:curve_match}{\small NIR spectra of NEOs 163348,  2006 XY, 2012 ER14, 2014 WN4, 2015 TB25, 2017 BW and 2017 FU64. Also shown the spectra 
of the CM2 carbonaceous chondrite Mighei (RELAB sample IDs cmms01 and ccms01), CM2 carbonaceous chondrite Murchison (RELAB sample ID cnms02), 
troilite from the Canyon Diablo iron meteorite (RELAB sample ID cae01), enstatite achondrite (aubrite) Khor Temiki (RELAB sample ID latb48), calcium-aluminum-rich 
inclusions (CAIs) from Allende (RELAB sample ID c1tm05) and CM2 carbonaceous chondrite ALH84033 (RELAB sample ID c1mp14).}}

\end{center}
\end{figure*}

As explained earlier, our spectroscopic data do not cover most of the visible wavelength range. This is particularly problematic for the compositional analysis of those asteroids 
in the C/X-complex whose spectra show diagnostic features in the visible. NEOs that were classified as C/X-complex include 2002 LY1, 2006 XY, 2014 WN4, 
2015 LK24, 2015 TF, 2016 CO247, 2016 EB1 and 2017 OP68. For objects like these, with weak features or featureless spectra, we used the Modeling for Asteroids (M4AST) 
online tool \citep{2012A&A...544A.130P} to look for possible meteorite analogs. If a good spectral match was found, it is shown along with the spectrum of the asteroid in 
Figure \ref{f:curve_match}. For NEOs 2015 LK24, 2015 TF and 2016 EB1 no good meteorite analogs were found. However, the thermal excess and low albedos calculated for 
these asteroids suggest that they might be composed of carbonaceous chondrite-like material. We also noticed that the albedos of these NEOs, which range from 0.015 to 
0.025 are closer to the albedo of P-type asteroids, which have a mean value of 0.03$\pm$0.01 \citep{2022AJ....163..165M}.
 
The spectra of 2002 LY1, 2006 XY, 2014 WN4 and 2016 CO247 are featureless, with relatively neutral spectral slopes and do not show signs of a thermal excess at 
wavelengths $>$2 $\mu$m. The spectrum of a dark NEO that is close to its perihelion would typically display a thermal tail. However, sometimes, if the object is faint, this 
thermal tail could be difficult to detect. Thus, for these four NEOs we cannot rule out a carbonaceous chondrite-like composition. Another alternative that could explain their 
spectral characteristics is the presence of iron-free enstatite. The NIR spectra of some enstatite chondrites and achondrites (aubrites), which have a high content of this 
mineral, are featureless and can have neutral or negative slopes depending on the grain size \citep[e.g.,][]{2009Icar..202..477V, 2016AJ....152..162R}. Therefore, a 
composition dominated by iron-free enstatite could also be compatible with the spectral characteristics of these asteroids. For 2006 XY and 2014 WN4 we 
found a good spectral match with the CM2 carbonaceous chondrite Murchison and the aubrite Khor Temiki, respectively (Figure \ref{f:curve_match}).  

Only one of the NEOs classified as C/X-complex, 2017 OP68, shows weak features in the NIR, with one possible absorption band centered at 0.932$\pm$0.005 $\mu$m and 
another at 1.138$\pm$0.010 $\mu$m. The spectra of carbonaceous chondrites that have undergone aqueous alteration exhibit absorption bands due to the presence of 
phyllosilicates such as the serpentine and the saponite groups \citep{2012Icar..220..586C}. The serpentine-group shows absorption bands at $\sim$0.90-0.94 $\mu$m and 
$\sim$1.1-1.2 $\mu$m, whereas the saponite-group at $\sim$0.90 $\mu$m and $\sim$1.1-1.2 $\mu$m \citep{2011Icar..216..309C, 2012Icar..220..586C}. These particular 
absorption bands are caused by an octahedral Fe$^{2+}$ crystal field transition. Given the position of the absorption bands of 2017 OP68, both phyllosilicates could be 
present, although the serpentine-group might be dominant. Possible meteorite analogs for this NEO include CM and CI carbonaceous chondrites. 

The NEO 163348 was the only object classified as a B-type. Asteroids belonging to this taxonomic type have spectra that can exhibit either negative or positive NIR slopes 
\citep[e.g.,][]{2012Icar..218..196D} and have a mean geometric albedo of 0.09$^{+0.05}_{-0.04}$ \citep{2022AJ....163..165M}. The NIR spectrum of 163348 is 
featureless, it has a slightly negative slope and shows a thermal excess at wavelengths $>$2 $\mu$m. These spectral characteristics are similar to some B-types, 
although the albedo derived for this object (0.015) is more consistent with the mean value estimated for P-type 
asteroids \citep{2011AJ....142...85T, 2022AJ....163..165M}. \cite{2023PSJ.....4..177C} studied grain size effects on the spectra of carbonaceous chondrites 
and found that grains $>$150 $\mu$m could, in some cases, turn a Ch-type object into a B-type. Thus, it is also possible that 163348 is being classified 
as a B-type due to larger grains on the surface. The spectral match with meteorite Mighei (Figure \ref{f:curve_match}) along with the low albedo suggest that 
carbonaceous chondrites could be good meteorite analogs for this asteroid.

\subsubsection{Xk- and Xn-type asteroids}

The NIR spectra of Xk- and Xn-types show weak absorption bands and no thermal excess, which allow us to separate them from the broader C/X-complex category for 
the compositional analysis. NEOs classified as Xk-types include 85990, 2015 BC and 2020 SN. The spectra of 85990 and 2015 BC show red spectral slopes and weak 
absorption bands at $\sim$0.93 $\mu$m and $\sim$2.0 $\mu$m due to the presence of pyroxene. Since both NEOs are also located in the basaltic achondrites region 
(Figure \ref{f:BIC_BAR}), the compositional analysis was done following the same procedure used with the V-types. We found that the pyroxene chemistry of 85990 is 
consistent with non-cumulate eucrites, whereas 2015 BC is more similar to howardites (Figure \ref{f:Wo_Fs}). It is important to note that given the relatively weak absorption 
bands of these objects, pyroxene is likely not the main mineral present on the surface. For example, some asteroids with similar spectral characteristics have been found to have a high metal content on their surface based on their radar albedos \citep[e.g.,][]{2011M&PS...46.1910H, 2014Icar..238...37N, 2021PSJ.....2..205S}. Thus, possible meteorite analogs could also include silicate-bearing NiFe meteorites, stony-iron meteorites and metal-rich carbonaceous chondrites.

The NIR spectrum of 2020 SN has a curved downward shape and a weak absorption band centered at $\sim$0.89 $\mu$m. The spectra of HEDs typically have 
Band I centers in the range of $\sim$0.92-0.95 $\mu$m, making the spectral calibrations derived by \cite{2009MandPS...44.1331B} not suitable for the compositional analysis 
of this asteroid. Nevertheless, the position of the band center suggests the presence of Fe-poor pyroxene on the surface of 2020 SN. Enstatite chondrites or aubrites containing 
traces of Fe$^{2+}$ are possible meteorite analogs for this asteroid. 

The Xn class was not originally included in the Bus-DeMeo taxonomy but was introduced by \cite{2019Icar..324...41B} to classify a 
small number of NEOs whose spectra are relatively flat and show a narrow feature centered at $\sim$0.9 $\mu$m. NEOs in our sample that were classified as 
Xn include 2015 TC25 and 2016 GU. 

The physical and compositional properties of 2015 TC25 have been already studied in great detail by \cite{2016AJ....152..162R}. The NIR spectrum of this object 
shows a negative slope and a narrow absorption band centered at $\sim$0.91 $\mu$m. Based on its high geometric albedo and spectral characteristics, 
\cite{2016AJ....152..162R} determined that 2015 TC25 was a “Nysa-like” E-type asteroid. They also found that the spectrum of this NEO could be reproduced with a 
mixture of 7\% orthopyroxene and 93\% aubrite.

The spectrum of 2016 GU has a neutral spectral slope an a weak and narrow absorption band centered at 0.91 $\mu$m. Like in the case of 2020 SN we did no use 
the equations of \cite{2009MandPS...44.1331B} for the analysis of this asteroid, as its Band I center is below the typical range measured for HEDs. The position of the 
Band I center, however, suggests the presence of Fe-poor pyroxene on the surface. Because this object does not have a red spectral slope, silicate-bearing NiFe 
meteorites are probably not good meteorite analogs. Instead, this NEO could be more similar to enstatite chondrites or aubrites containing traces of Fe$^{2+}$.

\subsubsection{D-type asteroids}

NEOs that were classified as D-types include 2012 ER14, 2015 AP43, 2019 RC and 2017 FU64. All these objects have featureless spectra with very 
steep slopes. The spectra of 2015 AP43 and 2017 FU64 also show a thermal excess, which allowed us to estimate albedos of 0.05 and 0.02, respectively for these 
objects. These values are within the range found for D-type asteroids ($\sim$0.02-0.07) by previous studies \citep{2011AJ....142...85T, 2022AJ....163..165M}. 
\cite{2022Icar..38014971D} noticed that a large percentage of D-type asteroids matched the spectra of iron meteorites. We obtained similar results when we looked for meteorite analogs for these NEOs. Figure \ref{f:curve_match} shows an example of such spectral match, in this case between 2012 ER14 and troilite from the Canyon 
Diablo iron meteorite. This is somehow expected, as the spectra of iron meteorites share the main characteristics of D-types, i.e., very steep slopes and lack of 
absorption features. Despite the good spectral match between the D-types and the iron meteorites, these asteroids, in general, are considered to be primitive bodies 
and have been associated with carbonaceous chondrites such as the Tagish Lake meteorite \citep[e.g.,][]{2001Sci...293.2234H, 2018MNRAS.476.4481B, 2021Icar..36114349G}. In the case of 2015 AP43 and 2017 FU64, their low albedo are consistent with a carbonaceous chondrite-like composition. For 2017 FU64, 
we found a good spectral match with the CM2 carbonaceous chondrite ALH84033 (Figure \ref{f:curve_match}). The spectra of 2012 ER14 and 2019 RC did not show 
a thermal tail, but, as discussed earlier, a thermal tail can be difficult to detect in faint objects. As a result, it is also possible that these two NEOs are primitive bodies.

\subsubsection{K-, L- and Indeterminate asteroids}

The only NEO in our sample that was classified as a K-type is 2014 PR62. The NIR spectrum of this object shows an absorption band centered at $\sim$1 $\mu$m 
and no signs of a 2-$\mu$m band. Given the scattering in the data at wavelengths $>$ 1.5 $\mu$m, it is not clear if the 2-$\mu$m band is really absent or too weak to 
be detected. The compositional analysis of 2014 PR62 was done using the Band I center and the equations of \cite{2020AJ....159..146S} (see section 2.4.1). The 
asteroid was found to have olivine and pyroxene chemistries similar to LL-chondrites (Figure \ref{f:Nakamura}). Some K-type asteroids in the main belt, e.g., 221 Eos 
and members of its family, have been linked to CO, CV, CK and R chondrites \citep{1998Icar..131...15D, 2001MandPS...36..245B, 2008Icar..195..277M}. We 
were not able to find a good meteorite analog for 2014 PR62 based on curve matching; however, considering its spectral characteristics, an affinity with any of those 
meteorites could be possible.

Four NEOs were classified as L-types, including 2014 VH2, 2015 TB25, 2017 BW and 2017 CR32. Previous work have found that some 
carbonaceous chondrites and calcium-aluminum-rich inclusions (CAIs) share similar spectral characteristics with L-type asteroids 
\citep{1992Metic..27..424B, 2008Sci...320..514S, 2018Icar..304...31D}. The spectra of the four NEOs show an increase in reflectance with increasing wavelengths from 
$\sim$0.75 to 1.5 $\mu$m and then become relatively flat at longer wavelengths. For 2014 VH2 we did not find a good spectral match with meteorite samples, but given its 
taxonomic classification, possible meteorite analogs could include carbonaceous chondrites. The NIR spectrum of 2015 TB25 shows a slightly negative slope at wavelengths 
$>$1.5 $\mu$m and the possible presence of a 2-$\mu$m band. For this NEO we found a good spectral match with CAIs from the CV3 carbonaceous chondrite Allende 
(Figure \ref{f:curve_match}). The spectrum of the CAI exhibits a characteristic absorption band centered at $\sim$2 $\mu$m due to the presence of spinel (MgAl$_{2}$O$_{4}$). 
For asteroid 2017 BW, the best meteorite analog was found to be the CM2 carbonaceous chondrite Mighei. No good meteorite analogs were found for 2017 CR32, however, 
as in the previous cases, a link with carbonaceous chondrites could be possible. 

There was one asteroid (2015 FL) whose NIR spectrum did not match any of the spectra in the Bus-DeMeo taxonomy and therefore its classification was 
labeled as "indeterminate". The spectrum of this asteroid shows a very red slope (comparable to a D-type) from $\sim$0.7 to 1.4 $\mu$m and then becomes flat at 
longer wavelengths. No absorption bands are visible, but given the scattering of the data beyond 1.4 $\mu$m we cannot rule out the presence of the 2-$\mu$m band. 
No good meteorite analogs were found for this NEO. 

\subsection{Space weathering and resurfacing processes} \label{sec:SW}

Space weathering refers to any process that modifies the optical properties and physical structure of the surface of airless bodies. It is characterized by producing an increase 
of the spectral slope and suppression of the absorption bands in the NIR spectra of S-complex objects \citep[e.g.,][]{2000M&PS...35.1101P, 2001JGR...10610039H, 2002aste.book..585C,  2010Icar..209..564G}. For asteroids in the S-complex, S-, Sq-, and Q-types are thought to represent a weathering gradient, with Q-types having relatively fresh surfaces and 
Sq- and S-types having more space-weathered surfaces \citep[e.g.,][]{2010Natur.463..331B, 2019Icar..324...41B}. 

Different processes have been proposed to explain the presence of fresh material on the surface of asteroids, including thermal fatigue fragmentation, YORP spin-up 
and planetary encounters. Thermal fragmentation caused by diurnal temperature variations can create fresh regolith by breaking down rocks and exposing their 
unweathered interiors \citep{2014Natur.508..233D}. YORP spin-up, on the other hand, can accelerate the rotation of an asteroid enough to displace weathered 
particles, refreshing with this the surface of the object \citep[e.g.,][]{2018Icar..304..162G}. Similarly, tidal stress caused during close encounters with planets can 
produce landslides exposing fresh subsurface material \citep[e.g.,][]{2010Icar..209..510N, 2010Natur.463..331B, 2019Icar..324...41B}. 

\cite{2019Icar..324...41B} investigated the efficiency of the different resurfacing processes as a function of the perihelion distance. For this they estimated 
the degree of space weathering experienced by the NEOs calculating the Space Weathering Parameter $\Delta\eta$, which uses the principal components 
PC1' and PC2' obtained from the taxonomic classification. This parameter is given by the scalar magnitude of the space weathering vector defined in the 
principal component space of the Bus-DeMeo taxonomy \citep{2010Natur.463..331B, 2019Icar..324...41B}: 

\begin{equation}
\Delta\eta = \frac{-\frac{1}{3}PC2'+PC1'+0.5}{1.0541}
\end{equation}

The analysis done by \cite{2019Icar..324...41B} 
showed that the amount of NEOs with fresh unweathered surfaces increases as the perihelion distance decreases. This is because at short perihelion 
distances there are more resurfacing processes operating simultaneously. At perihelion distances $<$1.0 AU, thermal fragmentation, YORP spin-up and 
planetary encounters with the Earth and Venus are all acting together, whereas at greater perihelion distances, YORP spin-up and encounters with Mars 
are the most dominant. 

\begin{figure*}[!ht]
\begin{center}
\includegraphics[height=9cm]{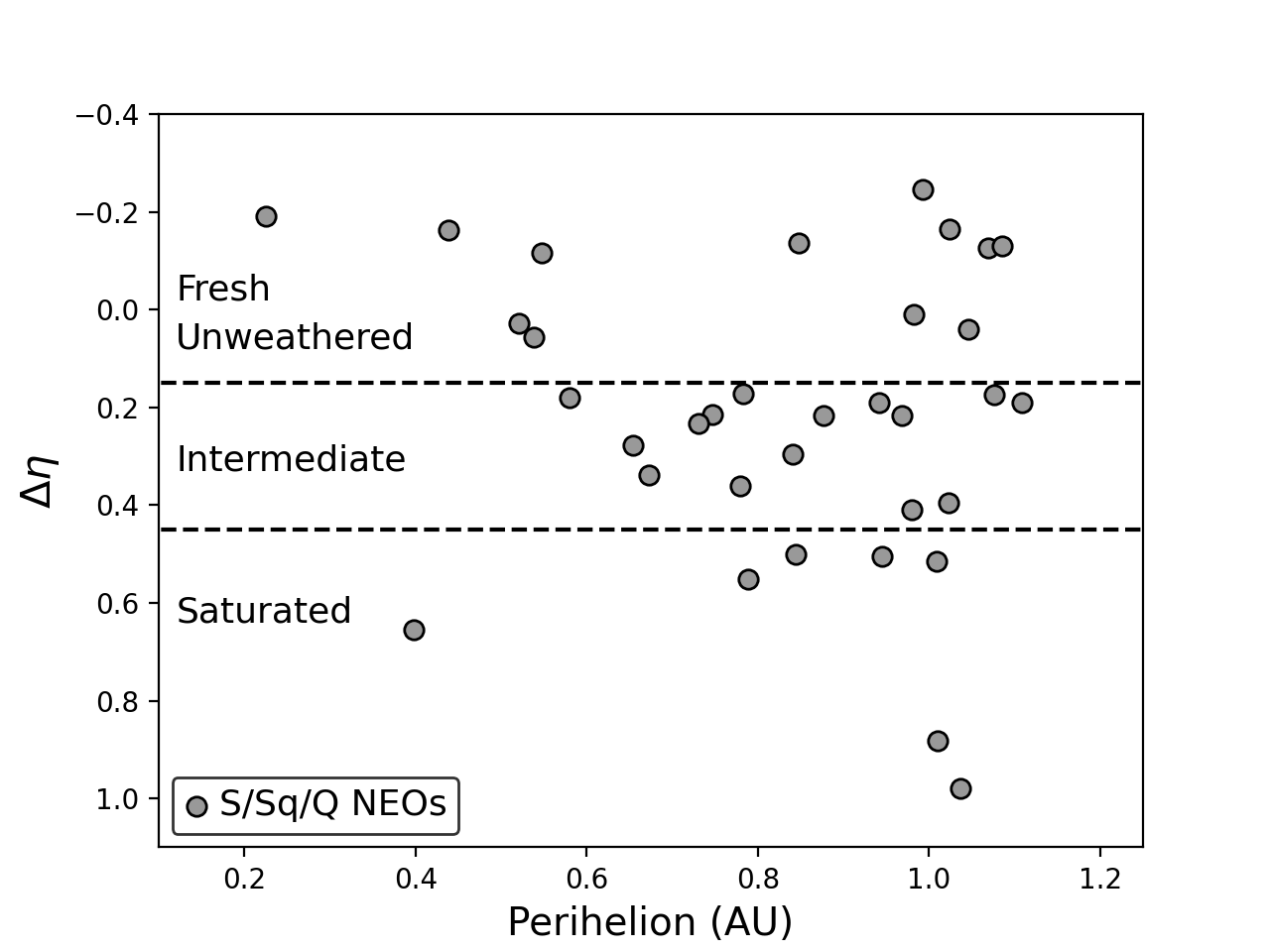}

\caption{\label{f:SW_q}{\small Space Weathering Parameter $\Delta\eta$ as a function of the perihelion distance for S-, Sq-, and Q-types. Regions corresponding 
to the different regimes of space weathering (fresh, intermediate and saturated) from \cite{2019Icar..324...41B} are indicated. Figure adapted from 
\cite{2019Icar..324...41B}.}}

\end{center}
\end{figure*}

We performed a similar analysis for the S-, Sq-, and Q-types in our sample. Spectra obtained with the 0.8 $\mu$m dichroic were not included. Figure \ref{f:SW_q} 
shows $\Delta\eta$ as a function of the perihelion distance for the NEOs. Most objects were found to fall within the fresh or intermediate regimes as defined by 
\cite{2019Icar..324...41B} and only a few fall in the saturated regime. In the context of space weathering, saturation occurs when surface grains become uniformly weathered after 
multiple re-arrangements events followed by extended periods of exposure to the space environment \citep{2019Icar..324...41B}.

The trend in Figure \ref{f:SW_q} is less clear than what it was found by \cite{2019Icar..324...41B}, this is probably 
because the maximum perihelion distance of the NEOs in our analysis is $\sim$1.1 AU, whereas the data used by \cite{2019Icar..324...41B} include objects with 
perihelion distances that extend up to $\sim$1.7 AU. Nevertheless, we note that most of the NEOs that fall in the saturated regime have perihelion distances 
$>$0.8 AU, which is consistent with the results of \cite{2019Icar..324...41B}.

\cite{2019PASJ...71..103H} found that surface refreshening due to close encounters with planets could explain only $\sim$50\% of known Q-types. To account for 
the other Q-types, they proposed an alternative explanation where these objects have weathered surfaces like S-types, but their spectral characteristics are explained by the 
presence of large particles ($>$100 $\mu$m). Different mechanisms have been proposed to explain the release of particles smaller than 100 $\mu$m, 
a condition required for this hypothesis to work. These include solar radiation pressure and electrostatic acceleration \citep{2019PASJ...71..103H}.

\begin{figure} 
 \includegraphics[width=9.5cm,angle=0]{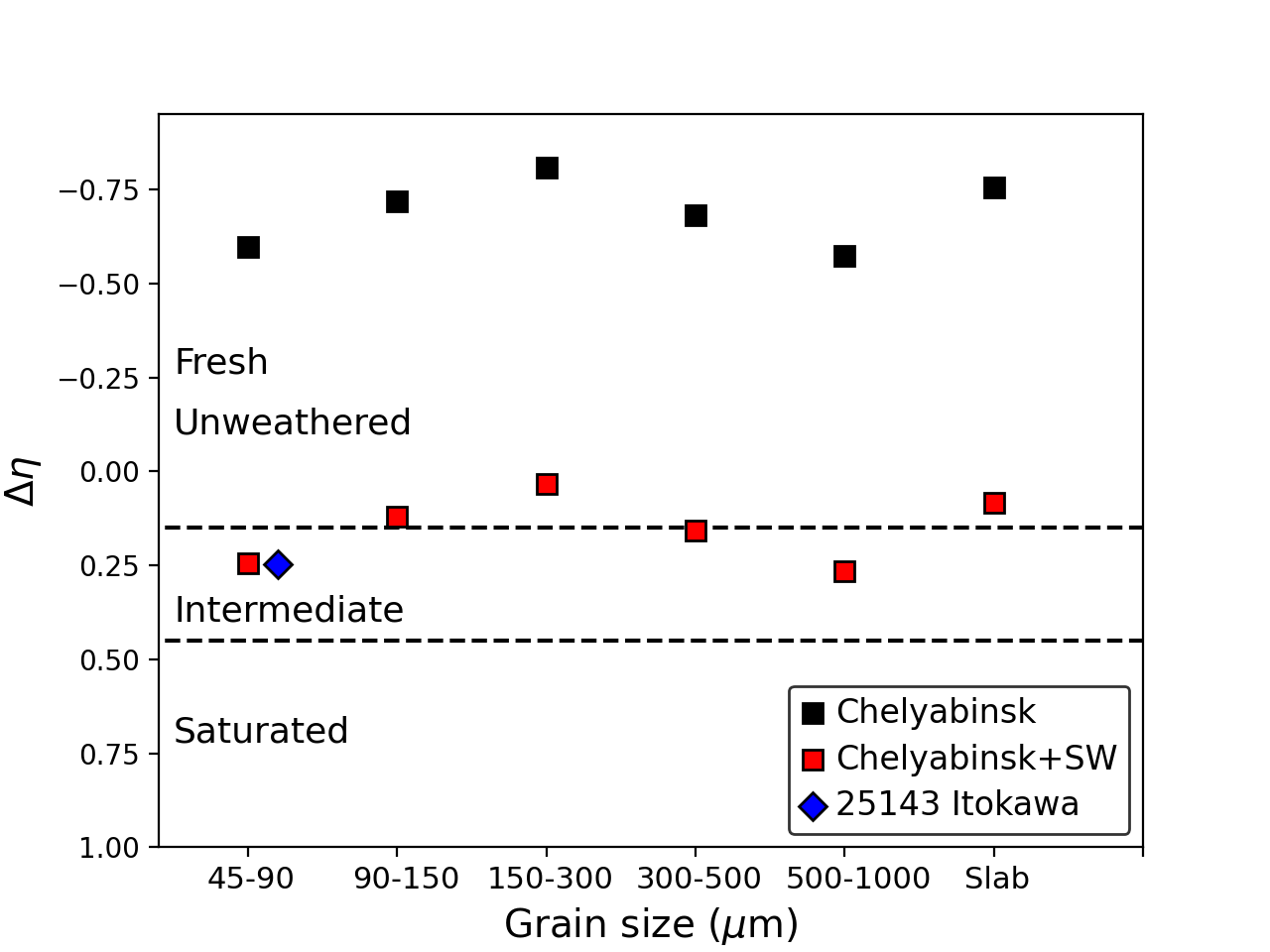}
 \hspace{-5mm}
 \includegraphics[width=9.5cm,angle=0]{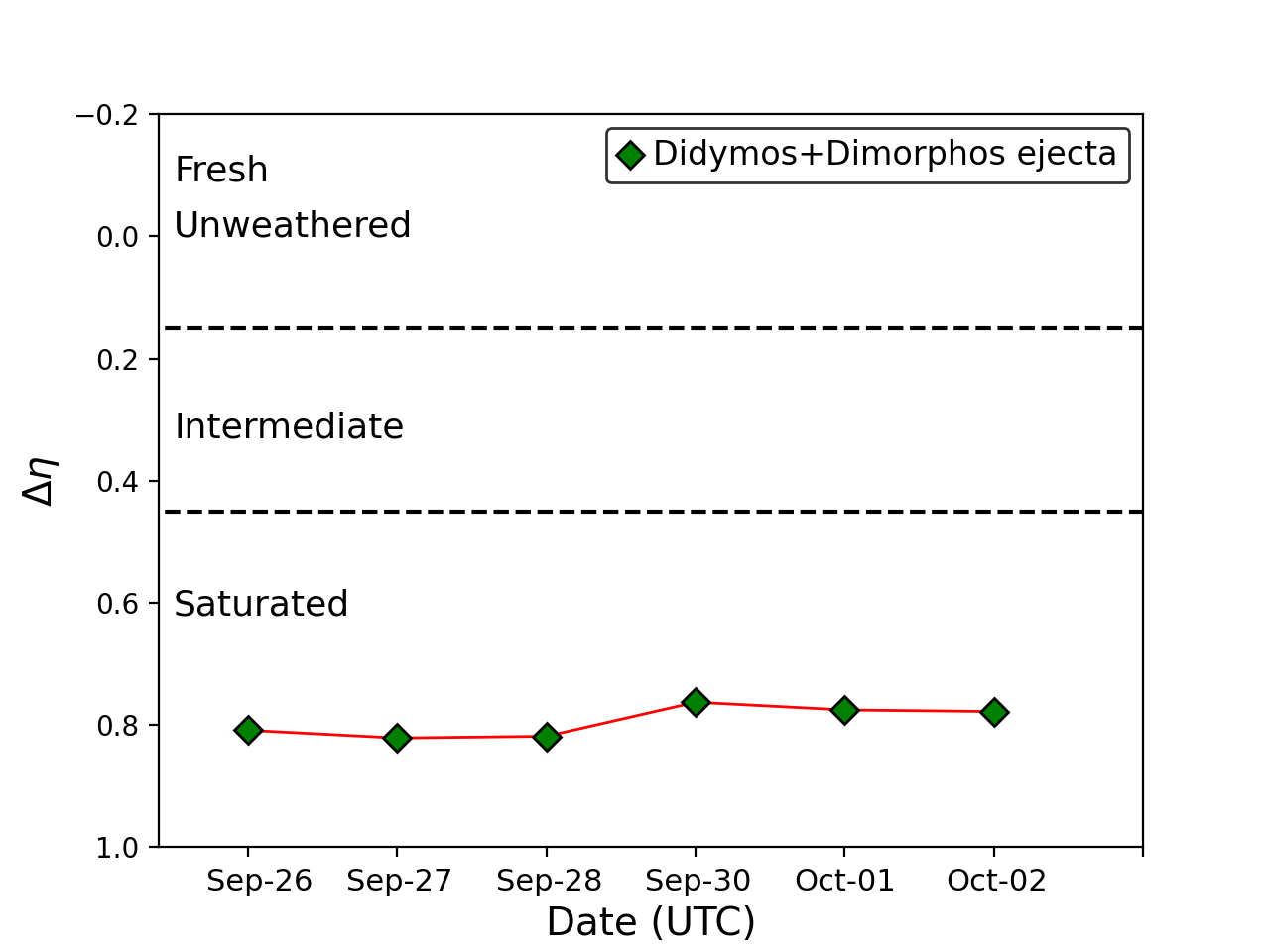}
 \caption{Left: Space Weathering Parameter $\Delta\eta$ vs. grain size for the LL-chondrite Chelyabinsk. Black squares represent the $\Delta\eta$ 
values calculated for different grain sizes of Chelyabinsk obtained by \cite{2023PSJ.....4...52B}. Red squares represent the $\Delta\eta$ values of 
Chelyabinsk plus space weathering (SW) after been offset to match the $\Delta\eta$ calculated for asteroid Itokawa (blue diamond) from the spectrum obtained by 
\cite{2001M&PS...36.1167B}. Right: Space Weathering Parameter $\Delta\eta$ vs. date for the Didymos system. $\Delta\eta$ has been calculated from the PC values reported by 
\cite{2023PSJ.....4..229P}. The location in the x-axis of the Didymos system corresponds to the observation dates carried out before (Sep 26) and after (Sep 27-Oct 02) the impact of the 
DART mission. For both figures, regions corresponding to the different regimes of space weathering (fresh, intermediate and saturated) from \cite{2019Icar..324...41B} are indicated.}

\label{f:SW_GS}
\end{figure}

We tested this hypothesis by comparing $\Delta\eta$ values obtained for different grain sizes of the LL-chondrite Chelyabinsk with the $\Delta\eta$ of asteroid 25143 
Itokawa. The $\Delta\eta$ values of Chelyabinsk were calculated from the spectra of samples with five different grain sizes (45-90, 90-150, 150-300, 300-500, 500-1000 
$\mu$m) obtained by \cite{2023PSJ.....4...52B} plus the spectrum of a slab obtained for this study. These values are depicted as black squares in Figure \ref{f:SW_GS} (left). The 
$\Delta\eta$ for Itokawa was calculated from the spectrum obtained by \cite{2001M&PS...36.1167B}. \cite{2006Sci...312.1334A} found that the average reflectance spectra of Itokawa were consistent with grain sizes $<$125 $\mu$m. 
Therefore, we matched the location of Itokawa in the x-axis with the 45-90 $\mu$m sample, since this grain size is probably the one that best represents the mean grain size of 
the asteroid. Then, we offset all the $\Delta\eta$ values calculated for Chelyabinsk (red squares in Figure \ref{f:SW_GS}), so that the $\Delta\eta$ corresponding to the 45-90 
$\mu$m sample overlaps with the $\Delta\eta$ of Itokawa. In this way we can visualize what would happen to the Space Weathering Parameter of Itokawa if we 
increase the grain size. 

As can be seen in Figure \ref{f:SW_GS} (left), increasing the grain size from 45-90 to 150-300 $\mu$m could, in theory, move Itokawa from the intermediate to the fresh (unweathered) 
regime. A similar result would be obtained if the surface of Itokawa were completely depleted of dust (slab), which is probably the case for very small asteroids of just a few meters in 
size \citep[e.g.,][]{2016AJ....152..162R}. Interestingly, if the grain size were in the range of $\sim$500-1000 $\mu$m, no 
change in $\Delta\eta$ would be seen compared to the 45-90 $\mu$m sample.

\begin{figure} 
 \includegraphics[width=9.5cm,angle=0]{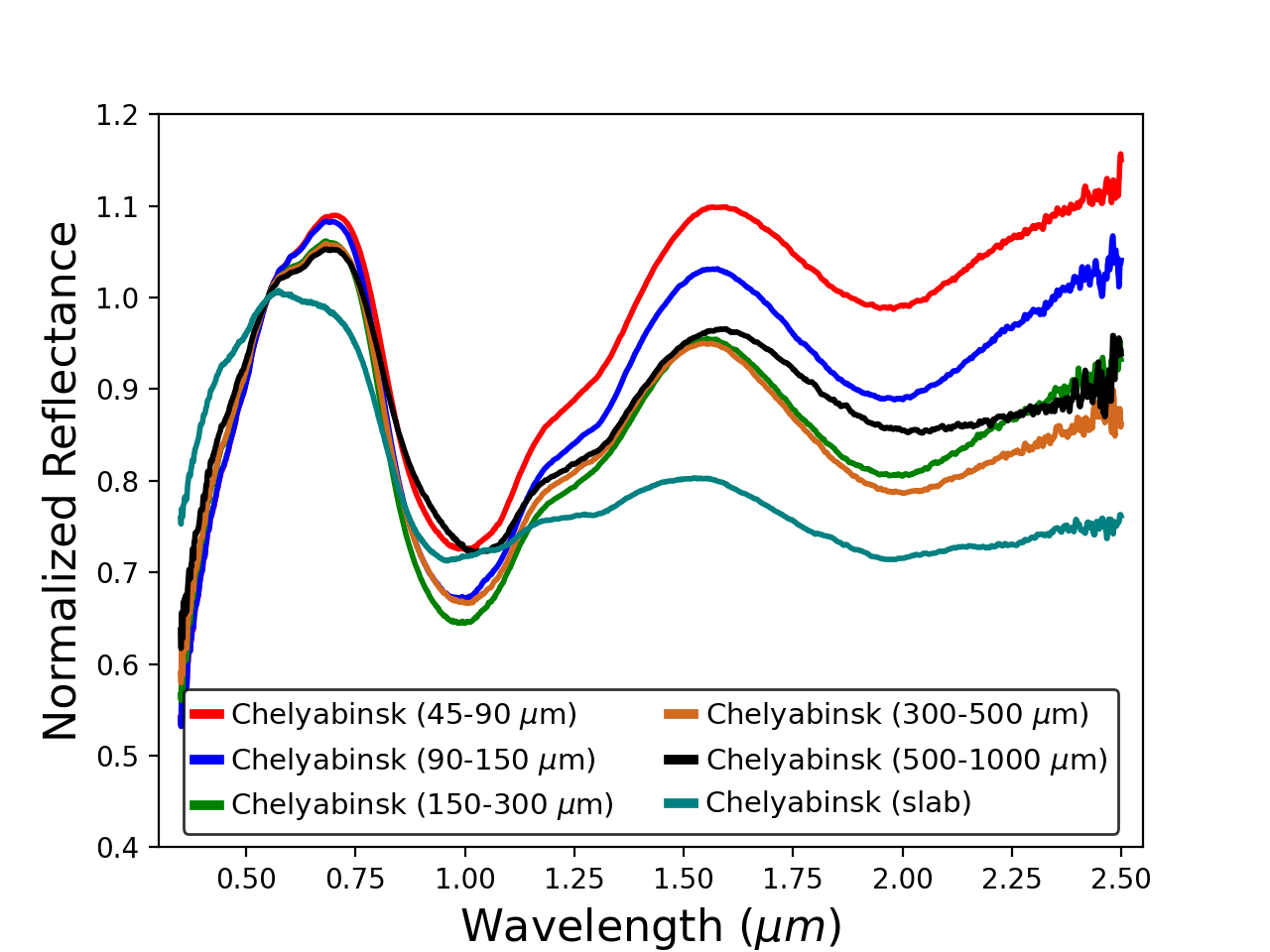}
 \hspace{-5mm}
 \includegraphics[width=9.5cm,angle=0]{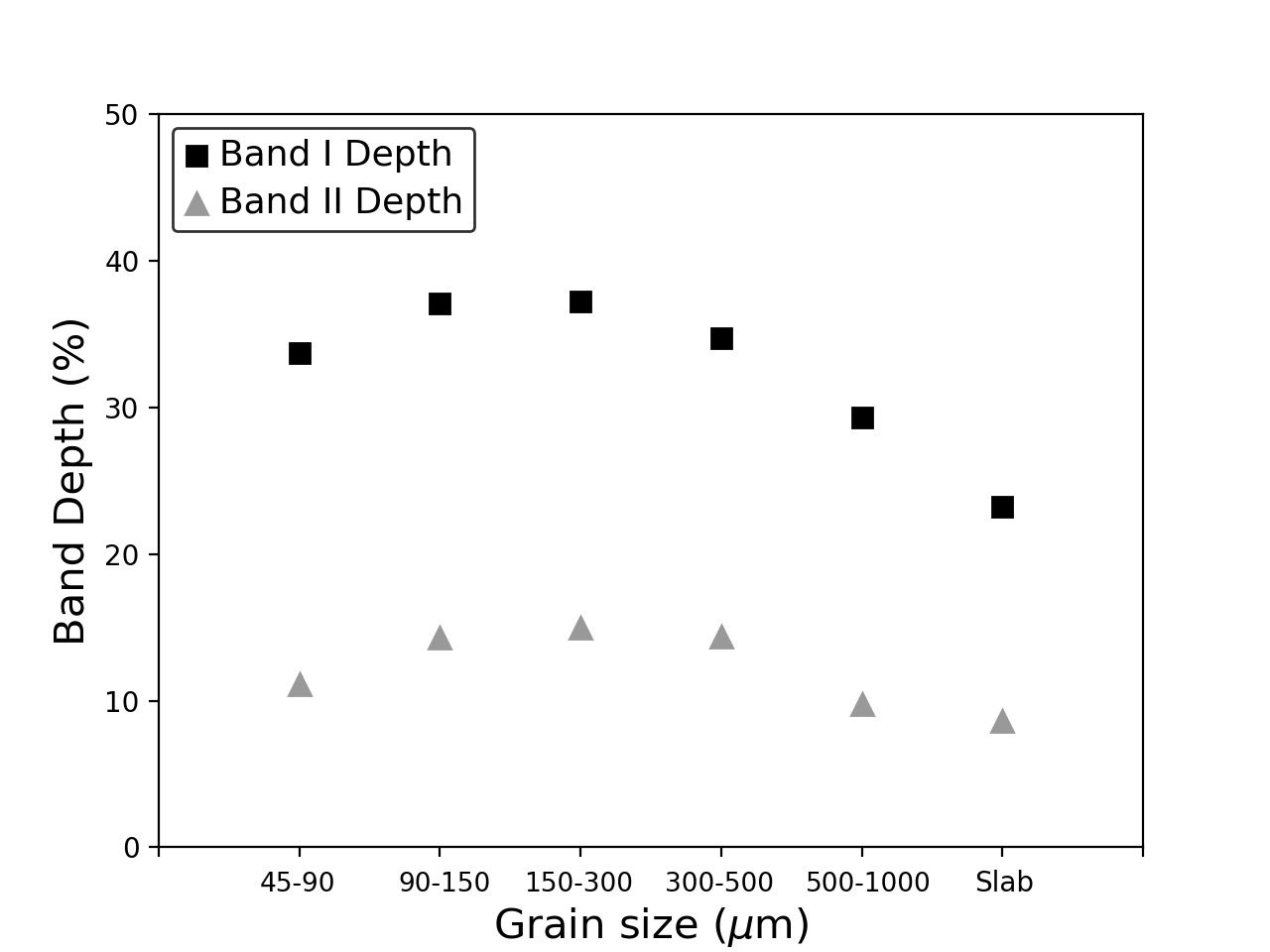}
 \caption{Left: Spectra of the LL-chondrite Chelyabinsk corresponding to five different grain sizes plus a slab from \cite{2023PSJ.....4...52B}. All spectra are normalized to unity 
 at 0.55 $\mu$m. Right: Band I and II depths as a function of grain size measured from the Chelyabinsk spectra. Uncertainties are smaller than the symbols.} 
\label{f:Chelyabinks_BD}
\end{figure}

Our results seem to confirm the findings of \cite{2019PASJ...71..103H}, but also impose an upper limit in the grain size for this mechanism to be effective. For the Chelyabinsk 
sample shown in Figure \ref{f:SW_GS}, this upper limit is $\sim$400 $\mu$m, for other meteorites with different compositions this limit might change. This 
constraint in grain size for which we see a meaningful increase of $\Delta\eta$ is related to the way light is absorbed within the grains. To illustrate this, it is
useful to see how the Chelyabinsk spectra change as the grain size increases. Figure \ref{f:Chelyabinks_BD} (left) shows the spectra corresponding to five 
different grain sizes plus a slab. The most noticeable change is an overall decrease in the spectral slope with increasing grain size, being particularly 
pronounced for the slab. Band depths, which were measured from the continuum to the band center \citep[e.g.,][]{2012Icar..220...36S}, show an 
increase from the 45-90 $\mu$m to the 150-300 $\mu$m sample and then a decrease for larger grain sizes (Figure \ref{f:Chelyabinks_BD} right). The initial increase in band 
depths happens because as the grain size increases, so does the mean optical path length, resulting in more absorption within the grain and deeper absorption bands 
\citep[e.g.,][]{1983JGR....88.9534P}. However, after reaching a maximum in band depth at a grain size of $\sim$225 $\mu$m, band saturation starts to occur (i.e., all photons that are not scattered are absorbed within the grains) and the absorption bands become shallower. Because $\Delta\eta$ is calculated from PC1' and PC2', which are 
sensitive to the intensity of the absorption bands, this parameter will closely follow the behavior of the band depths. It is important to note that composition also plays a role and that 
the pattern that we observe for the ordinary chondrite is probably not going to be the same for a meteorite with a different composition.

The obvious consequence of these results is that there will be some cases where the increase in grain size is too small (or too large) to produce a change in the 
perceived degree of space weathering of the asteroid. An example of such a case could be the 
Didymos system, the target of NASA’s Double Asteroid Redirection Test (DART) mission. \cite{2023PSJ.....4..229P} obtained NIR spectra of Didymos before and after the 
impact. They found that the ejecta cloud released from Dimorphos was the main contributor to the light of the system for $\sim$40 hours after the impact. During those hours, a decrease 
in spectral slope was observed and, as the ejecta cloud dispersed, the spectral slope returned to pre-impact level. However, no significant change in the absorption bands that could turn this S-type asteroid into a "fresh" Q-type was detected. This led to the conclusion that a negligible amount of unweathered material was ejected from Dimorphos and that the ejecta cloud 
was dominated by coarse debris $\geq$100 $\mu$m \citep{2023PSJ.....4..229P}.

We calculated the $\Delta\eta$ values for the Didymos system from the PC1' and PC2' values reported by \cite{2023PSJ.....4..229P}. As shown in Figure \ref{f:SW_GS} (right), 
there is little change in $\Delta\eta$ from Sep 26-28 and only a small decrease occurs for the following nights. This is consistent 
with the lack of unweathered material reported by \cite{2023PSJ.....4..229P}. Moreover, the small variation in $\Delta\eta$ combined with the decrease in spectral slope suggests two 
possibilities: 1) the increase in grain size was  small compared with the grains present before the impact or 2) the increase in grain size was large enough to start to produce band 
saturation. This example also highlights the fact that an asteroid that is well inside the saturated region (S-type) is unlikely to reach the intermediate (Sq-type) or unweathered (Q-type) 
regions even if the grain size increases to the point where $\Delta\eta$ reaches its lowest value.

\subsection{Possible evidence of metal or shock-darkening} \label{sec:SD}

When the results obtained from the taxonomic classification and the compositional analysis are combined it is possible to obtain further insights about 
particular characteristics of the NEOs. This is the case for the Sx-type objects, whose weak absorption bands could lead to an ambiguous classification in the C- or 
X-complex, even though their composition is similar to ordinary chondrites. In this section we investigate the possible causes behind these results.

\begin{figure}[!h]
 \includegraphics[width=9.5cm,angle=0]{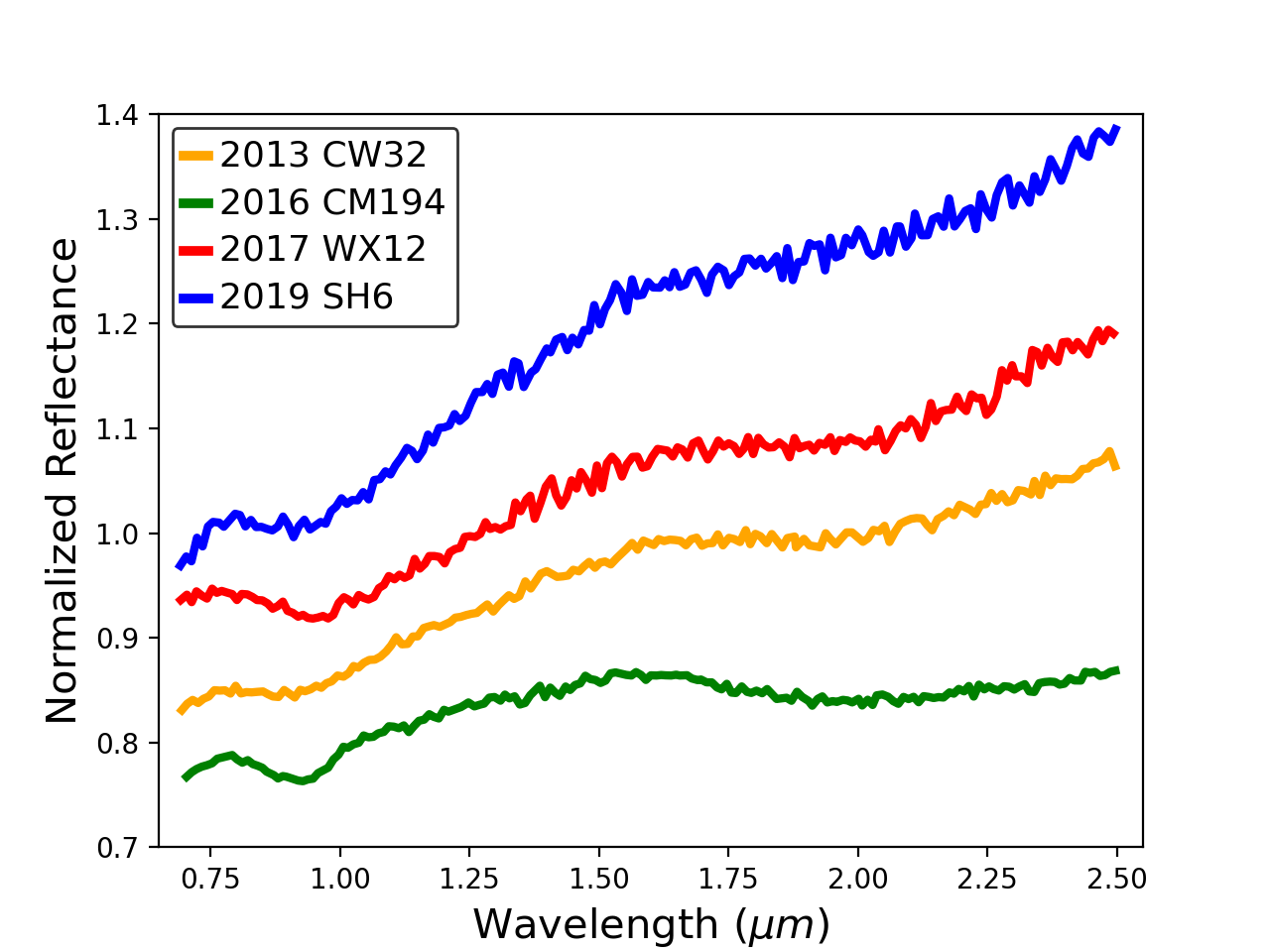}
 \hspace{-5mm}
 \includegraphics[width=9.5cm,angle=0]{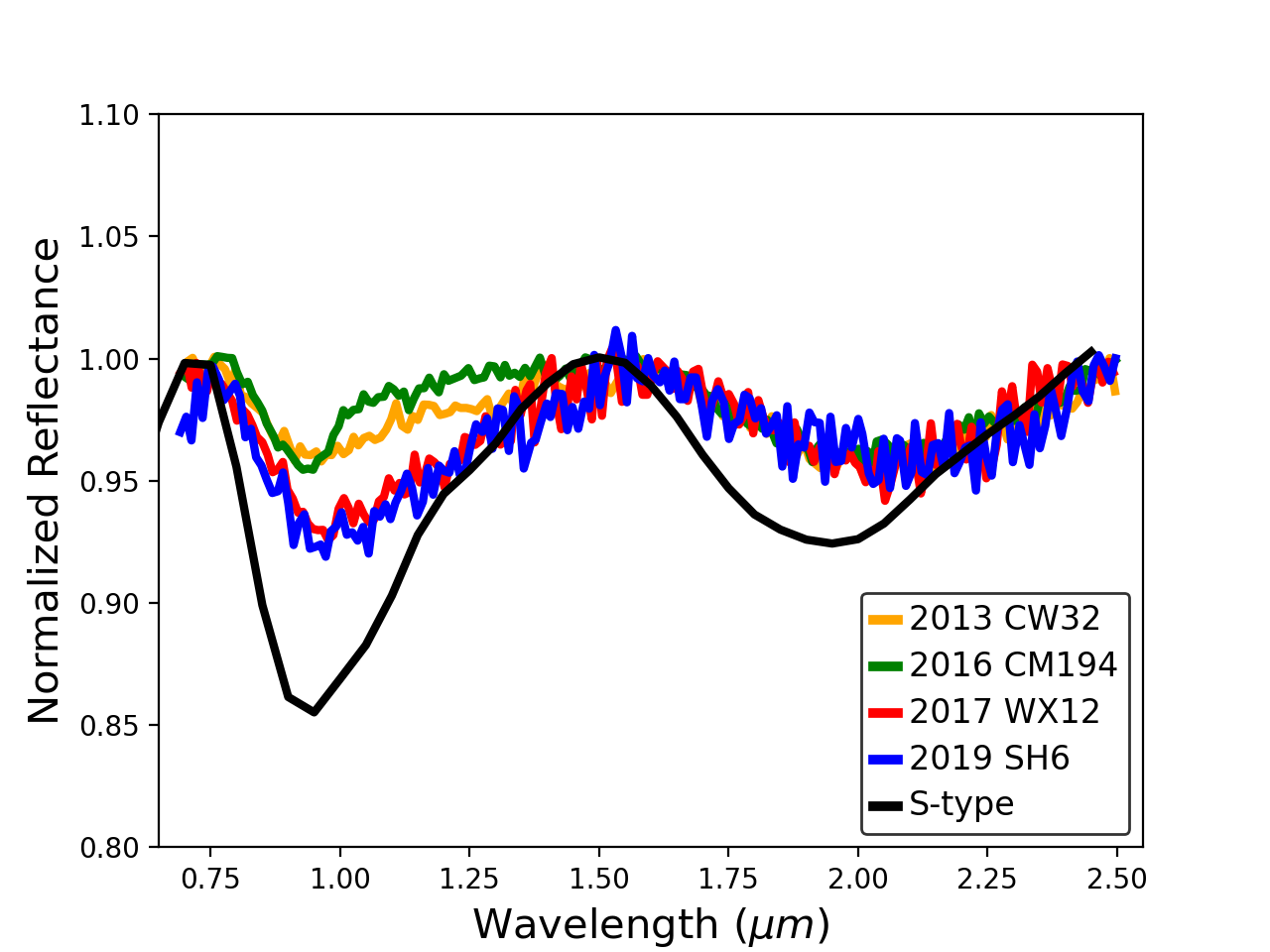}
 \caption{\label{f:SD_spectra}Left: spectra of NEOs 2013 CW32, 2016 CM194, 2017 WX12, 2019 SH6. All the spectra show absorption bands centered at 
$\sim$0.9 and 2.0 $\mu$m due to minerals olivine and pyroxene. Spectra have been offset for clarity. Right: continuum-removed spectra of the same objects. The mean spectrum of an S-type from \cite{2009Icar..202..160D} is also shown.}
\label{fig:abratio}
\end{figure}

As a reminder, four NEOs were found to share these characteristics, 2013 CW32, 2016 CM194, 2017 WX12 and 2019 SH6. The spectra of these objects are shown 
together in Figure \ref{f:SD_spectra} (left). At first glance, the spectral characteristics of these NEOs are typical of S-type asteroids showing absorption bands 
centered at $\sim$0.9 and 2.0 $\mu$m due to the presence of olivine and pyroxene. However, compared to the mean spectrum of an S-type, their absorption bands 
are much weaker (Figure \ref{f:SD_spectra}, right). 

\begin{figure*}[h]
\begin{center}
\includegraphics[height=10cm]{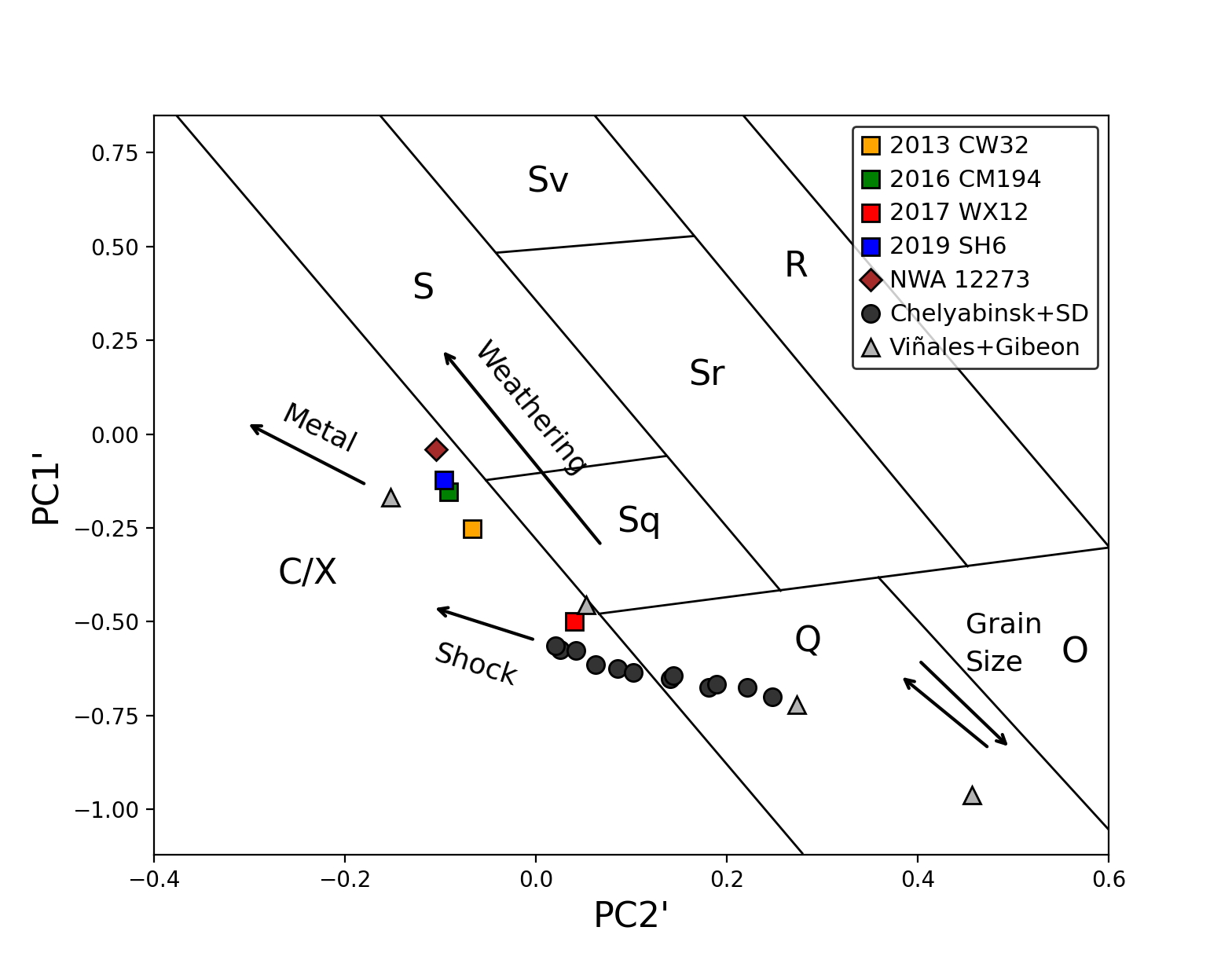}

\caption{\label{f:PC1_PC2}{\small PC2’ vs PC1’ diagram from \cite{2009Icar..202..160D}. PC values calculated for 2013 CW32, 2016 CM194, 2017 WX12, 2019 SH6 and the metal-rich chondrite 
NWA 12273 \citep{2019LPI....50.2212R} are depicted with different colors. PC values for intimate mixtures of the L-chondrite Viñales and the iron meteorite 
Gibeon and intimate mixtures of Chelyabinsk and shock-darkened (SD) material from \citep{2014Icar..237..116R} are also shown. Arrows indicate the direction in which the grain size or the content of metal and shock darkening increases. The first data point for the mixture of Viñales and Gibeon (lower right corner) corresponds to a mixture of 40 wt.\% Gibeon and 60 wt.\% Viñales. The following data points represent increments of 20 wt.\% Gibeon. The space weathering vector from \cite{2019Icar..324...41B} is also indicated.}}

\end{center}
\end{figure*}

NEO 2013 CW32 was found to have an L chondrite-like composition. Its NIR spectrum shows a red spectral slope with a Band I depth of 4.1$\pm$0.1\% and a Band II depth of 3.1$\pm$0.1\%. For comparison, the mean spectrum of an S-type asteroid from \cite{2009Icar..202..160D} has a Band I depth of 
13.0$\pm$0.2\% and a Band II depth of 5.5$\pm$0.1\%. The composition of 2016 CM194 was also found to be similar to L-chondrites, although an affinity with primitive achondrites is also possible. This object has the less steep spectral slope of the four NEOs. The Band I depth is 4.4$\pm$0.1\% and the Band II depth is 2.6$\pm$0.1\%. 
2017 WX12, on the other hand, has a composition consistent with LL-chondrites. The spectrum of this asteroid has a red spectral slope with a Band I depth of 
7.6$\pm$0.1\% and a Band II depth of 4.8$\pm$0.6\%. Similarly, the composition of 2019 SH6 is also consistent with LL-chondrites. The spectral slope of this NEO 
is the steepest of the four objects discussed in this section. The spectrum has a Band I depth of 7.8$\pm$0.2\% and the Band II depth is 3.8$\pm$0.5\%.

In the Bus-DeMeo taxonomy the PC1’ and PC2’ values are sensitive to the presence and intensity of the absorption bands. As a 
result, as the absorption bands become shallower, PC2’ moves towards negative values and PC1’ towards positive values, eventually crossing the line 
$\alpha$, which separates the C/X-complex, whose spectra are relatively featureless, from other taxonomic types such as those in the S-complex and V-types. The 
PC1’ and PC2’ values for the four NEOs are shown in Figure \ref{f:PC1_PC2}. Several factors are known to produce changes in the band depths, among them the most 
relevant are: space weathering, grain size, phase reddening, presence of metal and shock-darkening.  

As explained in the previous section, space weathering produces suppression of the absorption bands. As the degree of space weathering increases, an 
asteroid with an ordinary chondrite-like composition will move from the Q-type to the S-type taxonomy. In the PC2’ vs PC1’ diagram this transition is 
represented by a space weathering vector that moves parallel to the line $\alpha$, but that does not cross it (Figure \ref{f:PC1_PC2}). This suggests that space 
weathering alone is not responsible for the weak absorption bands of these asteroids.

As we have already seen, grain size variations can produce changes in the intensity of the absorption bands and the spectral slope. \cite{2023PSJ.....4...52B} studied the 
effects of grain size on spectral band parameters, composition and taxonomic classification. For this, they analyzed the spectra of ordinary chondrites for five different 
grain size groups (45-90, 90-150, 150-300, 300-500, and 500-1000 $\mu$m). In the PC2’ vs PC1’ diagram most samples were found to move away from the line $\alpha$ for 
grain size groups 45-90 to 150-300 $\mu$m, but then for the largest grain sizes, 300-500 and 500-1000 $\mu$m, the opposite behavior was observed and the PC values 
ended up close to where they started (Figure \ref{f:PC1_PC2}). This behavior is explained by the saturation in the large grains previously discussed. Considering this, it seems 
unlikely that grain size is the main responsible for the attenuation of the absorption bands observed in the NEOs.

Phase reddening produces an increase of the spectral slope and variations in the band depths as the phase angle increases \citep{2012Icar..220...36S}. 
This effect is particularly relevant for NEOs, since they are normally observed at high phase angles. \cite{2012Icar..220...36S} analyzed NIR spectra of 
S-complex NEOs and ordinary chondrites obtained at different phase angles. They found that the increase in spectral slope starts to become evident at 
phase angles $>$30$^{o}$ and absorptions bands become deeper reaching a maximum depth at phase angles $\sim$60$^{o}$. Many of the NEOs in this study were 
observed at phase angles $>$30$^{o}$, thus are likely affected by phase reddening. However, we note that while phase reddening can explain, at least in part, the 
red spectral slopes of some of the NEOs, the increase in band depths produced by phase reddening is inconsistent with the attenuation of the absorption 
bands of the NEOs discussed in this section.

The presence of metal can also suppress absorption bands. Radar observations have identified metal on the surface of many asteroids 
\citep[e.g.,][]{1991plas.rept..174O,  2010Icar..208..221S, 2015Icar..245...38S}. The NIR spectra of these objects typically show red slopes and, in some 
cases, weak absorption bands at $\sim$0.9 and 1.9 $\mu$m \citep[e.g.,][]{2010Icar..210..674O, 2011M&PS...46.1910H, 2014Icar..238...37N,  2019LPI....50.2212R, 2017AJ....153...29S, 2021PSJ.....2..205S}. Samples from these bodies are likely represented by several classes of meteorites found 
on Earth, including iron meteorites, stony-iron meteorites such as pallasites and mesosiderites, metal-rich carbonaceous chondrites and anomalous 
metal-rich chondrites.

\begin{figure*}[h]
\begin{center}
\includegraphics[height=9cm]{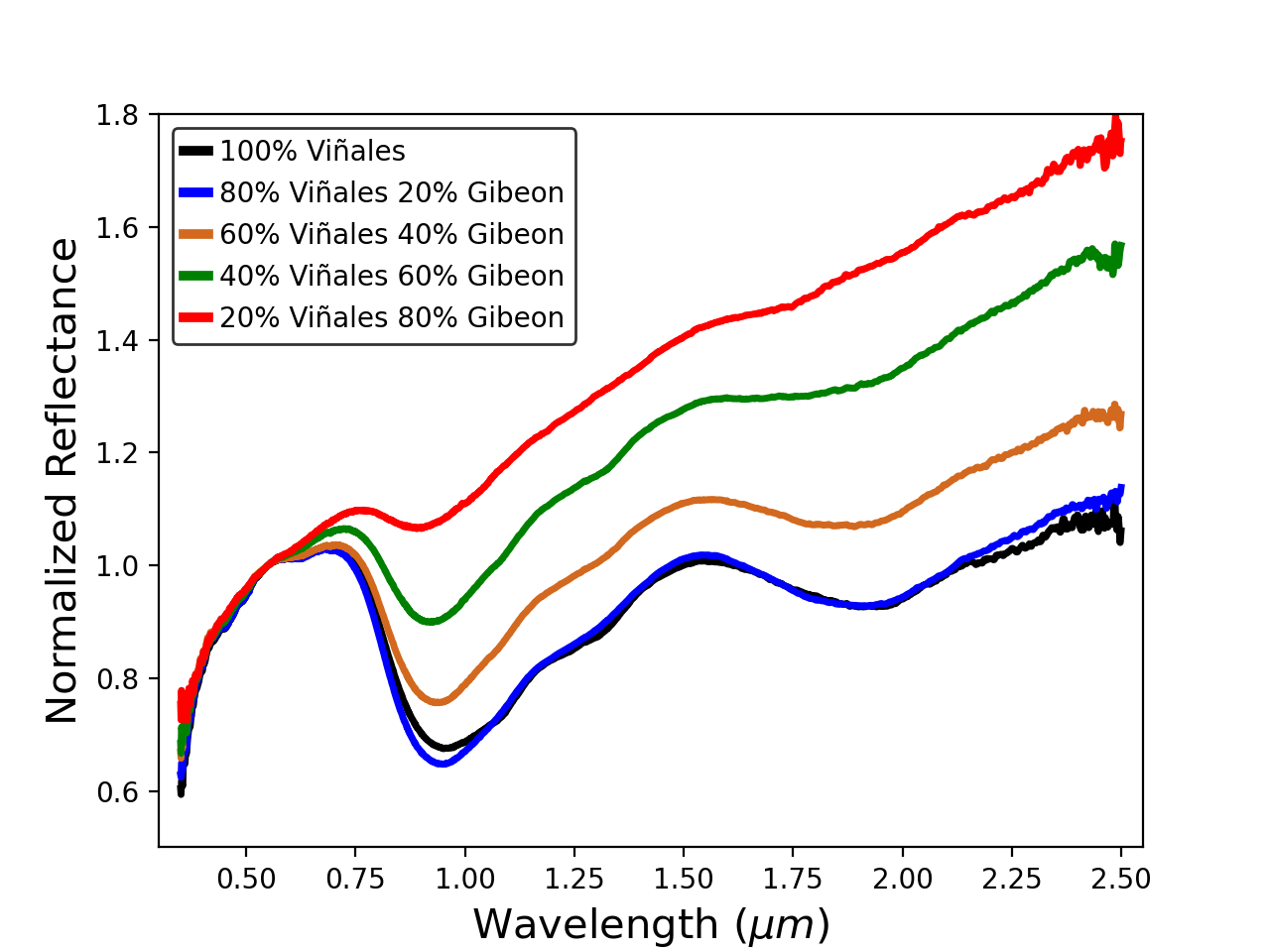}

\caption{\label{f:Vinales_Gibeon}{\small Spectra of intimate mixtures of the L-chondrite Viñales and the iron meteorite Gibeon. All spectra are normalized to unity at 0.55.}}

\end{center}
\end{figure*}

Evidence of metal-rich chondrites with affinity to ordinary chondrites has been recently found in meteorites NWA 12273 and NWA 12379 
\citep[e.g.,][]{2019LPI....50.1176A,  2019ChEG...79l5537J}. NWA 12273 is made up of $\sim$64\% Fe-Ni metal and $\sim$30\% chondrules 
\citep{2019LPI....50.1176A}. The olivine chemistry is consistent with L3/LL3-chondrites, whereas the low-Ca pyroxene is more consistent with H4-chondrites.  
 A NIR spectrum of NWA 12273 obtained by \cite{2019LPI....50.2212R} shows a red spectral slope and two weak absorption features at $\sim$0.9 and 2 $\mu$m 
 due to minerals olivine and pyroxene. NWA 12379 shares similar characteristics, it is made up of $\sim$70 vol.\% Fe-Ni metal and $\sim$25 vol.\% porphyritic 
 ferromagnesian chondrules \citep{2019ChEG...79l5537J}. The oxygen isotopic composition and mineralogical characteristics of this meteorite were found to be 
 similar to those of L3.8 ordinary chondrites \citep{2019ChEG...79l5537J}. According to \cite{2019ChEG...79l5537J}, NWA 12379 could have formed by a collision 
 between an ordinary chondrite-like body and a metal-rich body.

In order to investigate the effects of metal content on ordinary chondrites, we prepared intimate mixtures of the L-chondrite Viñales and the iron meteorite 
Gibeon. Viñales was crushed using an alumina mortar and pestle and sieved to a grain size of 45-90 $\mu$m. Metal shavings of Gibeon were sieved to a 
grain size of $<$300 $\mu$m. Four mixtures were prepared starting with a metal content of 20 wt.\% and increasing in 20 wt.\% metal intervals. A sample 
of 100 wt.\% Gibeon was also prepared. The visible and NIR spectra (0.35-2.5 $\mu$m) of the samples were obtained relative to a Labsphere Spectralon 
disk using an ASD Labspec4 Pro spectrometer at an incident angle i = 0$^{o}$ and emission angle e = 30$^{o}$ (Figure \ref{f:Vinales_Gibeon}). The Bus-DeMeo taxonomic classification 
was then applied to the spectra, the results are shown in Figure \ref{f:PC1_PC2}. The first data point, which falls in the Q-type region, corresponds to a mixture of 40 wt.\% 
Gibeon and 60 wt.\% Viñales, the last data point corresponds to the sample of 100 wt.\% Gibeon. We found that increasing the metal content would 
cause an object classified as Q-type to progressively move up and to the left in the PC2’ vs PC1’ diagram, reaching a point where it crosses the line 
$\alpha$ and falls in the C/X-complex region. We applied the taxonomic classification to the spectrum of the metal-rich chondrite NWA 12273 obtained 
by \cite{2019LPI....50.2212R} and found that the PC1’ and PC2’ values also fall in the C/X-complex region. These results show that the presence of metal could explain the weak 
absorption bands of the NEOs.

The presence of shock-darkened or impact melt material can also have a significant effect on the spectral properties of asteroids and meteorite samples. 
Shock-darkening occurs at pressures of $\sim$40-50 GPa, whereas impact melt requires pressures of over 90-150 Gpa 
\citep[e.g.,][]{ 2020A&A...639A.146K}. \cite{2014Icar..228...78K} found that the effects of shock-darkening and impact melt on the spectrum of 
an LL-chondrite are essentially the same, i.e., increasing the amount of these lithologies will produce a decrease in absolute reflectance and suppression 
of the absorption bands. Since the spectral effects of the two lithologies are indistinguishable, in the present work we will refer to them as just 
shock-darkening. 

\cite{2014Icar..237..116R} suggested that shock-darkening could be responsible for the low albedo and subdued absorption bands seen among asteroids of 
the Baptistina family. More recently, \cite{2022PSJ.....3..226B} found compelling evidence for the presence of shock-darkening in the NEO (52768) 1998 
OR2, that could explain its weak absorption bands and its classification as an Xn-type.

Figure \ref{f:PC1_PC2} (black circles) shows the results of applying the Bus-DeMeo taxonomic classification to the spectra of intimate mixtures of Chelyabinsk (light 
colored lithology) and the shock-darkened material from \cite{2014Icar..237..116R}. The results show how shock-darkening can 
effectively turn a Q-type asteroid into a C/X-type object when the amount of shock-darkened material is $>$50\%. 

To summarize, in this section we have discussed the factors that could be responsible for the weak absorption bands of four NEOs whose compositions are 
similar to those of ordinary chondrites. These objects represent $\sim$5\% of the entire sample and $\sim$7\% of the objects with ordinary chondrite-like composition 
studied in this work. We suggest that the presence of metal or shock-darkening could be responsible for the unusual spectral characteristics of these objects. Other posible 
explanations such as the presence of carbonaceous chondrite material or even a mixture of carbonaceous chondrite and metal could have a similar effect. Although the spectral 
characteristics of these NEOs do not seem to be consistent with this scenario, we leave this possibility open for a future investigation.  

Determining which asteroids have metal and which have shock-darkened material is a more complicated task, but in some cases it might 
be possible from the NIR spectrum. For example, the spectrum of 2016 CM194 has the less steep spectral slope of the four NEOs, if the presence of metal were 
responsible for the weak absorption bands of this object, we should also see a very steep spectral slope. This is evident in Figure \ref{f:Vinales_Gibeon} that shows 
the effects of adding meteoritic metal to an ordinary chondrite. Increasing the amount of metal will suppress the absorption bands and also increase the spectral slope. 
For the spectrum corresponding to the 100\% Viñales sample we measured Band I and II depths of 34\% and 11\%, respectively, but for the mixture corresponding to 
80\% Gibeon and 20\% Viñales the Band I depth decreased to 9\% and the Band II depth to $\sim$3\%. In the case of the spectral slope, which was measured as the 
slope of a linear fit performed from the reflectance maximum at $\sim$0.7 $\mu$m to the reflectance maximum at $\sim$1.48 $\mu$m, the value changed from 
-0.02621 $\mu$m$^{-1}$ (100\% Viñales) to 0.4112 $\mu$m$^{-1}$ (80\% Gibeon and 20\% Viñales). Thus, for 2016 CM194 we favor shock-darkening, which can 
suppress the absorption bands, but has little effect on the spectral slope \citep{2014Icar..237..116R}.

The spectra of NEOs 2013 CW32, 2017 WX12 and 2019 SH6 show red (steep) spectral slopes, which makes it more difficult to rule out metal just by looking 
at the spectra as in the case of 2016 CM194. Apart from metal, both space weathering and phase reddening also increase the spectral slope, introducing 
further complications to the analysis. NEOs 2013 CW32 and 2017 WX12 were observed at phase angles of 10$^{o}$ and 21$^{o}$, respectively and are probably not very 
affected by phase reddening \citep{2012Icar..220...36S}, but they might have experienced some degree of space weathering. 2019 SH6, on the other hand, 
was observed at a phase angle of 53$^{o}$ and its spectral slope is likely affected by phase reddening. The surface of this object could also be affected by 
space weathering. In cases like these, radar data could be useful to rule out or confirm the presence of metal, since high radar albedos are typically 
associated with metal-rich asteroids.

For NEO 2013 CW32, \cite{2022PSJ.....3..222V} reported a radar albedo ($\hat{\sigma}_{OC}$) in the range of 0.23-0.31. This range is higher than the 
mean $\hat{\sigma}_{OC}$ found by these authors for S- and Q-type NEOs (0.19$\pm$0.06), which could indicate the presence of some metal mixed with 
ordinary chondrite-like material, perhaps similar to meteorites NWA 12273 and NWA 12379. Unfortunately, for 2017 WX12 and 2019 SH6 there is no radar data available and the limited information that we have for these objects is insufficient to further constrain their composition.

\subsection{Near-Earth Object source regions} \label{sec:source}

An important part in the study of NEOs is to determine their source regions, as this allow us to establish which of those regions contribute 
the most to the influx of asteroids to the near-Earth space, as well as identifying the possible parent bodies of NEOs and meteorites that fall on Earth.
In order to determine the likely source regions for the NEOs in our sample, we used the NEO model described 
by \cite{2017A&A...598A..52G, 2018Icar..312..181G}. This model yields the probability that an asteroid escaped out of one of seven different regions, including 
the $\nu_{6}$ secular resonance; the 3:1, 5:2, and 2:1 mean motion resonances (MMR) with Jupiter; and the Hungaria, Phocaea, and Jupiter family comet 
regions. 

\begin{figure*}[!ht]
\begin{center}
\includegraphics[height=9cm]{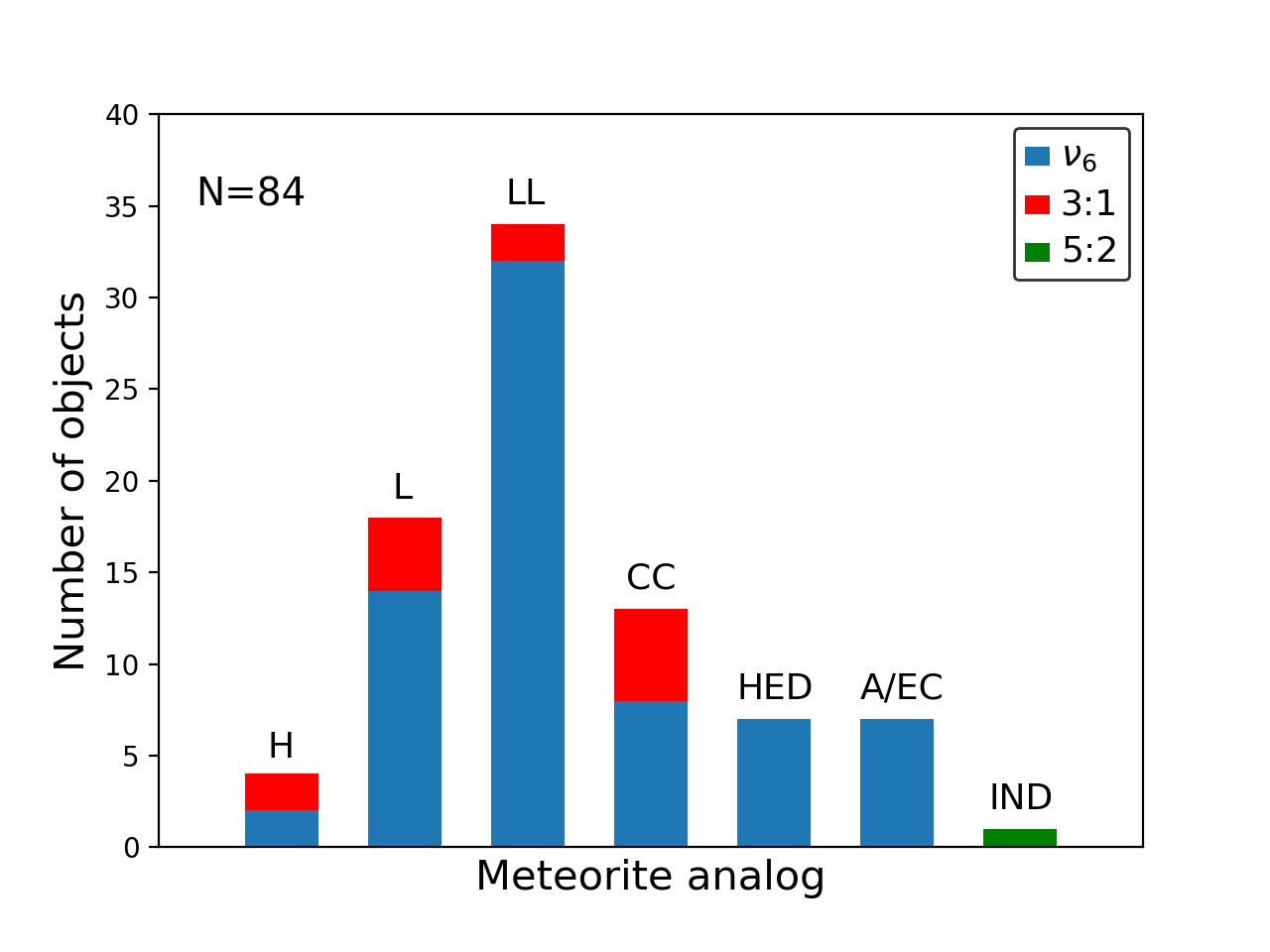}

\caption{\label{f:met_analog_bar}{\small Distribution of meteorite analogs found in the present study. Meteorite abbreviations are: H-, L- and LL-ordinary chondrites, 
carbonaceous chondrites (CC), howardites, eucrites and diogenites (HED), aubrites and enstatite chondrites (A/EC), indeterminate (IND). The source regions of the 
NEOs are indicated with different colors.}}

\end{center}
\end{figure*}

The dynamical modeling showed that the $\nu_{6}$, the 3:1 and the 5:2 resonances have the highest probabilities of escape for the studied NEOs. In 
particular we found that 83\% of the objects escaped from the $\nu_{6}$, 16\% from the 3:1 and just 1\% from the 5:2. The number of objects from each 
region corresponding to each taxonomic type and meteorite analog are shown in Figures \ref{f:Tax_Hist} and \ref{f:met_analog_bar}, respectively.
In total, 50\% of the NEOs with an H chondrite-like composition escaped from the $\nu_{6}$ resonance and 50\% from the 3:1 resonance. For those objects 
with an L chondrite-like composition, the major contribution was from the $\nu_{6}$ (78\%) followed by the 3:1 resonance (22\%). In the case of the NEOs with 
an LL chondrite-like composition, 94\% escaped from the $\nu_{6}$ resonance and only 6\% from the 3:1 resonance. Approximately 62\% of the objects with 
an affinity to carbonaceous chondrites escaped from the $\nu_{6}$ and the rest from the 3:1 MMR. All the objects whose meteorite analogs were found 
to be HEDs, aubrites or enstatite chondrites were delivered to the near-Earth space through the $\nu_{6}$ resonance. The only asteroid whose taxonomic 
type was indeterminate was found to escape from the 5:2 MMR. 

The results presented in this study are consistent with previous work that found that most NEOs originate in the innermost region of the 
main belt, with the $\nu_{6}$ resonance being the major contributor \citep[e.g.,][]{2013Icar..222..273D, 2019Icar..324...41B}. We notice that H-chondrites, 
the less common type of ordinary chondrites in our sample, also show the lowest contribution from the $\nu_{6}$ resonance compared to the L- and LL-chondrites. This 
result is similar to the findings of \cite{2019Icar..324...41B} that showed that the 3:1 resonance, the Phocaea region and the 5:2 resonance contribute the most to the delivery 
of H-chondrites from the main belt. This could explain the relatively low number of H-chondrites in our sample, as only 14\% of all NEOs with an ordinary-chondrite like 
composition escaped from the 3:1 resonance, and none from the Phocaea region nor the 5:2 resonance. The lack of objects coming from the Phocaea region and 
the 5:2 resonance is to be expected, since the model of \cite{2018Icar..312..181G} shows a negligible contribution of small objects from these source regions.

\section{Photometric Study} \label{sec:photostudy}

\subsection{Observations and analysis} \label{sec:photoobs}

As explained in Section 2.1, for those observations where the asteroids were faint, MORIS was used for guiding and images were saved to simultaneously obtain the 
lightcurves. MORIS images were taken using an LPR600 filter, exposure times were selected according to the object brightness and weather conditions to maximize the 
target's S/N. Typically, the exposure times were between 5 and 15 seconds for an error between 0.03 to 0.08 mag the photometric data. Due to the small field of view of the MORIS imager and the fast sky motion of the observed NEOs, observations were carried out by tracking the targets, so 
there is no reference star in the field of view for the photometry. Therefore, we report only instrumental photometry \citep{2013PhDT.......246T}. Using 
the \cite{1976Ap&SS..39..447L} procedure, we searched for periodicity in the time-series photometry of each NEO to infer its lightcurve. 

Since MORIS was not used for all the spectroscopic observations, the number of asteroids for which we obtained photometric data is smaller than the number of asteroids with spectroscopic data. For this reason, lightcurves are not available for all the NEOs presented in Section 2. It is also important to note that there were some 
cases where the S/N of the final spectrum of the asteroid was deemed too low to be published. As a result, not all the NEOs presented in this section have a 
NIR spectrum (see Table 1).  

\subsection{Lightcurves and rotational periods} \label{sec:photoobs}

\begin{figure*}[h]
\begin{center}
\includegraphics[height=9.5cm]{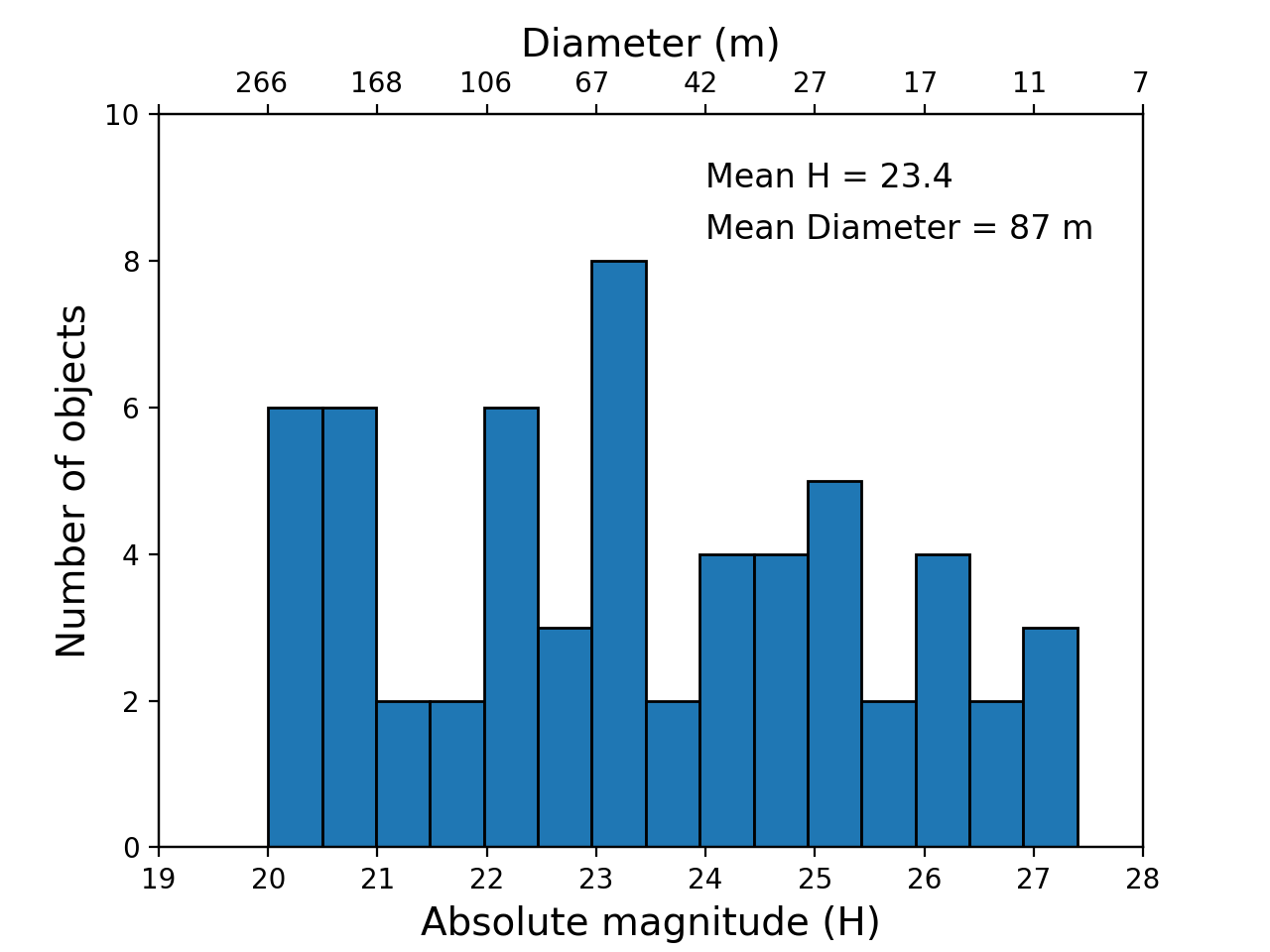}

\caption{\label{f:H_D_Histogram_photo}{\small Distribution of absolute magnitudes (H) and diameters for the NEOs included in the photometric study. Diameters were 
calculated from the absolute magnitudes and the mean geometric albedo of the photometric sample ($P_{V}$=0.25).}}

\end{center}
\end{figure*}

The sample for the photometric study consist of 59 NEOs with H ranging from $\sim$20 to 27.4~mag (Figure \ref{f:H_D_Histogram_photo}). 
Our target sample is probing the Aten, Amor, and Apollo dynamical classes, but the sample is dominated by the Apollo class with 59\% of our targets in this group while 27\% are Amor NEOs and 14\% Aten NEOs. 

\begin{figure*}[h]
\begin{center}
\includegraphics[height=8cm]{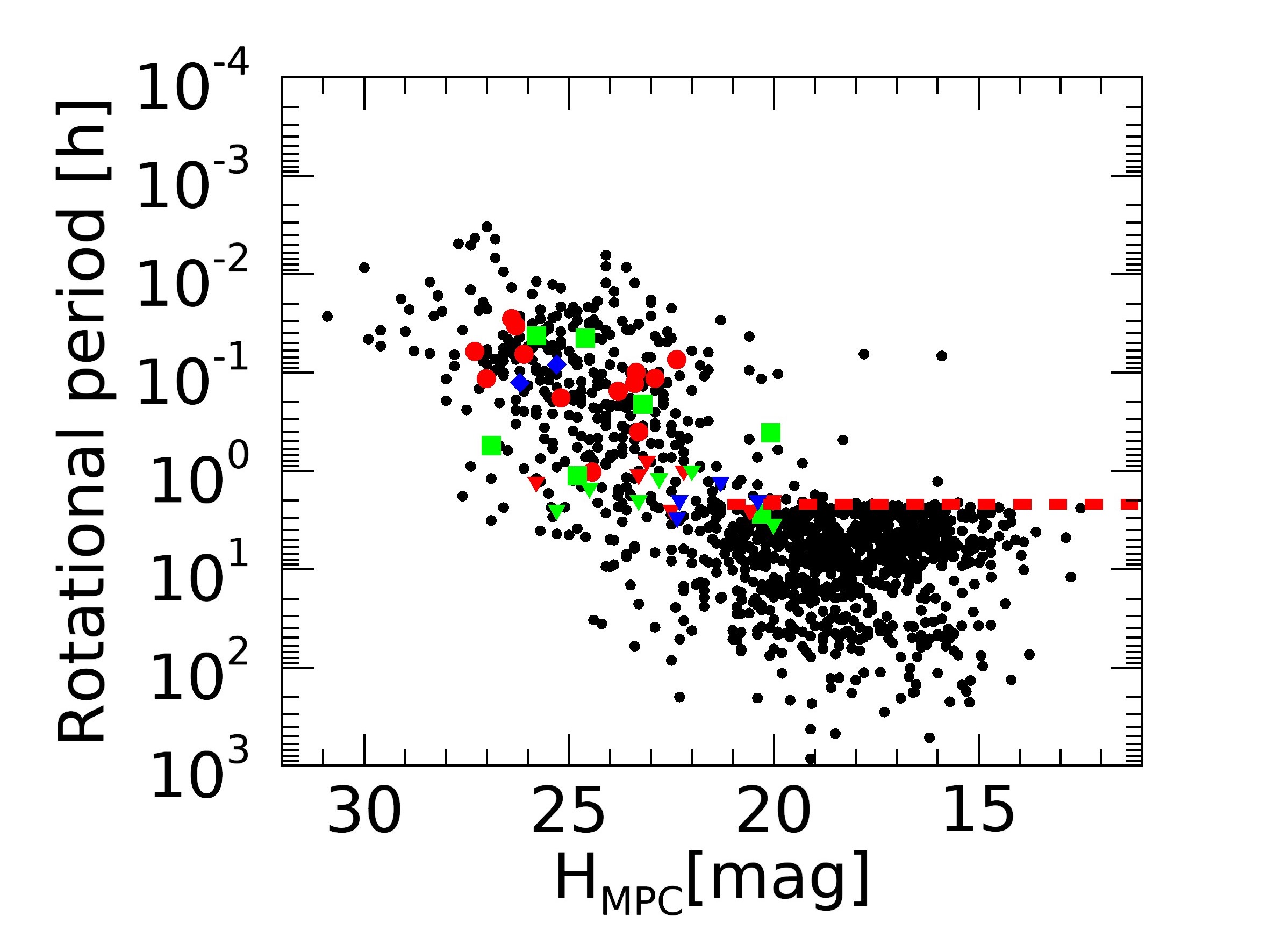}

\caption{\label{f:P_H}{\small Rotational periods versus absolute magnitudes for all NEOs with published lightcurves (black dots). NEOs with a full lightcurve presented in this study are depicted with red circles (Apollos), green squares (Amors) and blue diamonds (Atens). NEOs with partial lightcurves (i.e., a lower limit for their periods) are indicated with a triangle and the same color code as mentioned before. The red dashed line is the spin barrier at 2.2 h.}}

\end{center}
\end{figure*}

The amount of time dedicated to obtain the photometric data was limited by the amount of time needed to obtain the asteroid spectra. Because of this, for most NEOs 
the observing block was $\sim$2 h. This, in general, is enough time for most of the objects with H$>$22, since they 
typically have a rotational period of less than 2 h \citep{2009Icar..202..134W, 2016AJ....152..163T}. However, such an observing strategy created a bias against slow 
rotators as we are not able to report full lightcurves with rotational periods longer than our observing block. For NEOs with H$<$22, our 
observing strategy is an issue as most of the NEOs in this size range have rotational periods longer than 2 h (Figure \ref{f:P_H}). 

\startlongtable
\begin{deluxetable*}{lccc|cccc}

\tablecaption{\label{tab:Summary_photo} For each observed NEO we report its rotational period and lightcurve amplitude (Period and $\Delta m$($\alpha$)). Lightcurve amplitudes are not corrected from phase angle effects, thus we report the lightcurve amplitude at a phase angle $\alpha$. We also include rotational periods and lightcurve amplitudes found in the literature (Period$_{lit}$ and $\Delta m$$_{lit}$) and the corresponding references.}
\tablewidth{0pt}
\tablehead{Number&Designation  &  Period &  $\Delta m$($\alpha$) & Period$_{lit}$ &  $\Delta m$$_{lit}$  & Reference \\
 &                           &  (h)  & (mag)   &   (h)  & (mag)      &  }
\startdata
412995& 1999 LP28& 0.409 & 0.34$\pm$0.07 & - & - & -\\
459872& 2014 EK24 & 0.0997  & 0.98$\pm$0.03 & 0.0996$\pm$0.0002 & 0.56$\pm$0.02 & \citet{2016AJ....152..163T}\\
&  &  &  & 0.0998$\pm$0.0002 & 1.26$\pm$0.01&\citet{2016AJ....152..163T}\\
&  &   &  & 0.0998$\pm$0.0001 & 0.83$\pm$0.15&\citet{2016MPBu...43..156G}\\
&  &   &  & 0.0999$\pm$0.0001 & 	0.76$\pm$0.04& \citet{2018MPBu...45...57T}\\
&  &   &  & 0.0998$\pm$0.0005  &  0.45$\pm$0.03 & \citet{2018PSS..164...54M}\\
469737& 2005 NW44 & $>$2 & $>$0.15 & 31.5$\pm$0.2 & 0.13$\pm$0.02 & \citet{2018MPBu...45..366W}\\
496816& 1989 UP  & $>$2.5 & $>$0.65  &6.98$\pm$0.02  & 1.16$\pm$0.02& \citet{1997Icar..126..395W}\\
501647& 2014 SD224 & $>$3 & $>$0.25 &-& - & -\\
515742&2015 CU & ... &...  &-  &  -  &  - \\
515767& 2015 JA2 & ... &  ... &0.9801$\pm$0.0003 &0.10$\pm$0.02& Unpublished$^{b}$\\
&2001 YV3  &  $>$2.5 & $>$0.45  & - & - & -\\
&2002 LY1  &  $>$2& $>$1.6 & 3.204$\pm$0.005 & 1.24$\pm$0.05 & \citet{2016MPBu...43..311W}\\
&2005 NE21 & $>$1.3 & $>$0.25 & - & - & -\\
&2005 TF & 2.74  & 0.22$\pm$0.05 & 2.724$\pm$0.005 & 0.22$\pm$0.04 & \cite{2017EMP..120...41V} \\
&  &    &  & 2.57$\pm$0.01 & 0.32$\pm$0.03 & \citet{2017MPBu...44...22W}\\
&  &    &  & 2.630$\pm$0.014 & 0.20$\pm$0.03 & \citet{2018MPBu...45....6C}\\
&  &    &  & 2.57$\pm$0.05 & 0.18$\pm$0.05 & \citet{2018PASJ...70..114H}\\
&2013 RS43$^{a}$ & 0.1156 & ... & - &- & -\\
&2013 XA22 & 0.1149  & 0.40$\pm$0.05 & 2.2912$\pm$0.0008 & 0.26$\pm$0.03 & \citet{2020MPBu...47..290W}\\    
&2014 QH33 & 1.03 & 0.47$\pm$0.03 &-& - & -\\
&2014 QL32 & $>$1.2 & $>$0.5 & - & - & -\\
 &2014 QZ265 & 0.0831 & 0.30$\pm$0.05 &-& - & -\\
&2014 SF304 &0.0610 & 0.45$\pm$0.07 &-& - & -\\
&2014 UV210 &0.5553   & 0.96$\pm$0.05 &0.5559$\pm$0.0002& 0.91$\pm$0.04& \citet{2016AJ....152..163T}\\
&2014 VH2$^{a}$ & ... & ... &38.9$\pm$0.5 & 0.931$\pm$0.04& \citet{2015MPBu...42..115W}\\
&2014 WN4  & $>2$ & $>$0.3 &- &-& -\\
&2014 WO4  & ... & ... &- &-& -\\
&2014 WO7  & ... & ... &- &-& -\\
&2014 WP4  & ... & ... &- &-& -\\
&2014 WY119  & 0.0337  & 0.68$\pm$0.09 &- &-& -\\
&2014 XB6$^{a}$  & ... & ... &- &-& -\\
&2015 AP43  & $>$3.5 & $>$0.25 &- &-& -\\
&2015 BK509 & 0.07406 & 0.51$\pm$0.08 &0.074114$\pm$0.000007 &0.39$\pm$0.03& Unpublished$^{b}$\\
&2015 CA40$^{a}$ &  0.4440 & ... & - &-  & -  \\
&2015 CN13  & ... & ... & 22.7$\pm$0.3 & 0.60$\pm$0.05  & \citet{2015MPBu...42..196W} \\
&2015 GY & ... & ... &-  &  -  &  - \\
&2015 HE10 & 0.0652 & 0.59$\pm$0.02 &-  &  -  &  - \\
&2015 HW11 &$>$1.1  & $>$0.2 &-  &  -  &  - \\
 &2015 JW & 0.0422  & 0.51$\pm$0.03  &- &-&-\\
 &2015 KA$^{a}$& 0.1274 & ...  &- &-&-\\
 &2015 MC& ... & ... &- &-&-\\
& 2015 SA$^{a}$ & ... & ... &- &-&-\\
 &2015 SE & 0.0283 & 1.08$\pm$0.05 &- &-&-\\
 &2015 SZ$^{a}$&  ...& ... &$>$4&$>$0.03&\citet{2016AJ....152..163T}\\
&  &   &   & 41.0$\pm$1.0  &  1.33$\pm$0.10 & \citet{2016MPBu...43..143W}\\
 &2015 TB25$^{a}$&$>$1.5 & ... &- &-&-\\
& 2015 TE &  $>$2.5 & $>$0.5 &1.68$\pm$0.05 &0.19$\pm$0.03&Unpublished$^{c}$\\
 &2015 TF &  $>$1 & $>$0.1   &- &-&-\\
& 2015 WF13&  0.2111 & 0.19$\pm$0.05 &0.21194$\pm$0.00005 &0.23$\pm$0.05 & \citet{2016MPBu...43..143W}\\
 &2015 XC$^{a}$&0.1814   & ... &0.181099$\pm$0.000006&0.53$\pm$0.05 & \citet{2016MPBu...43..143W}\\
&  &    &  & 0.2767$\pm$0.0001  &  0.39$\pm$0.04 & \citet{2016MPBu...43..160C}\\
& 2016 CO247& ... & ... &-&- &-\\
 &2016 EV27$^{a}$& 0.1296 & ... &61$\pm$1&0.45$\pm$0.05 & \citet{2016MPBu...43..240W}\\
 &2016 FV13&$>$1.3 & $>$0.4 &-&-& -\\
 &2016 JD18&  - & -  &$>$0.5&$>$1.3&\citet{2018ApJS..239....4T}\\
&2016 LG&   $>$2.5 & $>$0.6  &4.39$\pm$0.01&0.58$\pm$0.05&\citet{2016MPBu...43..311W}\\  
& 2017 BY93$^{a}$&$>$0.8   & $>$0.15 &0.82255$\pm$0.00008&0.14$\pm$0.01& Unpublished$^{b}$\\
&  &   &  & $>$1.75& $>$0.14 & \citet{2017AJ....154..162E}\\
& 2017 FU64&0.1553  & 0.29$\pm$0.04 &-&-&-\\
&2017 RR15& ... & .... &-&-&-\\   
& 2018 XG5& ...  & ... &2.6594$\pm$0.0003&0.26$\pm$0.03&\citet{2019MPBu...46..423W}\\
& 2018 XS4$^{a}$&  ... & ...  &-&-&-\\
& 2019 YM3& 0.4027  & 0.27$\pm$0.03 &-&-&-\\
& 2020 KC5&...   &...   &	-&-& -\\
& 2020 RO6& ...  & ...  &	-&-& -\\
& 2020 SN& 1.117  & 0.25$\pm$0.05 &	1.193$\pm$0.003&0.19$\pm$0.04&\citet{2021MPBu...48...30W}\\
& 2020 ST1$^{a}$&  $>$1 & $>$1.3  &2.879$\pm$0.005& 1.29$\pm$0.08 & \citet{2021MPBu...48..170W}\\
& 2020 YQ3$^{a}$& $>$2 & $>$0.3 &14.752$\pm$0.003&0.65$\pm$0.04& Unpublished$^{b}$\\
&  &   &  & 11.148$\pm$0.004&0.49$\pm$0.03 & \citet{2021MPBu...48..294W}\\
\enddata
\tablenotetext{a}{Objects with non principal axis rotation (i.e., tumblers).} \tablenotetext{b}{Unpublished lightcurve, but results are available at \url{https://www.asu.cas.cz/~ppravec/}}\tablenotetext{c}{Unpublished lightcurve, but results are available at \url{https://obswww.unige.ch/~behrend/page_cou.html}}
\end{deluxetable*}

Lightcurves obtained for this work are classified into four groups: (1) full lightcurves with a rotational period and lightcurve peak-to-peak amplitude estimates (tumblers are not included), (2) partial lightcurves with an increase/decrease in brightness and only a lower limit for the rotational period and peak-to-peak amplitude, (3) flat lightcurves without any clear trend of brightness variability, and (4) lightcurves of NEOs with a non-principal axis rotation also known as tumblers. Table~\ref{tab:Summary_photo} summarizes our results. 

\paragraph{Full lightcurves}  We derived the full lightcurve of 17 NEOs which corresponds to about 29\% of our sample (Appendix B.1). For each NEO, we plotted its Lomb periodogram and the lightcurve corresponding to the periodicity with the highest confidence level. For each lightcurve, a Fourier series is fitted to the photometric data and the order of the fit depends on the lightcurve morphology.

Full lightcurves of 2014 EK24, 2014 UV210, 2005 TF, 2015 WF13, and 2020 SN are in agreement with already published lightcurves (see references in Table~\ref{tab:Summary_photo}). However, our lightcurve of 2013 XA22 is inconsistent with the literature. \citet{2020MPBu...47..290W} inferred a rotational period of about 2.3 h but our photometric data are best fitted with a rotational period of 0.1149 h for 2013 XA22. Our data are incompatible with a rotational period of 2.3 h, so we cannot confirm the \citet{2020MPBu...47..290W} results. We note that both lightcurves present a high dispersion, therefore higher data quality would be useful to secure the rotational period of this object. Similarly, the lightcurve of 1999 LP28 is incomplete and the derived rotational period seems a bit too short for an object in this size range, therefore more data are required to confirm our results. 

All full lightcurves reported in this paper are asymmetric with both peaks (maxima or minima) not reaching the same relative magnitude. Several lightcurves, such as the one of 2020 SN, 2015 WF13, 2014 SF304, 2014 QH33, require a high fit order to match the observations as they display additional peaks/valleys inferring that these NEOs have a complex shape. 

\paragraph{Partial lightcurves} All partial lightcurves are plotted as relative magnitude versus Julian date (Appendix B.2). For these lightcurves, lower limits for the rotational period and the lightcurve amplitude are inferred based on the duration of our observing blocks. Large objects tend to rotate in more than 2 h 
and, thus, our observing blocks are too short to cover the (nearly) full object's rotation. As an example, we observed 2005 NW44 for about 2 h, but since its rotational 
period is $\sim$32 h, our observing block was too short to cover a significant amount of the object's rotation to retrieve such a long rotational period \citep{2018MPBu...45..366W}. 

We highlight one NEO due to its very large lightcurve amplitude: 2002 LY1 presents an amplitude larger than 1.6 mag over about 2 h of observations which is in 
agreement with \citet{2016MPBu...43..311W}. The lightcurve of 2016 JD18 is interesting due to its large amplitude and irregular morphology. Unfortunately, due to the 
limited observing block, we are not able to derive nor constrain the rotational period of this asteroid. In fact, this object may present a combination of complex shape 
and tumbling rotation. To our knowledge, only one partial lightcurve of this object is available in \citet{2018ApJS..239....4T}, which suggested that 2016 JD18 has a 
complex shape. 

\paragraph{Flat lightcurves} Fourteen NEOs display a flat lightcurve and so we cannot constrain their rotational properties (Appendix B.3). Most lightcurves 
present a large dispersion due to low data quality, and/or bad weather. One can appreciate that the dispersion for the lightcurve of 2020 RO6 is larger than the 
dispersion of the other lightcurves. But, in some cases, the observing block was too short to see any sign of variability, which is the case for the lightcurve of 
2015 CN13.    
 
\paragraph{Tumblers} We classified nine NEOs as tumblers$\footnote{Due to the nature of the lightcurve of a tumbler and because we are not able to retrieve the secondary periodicity, we do not report lightcurve amplitude. }$. For five of them, we were able to retrieve the primary rotational period and we plotted the corresponding lightcurve (Appendix B.4), but for four of them no rotational period was retrieved. 2015 XC and 2020 ST1 were already classified as tumblers by \citet{2016MPBu...43..143W, 2021MPBu...48..170W}, and our work confirmed their results. 2020 YQ3 is classified as tumbler by 
\citet{2021MPBu...48..294W}, but our data are insufficient to confirm such a conclusion. For 2016 EV27, we report a lightcurve with a shorter period than the one reported in the literature and we infer that this object is a tumbler. 

The peak-to-peak lightcurve amplitudes in Table~\ref{tab:Summary_photo} are not corrected from phase angle ($\alpha$) effects. Therefore, to correct 
these values and derive the axis ratio (a/b) of the observed NEOs, one has to take into account the phase angle correction as follows: 

\begin{equation}
\Delta m(\alpha=0^{o})=\frac{\Delta m(\alpha)}{1+s\alpha}
\end{equation}

\begin{equation}
\frac{a}{b}\geq 10^{(0.4 \Delta m(\alpha))/(1+s\alpha)}
\end{equation}

where $\Delta m(\alpha=0^{o})$ is the lightcurve amplitude at zero phase angle and s=0.03 mag/deg is the slope correlating the amplitude and the phase angle \citep{2016AJ....152..163T, 2006A&A...454..367G, 1990A&A...231..548Z}. The 
average axis ratio for our sample using only the full lightcurves is 1.33 while it is 1.31 if we consider the full and partial lightcurves. 
 
\begin{figure}[h] 
 \includegraphics[width=8.75cm,angle=0]{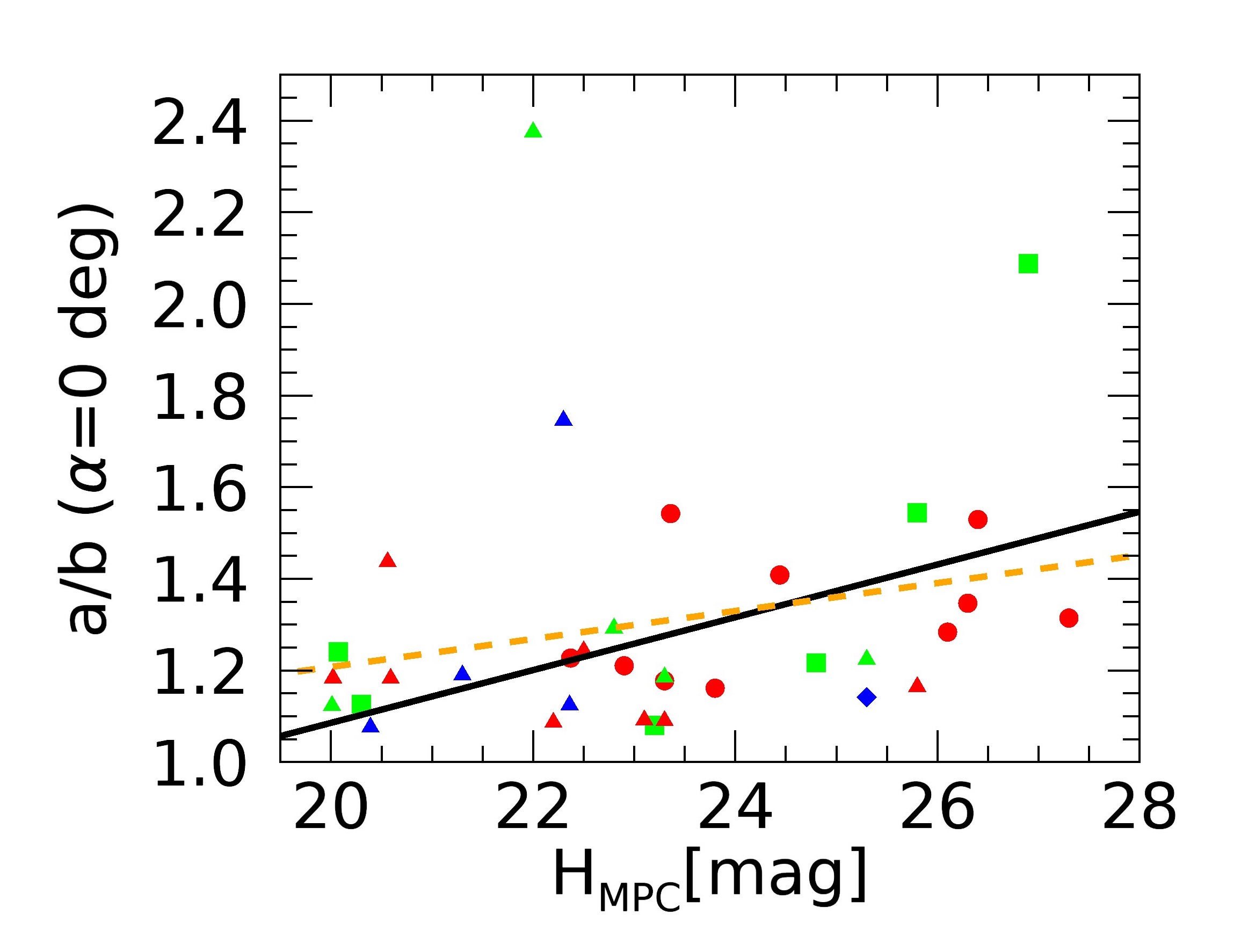}
 \hspace{-5mm}
\includegraphics[width=9cm,angle=0]{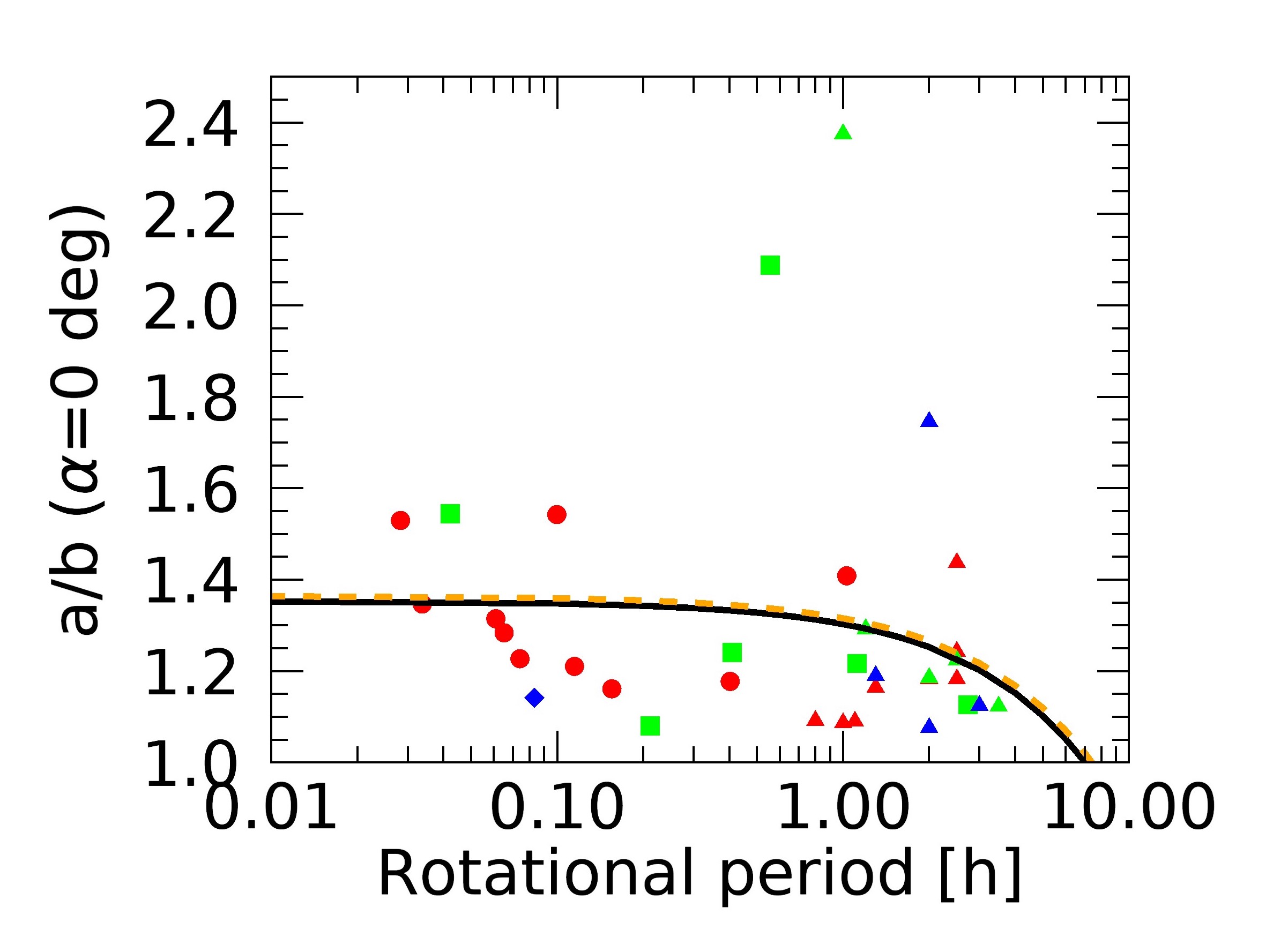}
 \caption{Axis ratio (a/b) vs. absolute magnitude (left) and rotational period (right). The same legend of Figure \ref{f:P_H} has been used. Orange and black 
lines are linear fits to the entire sample (full and partial lightcurves) and to the full lightcurves only, respectively.}
\label{fig:abratio}
\end{figure}

 \begin{figure*}[h]
\begin{center}
\includegraphics[height=8cm]{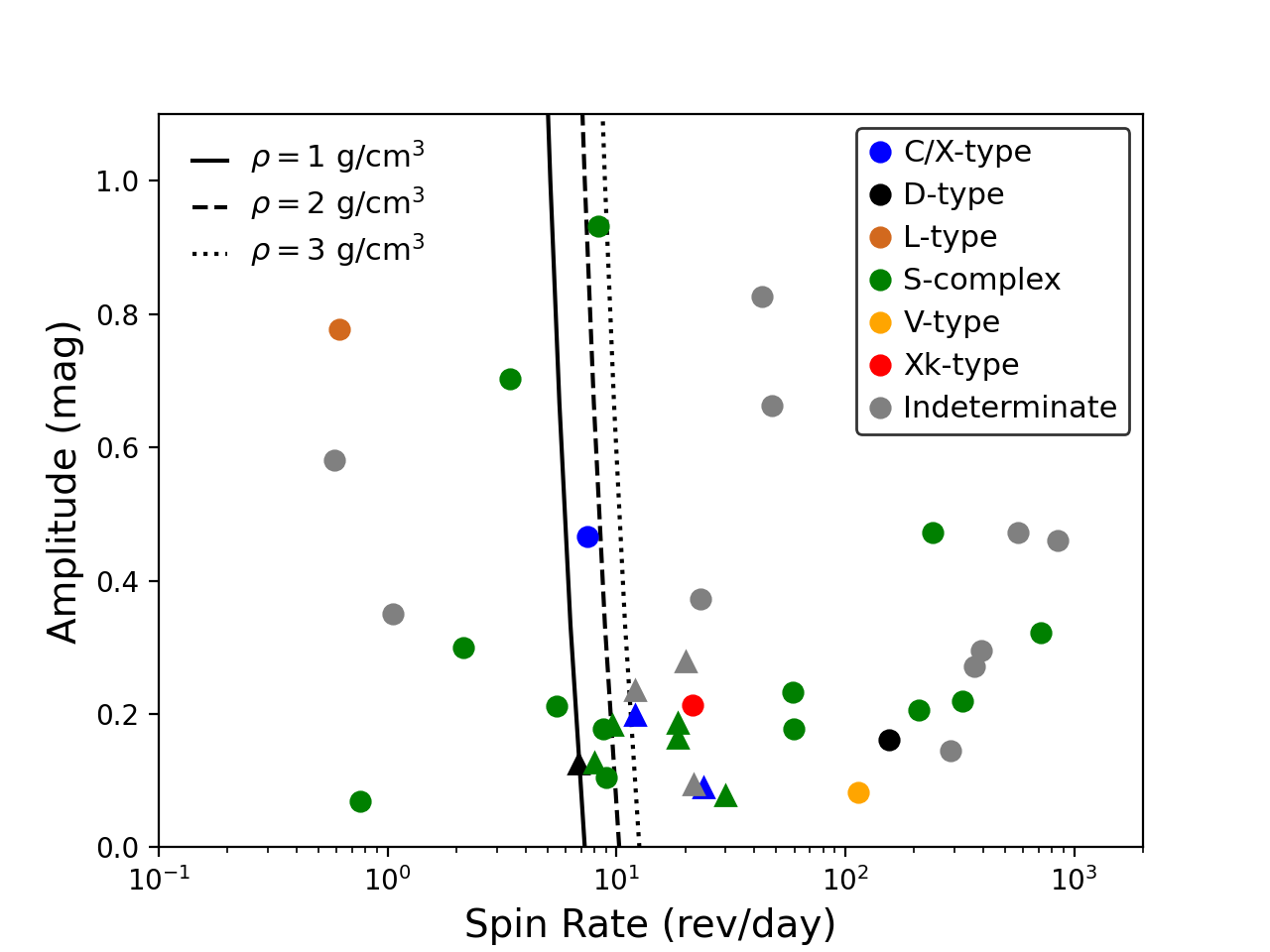}

\caption{\label{f:ampl_rate}{\small Amplitude vs. spin rate for the NEOs included in the photometric study. Taxonomic types are indicated with different colors. 
Objects whose amplitudes and spin rates were determined in this study (or previous work) are depicted with circles. Objects for which only a lower limit for the 
amplitude and spin rate was calculated are depicted with triangles. The curves represent the critical spin rate for bulk densities ($\rho$) of 1.0, 2.0, and 3.0 g/cm$^{3}$. 
Figure adapted from \cite{2000Icar..148...12P}.}}

\end{center}
\end{figure*}

In Figure~\ref{fig:abratio}, we plotted the axis ratio corrected from phase angle versus absolute magnitude and rotational period, the different dynamical classes 
have been highlighted. Unfortunately, due to the limited sample size of NEOs with partial and/or full lightcurves we can only report trends with a low confidence level. For 
the sample with only full lightcurves we obtained an R$^{2}$=0.2575, whereas for full+partial lightcurves this value falls to R$^{2}$=0.0541. Similarly, an anti-correlation 
between axis ratio and rotational period is also highly uncertain with an R$^{2}$=0.0199 for the sample with only full lightcurves and R$^{2}$=0.0322 for full+partial lightcurves.

Figure \ref{f:ampl_rate} shows the amplitude versus spin rate for the NEOs included in the photometric study. Most of the objects in this figure fall to the right side of the critical spin rate curves corresponding to 
bulk densities ($\rho$) of 1.0, 2.0, and 3.0 g/cm$^{3}$. This is consistent with the fact that the sample is dominated by small NEOs with H$>$22 and diameters $\lesssim$150 m. Objects in 
this part of the diagram cannot be held together only by self-gravitation and are often referred to as monoliths, while those falling to the left side of the critical spin rate curves are $
\gtrsim$150 m and likely rubble piles or shattered bodies \citep[e.g.,][]{2000Icar..148...12P}. The different taxonomic types of the NEOs are indicated in the 
figure, however, no obvious trend can be seen. The analysis of a larger sample could help determine whether there is a link between taxonomic type and 
the rotational properties of the asteroids.

\section{Summary} \label{sec:Summ}

We carried out a NIR spectroscopic and photometric survey of small NEOs in order to constrain their surface mineralogy and rotation rates. The 
spectroscopic study included 84 objects with a mean diameter of 126 m and the photometric study 59 objects with a mean diameter of 87 m. Thermal modeling 
was used to derive the albedo and diameter of those asteroids whose spectra showed a thermal excess at wavelengths $>$2 $\mu$m. A compositional 
analysis was performed and possible meteorite analogs were identified for most of the NEOs in our 
sample. For the S-, Sq-, and Q-types we estimated the degree of space weathering experienced by the objects and investigated the effect of grain size on the 
Space Weathering Parameter $\Delta\eta$. 

Our research revealed the existence of NEOs with spectral characteristics and compositions consistent with ordinary chondrites, but whose weak absorption bands could 
lead to an ambiguous classification in the C- or X-complex. For these objects we defined a new subclass within the S-complex called Sx-types. The source regions 
of all the NEOs in our sample were also determined. Rotational periods and lightcurve amplitudes were obtained from the photometric 
data and this information was used to derive the axis ratios of the observed NEOs. Overall, most of our results are consistent with previous studies and can be summarized as follows:

\begin{itemize}

\item The observed NEOs are dominated by S-complex asteroids, which comprise $\sim$66\% of the sample. Objects classified as C/X-complex represent 
$\sim$17\% of the sample and the other $\sim$17\% less common taxonomic types. The proportion of taxonomic types found in this study is similar to previous work 
that included kilometer-sized objects.

\item For asteroids in the S-complex, we found that 8\% were classified as H-, 31\% as L- and 61\% as LL-chondrites. These results agree with previous studies that 
showed that LL-chondrites are dominant among NEOs with ordinary chondrite-like compositions.

\item We confirmed that Q-type asteroids could have weathered surfaces and their spectral characteristics could result from the presence of large grains on the 
surface. However, for this mechanism to be effective, those grains cannot exceed the size at which absorption band saturation occurs. For asteroids with an 
LL chondrite-like composition, this size limit is $\sim$400 $\mu$m. 

\item We found that some NEOs with ordinary chondrite-like composition (the Sx-types) could be hidden within the C- or X-complex as the result of their weak absorption 
bands. Our analysis showed that the presence of metal or shock-darkening could be responsible for the attenuation of the absorption bands. 

\item The dynamical modeling showed that 83\% of the NEOs escaped from the $\nu_{6}$ resonance, 16\% from the 3:1 and just 1\% from the 5:2. The small fraction 
of NEOs coming from the 3:1 and 5:2 resonances and the lack of objects from the Phocaea region could explain the relatively low number of H-chondrites in our 
sample.

\item Full lightcurves were derived for 17 NEOs ($\sim$29\% of our sample) and partial lightcurves for 19 NEOs ($\sim$32\% of our sample). Flat lightcurves were 
obtained for 14 asteroids, which represents $\sim$24\% of the sample, whereas 9 NEOs were classified as tumblers ($\sim$15\%). 

\item No no clear trend between the axis ratio and the absolute magnitude or rotational period was found. Similarly, no correlation was observed between the taxonomic type 
and the rotational properties of the NEOs.

\end{itemize}

\begin{acknowledgments}

This research work was supported by NASA Near-Earth Object Observations Grant NNX17AJ19G and NASA Yearly Opportunities for Research in Planetary Defense Grant 
80NSSC22K0514 (PI: V. Reddy). We thank the IRTF TAC for awarding time to this project, and to the IRTF TOs and MKSS staff for their support. The authors wish to 
recognize and acknowledge the very significant cultural role and reverence that the summit of Maunakea has always had within the indigenous Hawaiian community. We are 
most fortunate to have the opportunity to conduct observations from this mountain. Taxonomic type results presented in this work were determined, in whole or in part, using a Bus-DeMeo Taxonomy Classification Web tool by Stephen M. Slivan, developed at MIT with the support of National Science Foundation Grant 0506716 and NASA Grant 
NAG5-12355. We thank the anonymous reviewers for useful comments that helped improve this paper.

\end{acknowledgments}

\begin{appendix}

\vspace{-6mm}

\textbf{Appendix A.}

\vspace{-3mm}

\begin{figure*}[!ht]
\begin{center}
\includegraphics[width=16cm,angle=0]{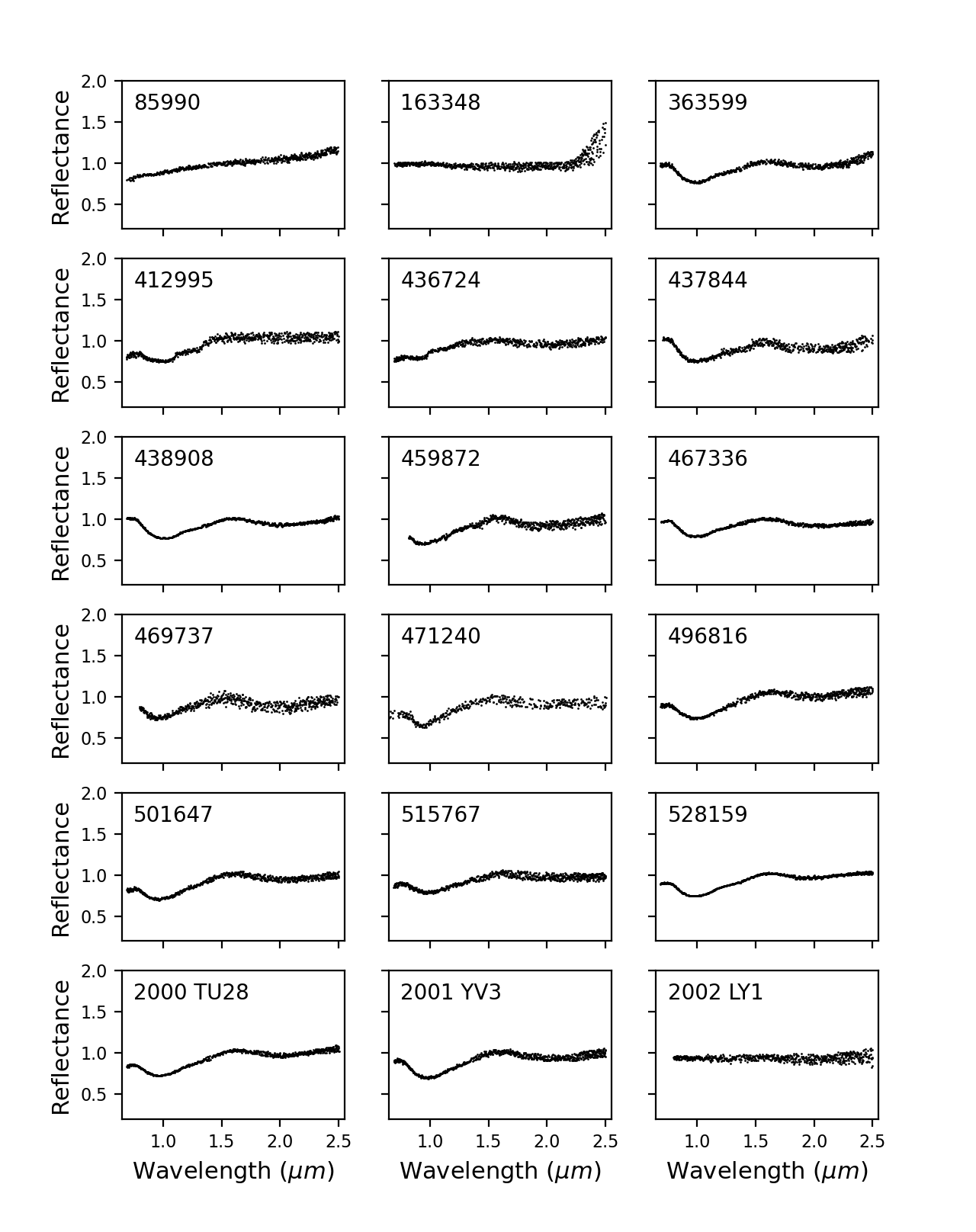}
\end{center}
\caption{Near-infrared spectra of NEOs included in the spectroscopic study.}
\label{fig:all_spectra}
\end{figure*}

\begin{figure*}[!ht]
\begin{center}
 \includegraphics[width=16cm,angle=0]{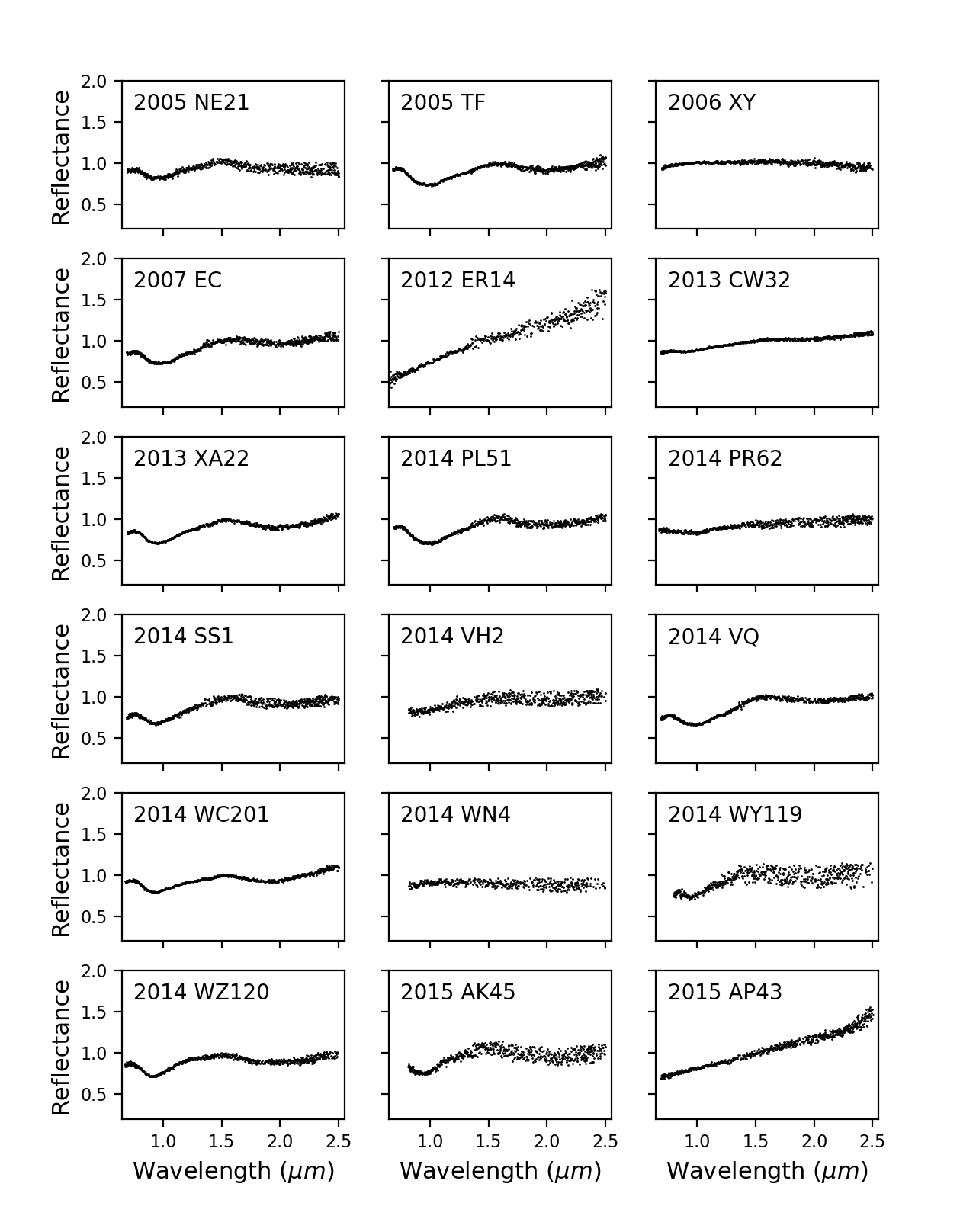}
 \end{center}
 \caption{Near-infrared spectra of NEOs included in the spectroscopic study.}
\label{fig:all_spectra}
\end{figure*}

\begin{figure*}[!ht]
\begin{center}
 \includegraphics[width=16cm,angle=0]{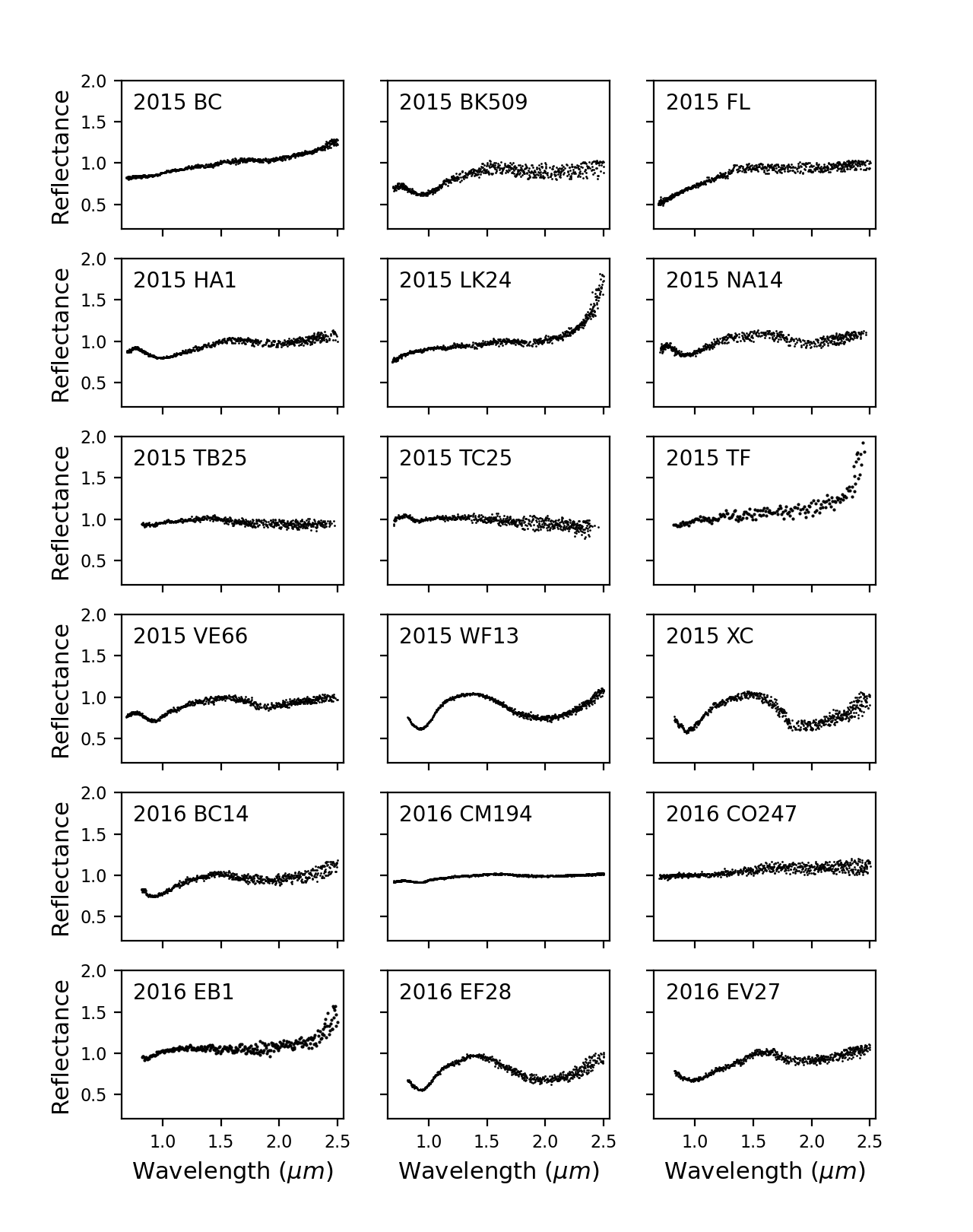}
 \end{center}
 \caption{Near-infrared spectra of NEOs included in the spectroscopic study.}
\label{fig:all_spectra}
\end{figure*}

\begin{figure*}[!ht]
\begin{center}
 \includegraphics[width=16cm,angle=0]{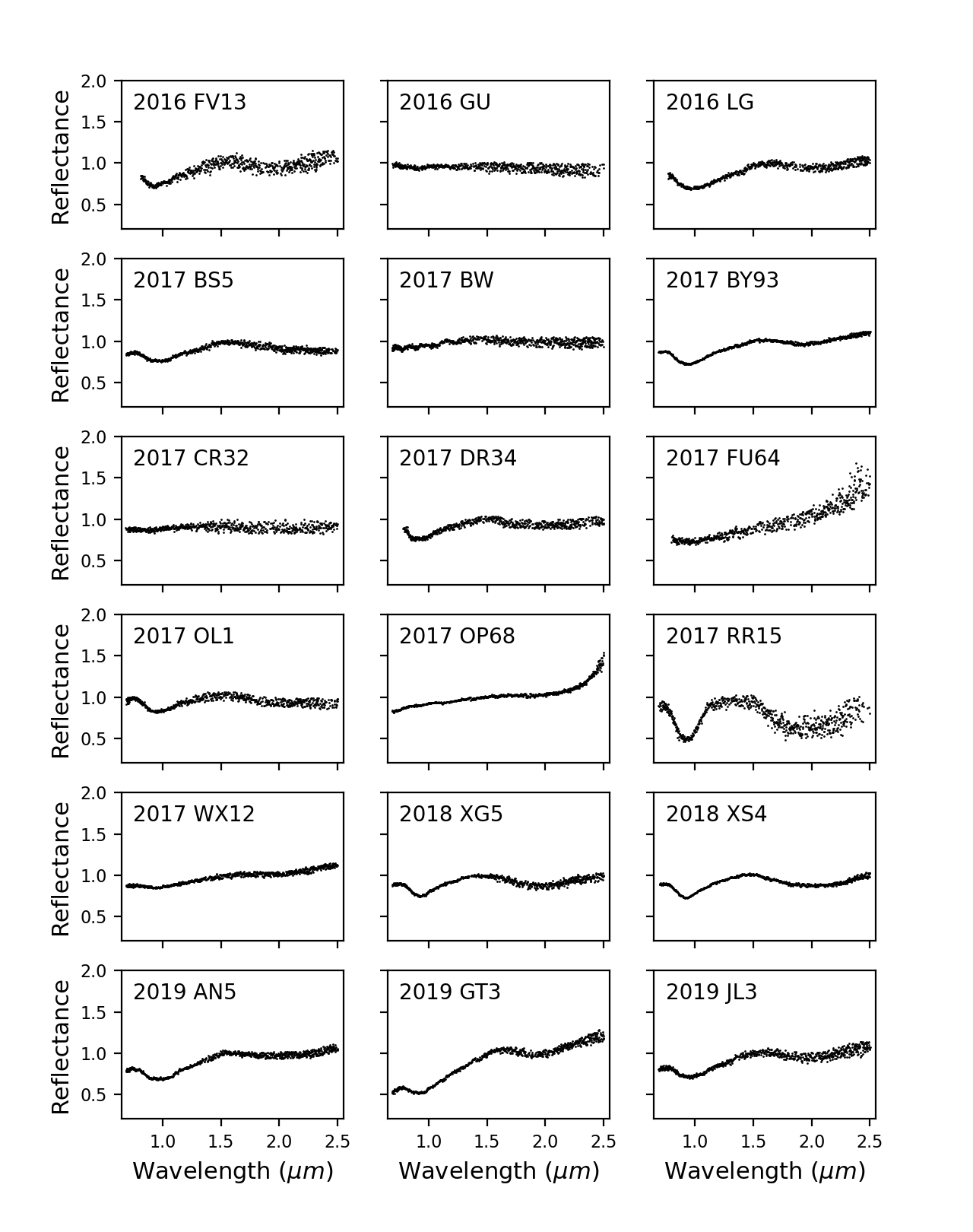}
 \end{center}
 \caption{Near-infrared spectra of NEOs included in the spectroscopic study.}
\label{fig:all_spectra}
\end{figure*}

\begin{figure*}[!ht]
\begin{center}
 \includegraphics[width=16cm,angle=0]{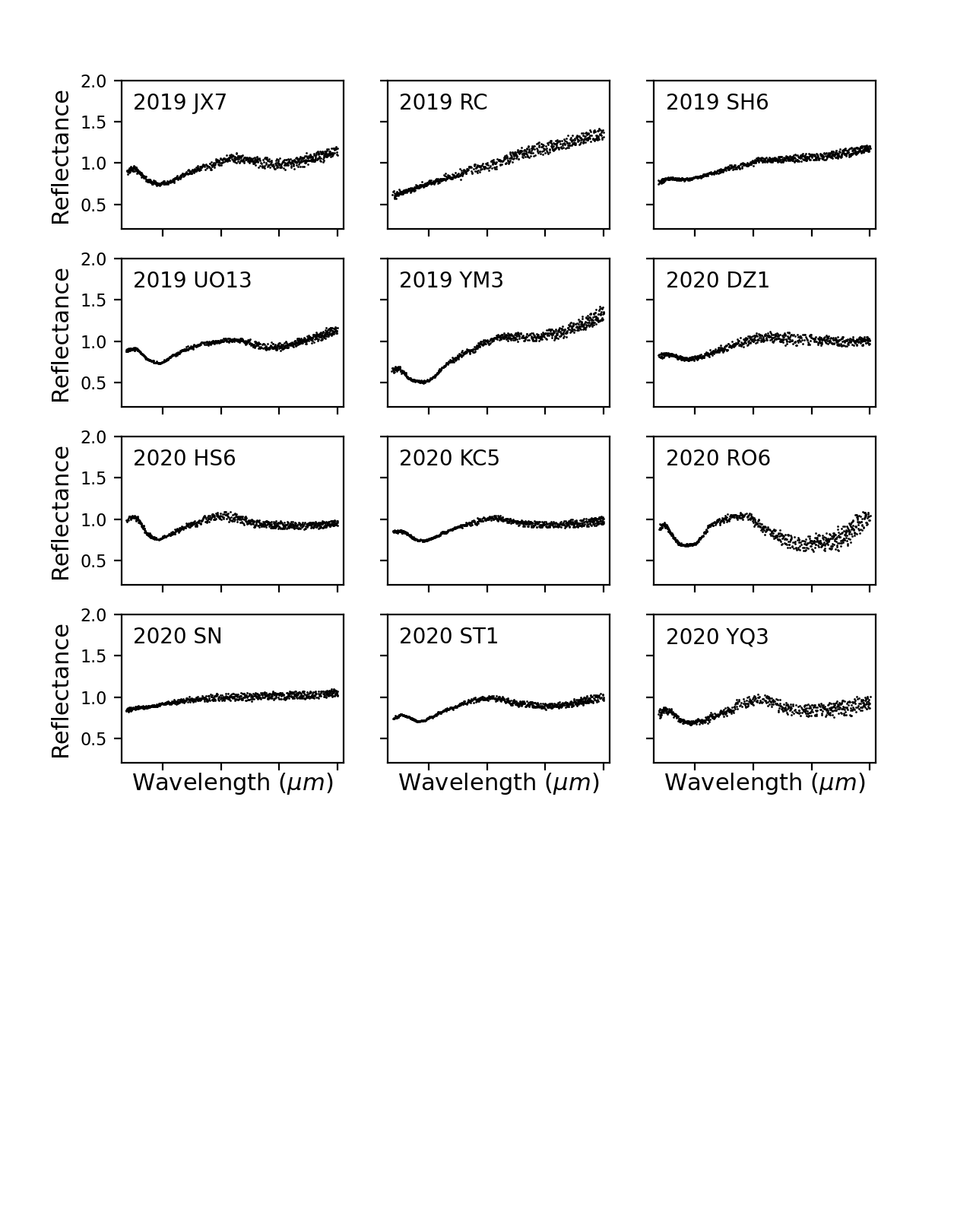}
 \end{center}
 \vspace{-65mm}
\caption{Near-infrared spectra of NEOs included in the spectroscopic study.}
\label{fig:all_spectra}
\end{figure*}

\clearpage

\textbf{Appendix B.1.}

\begin{figure*}[!h]
 \includegraphics[width=9cm,angle=0]{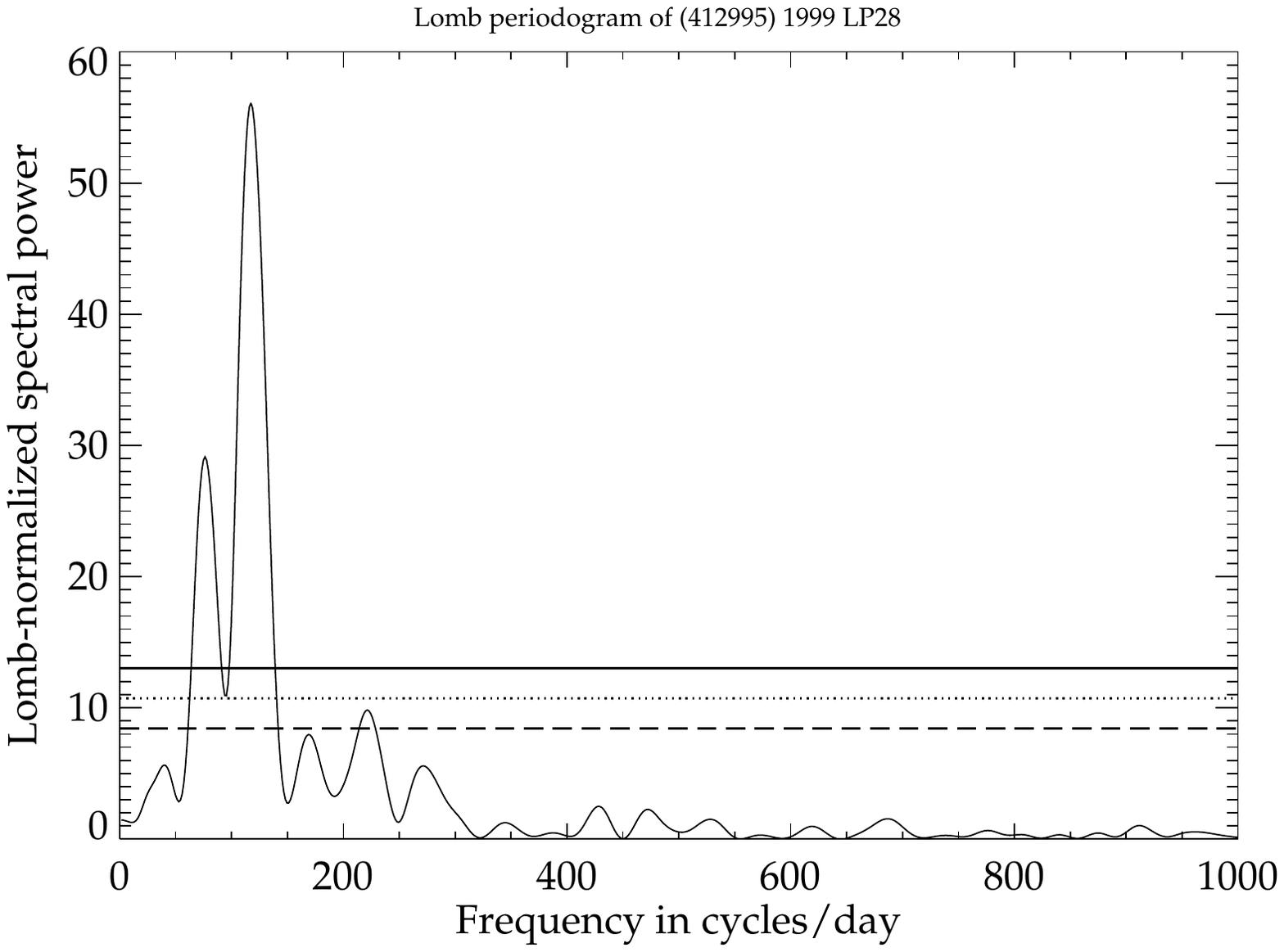}
 \includegraphics[width=9cm,angle=0]{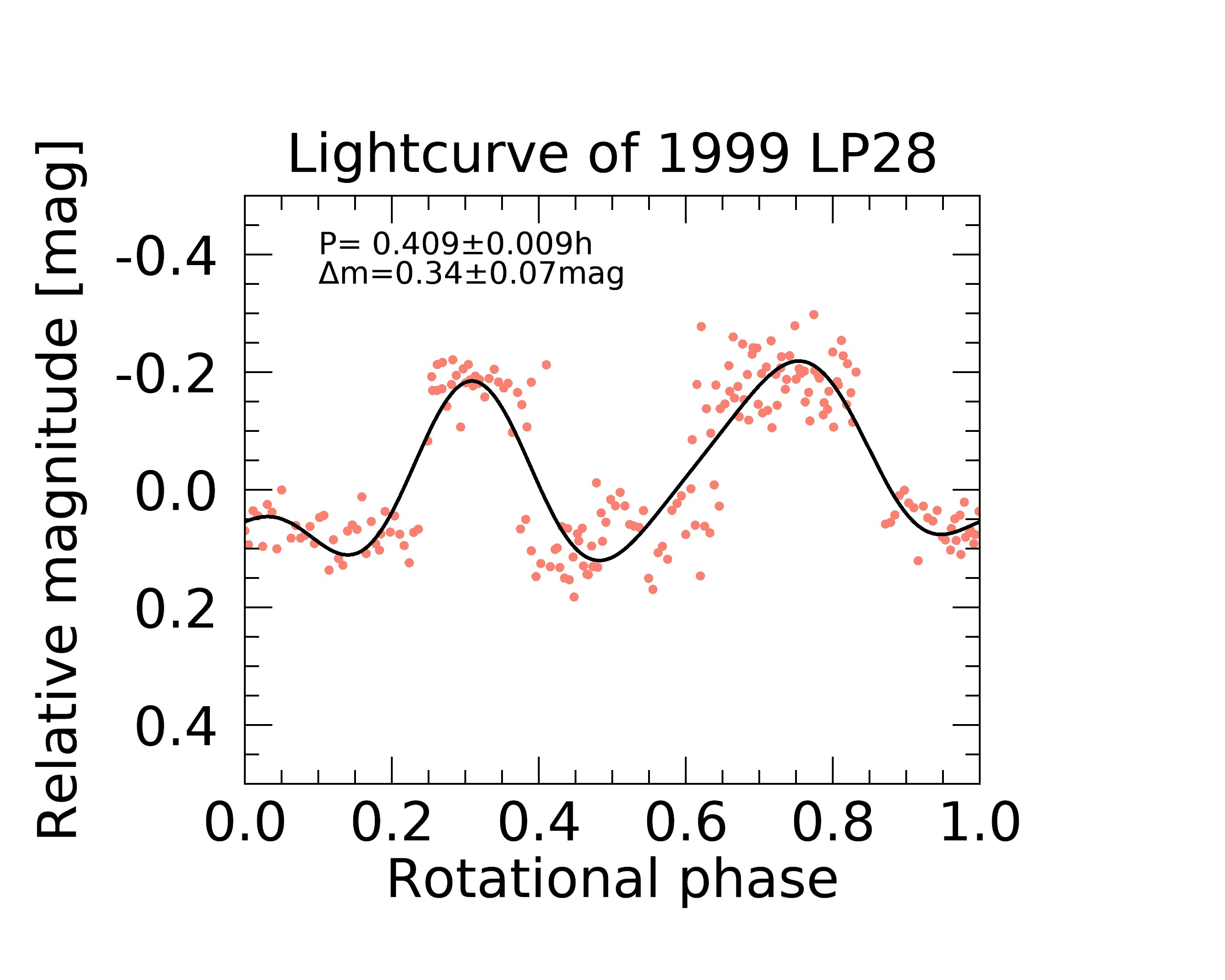}
 \includegraphics[width=9cm,angle=0]{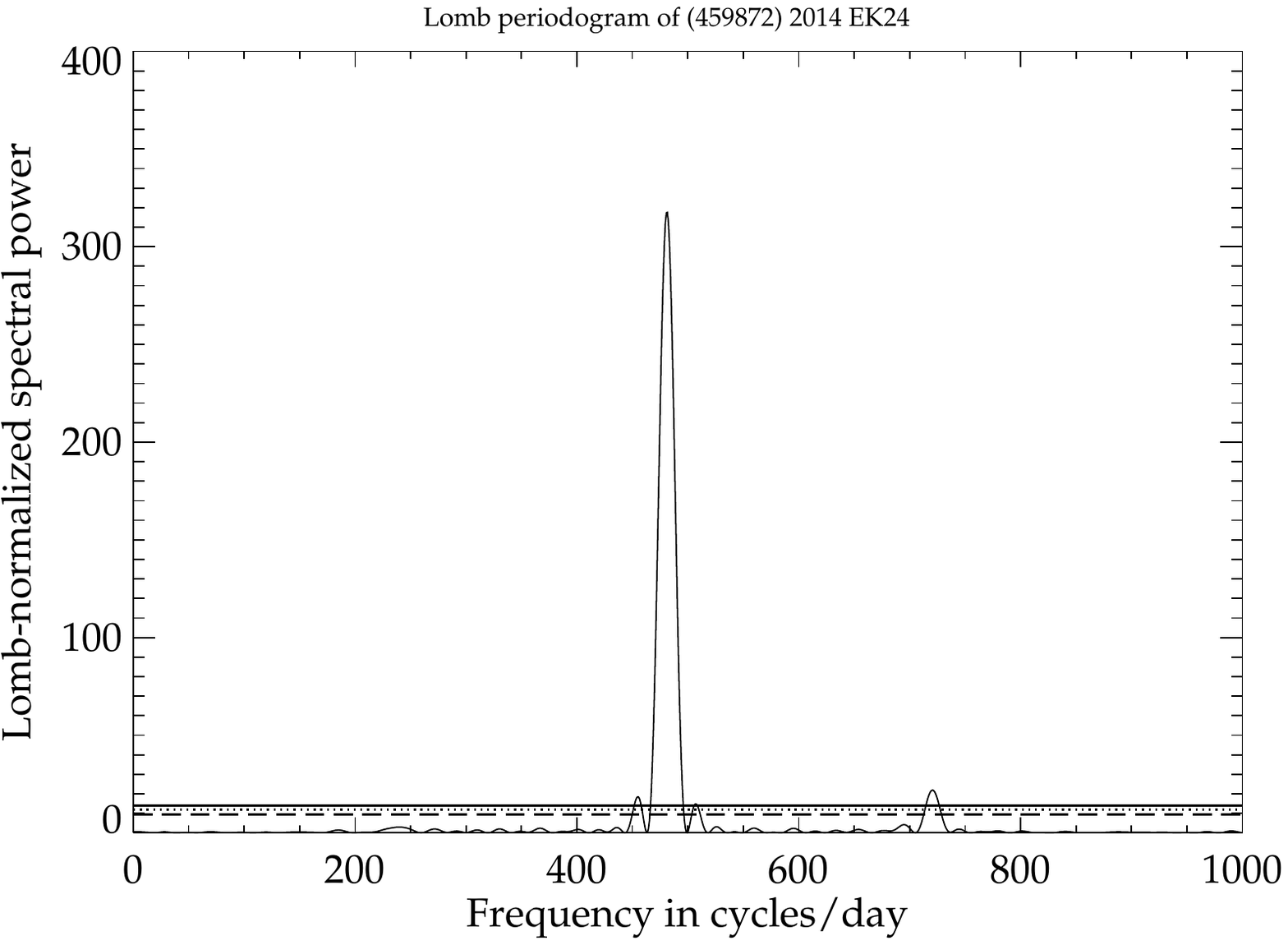}
 \includegraphics[width=9cm,angle=0]{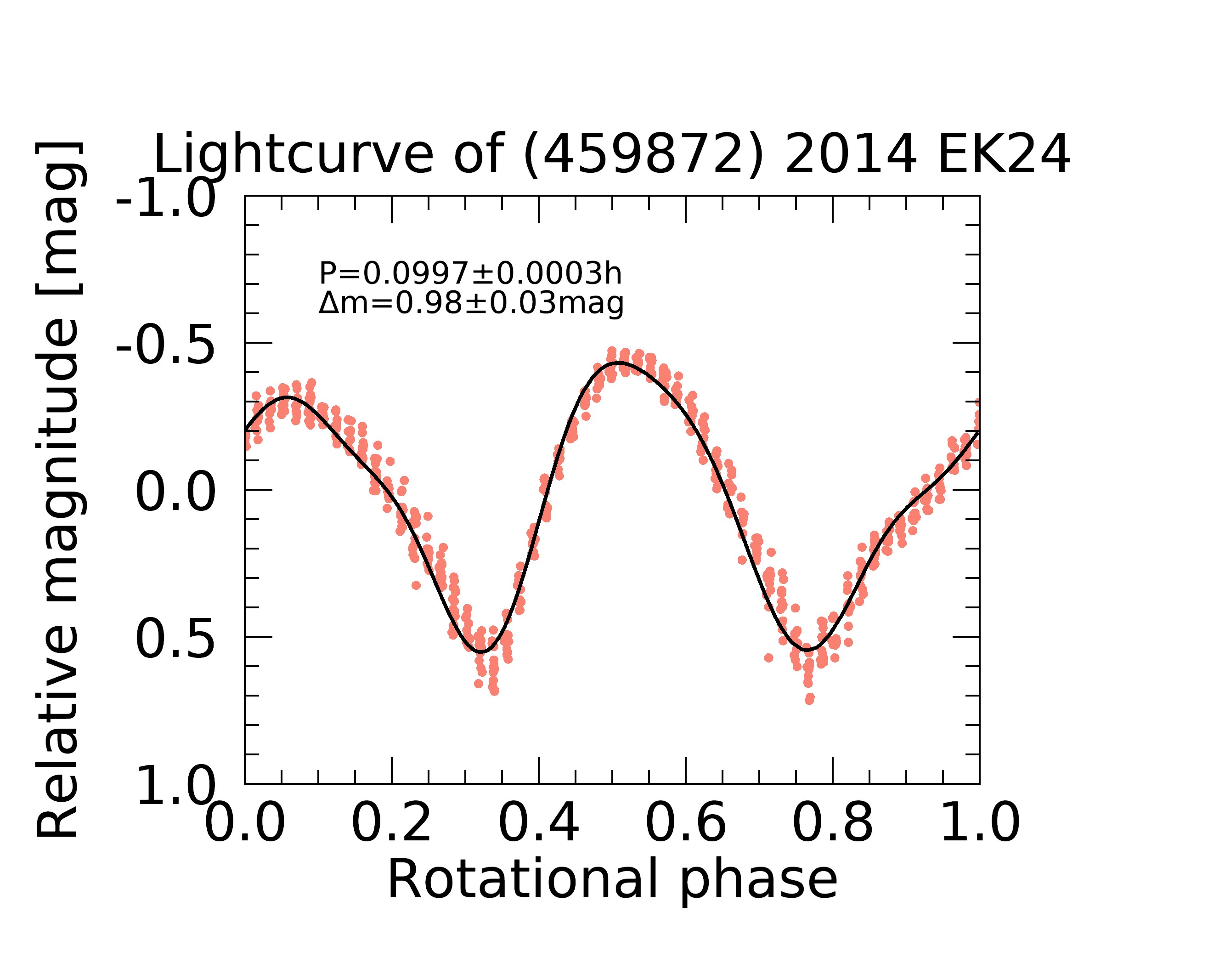}
 \caption{Full lightcurves of NEOs included in the photometric study. The highest peak of the Lomb periodogram is the single-peaked rotational period with the highest
confidence level. The 99.9\% confidence level is indicated with a continuous line while the confidence level at 99\% is the dotted line, and the dashed line corresponds to a 
confidence level of 90\%. On the right, the lightcurves corresponding to the highest confidence level peak are plotted. The lightcurves have been fitted with a Fourier series fit 
(black curves).}
\label{fig:Full_lightcurves}
 \end{figure*}

\begin{figure*} [!h]
 \includegraphics[width=9cm,angle=0]{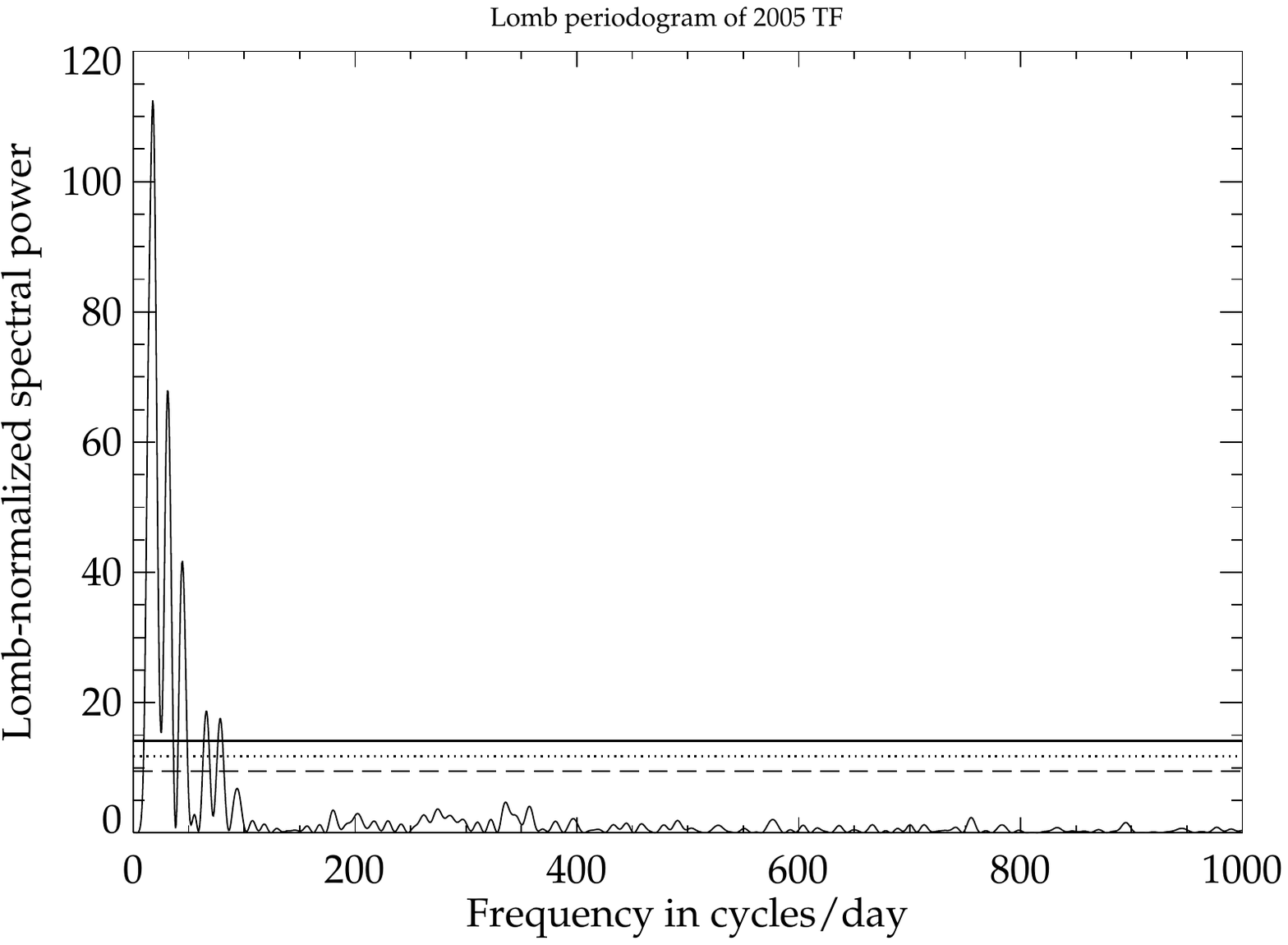}
 \includegraphics[width=9cm,angle=0]{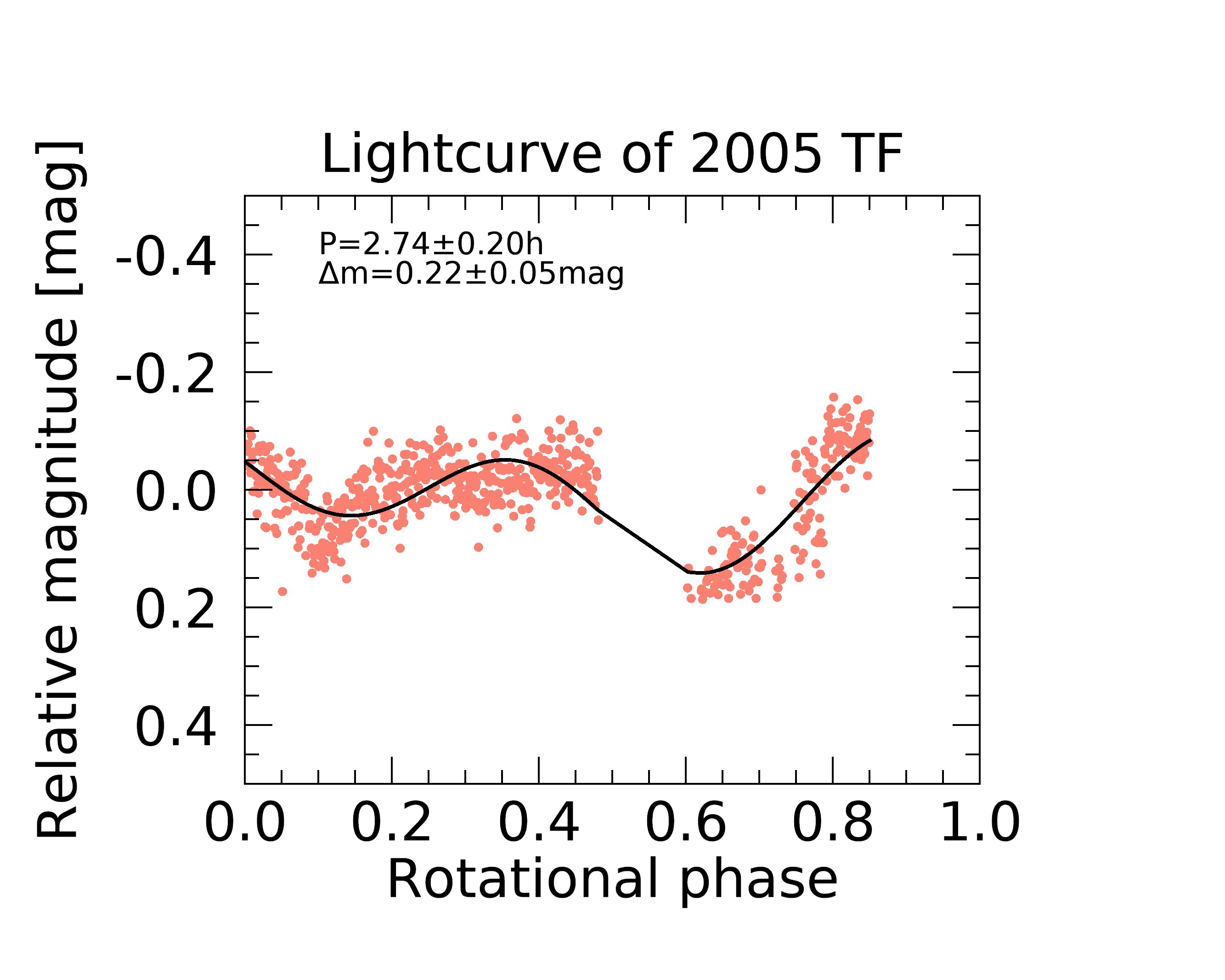}
 \includegraphics[width=9cm,angle=0]{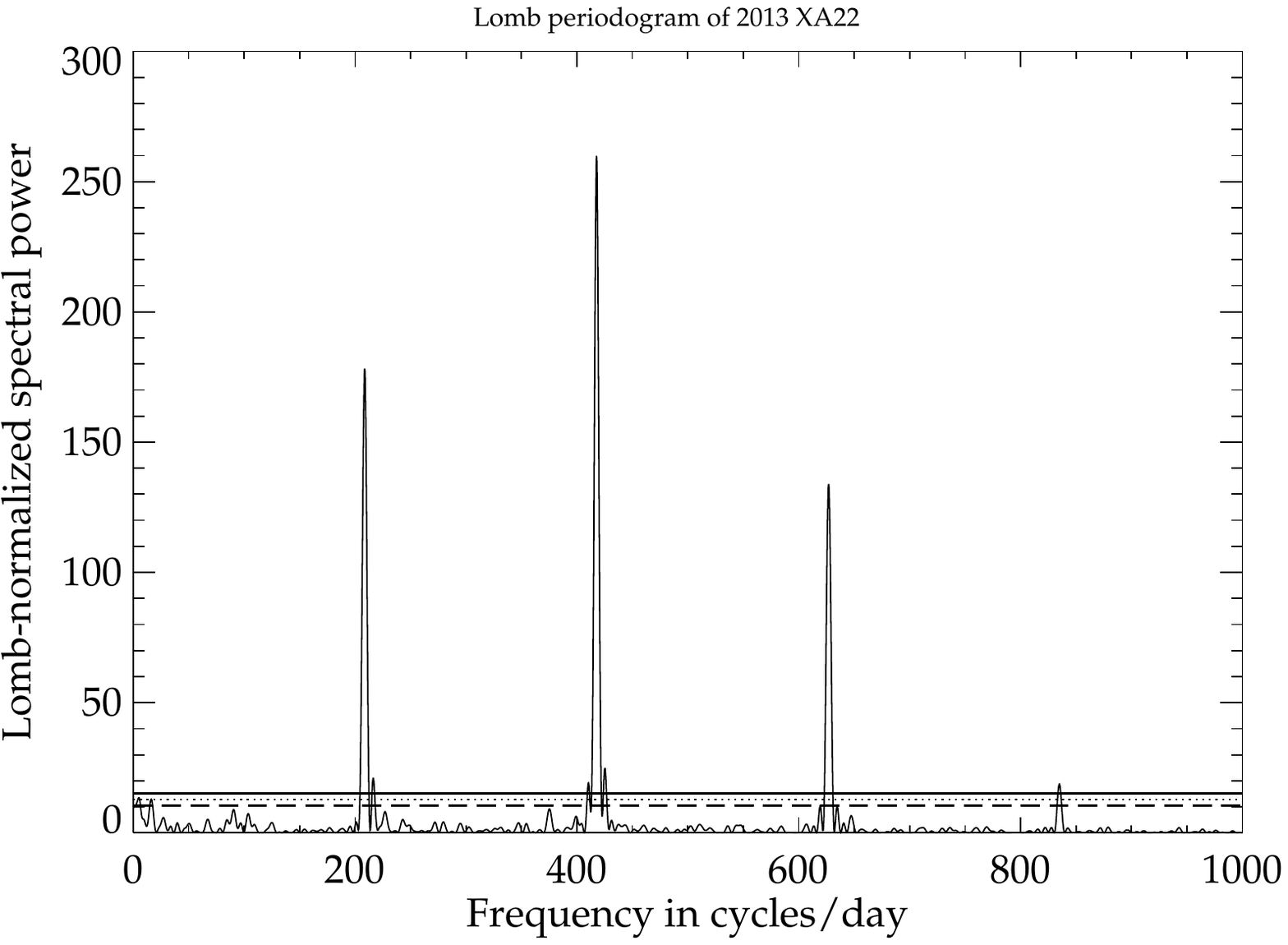}
 \includegraphics[width=9cm,angle=0]{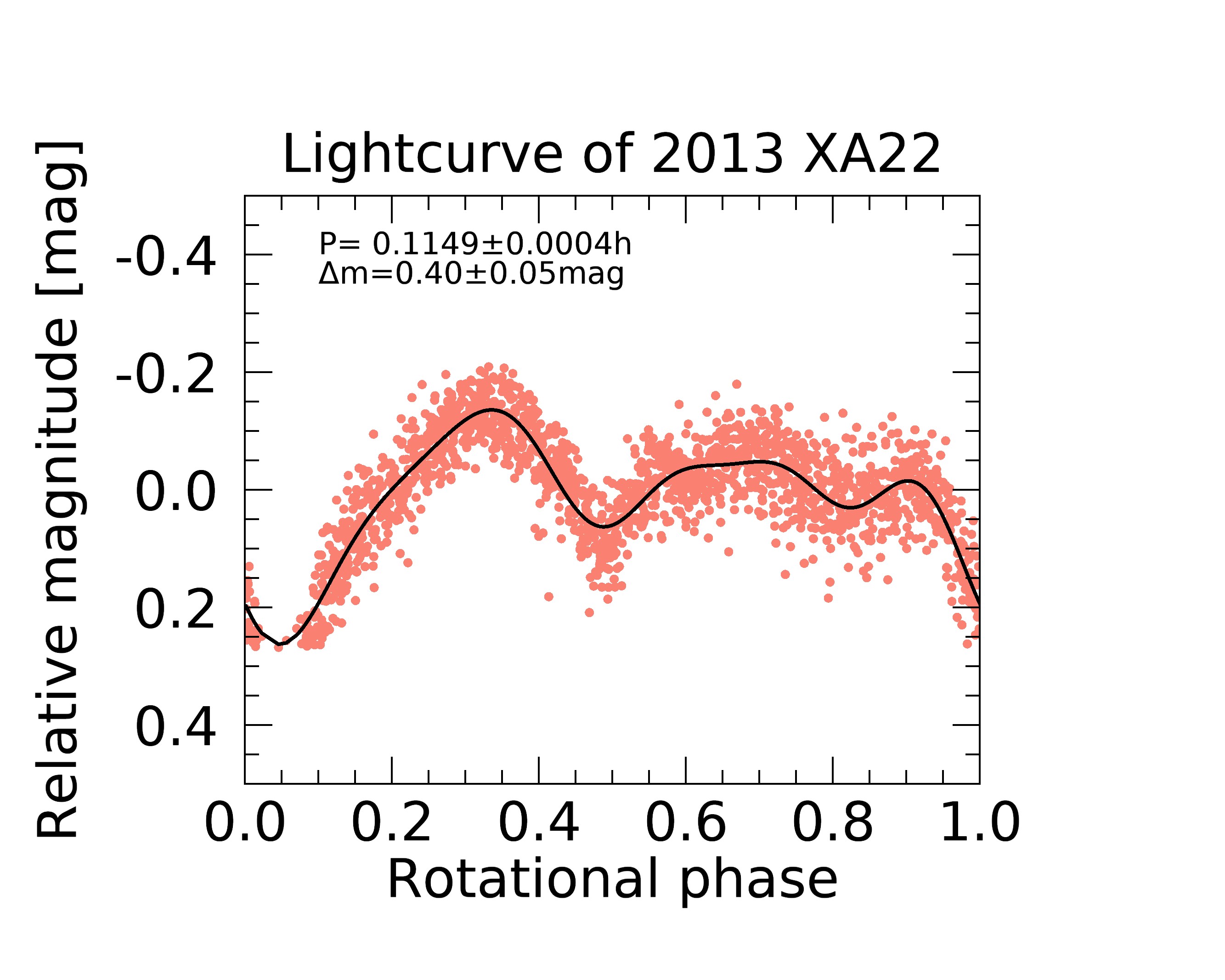}
\includegraphics[width=9cm,angle=0]{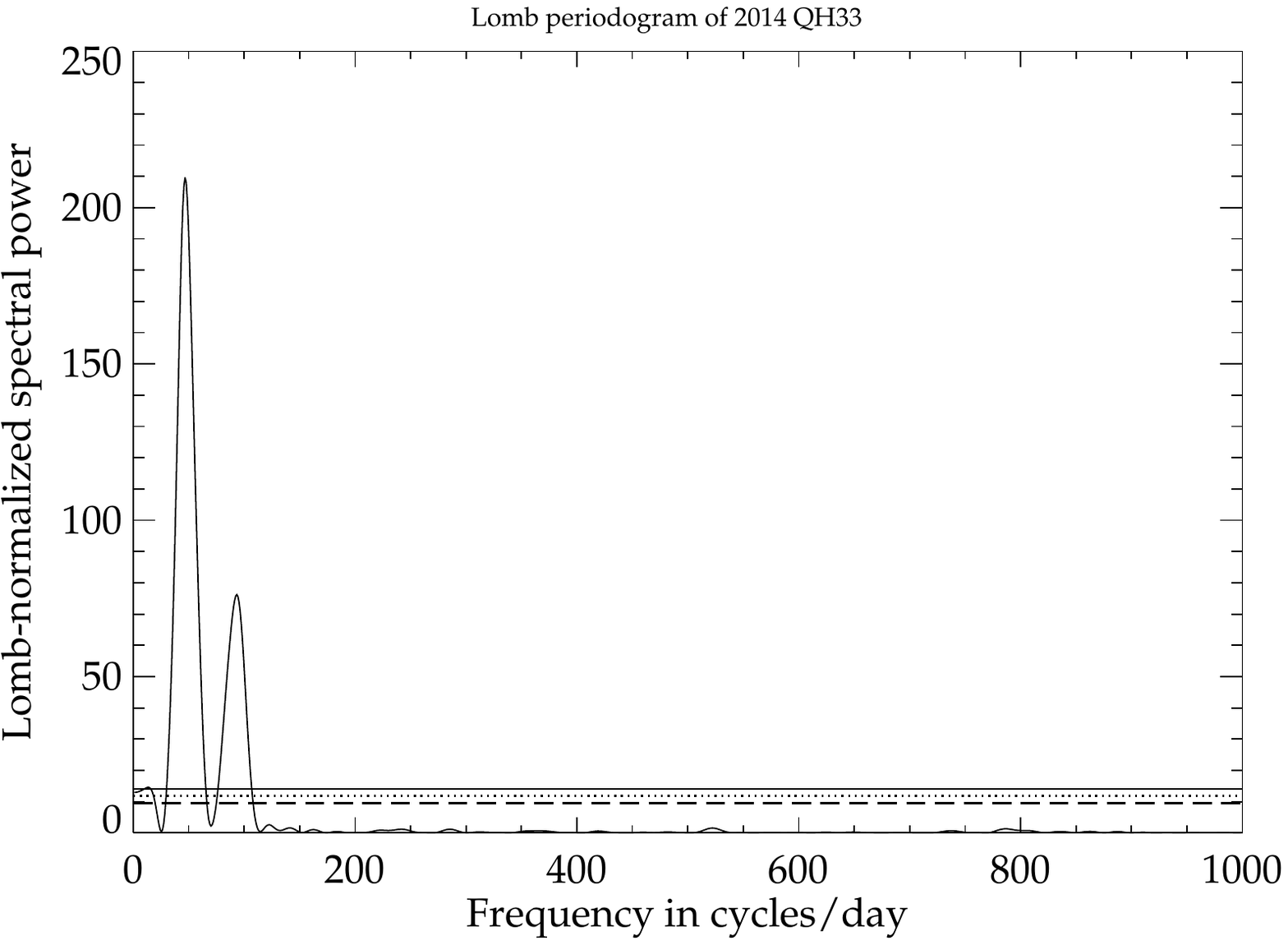}
 \includegraphics[width=9cm,angle=0]{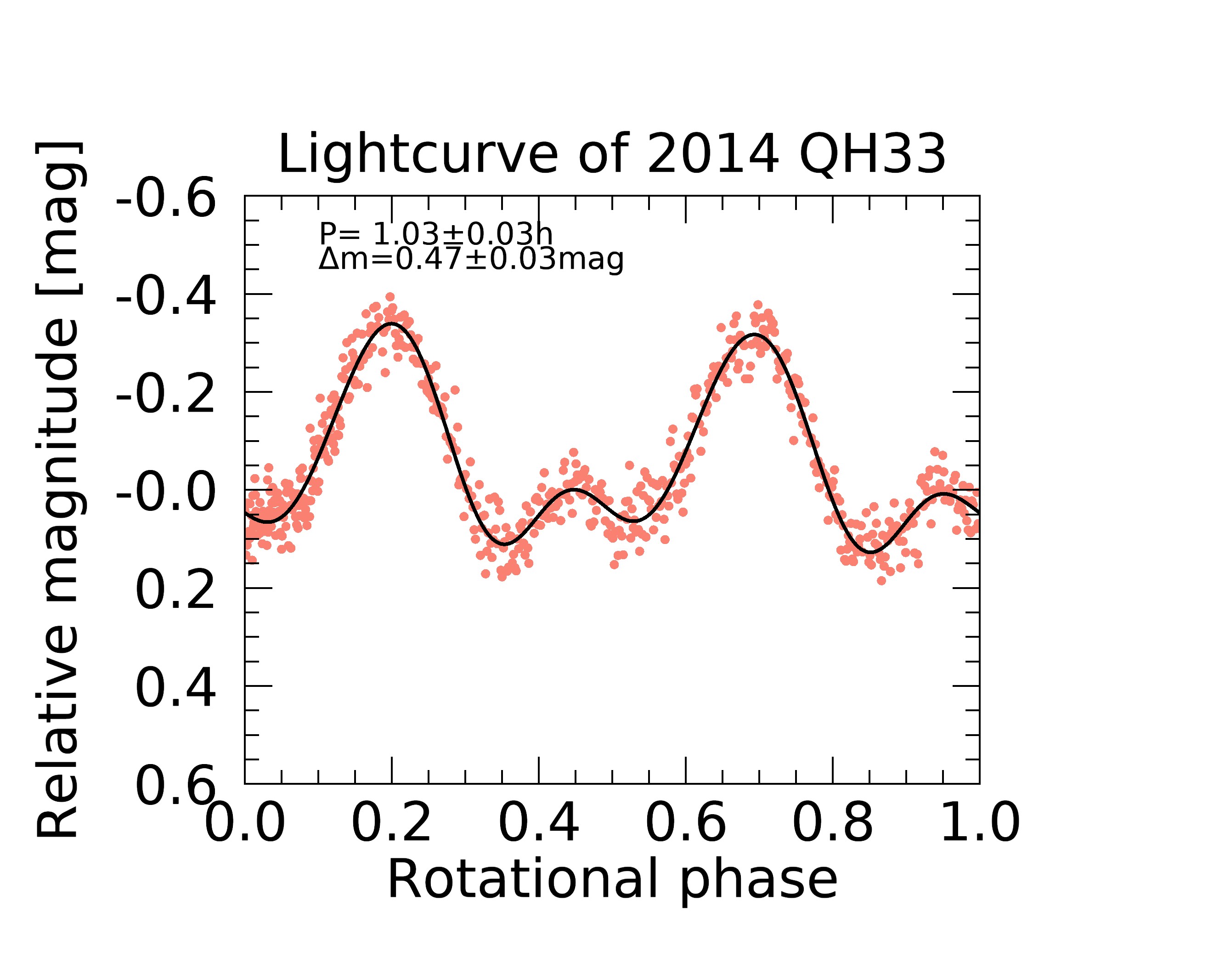}
 \caption{Full lightcurves of NEOs included in the photometric study. The highest peak of the Lomb periodogram is the single-peaked rotational period with the highest
confidence level. The 99.9\% confidence level is indicated with a continuous line while the confidence level at 99\% is the dotted line, and the dashed line corresponds to a 
confidence level of 90\%. On the right, the lightcurves corresponding to the highest confidence level peak are plotted. The lightcurves have been fitted with a Fourier series fit 
(black curves).}
\label{fig:Full_lightcurves}

 \end{figure*}

 \begin{figure*}
 \includegraphics[width=9cm,angle=0]{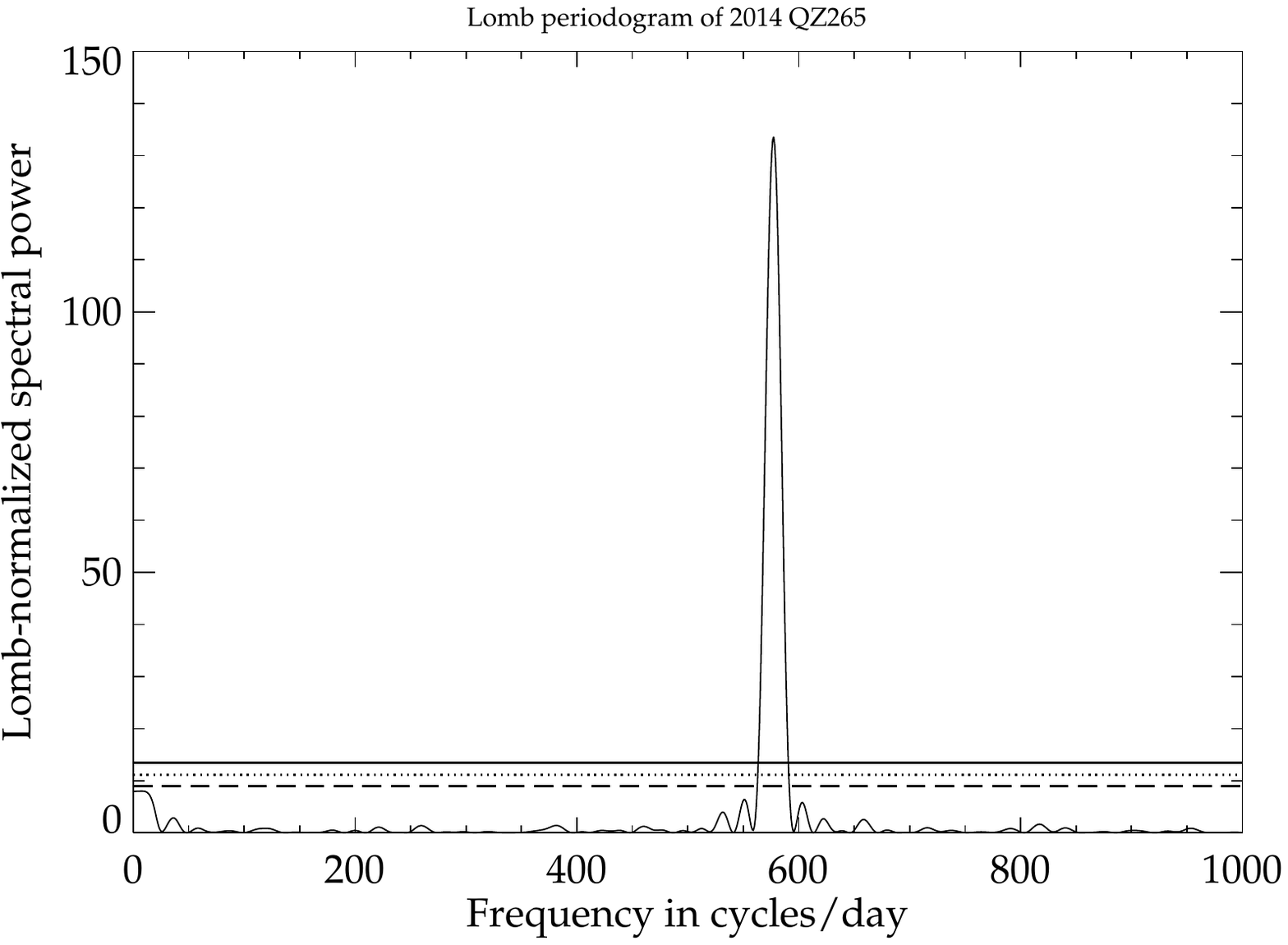}
 \includegraphics[width=9cm,angle=0]{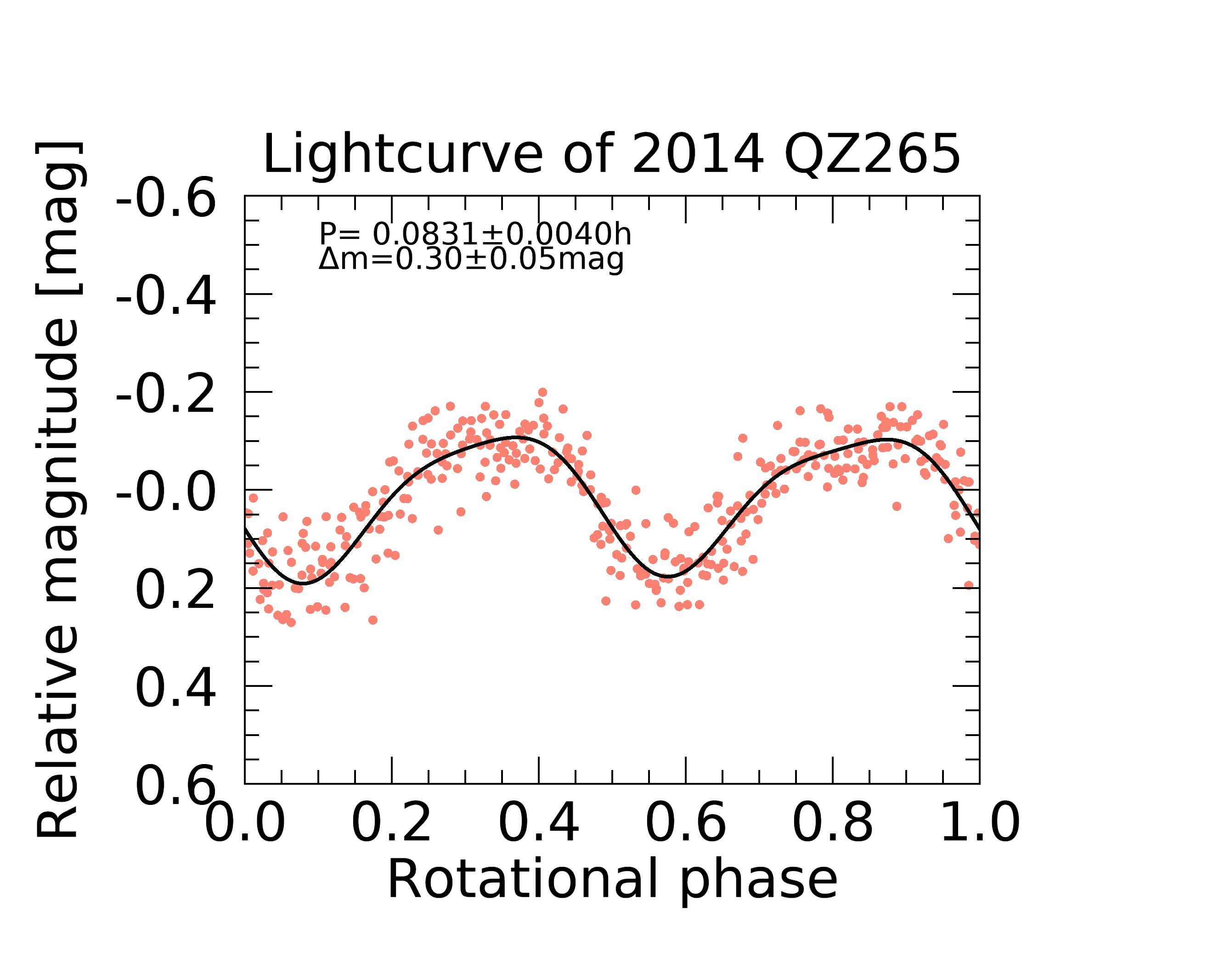}
 \includegraphics[width=9cm,angle=0]{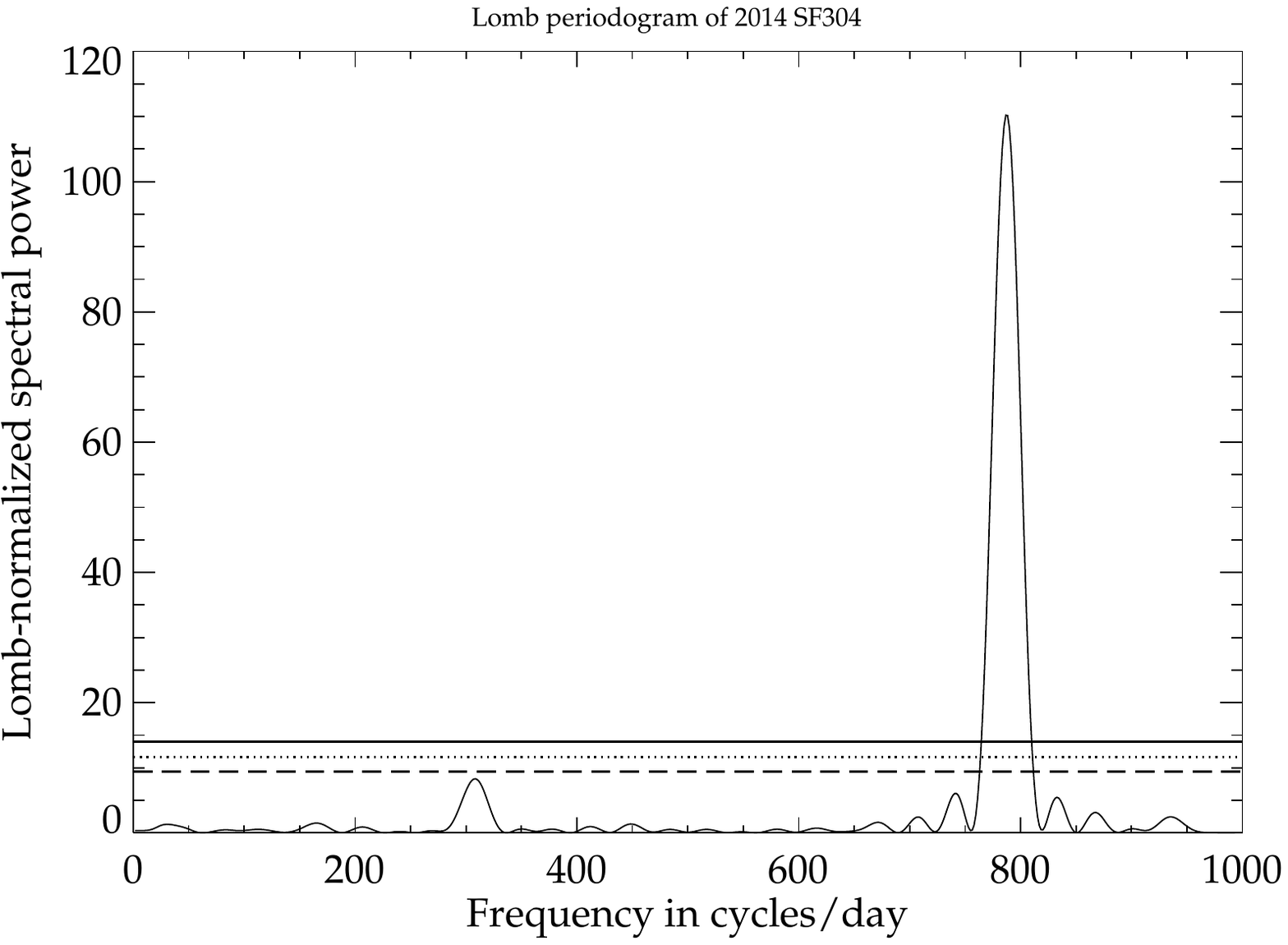}
 \includegraphics[width=9cm,angle=0]{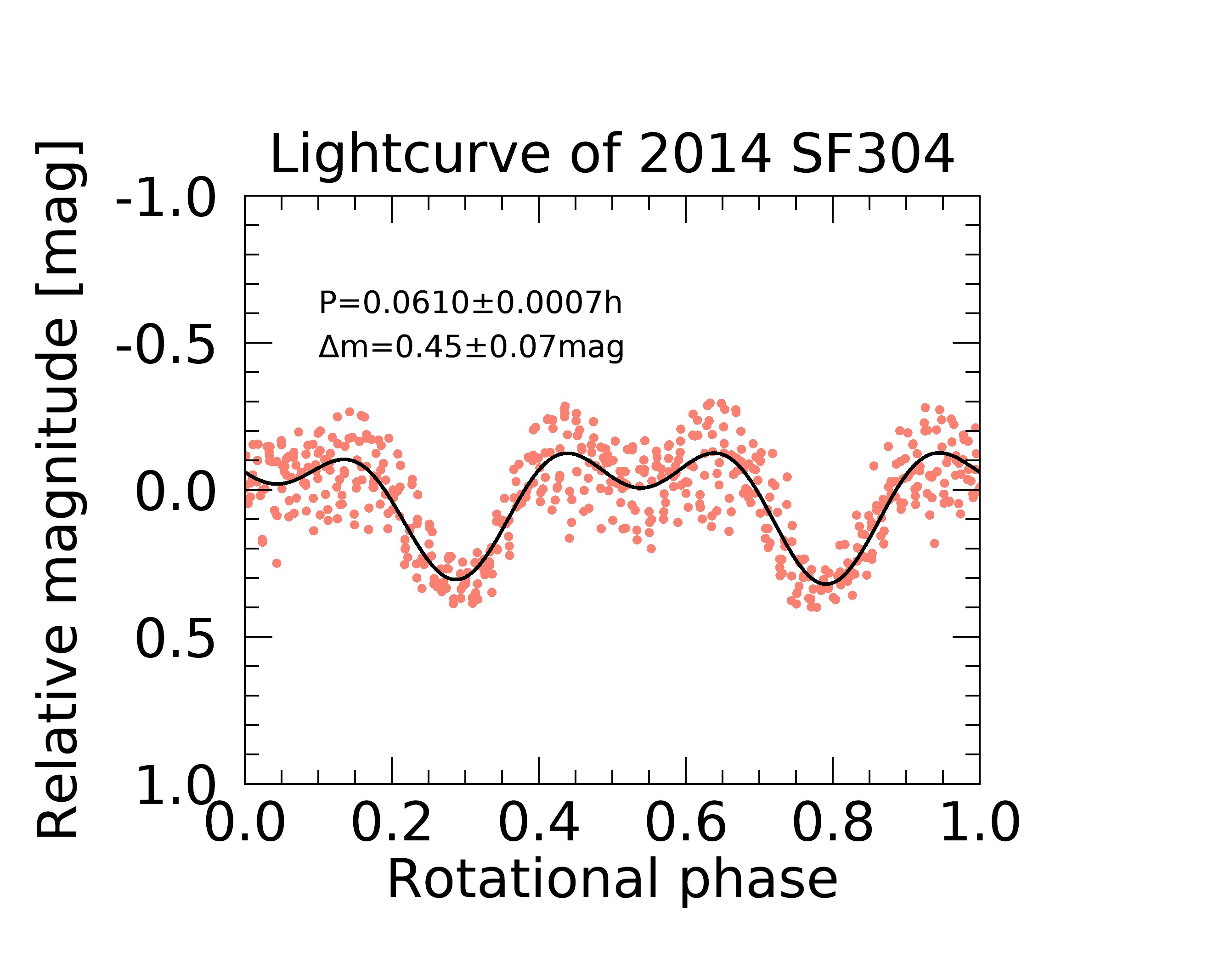}
 \includegraphics[width=9cm,angle=0]{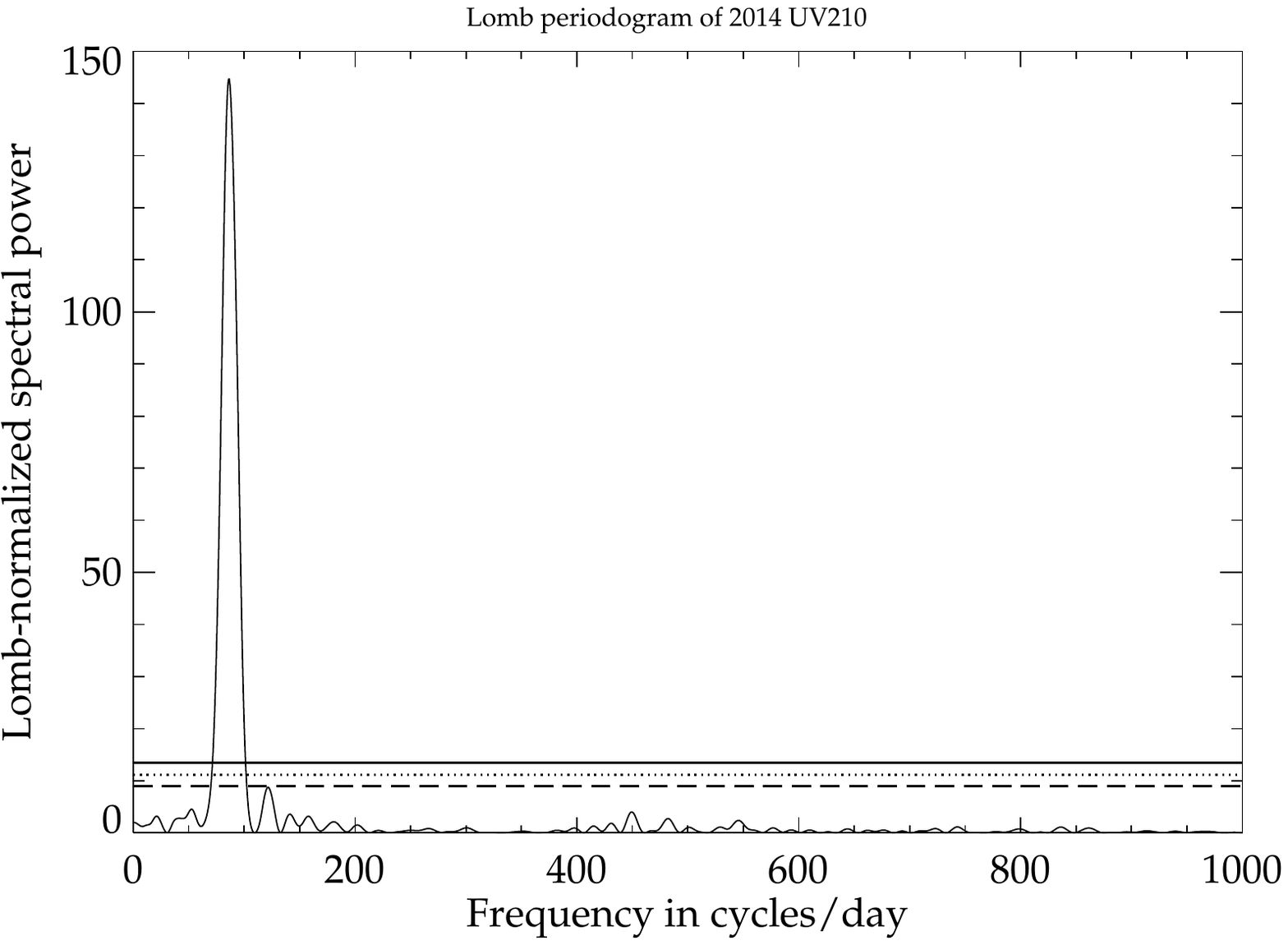}
 \includegraphics[width=9cm,angle=0]{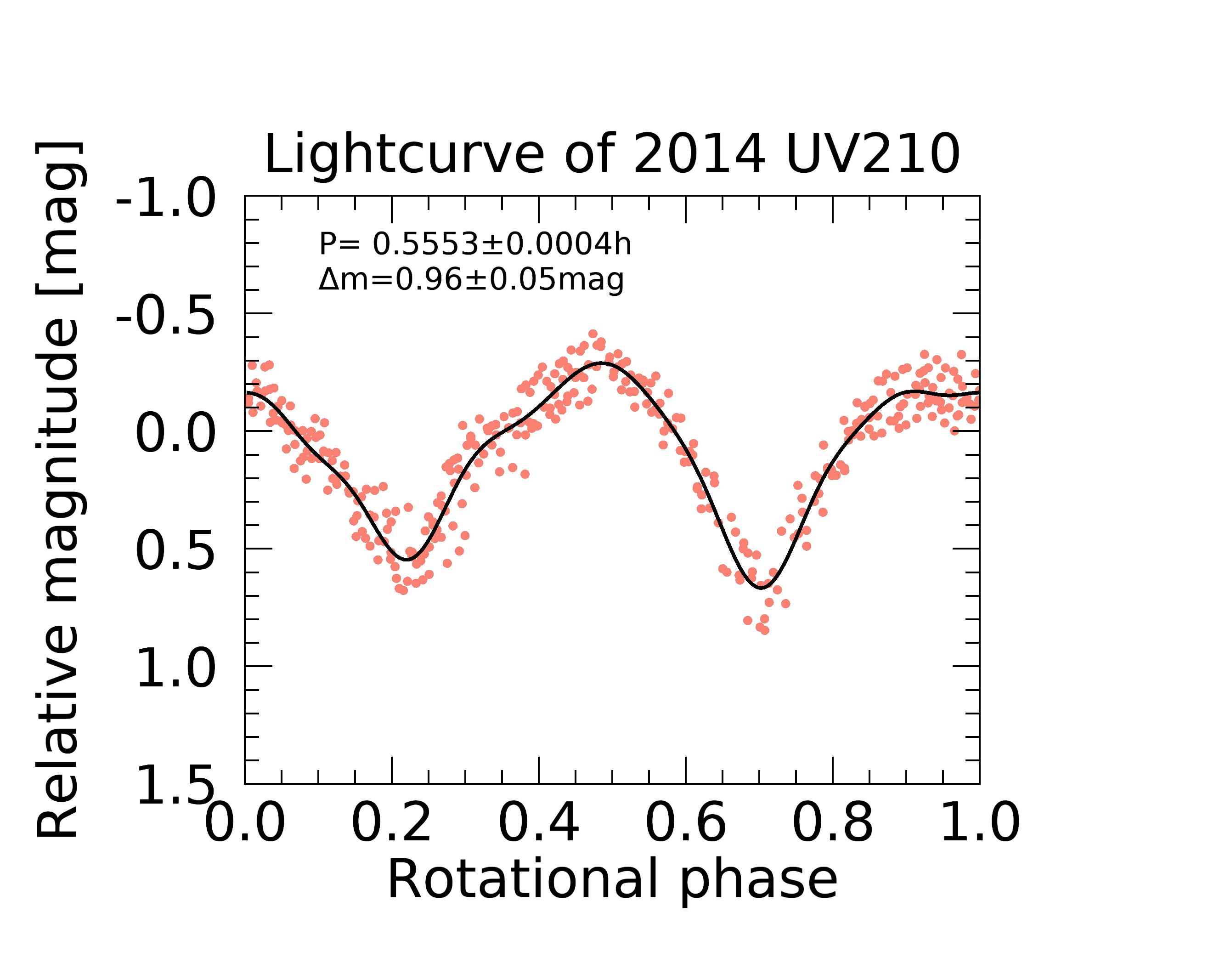}
 \caption{Full lightcurves of NEOs included in the photometric study. The highest peak of the Lomb periodogram is the single-peaked rotational period with the highest
confidence level. The 99.9\% confidence level is indicated with a continuous line while the confidence level at 99\% is the dotted line, and the dashed line corresponds to a 
confidence level of 90\%. On the right, the lightcurves corresponding to the highest confidence level peak are plotted. The lightcurves have been fitted with a Fourier series fit 
(black curves).}
\label{fig:Full_lightcurves}

 \end{figure*}

 \begin{figure*}
 \includegraphics[width=9cm,angle=0]{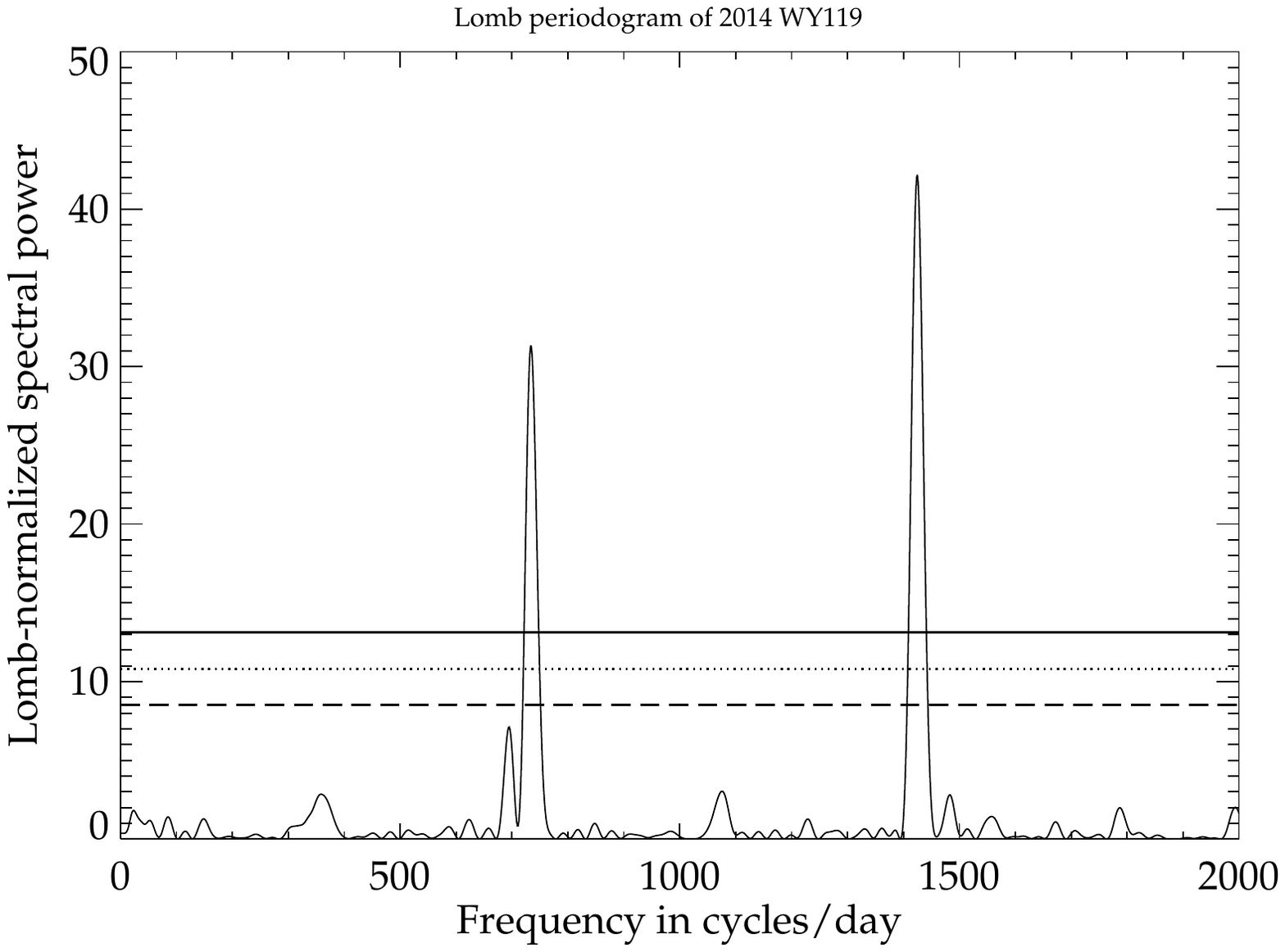}
 \includegraphics[width=9cm,angle=0]{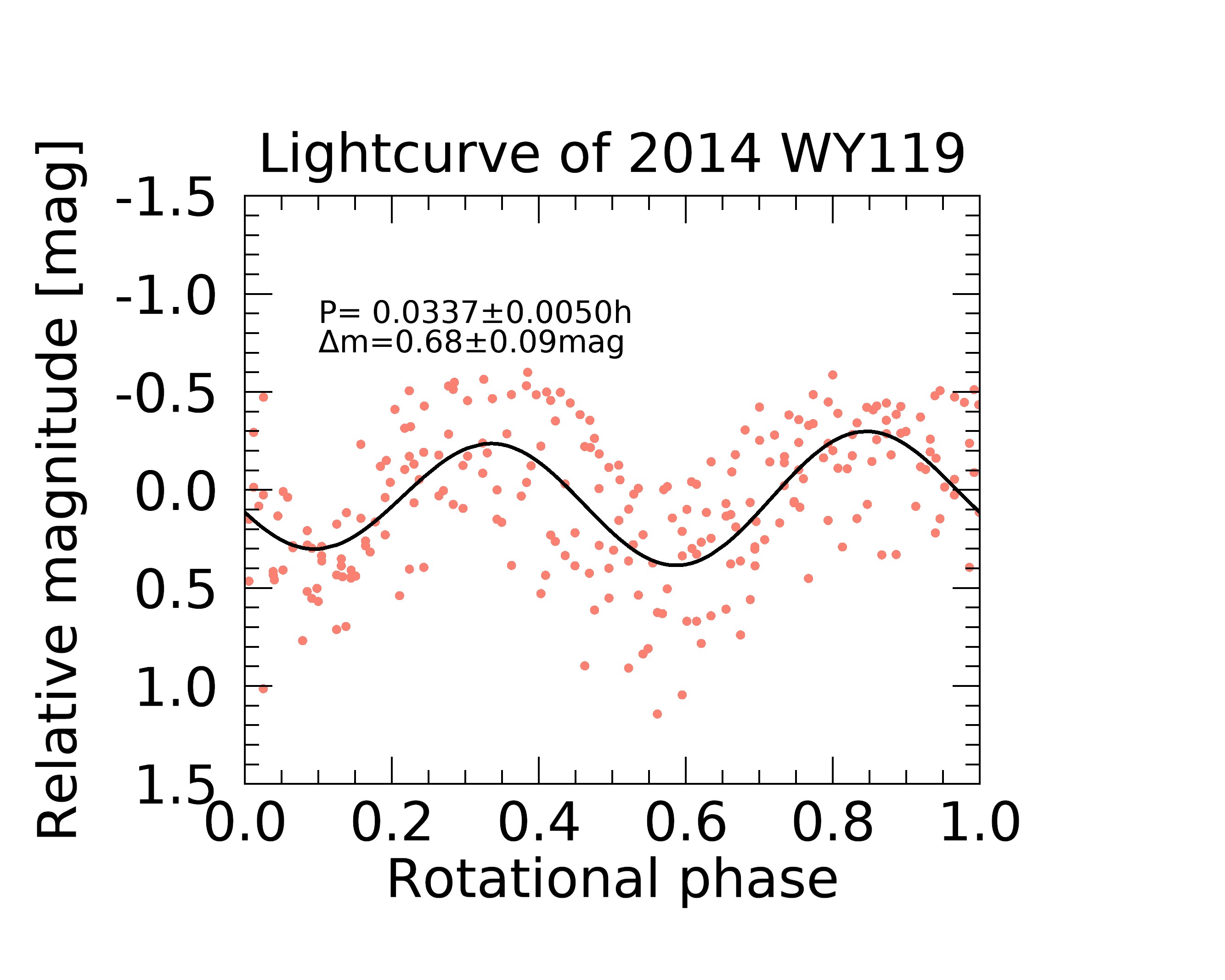} 
 \includegraphics[width=9cm,angle=0]{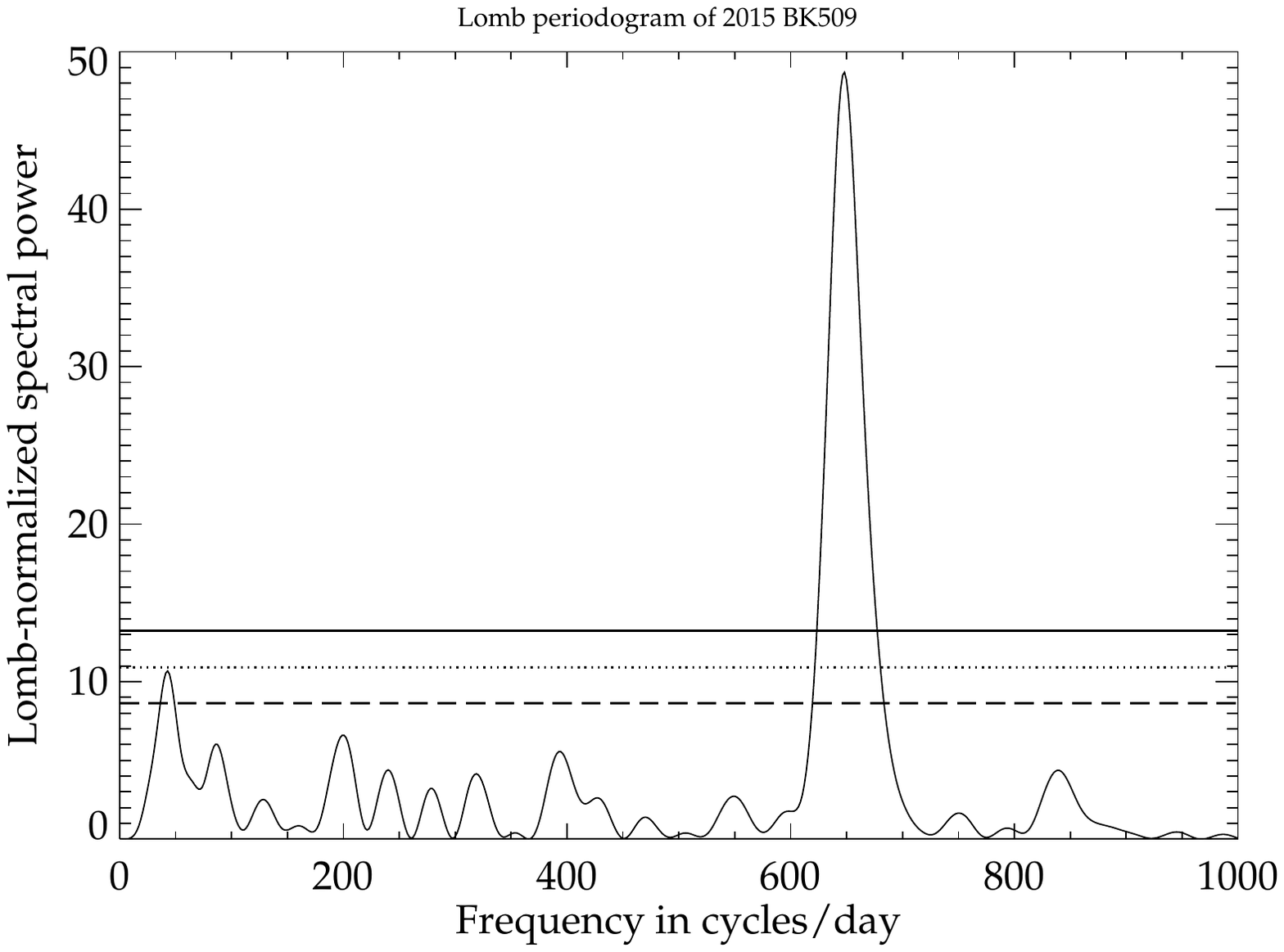}
 \includegraphics[width=9cm,angle=0]{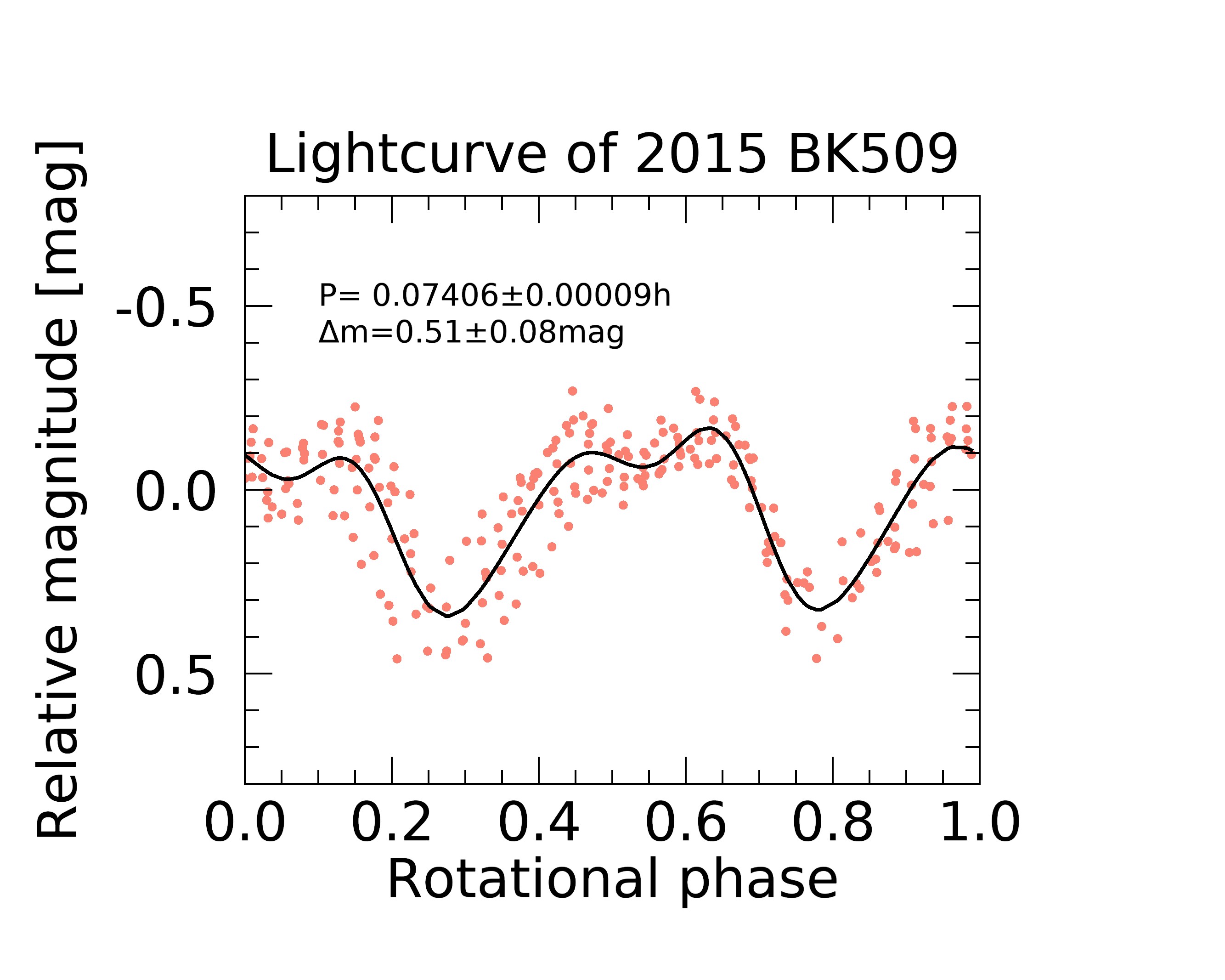} 
 \includegraphics[width=9cm,angle=0]{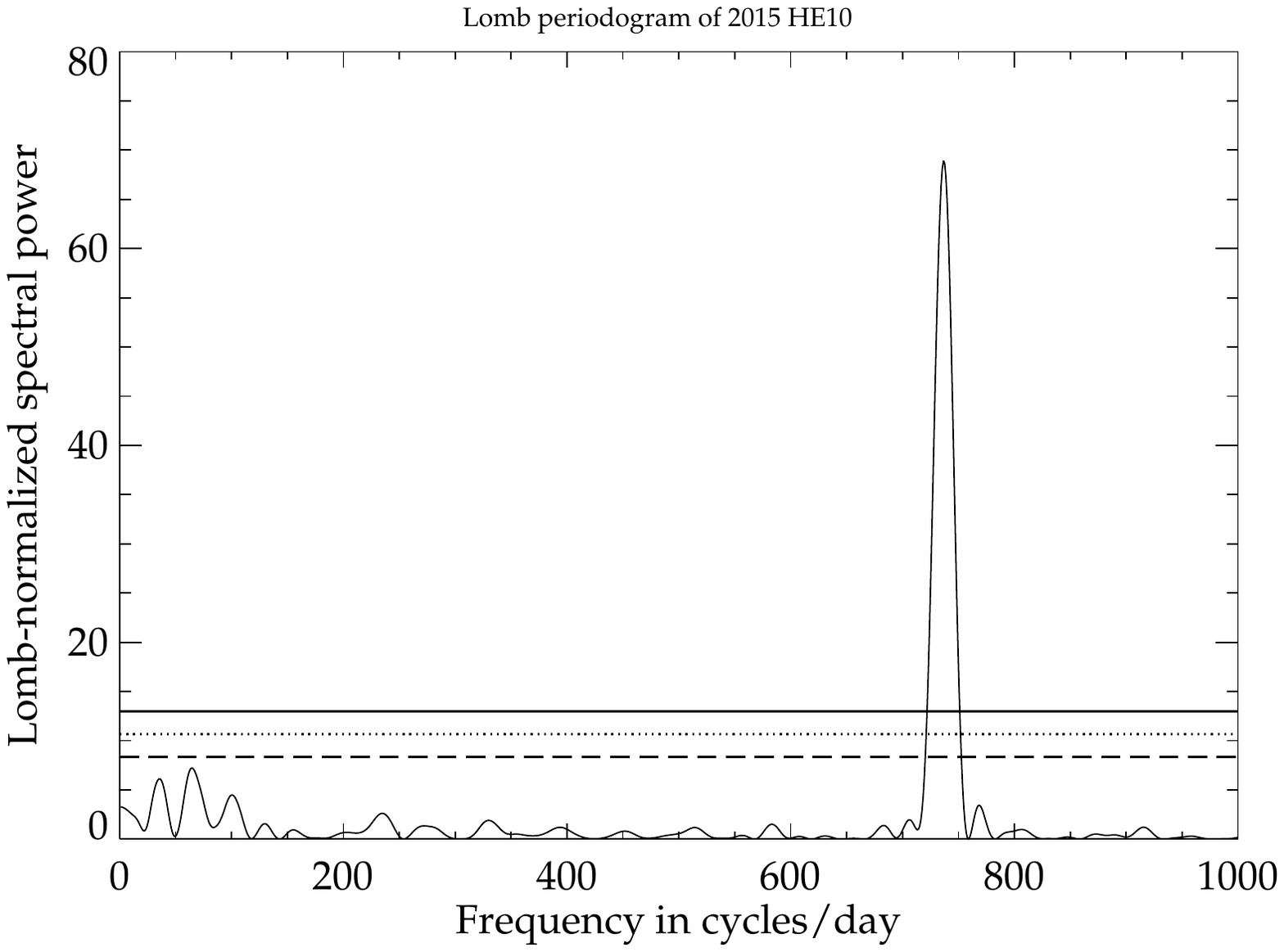}
 \includegraphics[width=9cm,angle=0]{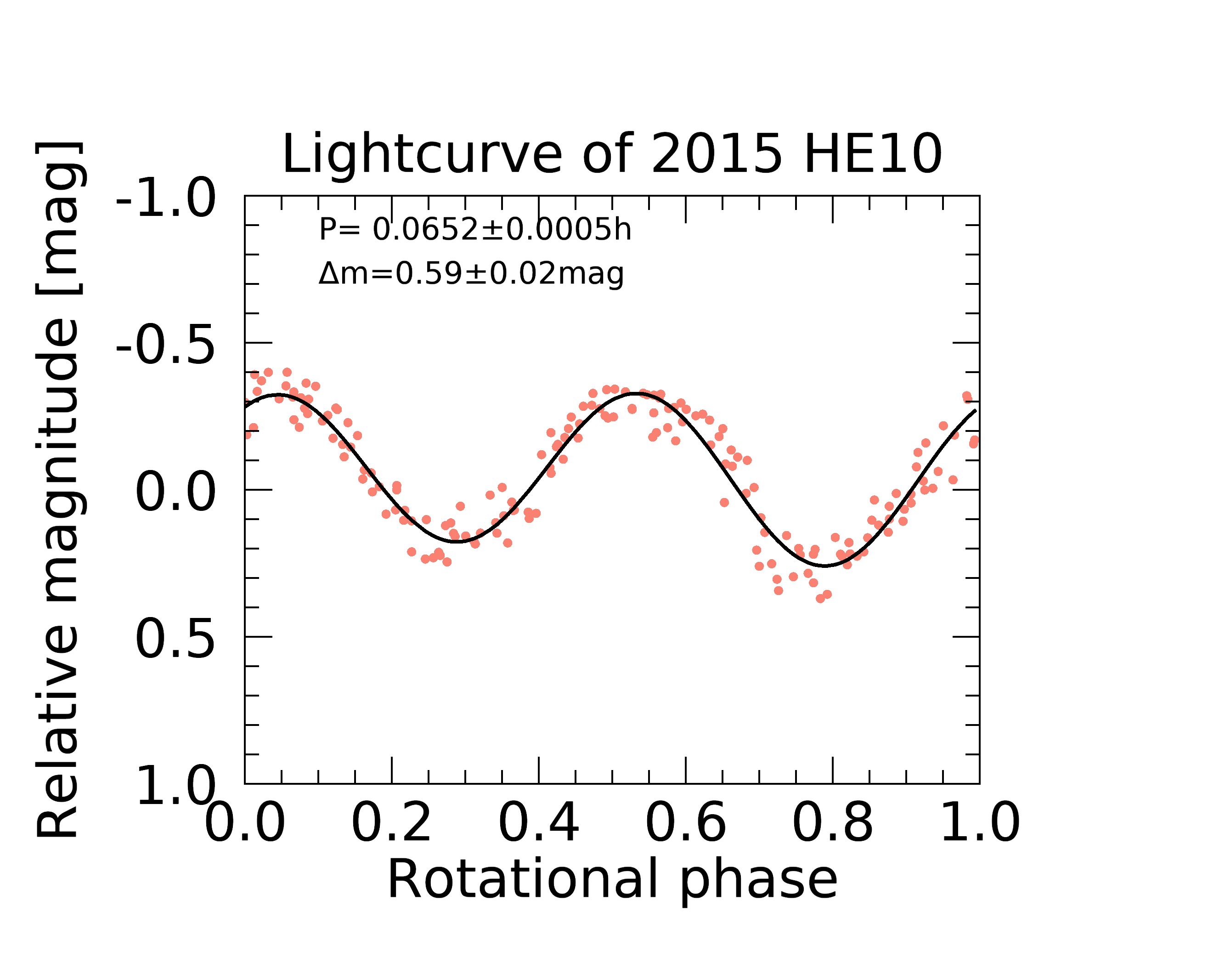} 
 \caption{Full lightcurves of NEOs included in the photometric study. The highest peak of the Lomb periodogram is the single-peaked rotational period with the highest
confidence level. The 99.9\% confidence level is indicated with a continuous line while the confidence level at 99\% is the dotted line, and the dashed line corresponds to a 
confidence level of 90\%. On the right, the lightcurves corresponding to the highest confidence level peak are plotted. The lightcurves have been fitted with a Fourier series fit 
(black curves).}
\label{fig:Full_lightcurves}

 \end{figure*}

 \begin{figure*}
 \includegraphics[width=9cm,angle=0]{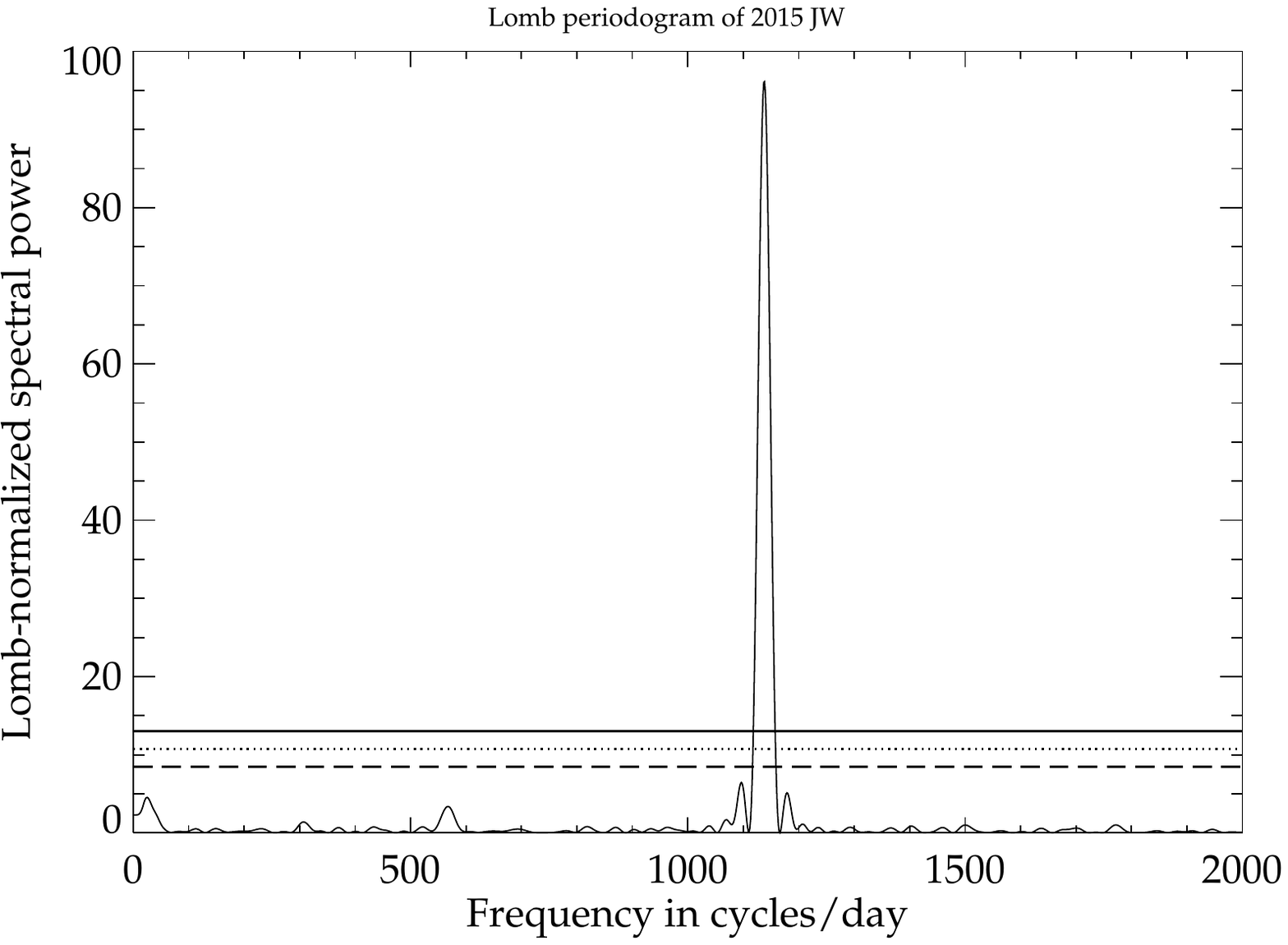}
 \includegraphics[width=9cm,angle=0]{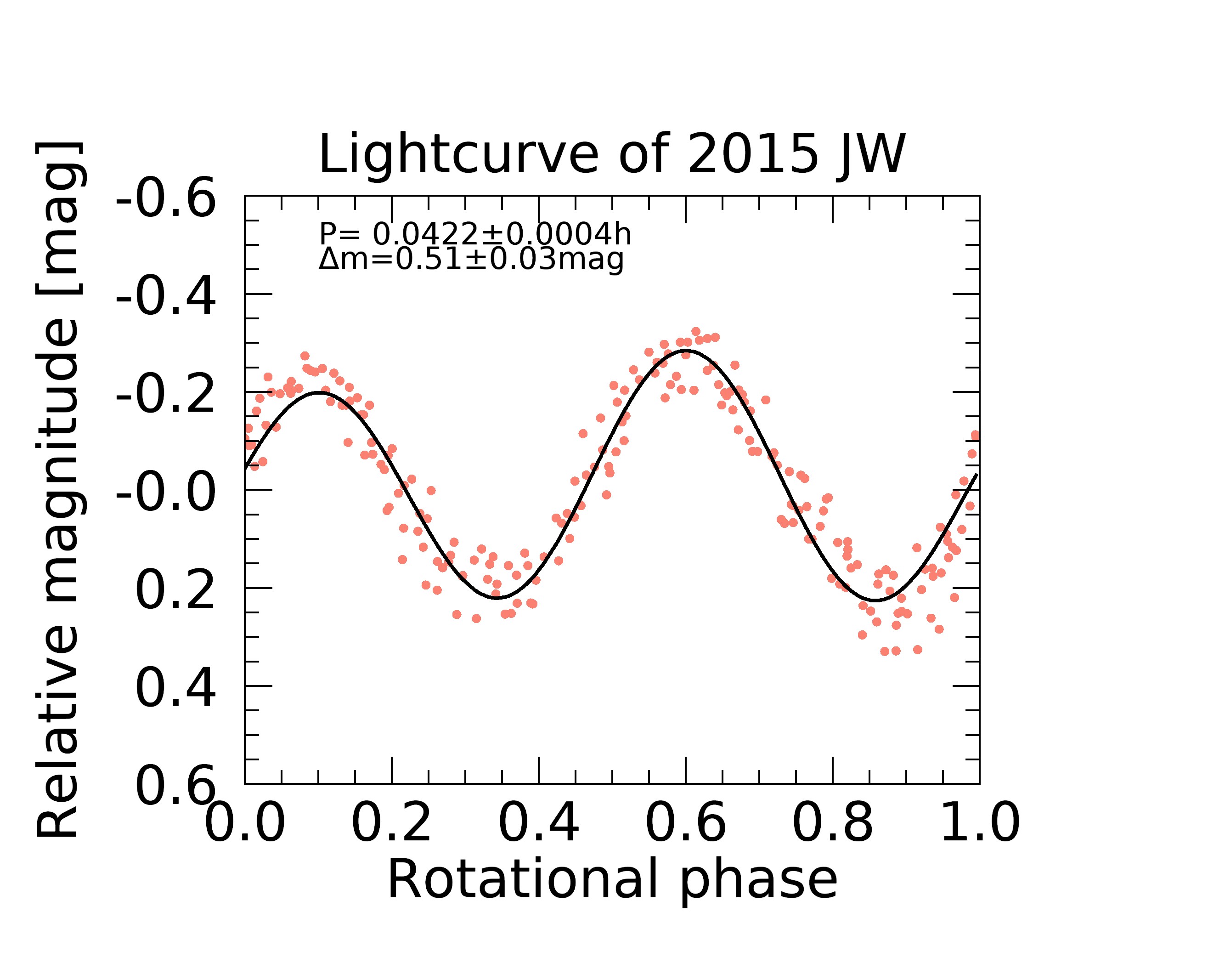} 
 \includegraphics[width=9cm,angle=0]{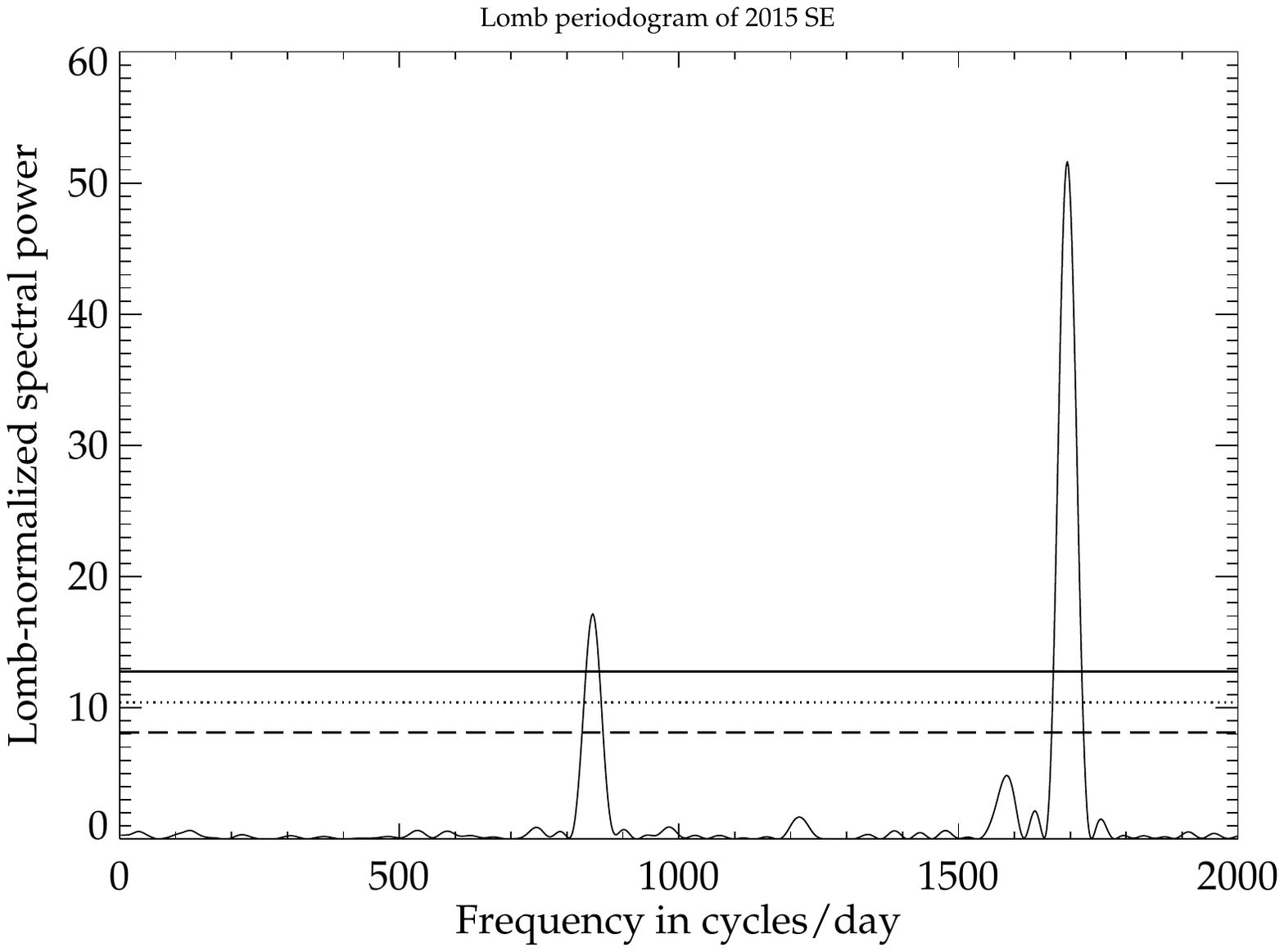}
 \includegraphics[width=9cm,angle=0]{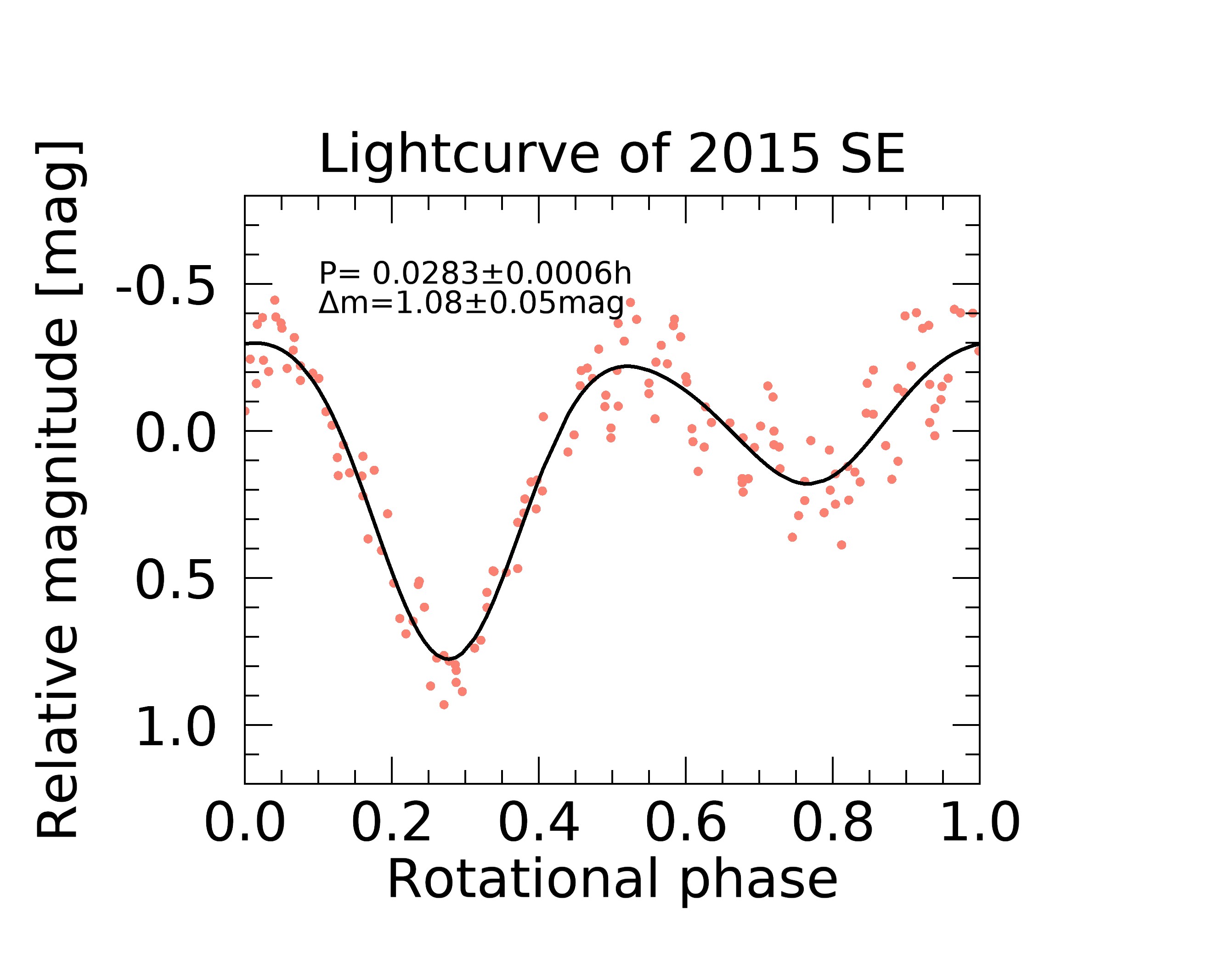} 
 \includegraphics[width=9cm,angle=0]{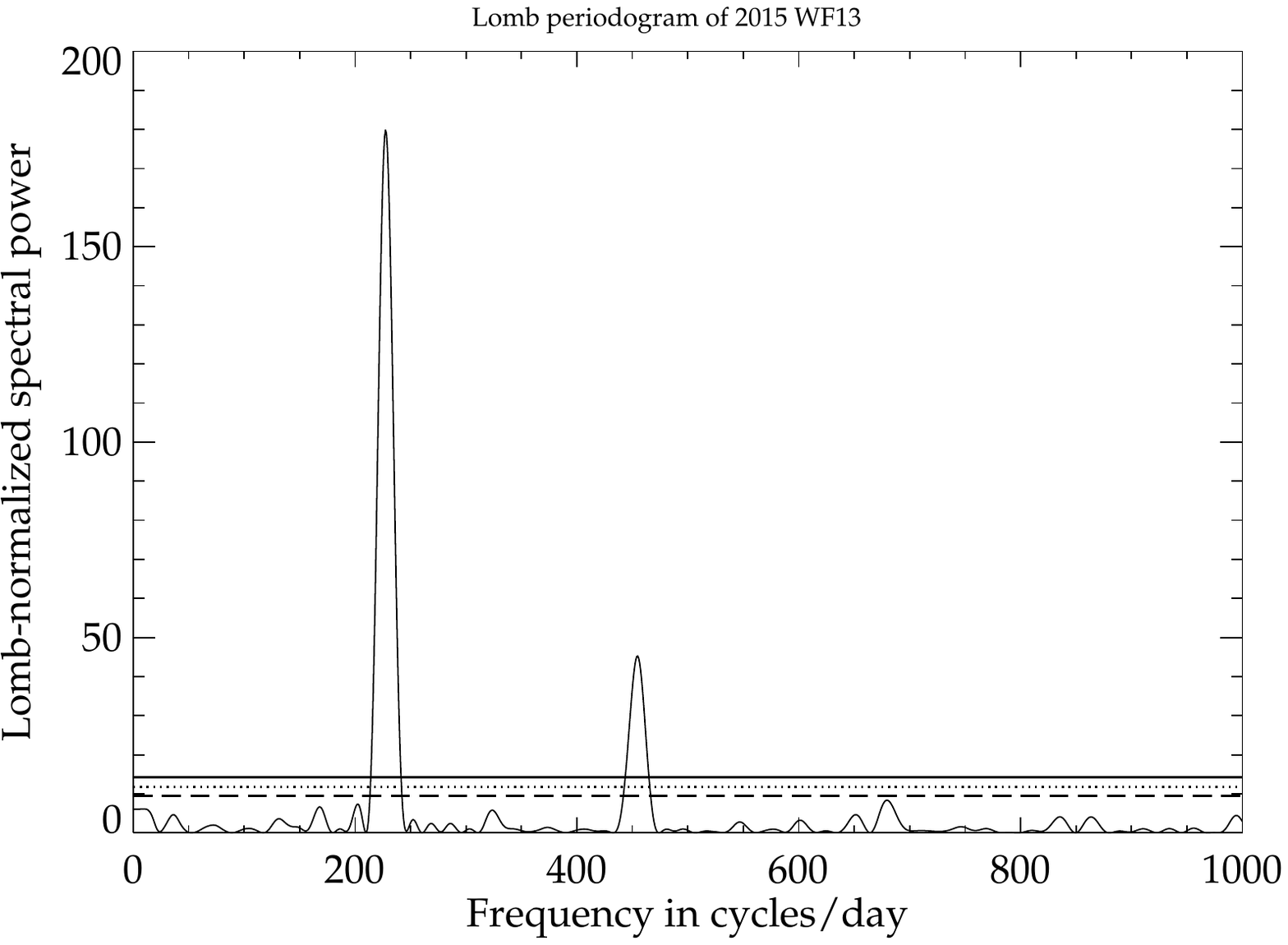}
 \includegraphics[width=9cm,angle=0]{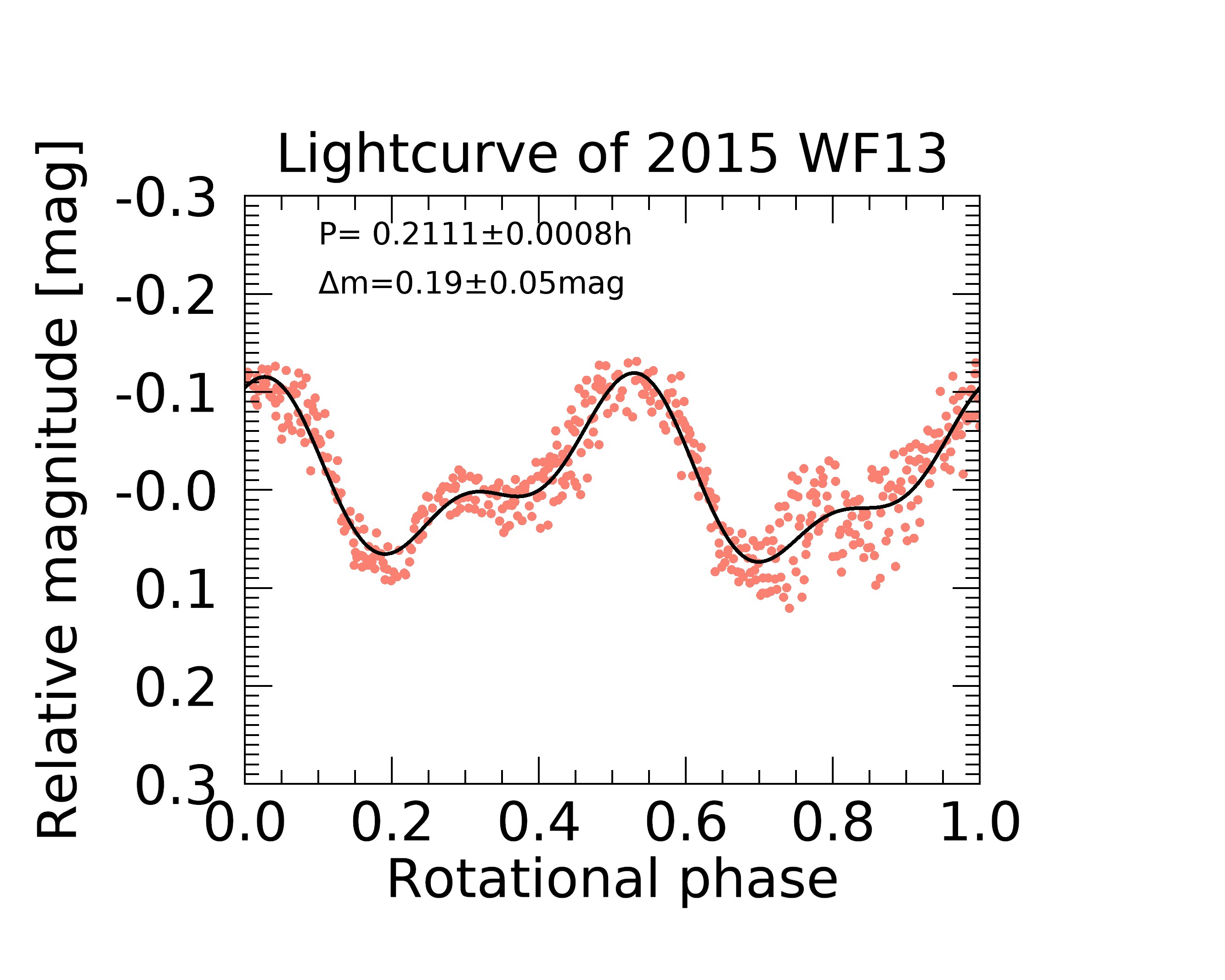} 
 \caption{Full lightcurves of NEOs included in the photometric study. The highest peak of the Lomb periodogram is the single-peaked rotational period with the highest
confidence level. The 99.9\% confidence level is indicated with a continuous line while the confidence level at 99\% is the dotted line, and the dashed line corresponds to a 
confidence level of 90\%. On the right, the lightcurves corresponding to the highest confidence level peak are plotted. The lightcurves have been fitted with a Fourier series fit 
(black curves).}
\label{fig:Full_lightcurves}

 \end{figure*}
 
 \begin{figure*}
 \includegraphics[width=9cm,angle=0]{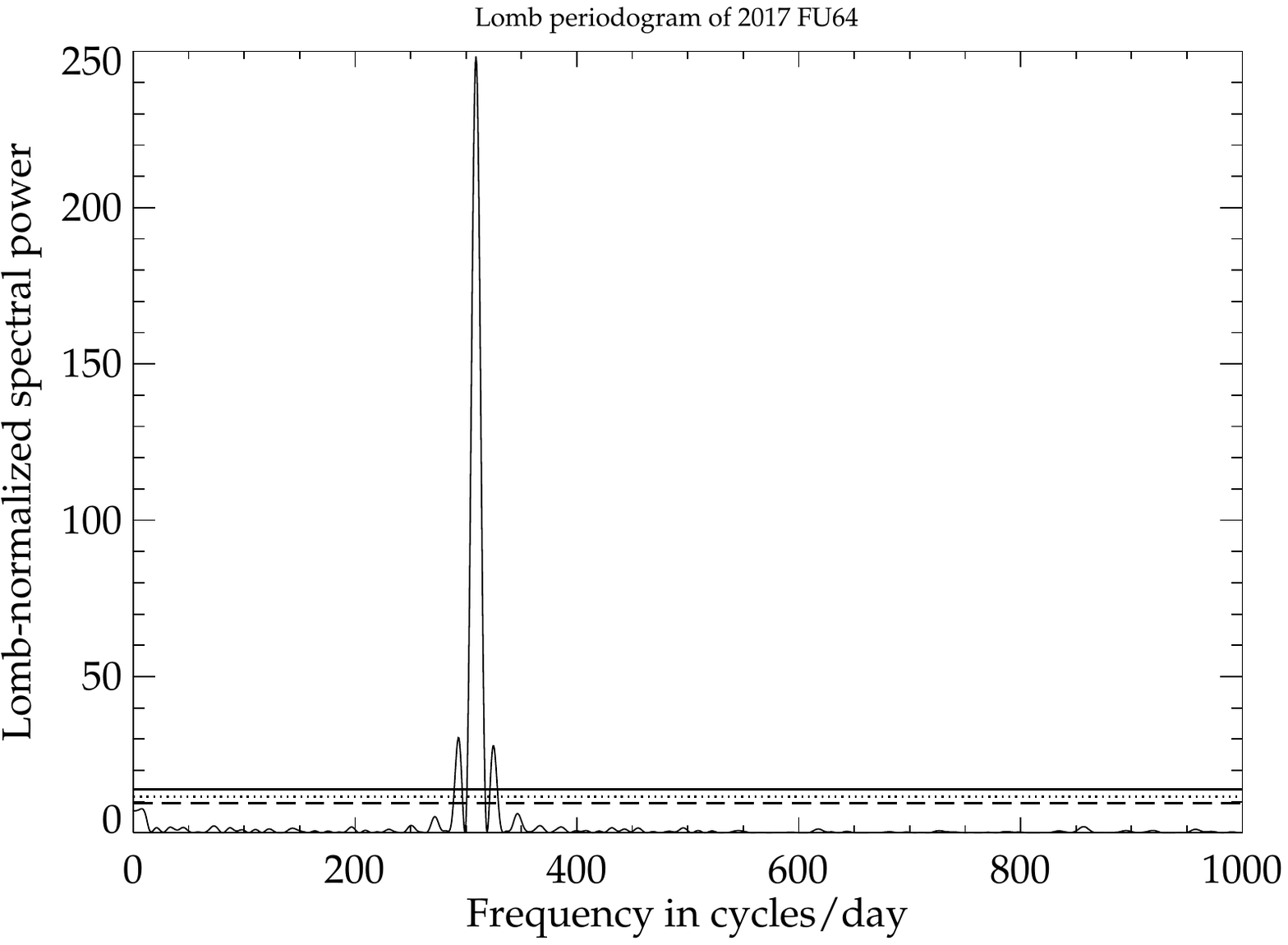}
 \includegraphics[width=9cm,angle=0]{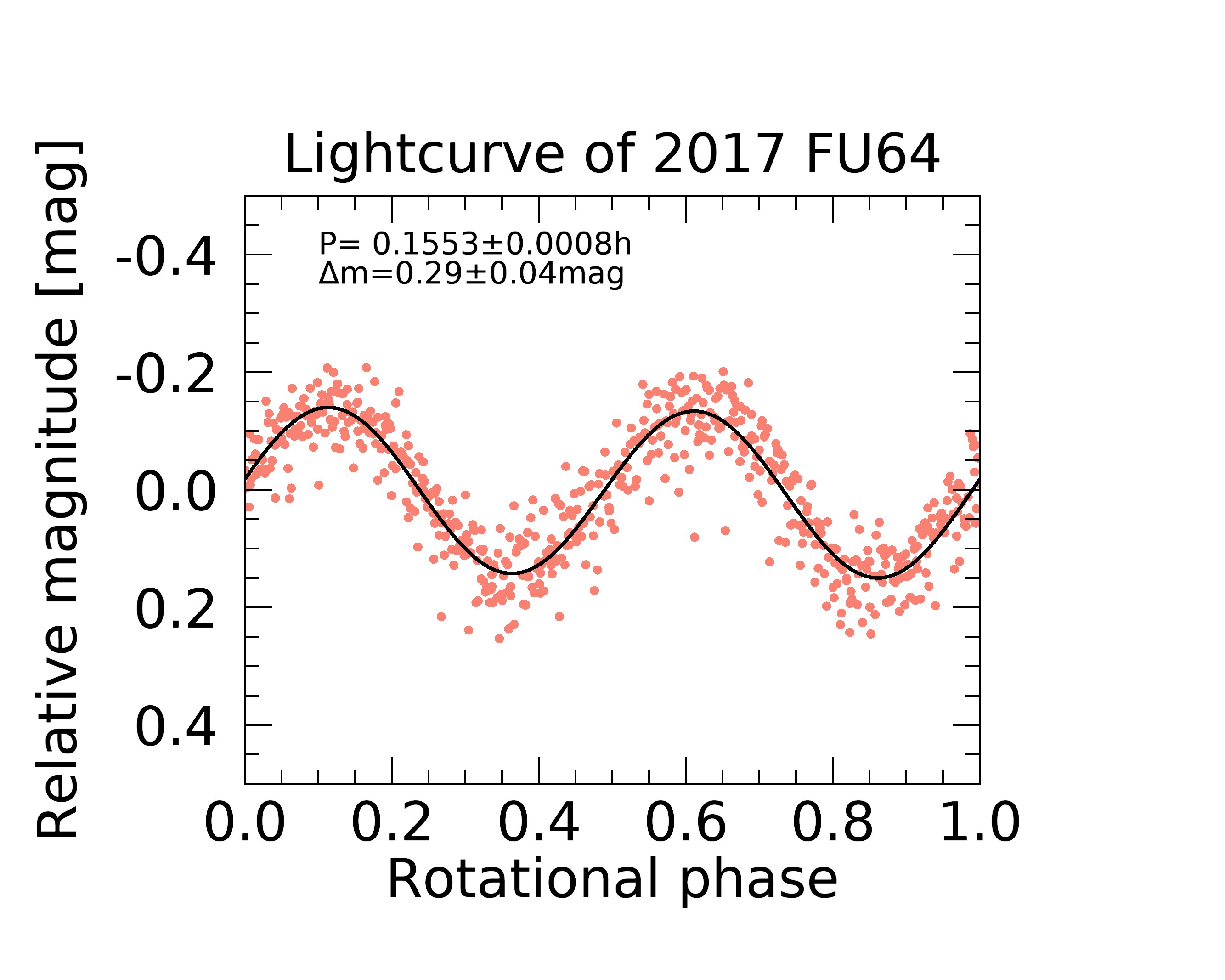} 
 \includegraphics[width=9cm,angle=0]{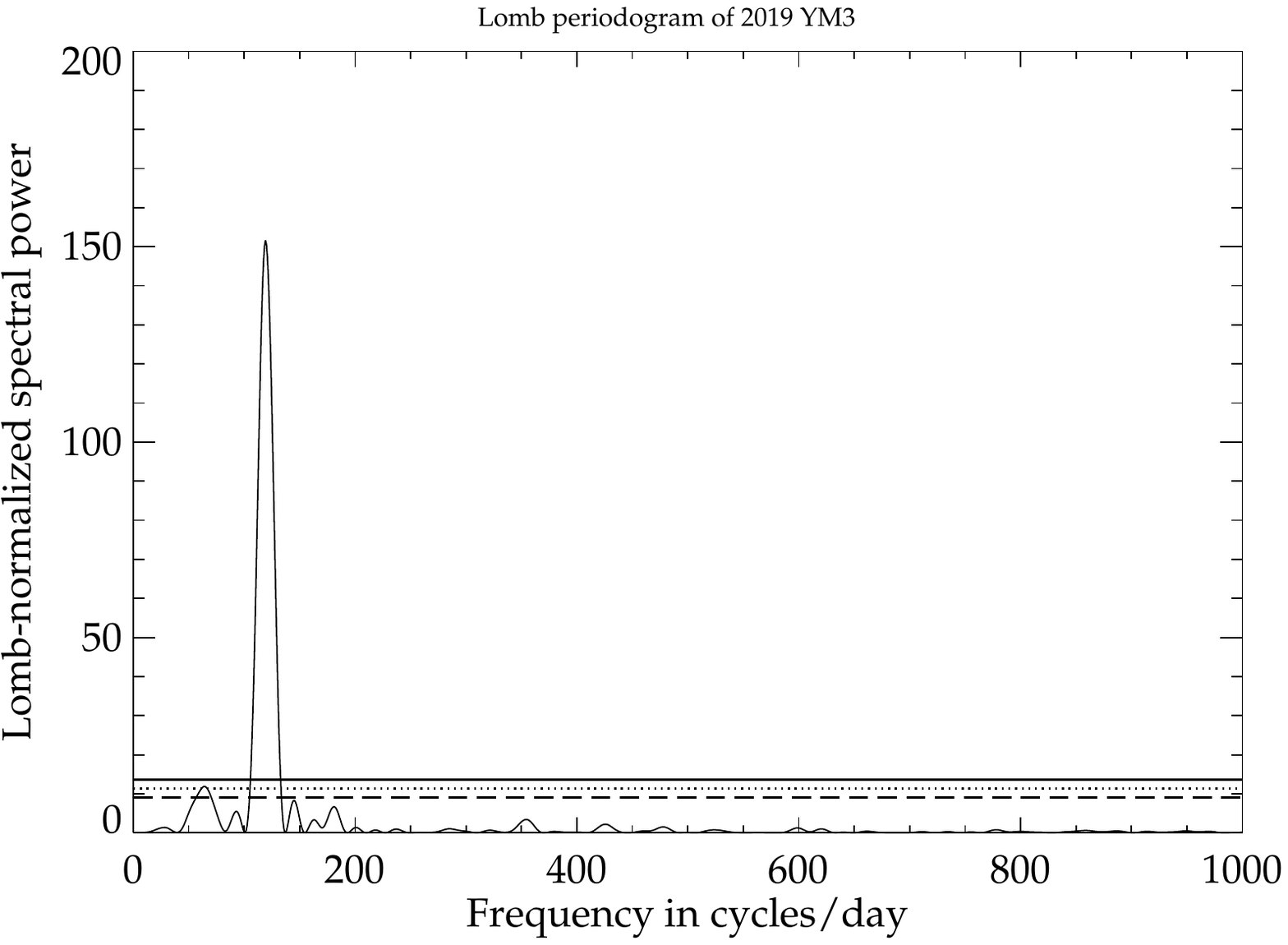}
 \includegraphics[width=9cm,angle=0]{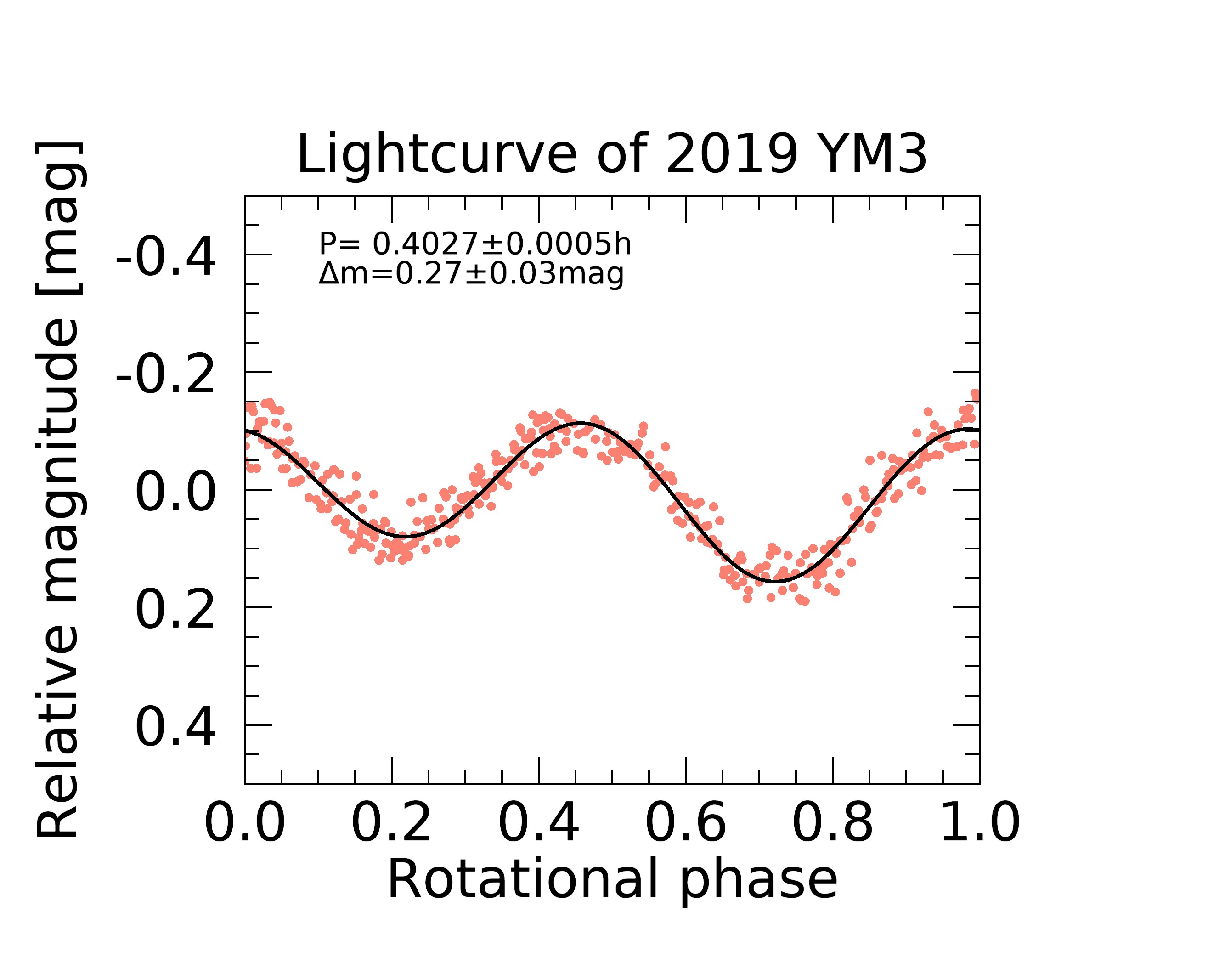} 
 \includegraphics[width=9cm,angle=0]{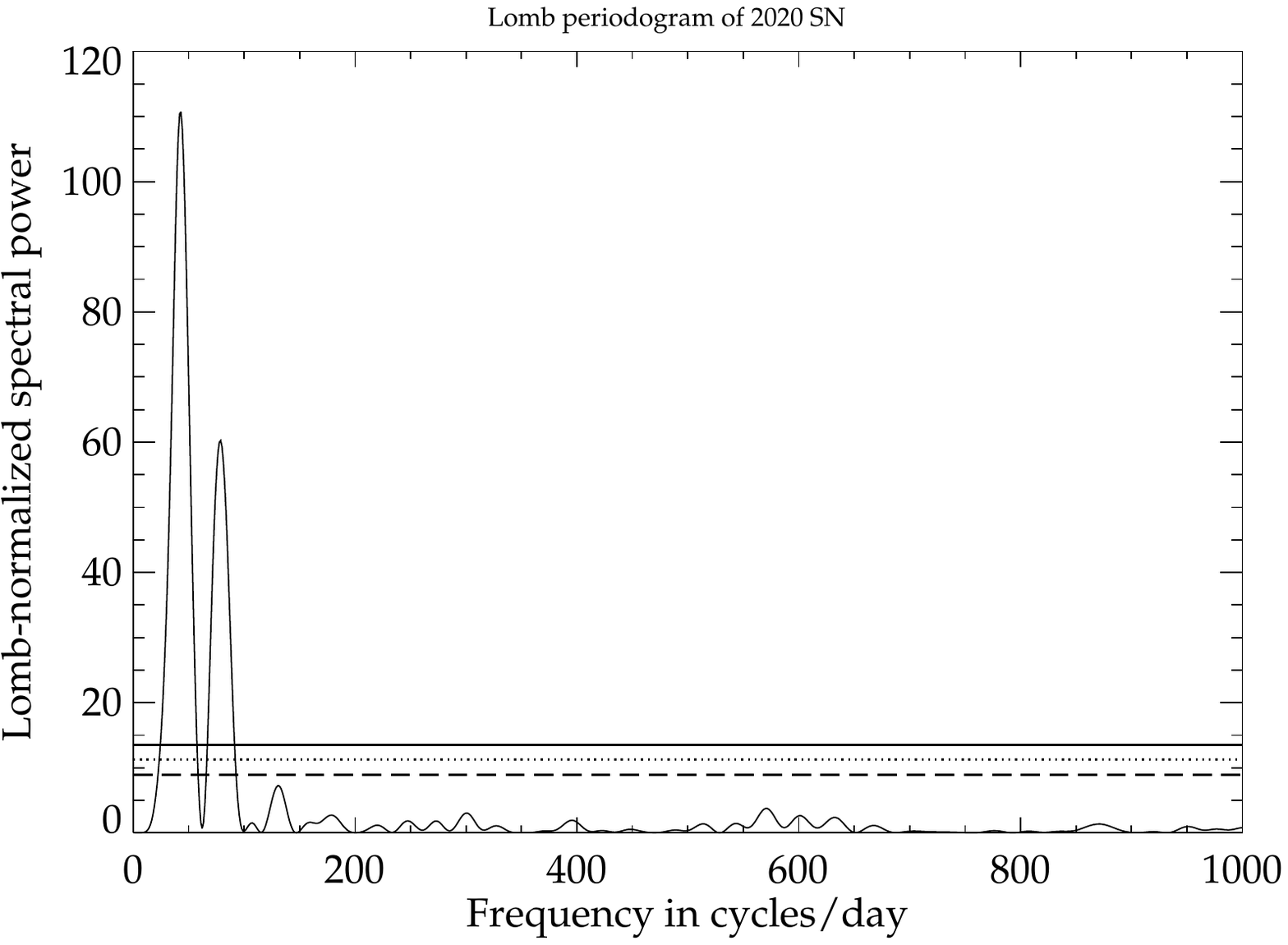}
 \includegraphics[width=9cm,angle=0]{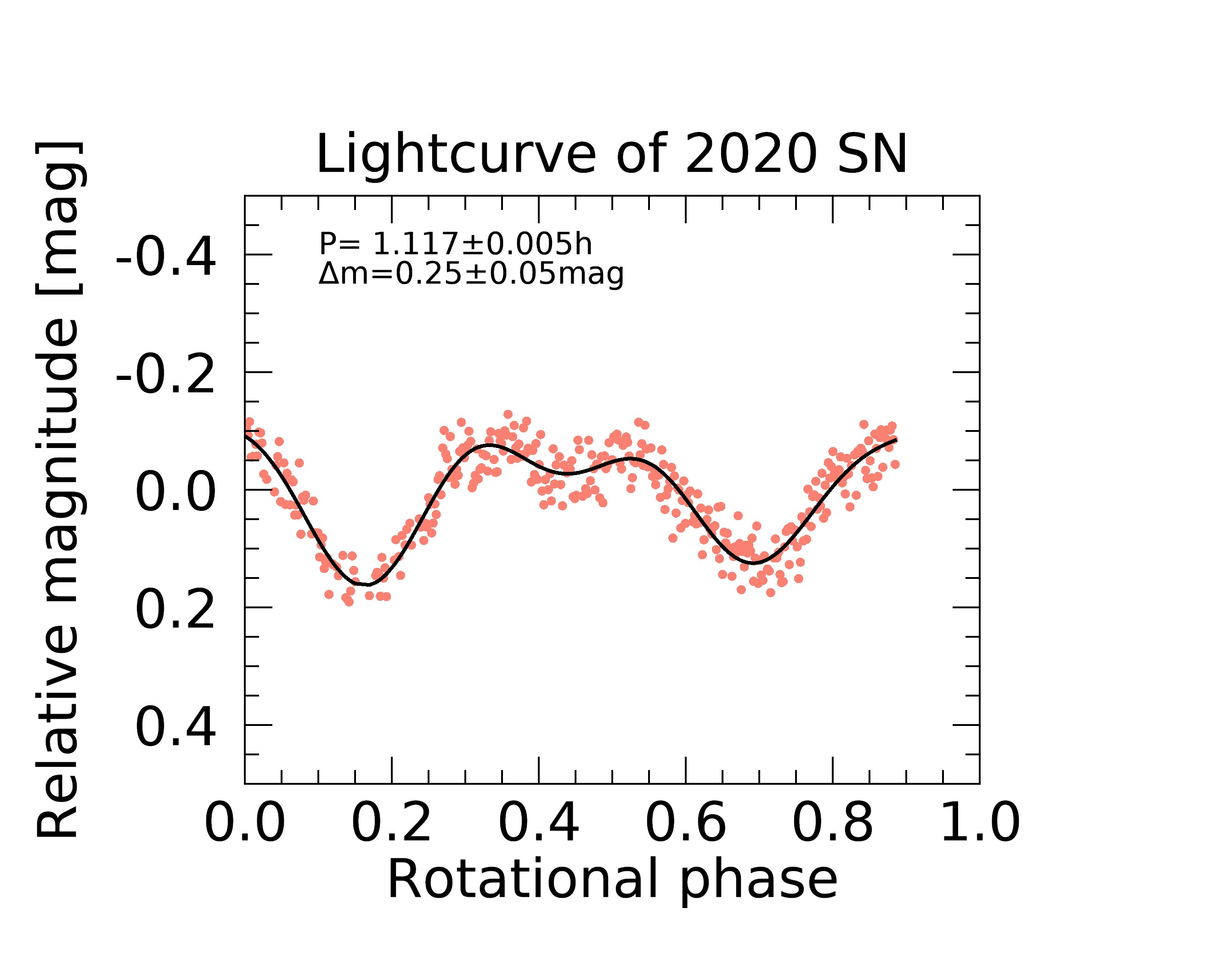} 
 \caption{Full lightcurves of NEOs included in the photometric study. The highest peak of the Lomb periodogram is the single-peaked rotational period with the highest
confidence level. The 99.9\% confidence level is indicated with a continuous line while the confidence level at 99\% is the dotted line, and the dashed line corresponds to a 
confidence level of 90\%. On the right, the lightcurves corresponding to the highest confidence level peak are plotted. The lightcurves have been fitted with a Fourier series fit 
(black curves).}
\label{fig:Full_lightcurves}
\end{figure*}

\clearpage

\textbf{Appendix B.2.}

\vspace{-5mm}

\begin{figure*}[!h]
 \includegraphics[width=9cm,angle=0]{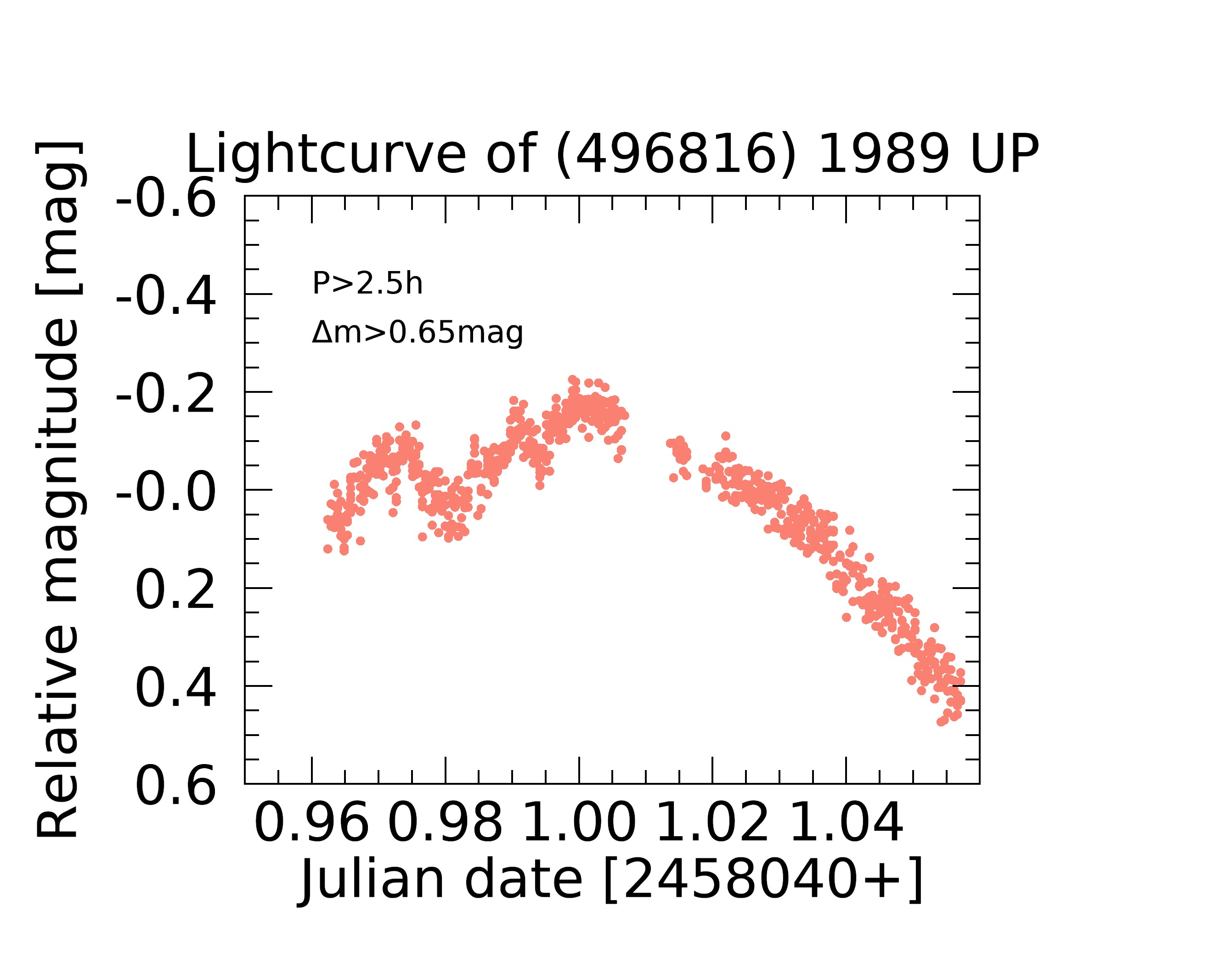}
  \includegraphics[width=9cm,angle=0]{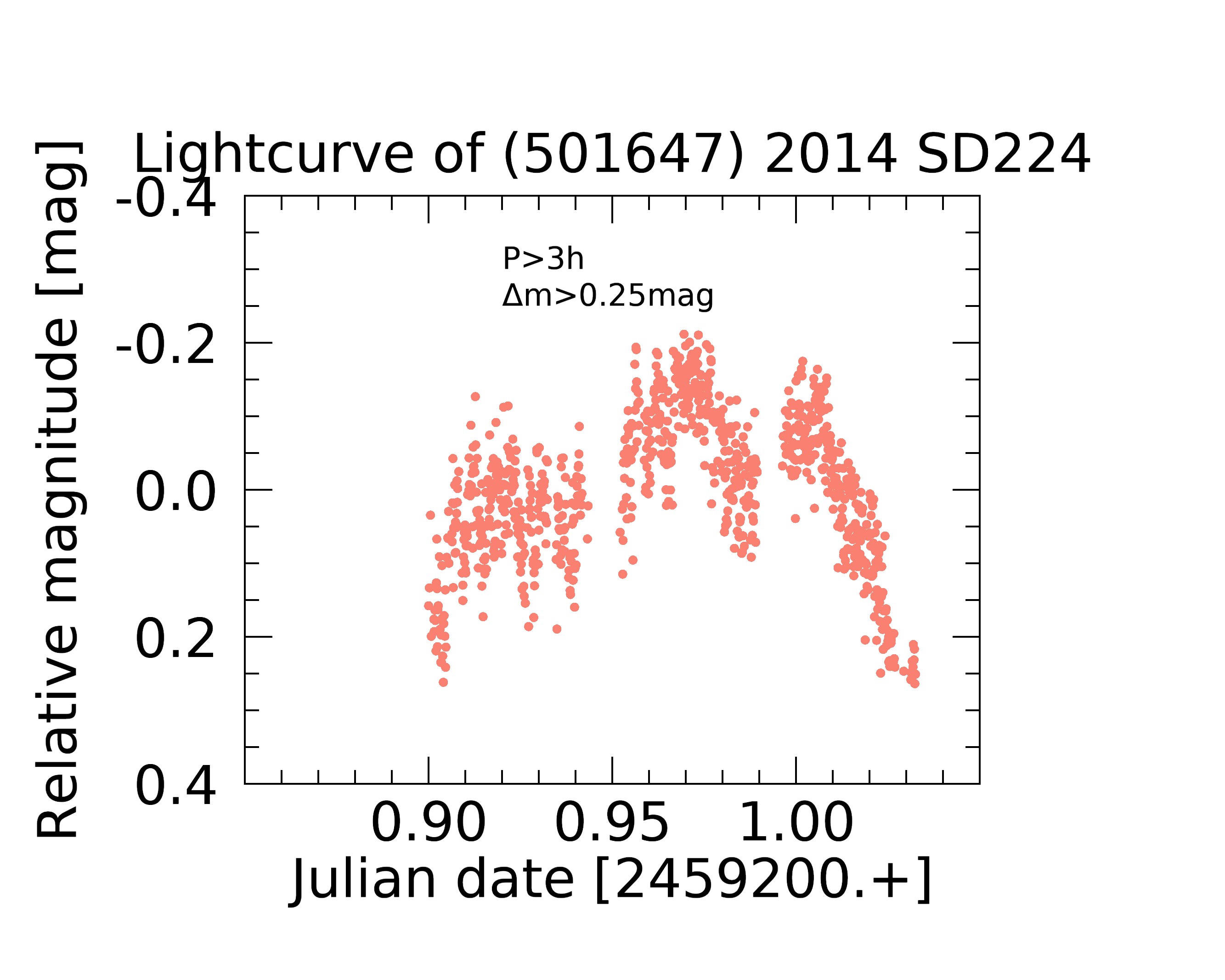}
 \includegraphics[width=9cm,angle=0]{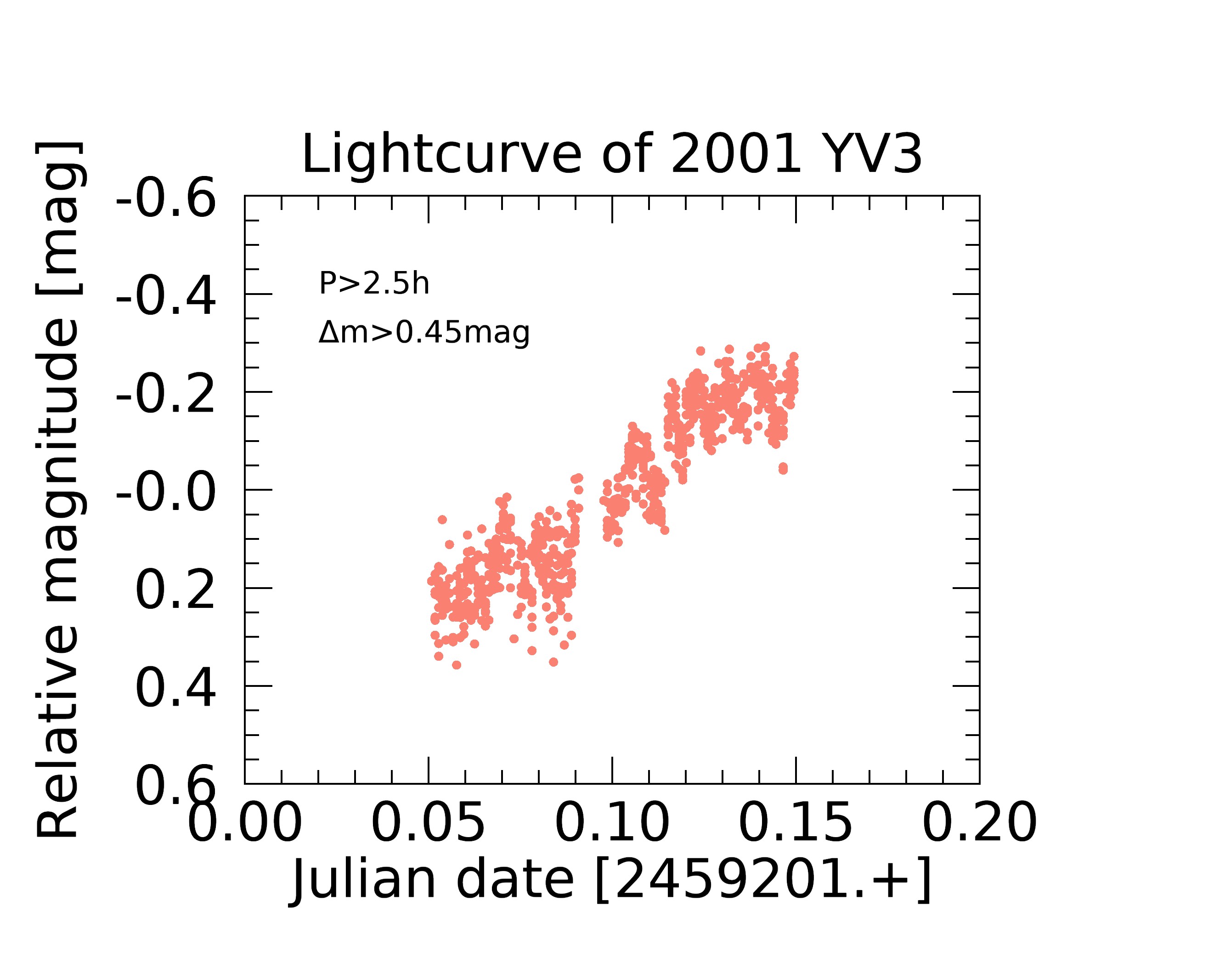}
 \includegraphics[width=9cm,angle=0]{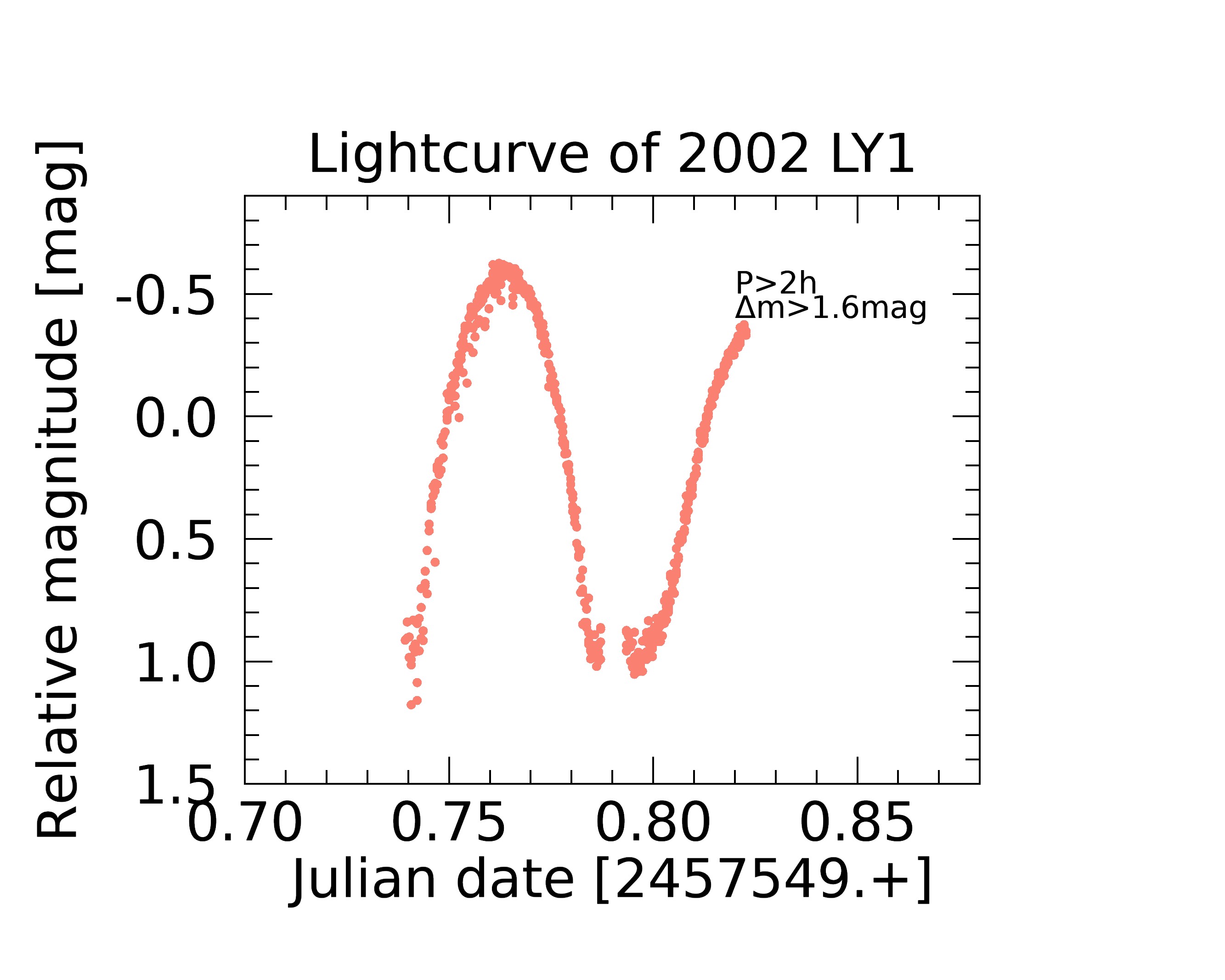}
 \includegraphics[width=9cm,angle=0]{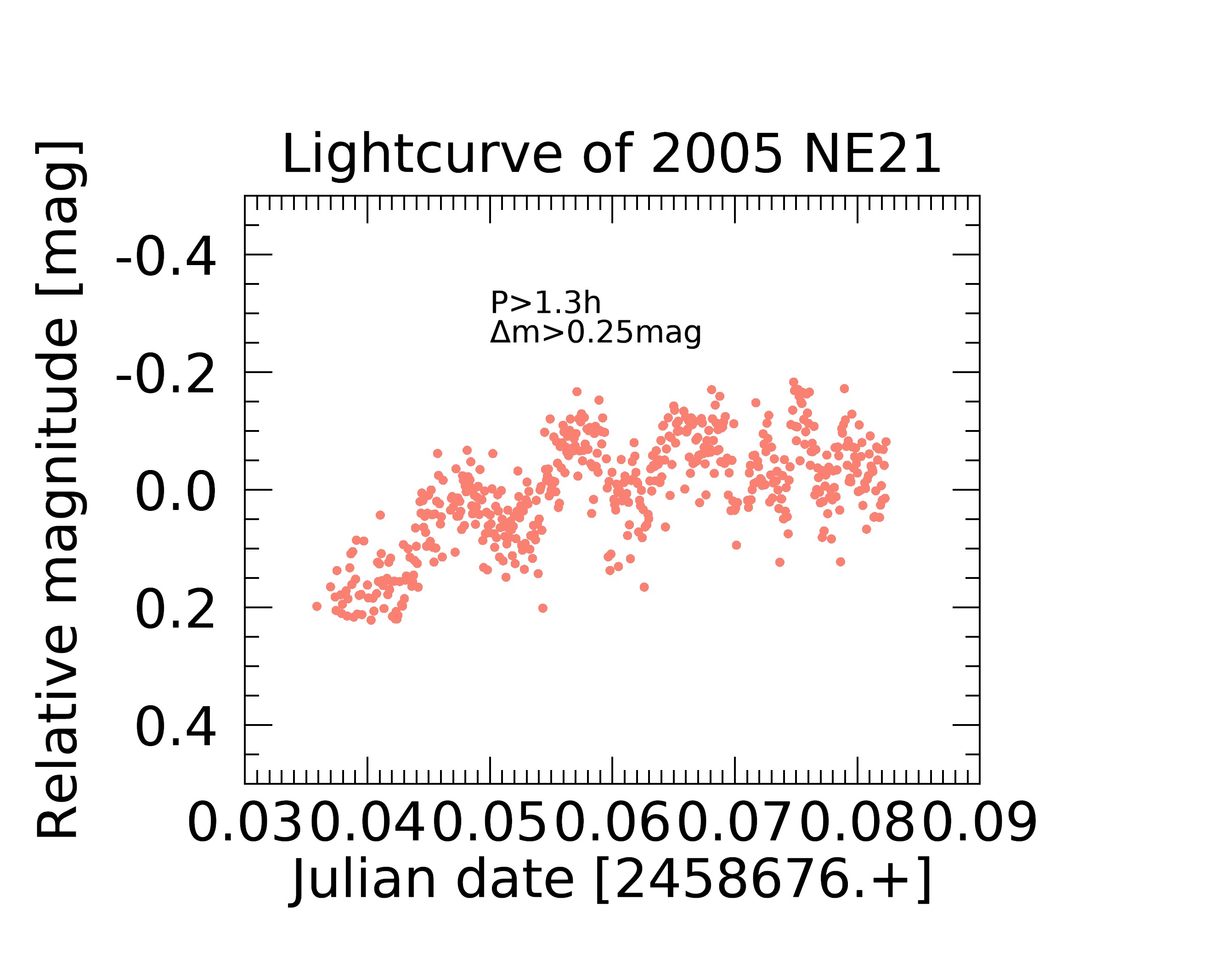}
 \includegraphics[width=9cm,angle=0]{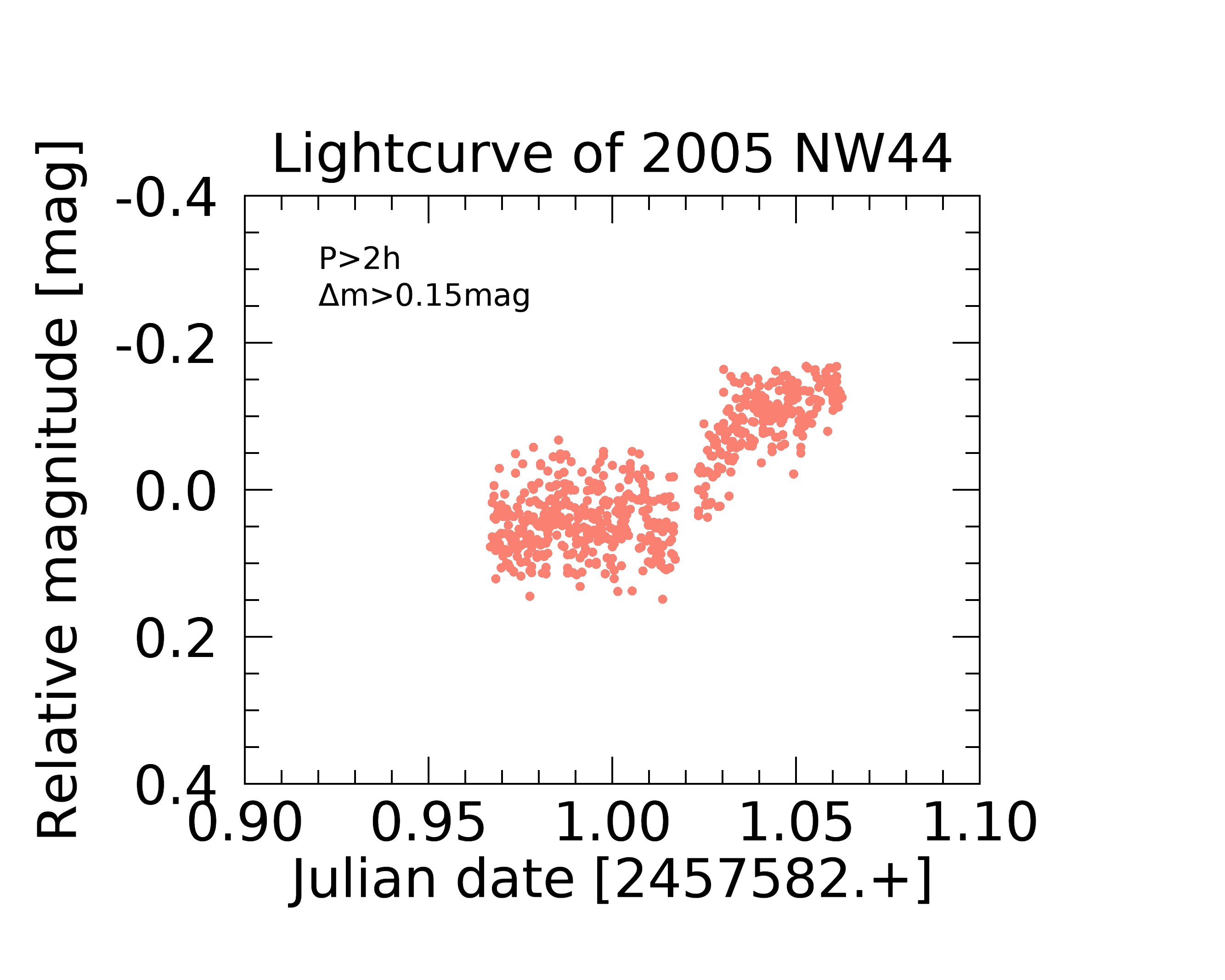}
\caption{Partial lightcurves of NEOs included in the photometric study.}
\label{fig:Partial_lightcurves}

\end{figure*}

 \begin{figure*}
 \includegraphics[width=9cm,angle=0]{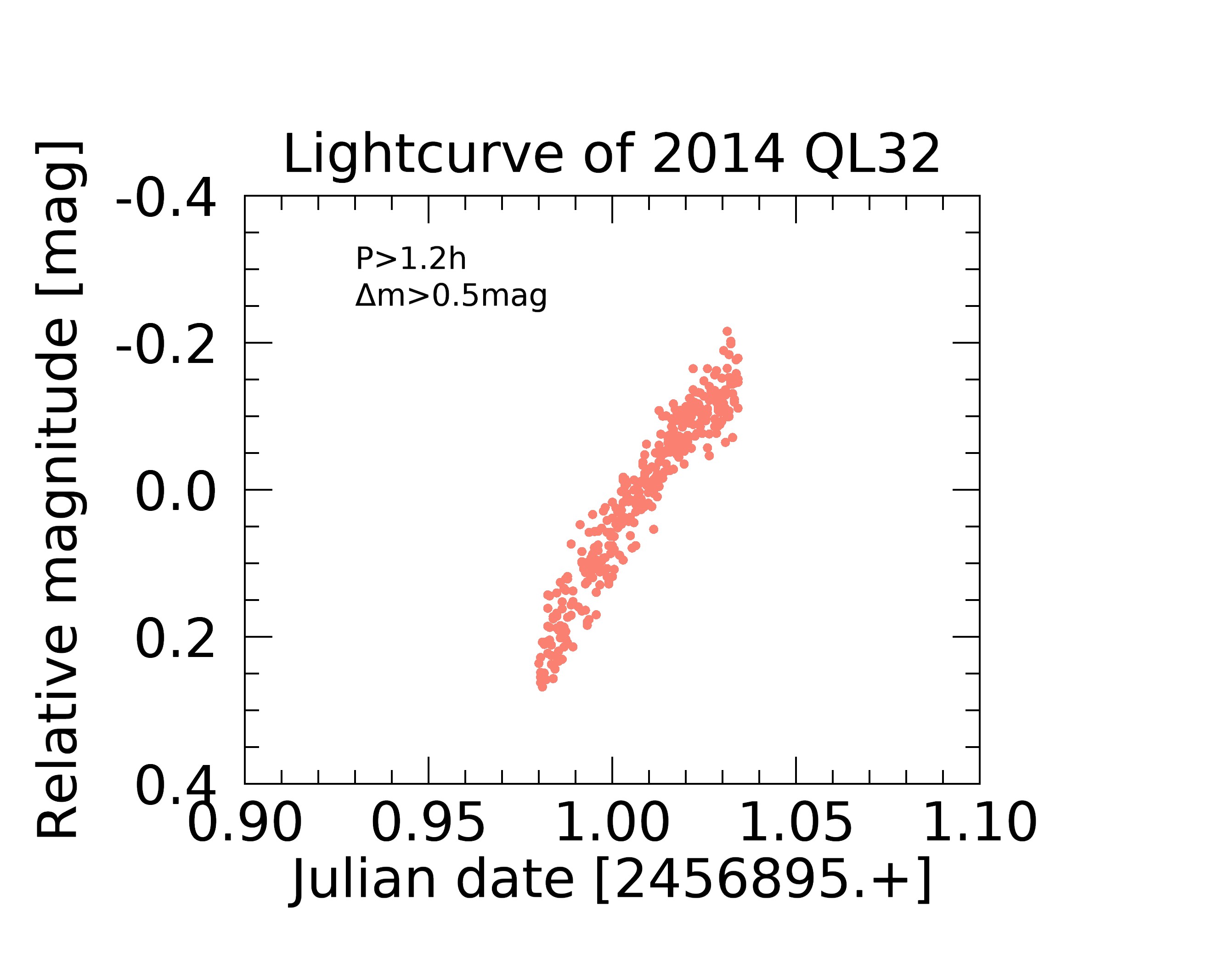}
 \includegraphics[width=9cm,angle=0]{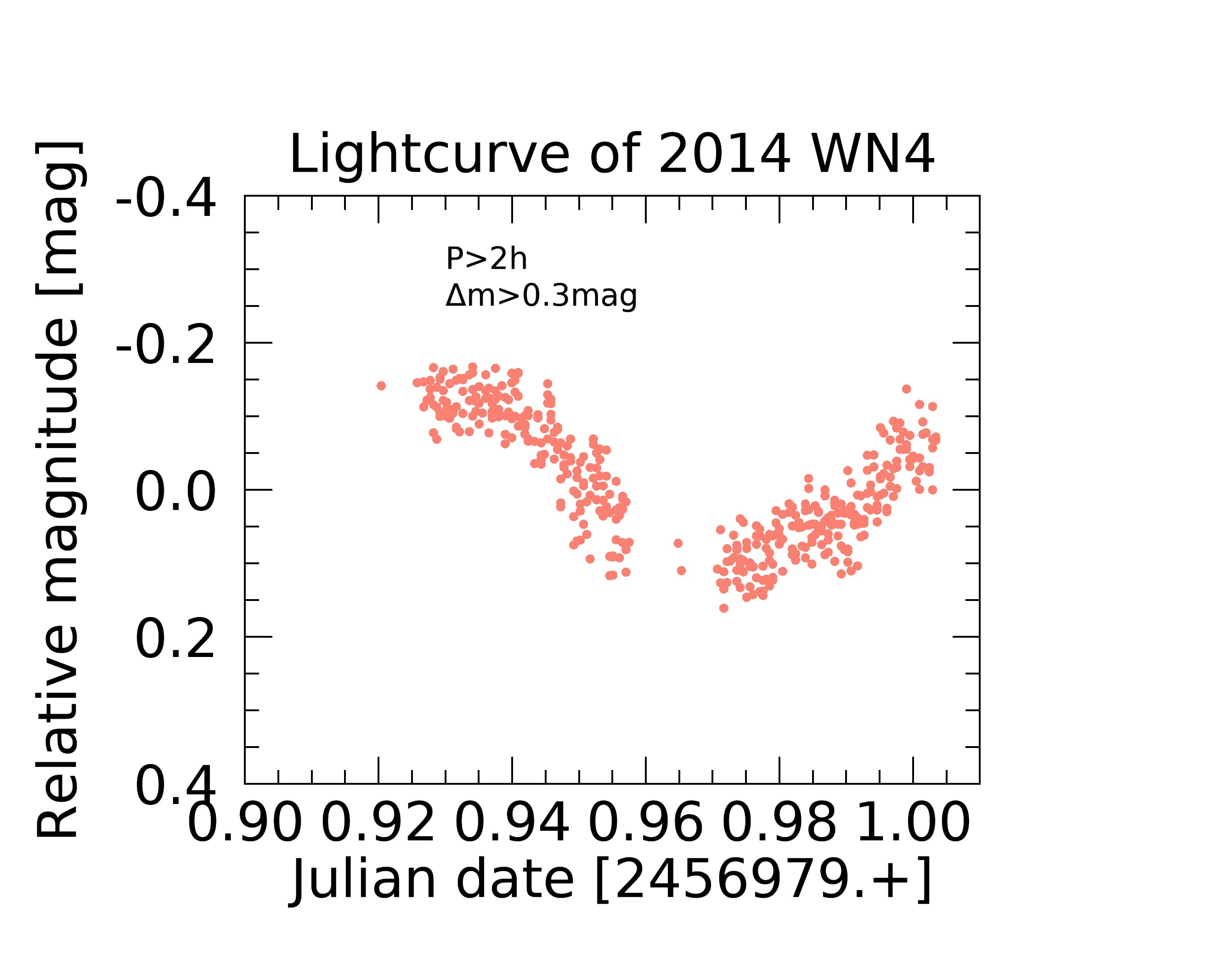}
 \includegraphics[width=9cm,angle=0]{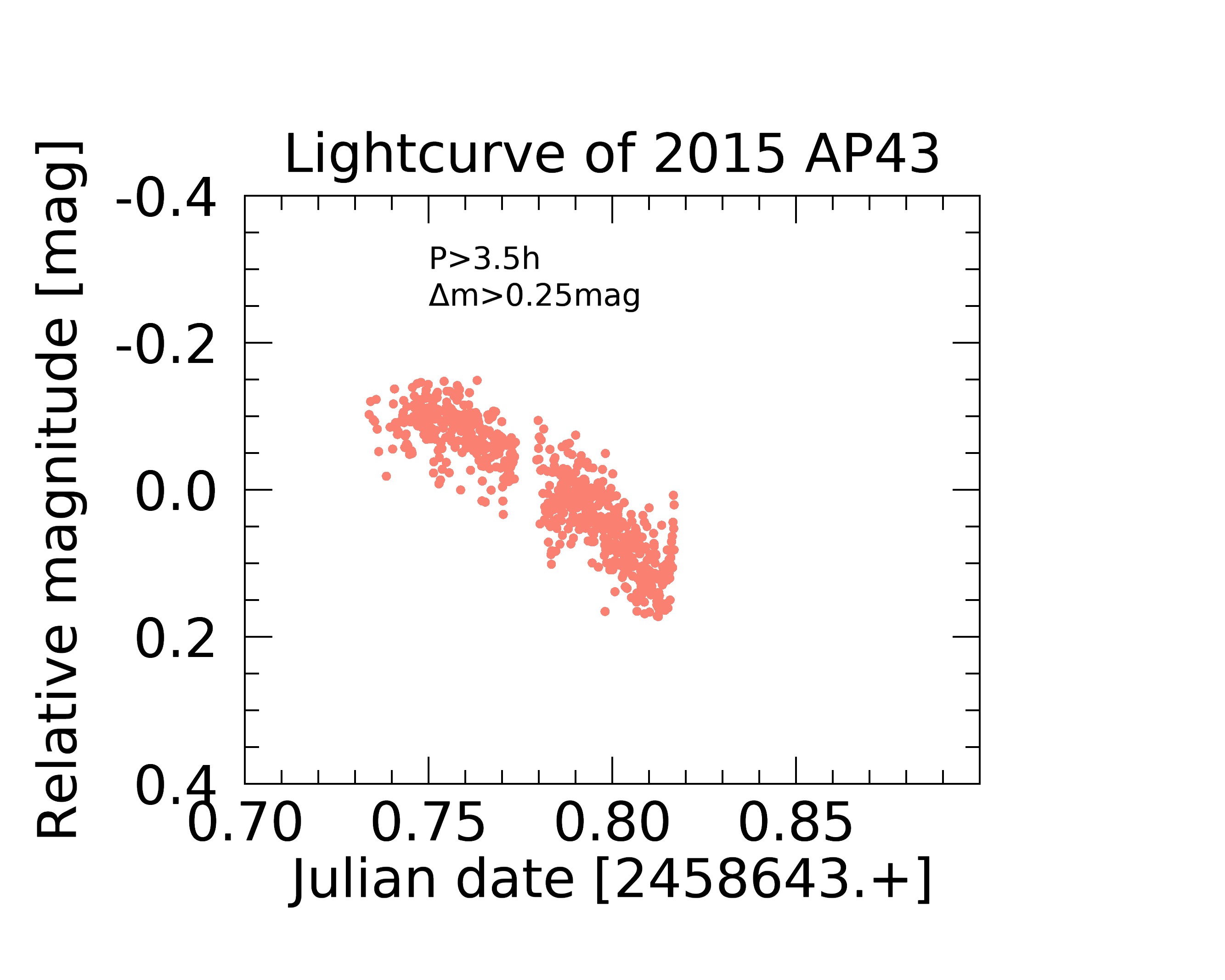}
 \includegraphics[width=9cm,angle=0]{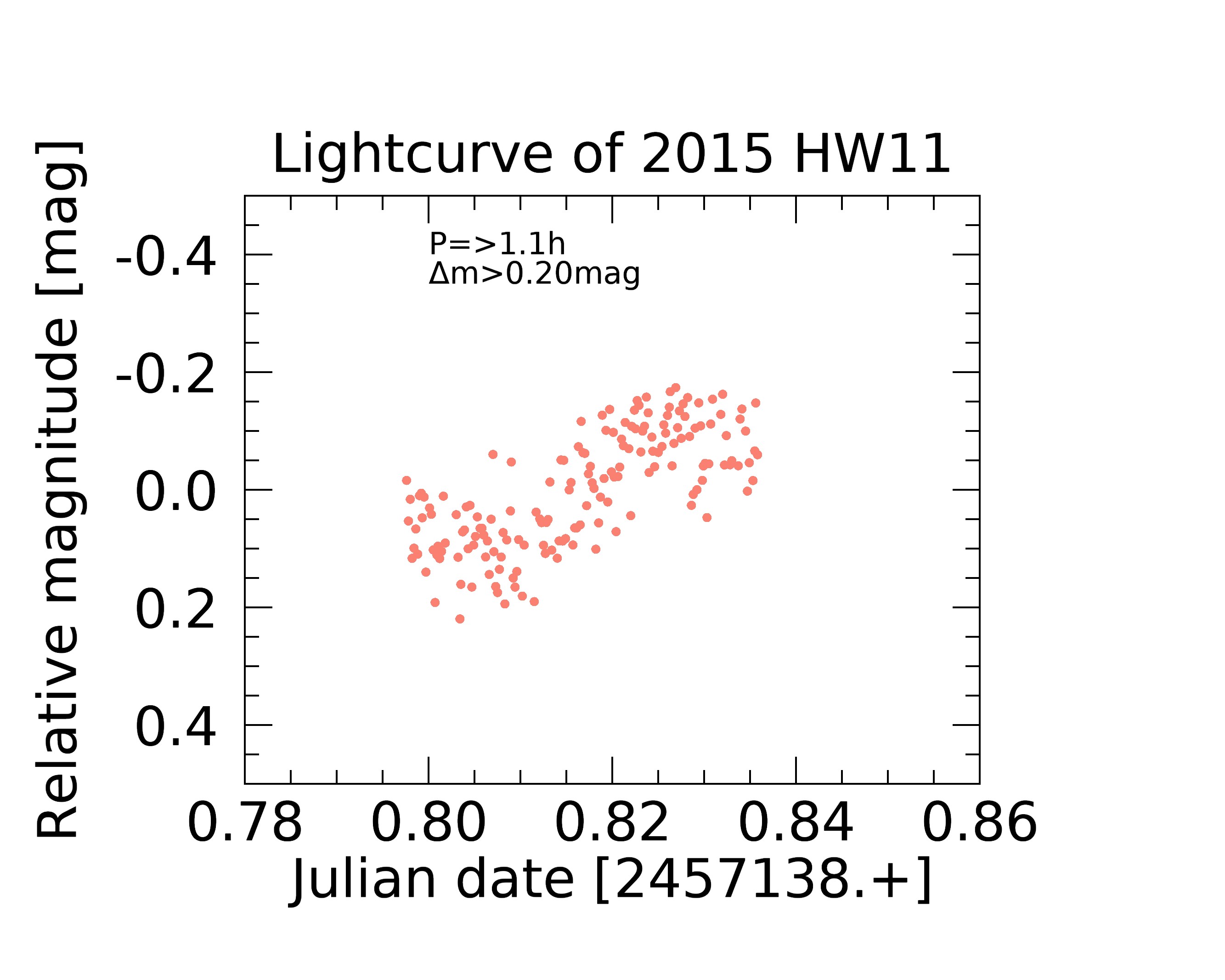}
 \includegraphics[width=9cm,angle=0]{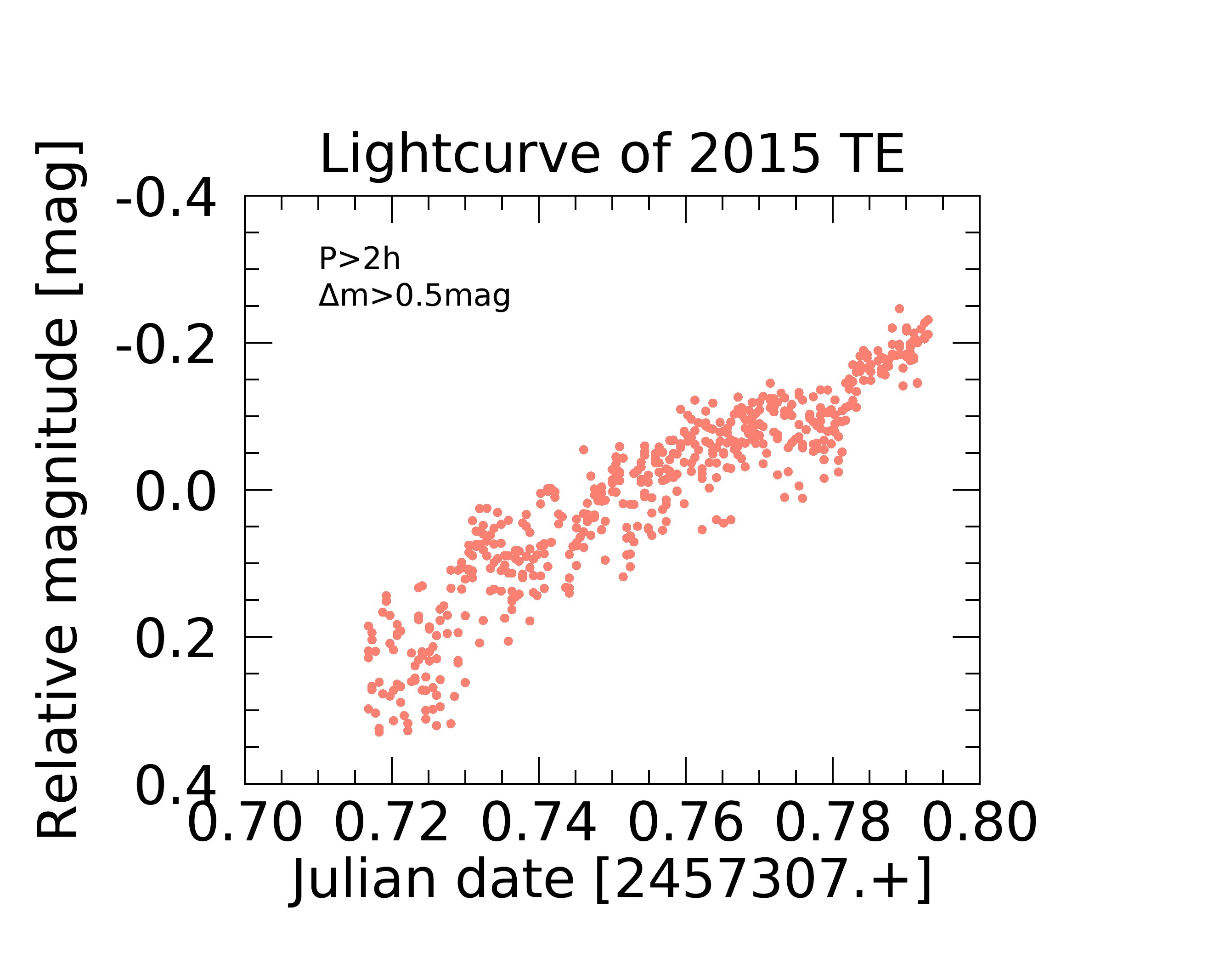}
 \includegraphics[width=9cm,angle=0]{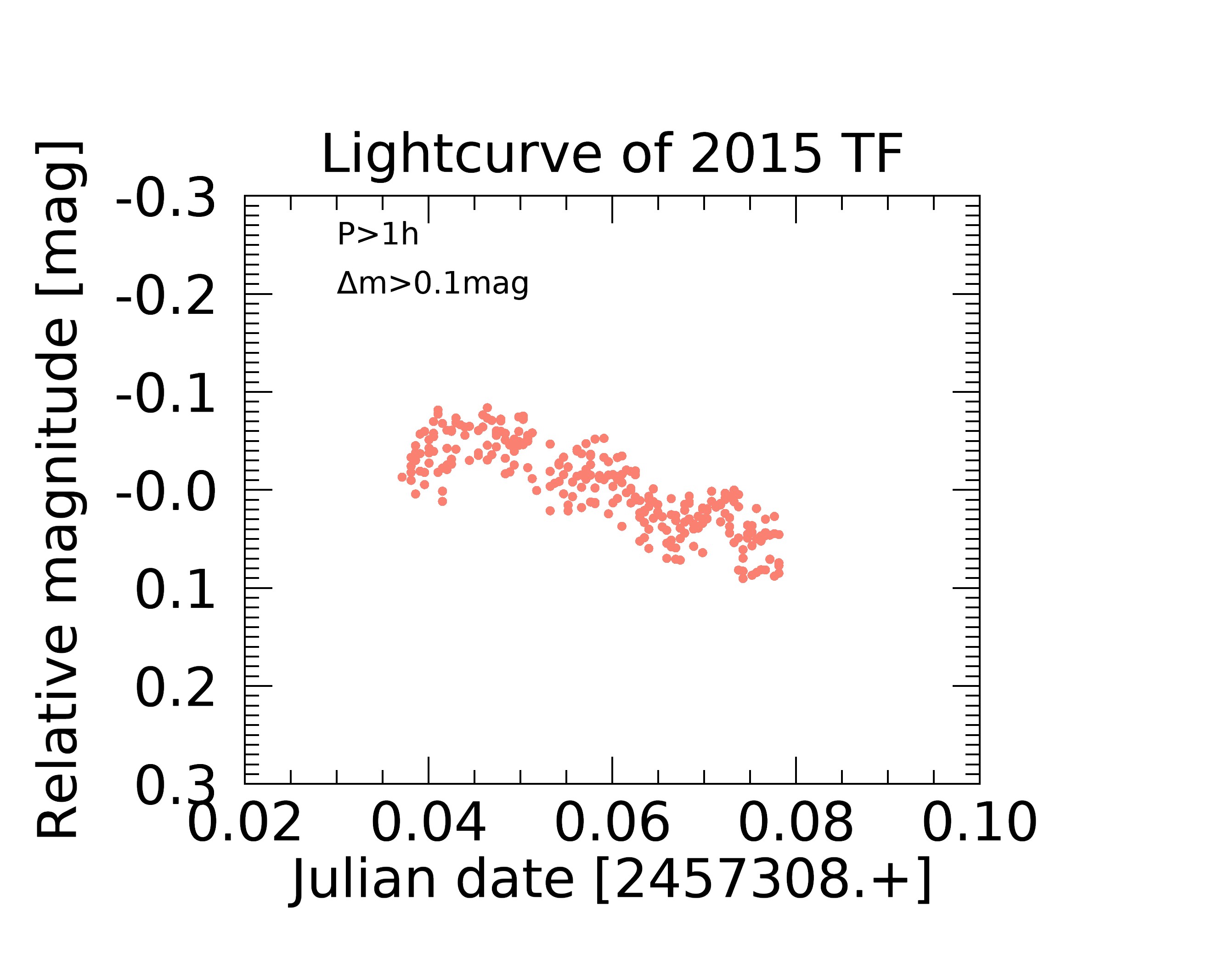}
\caption{Partial lightcurves of NEOs included in the photometric study.}
\label{fig:Partial_lightcurves}

\end{figure*}

 \begin{figure*}
 \includegraphics[width=9cm,angle=0]{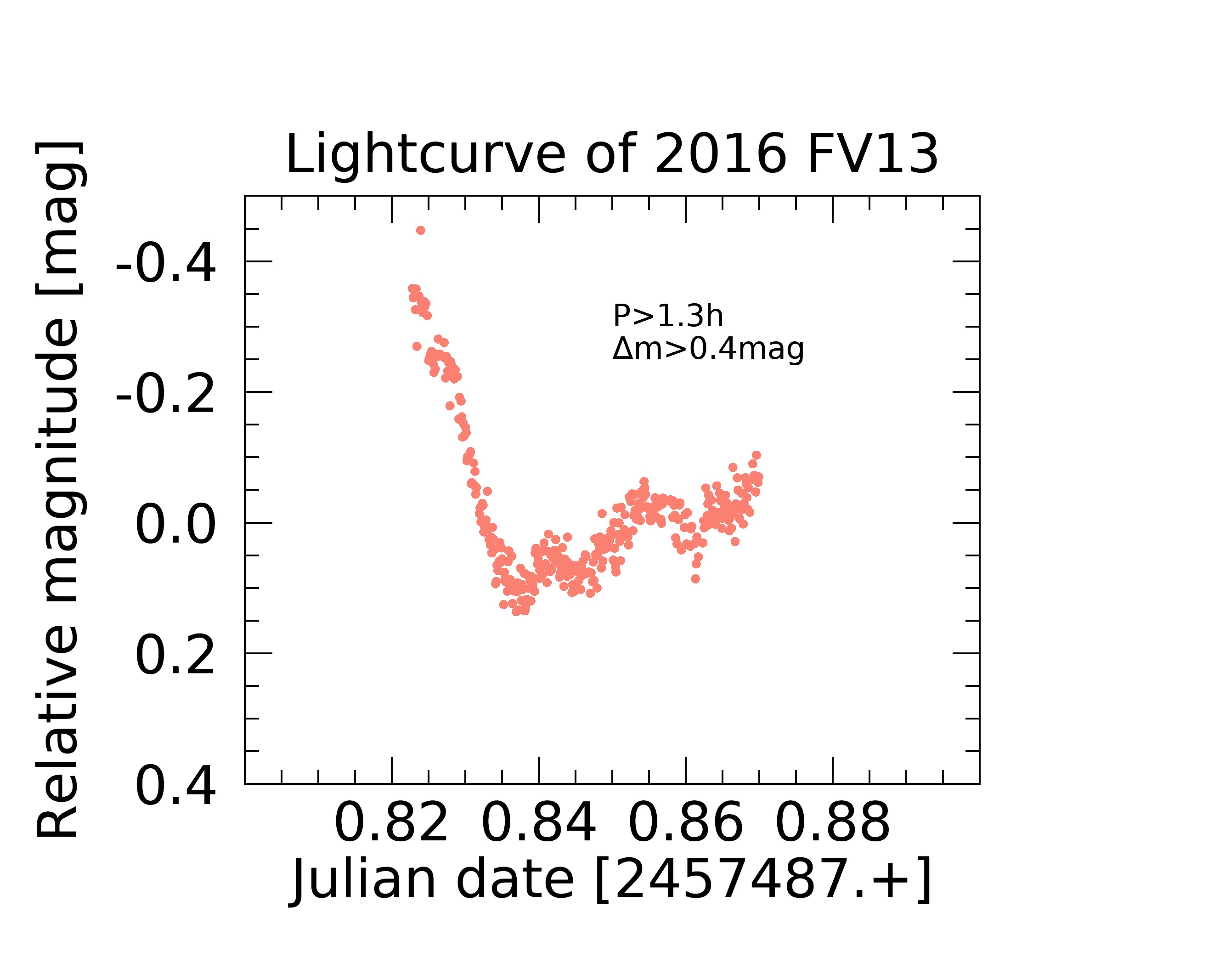}
 \includegraphics[width=9cm,angle=0]{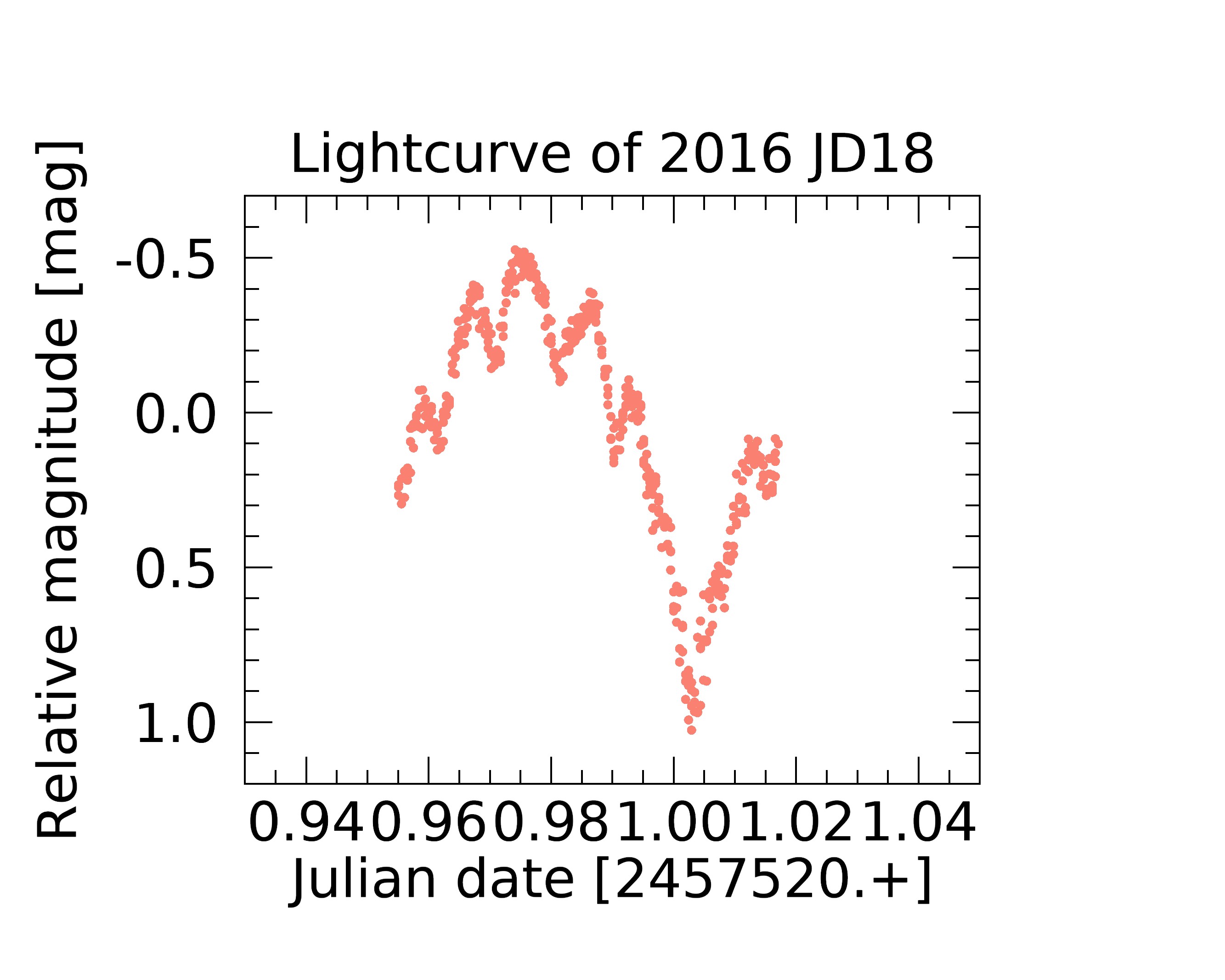}
 \includegraphics[width=9cm,angle=0]{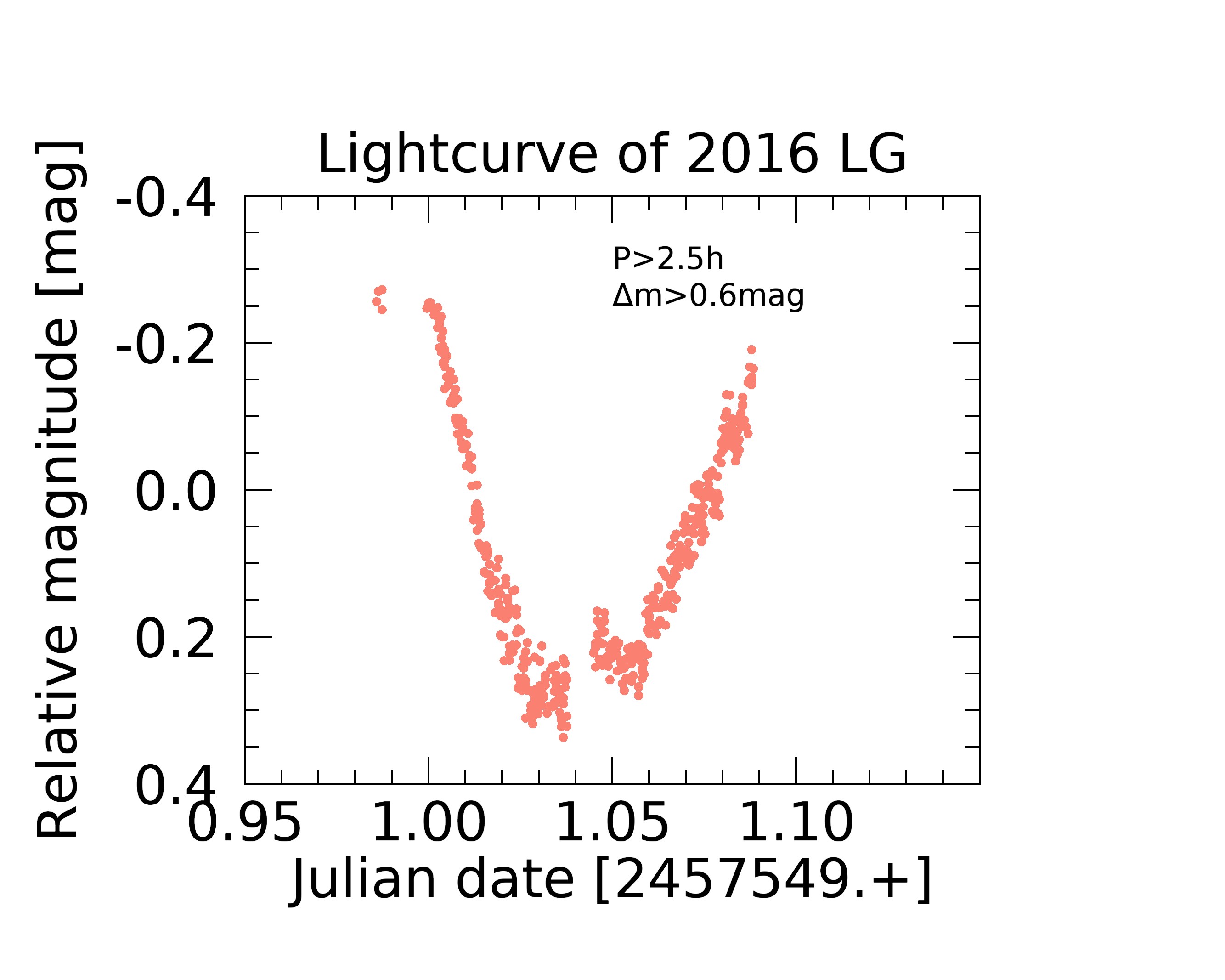}
 \includegraphics[width=9cm,angle=0]{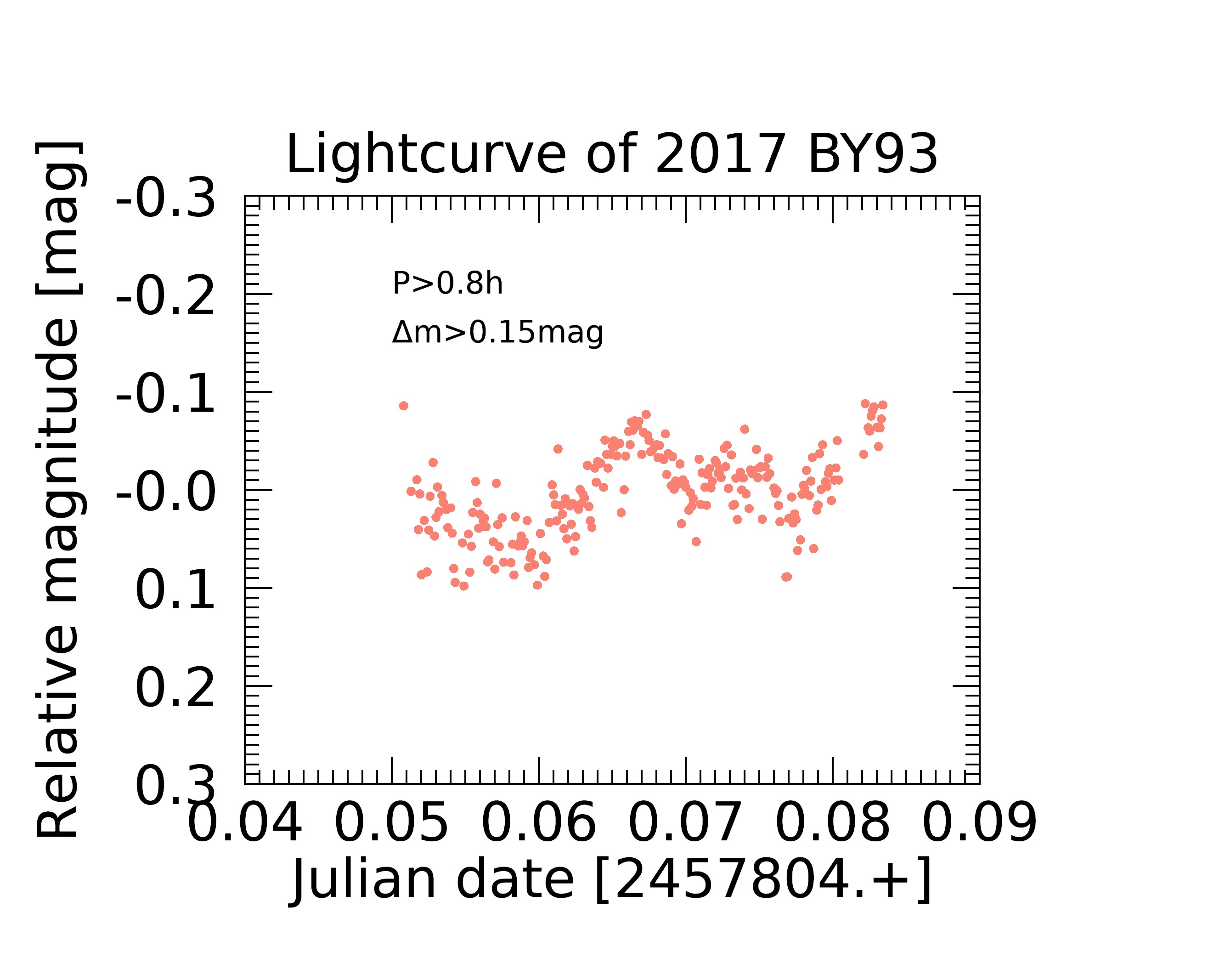}
 \includegraphics[width=9cm,angle=0]{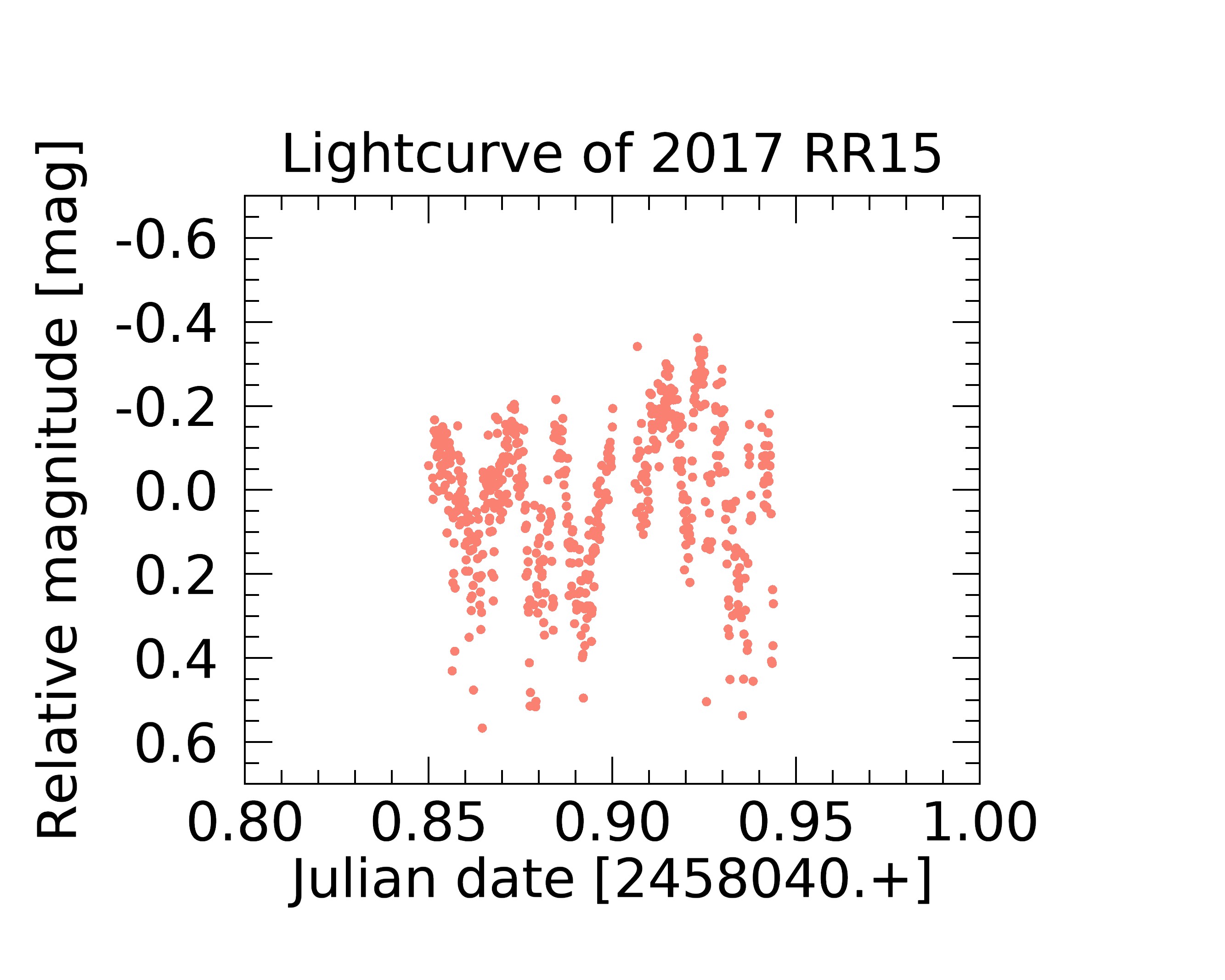}
\includegraphics[width=9cm,angle=0]{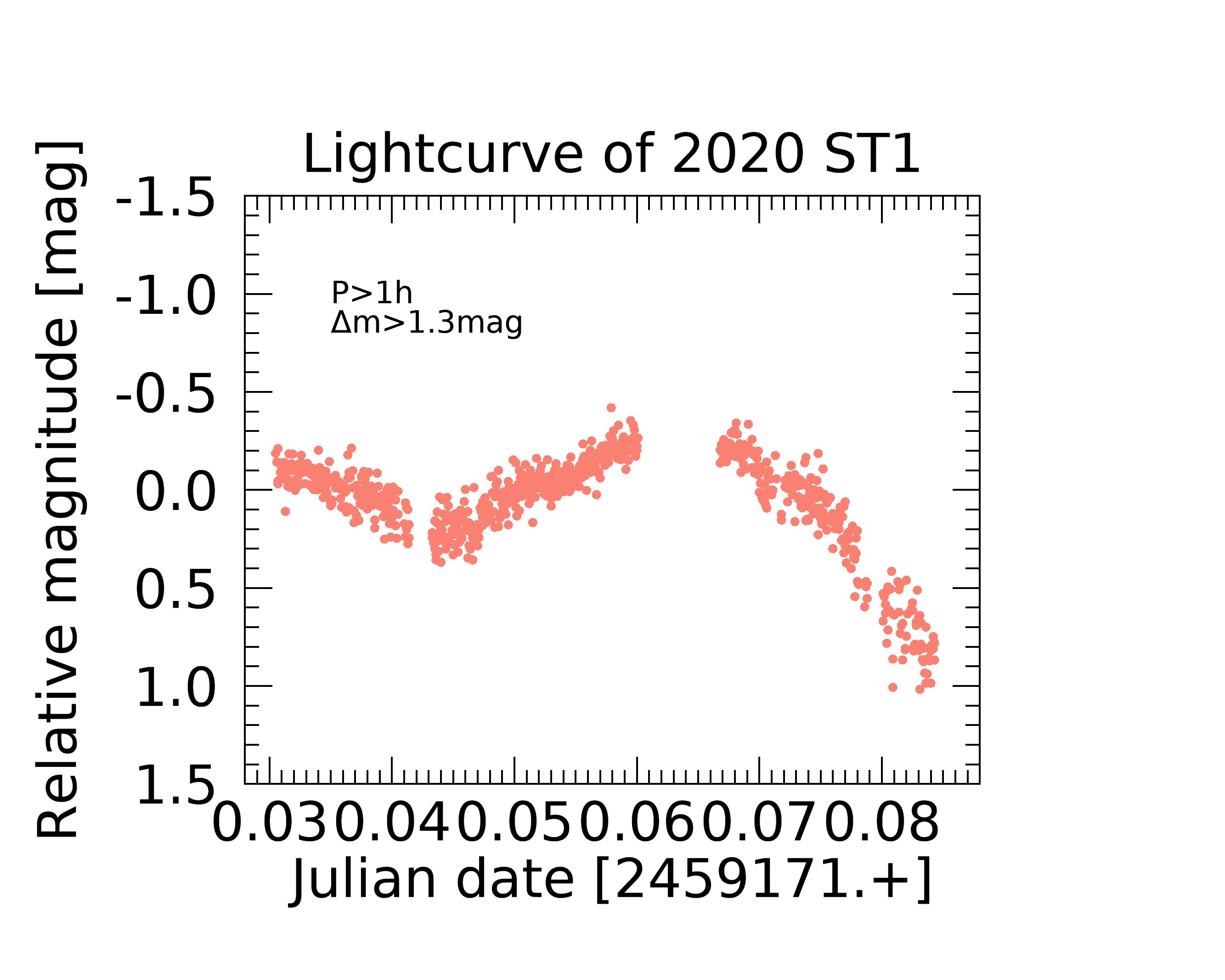}
\caption{Partial lightcurves of NEOs included in the photometric study.}
\label{fig:Partial_lightcurves}

\end{figure*}

 \begin{figure*}
  \includegraphics[width=9cm,angle=0]{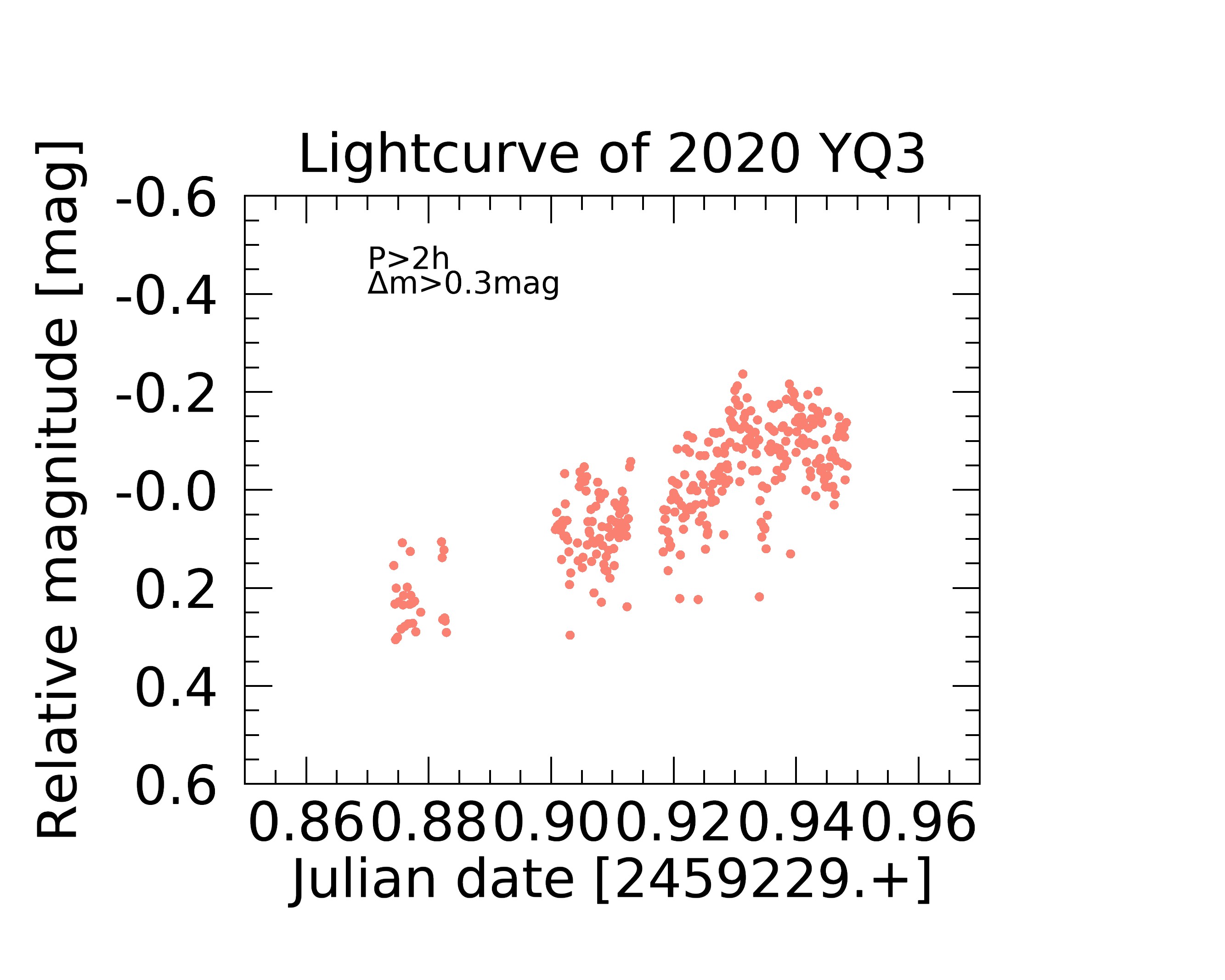}
 \caption{Partial lightcurves of NEOs included in the photometric study.}
\label{fig:Partial_lightcurves}
\end{figure*}

\clearpage

\textbf{Appendix B.3.}

\begin{figure*}[!h]
 \includegraphics[width=9cm,angle=0]{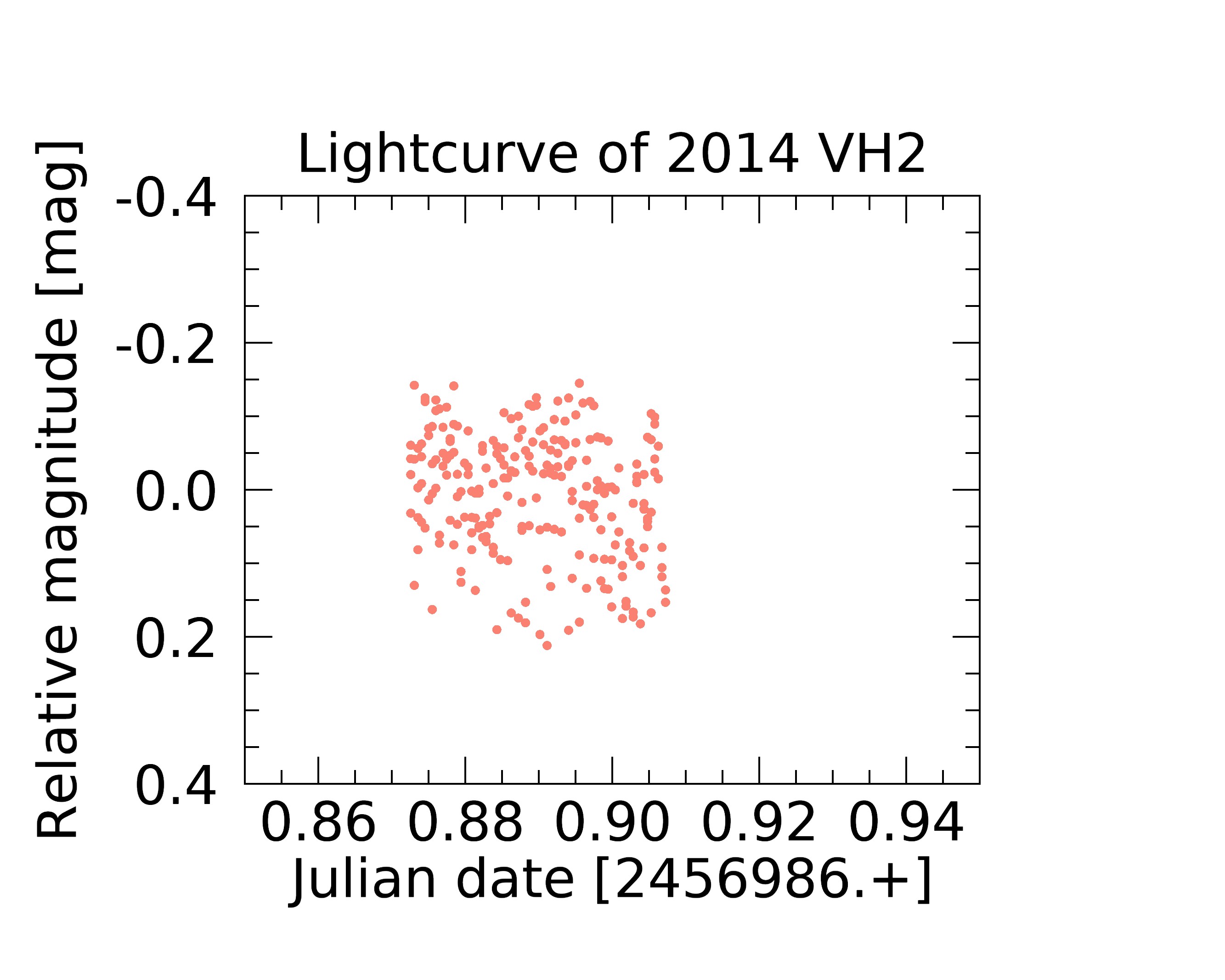} 
 \includegraphics[width=9cm,angle=0]{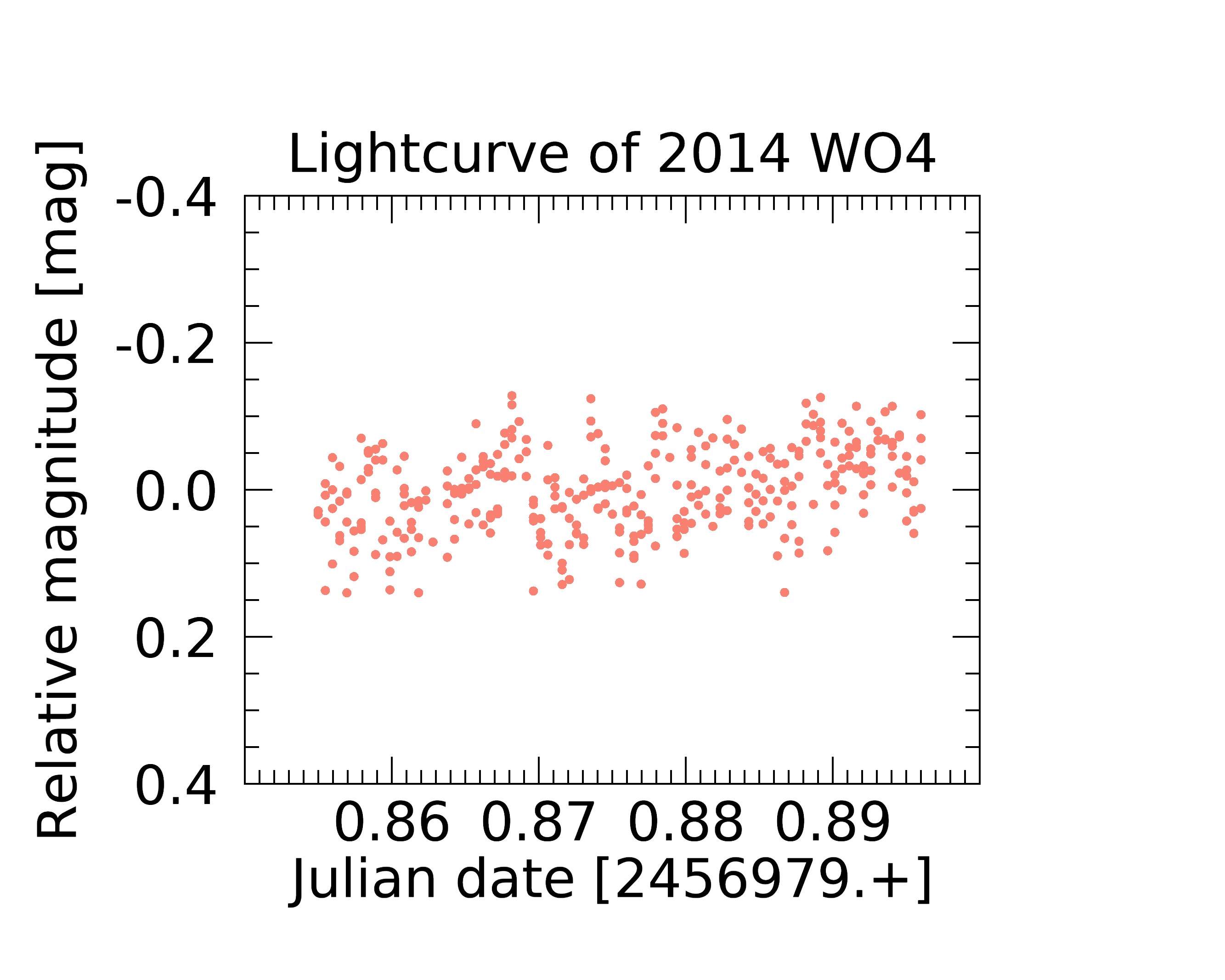} 
 \includegraphics[width=9cm,angle=0]{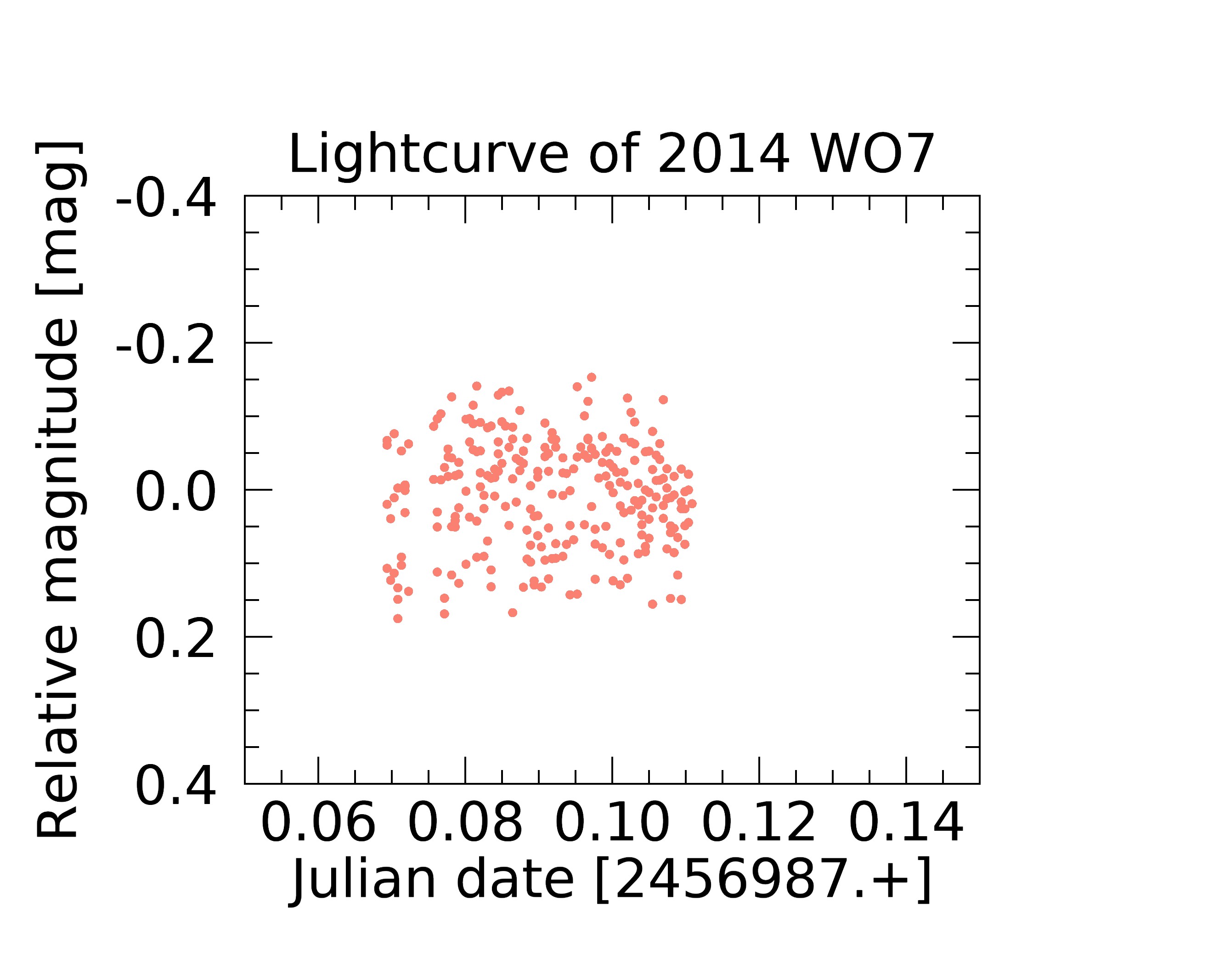} 
 \includegraphics[width=9cm,angle=0]{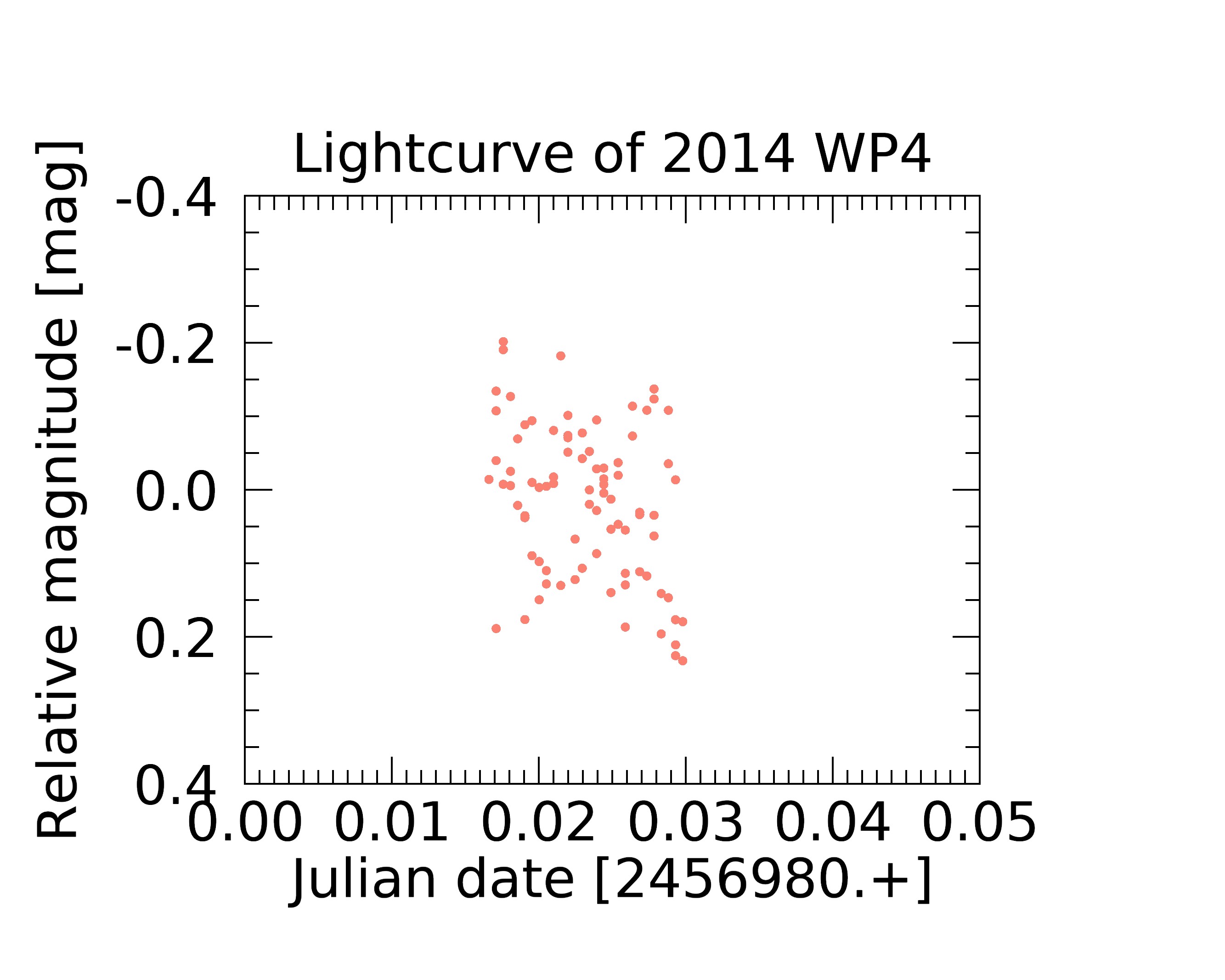} 
 \caption{Flat lightcurves of NEOs included in the photometric study.}
\label{fig:Flat_lightcurves}
 
 \end{figure*}
 
  \begin{figure*}
 \includegraphics[width=9cm,angle=0]{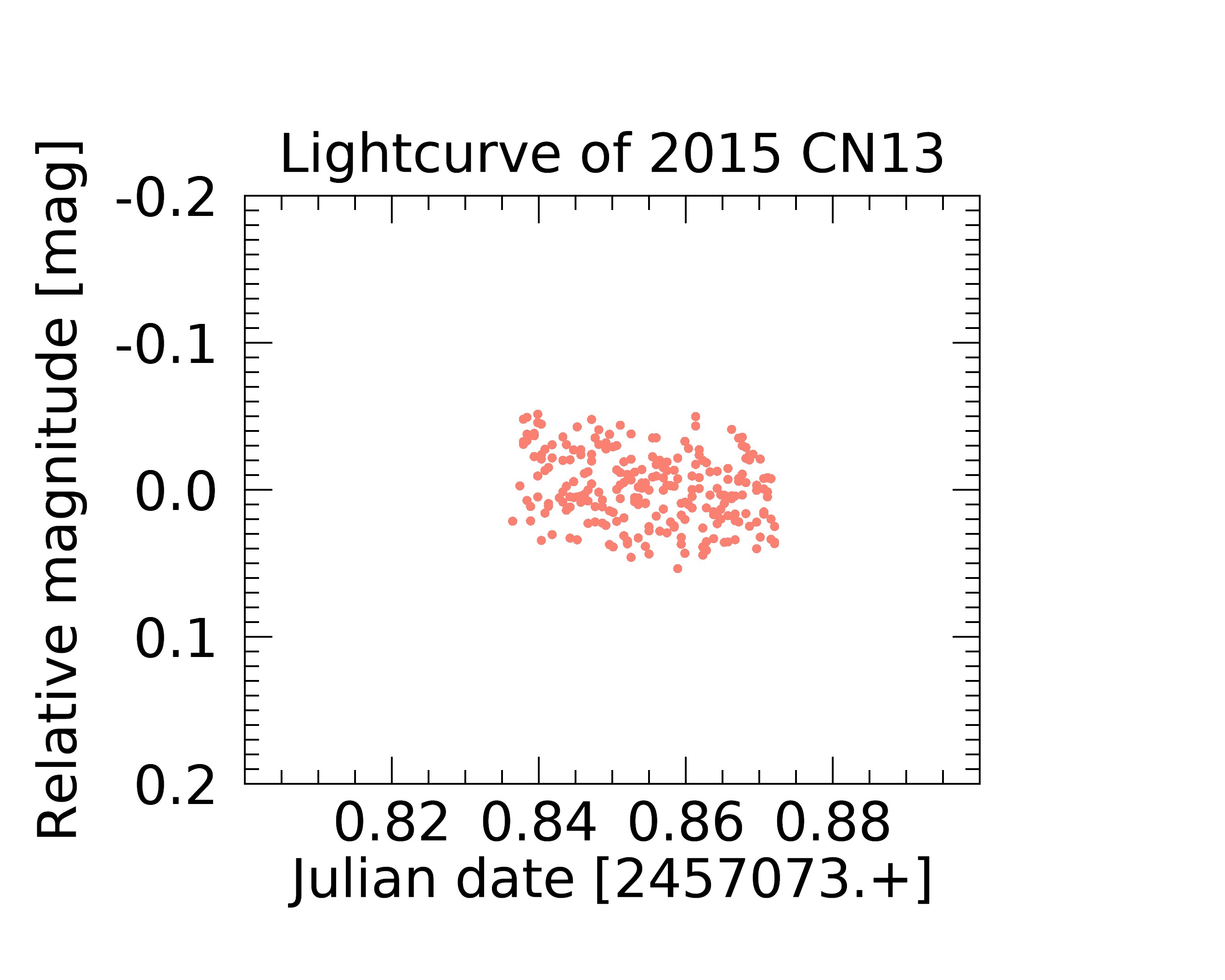} 
 \includegraphics[width=9cm,angle=0]{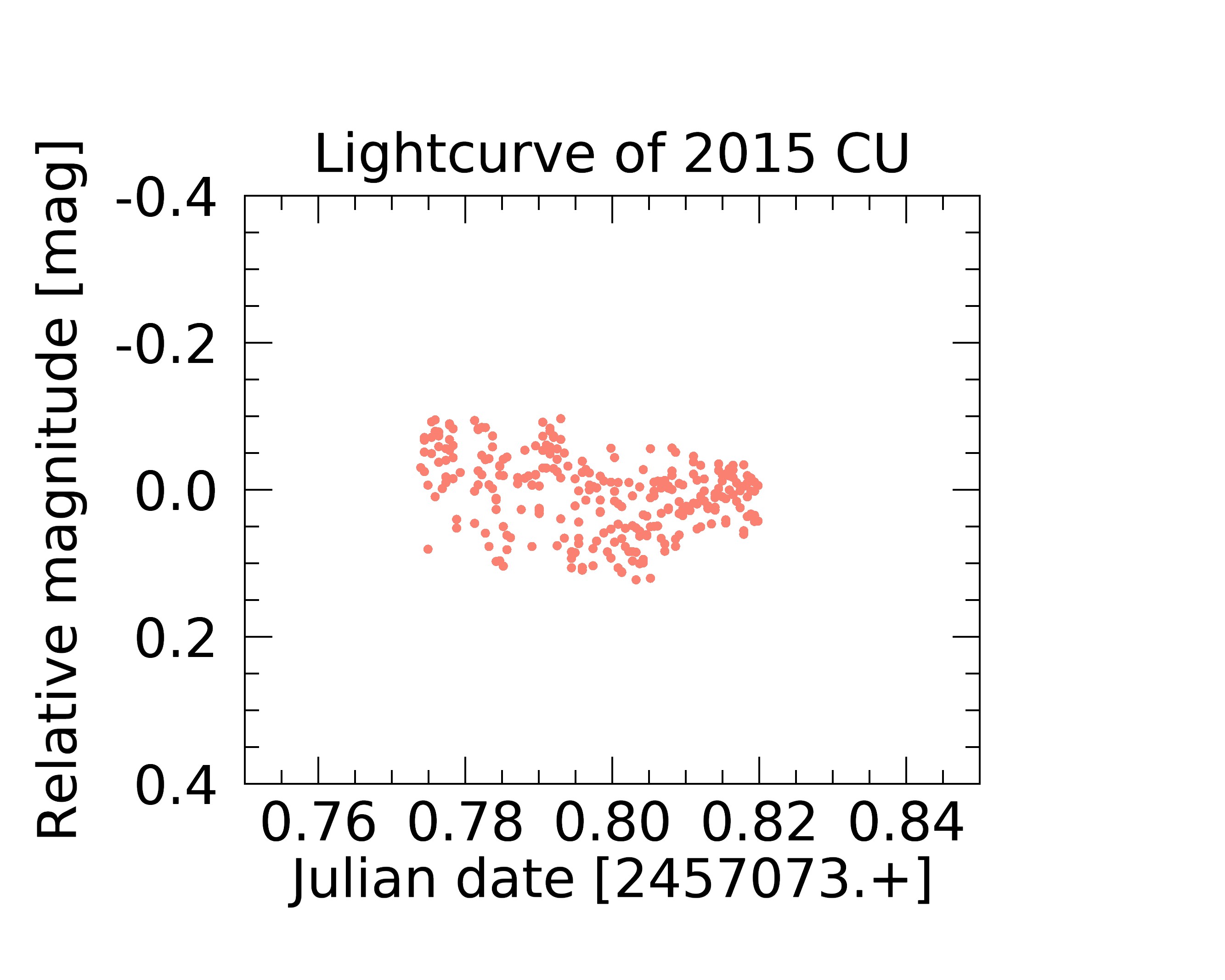} 
\includegraphics[width=9cm,angle=0]{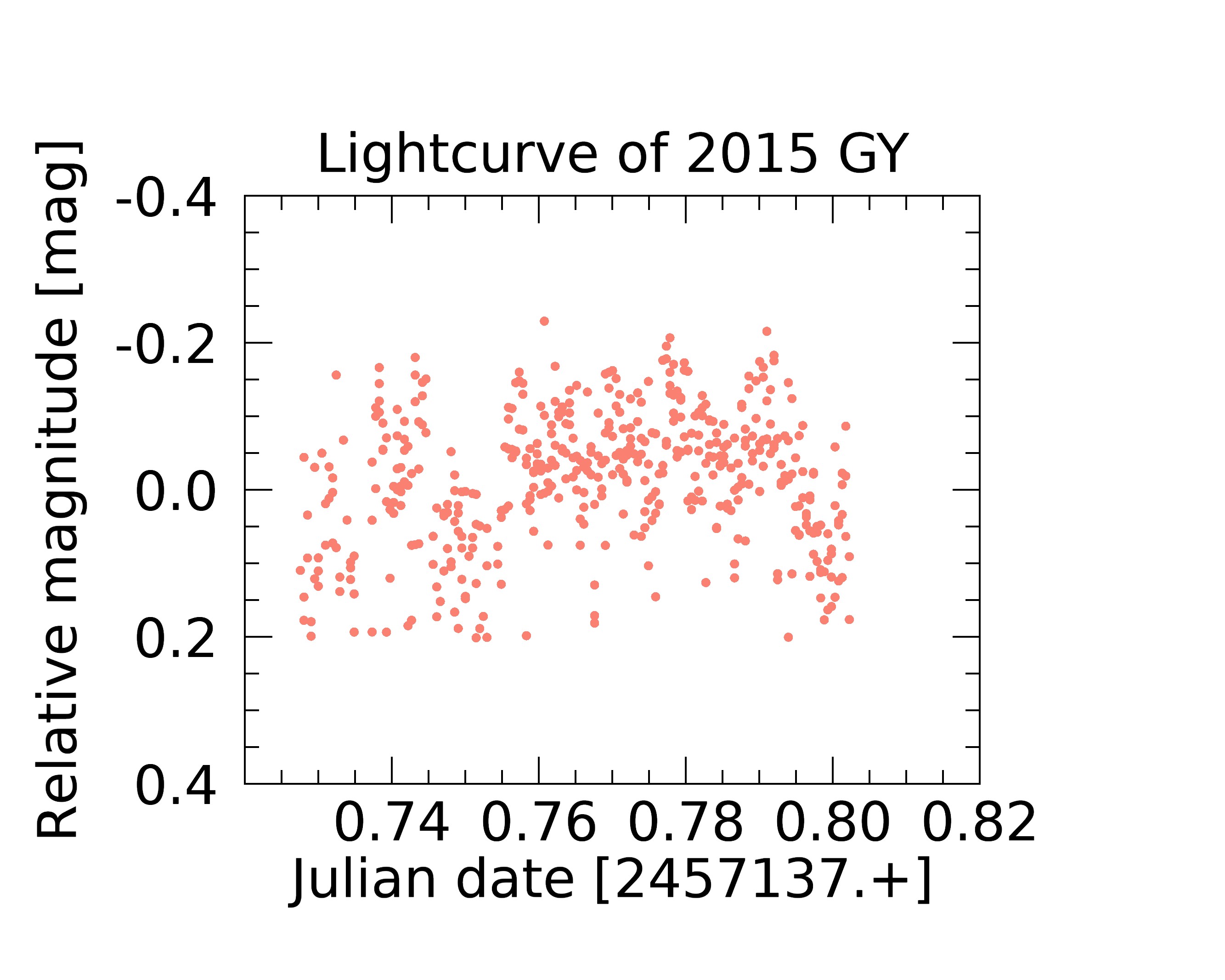} 
 \includegraphics[width=9cm,angle=0]{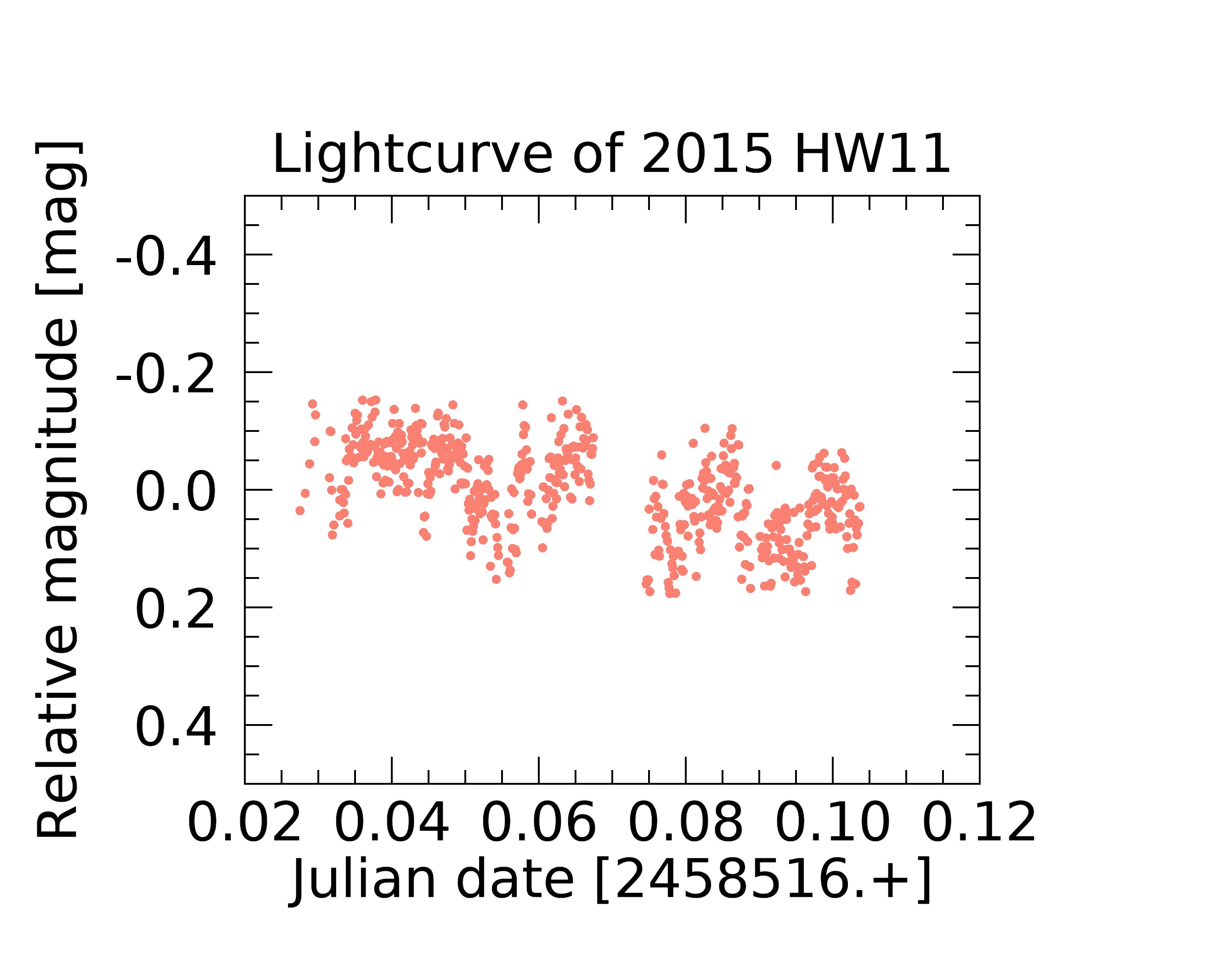} 
 \includegraphics[width=9cm,angle=0]{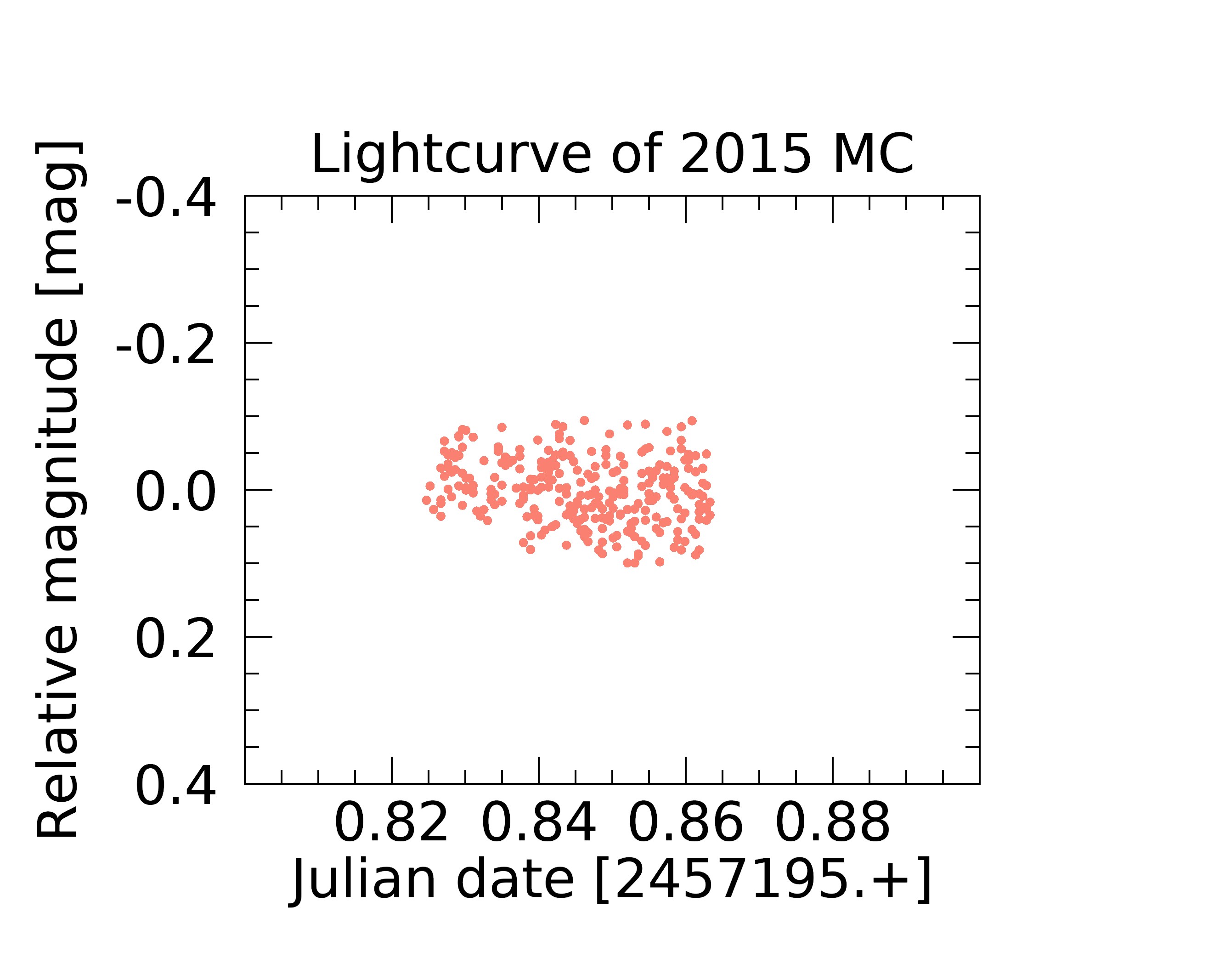} 
 \includegraphics[width=9cm,angle=0]{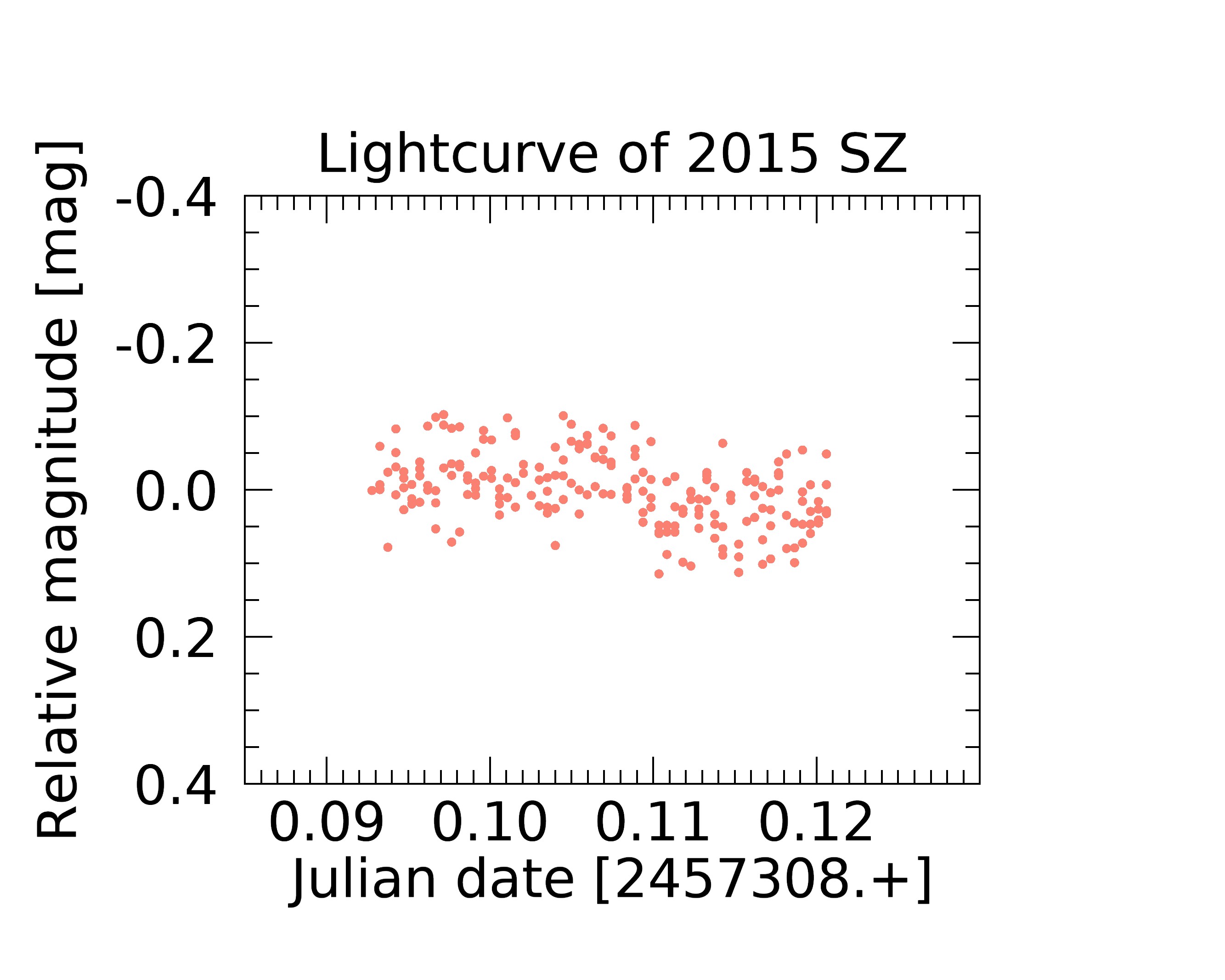}
 \caption{Flat lightcurves of NEOs included in the photometric study.}
\label{fig:Flat_lightcurves}
  
 \end{figure*}

 \begin{figure*}
 \includegraphics[width=9cm,angle=0]{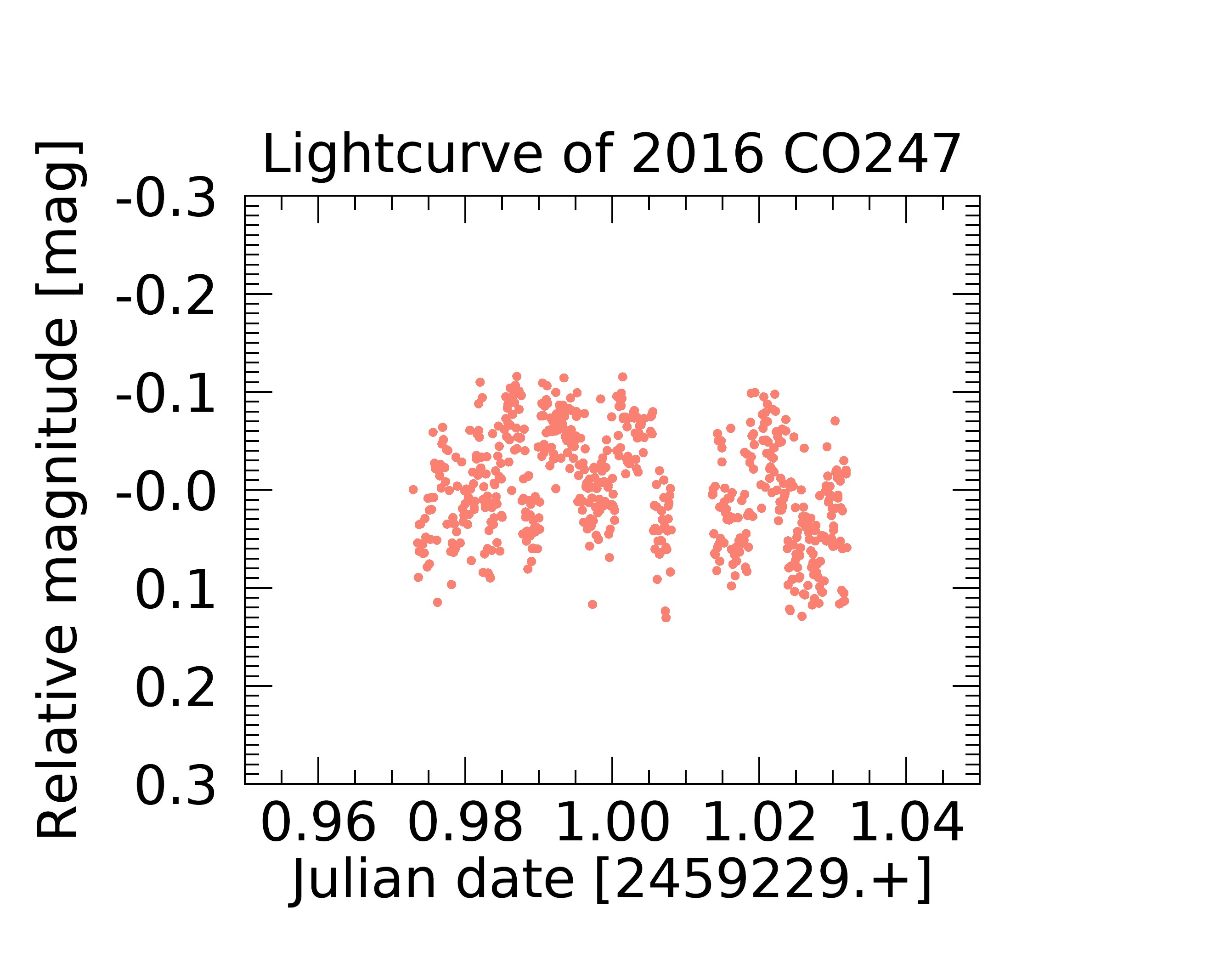}  
 \includegraphics[width=9cm,angle=0]{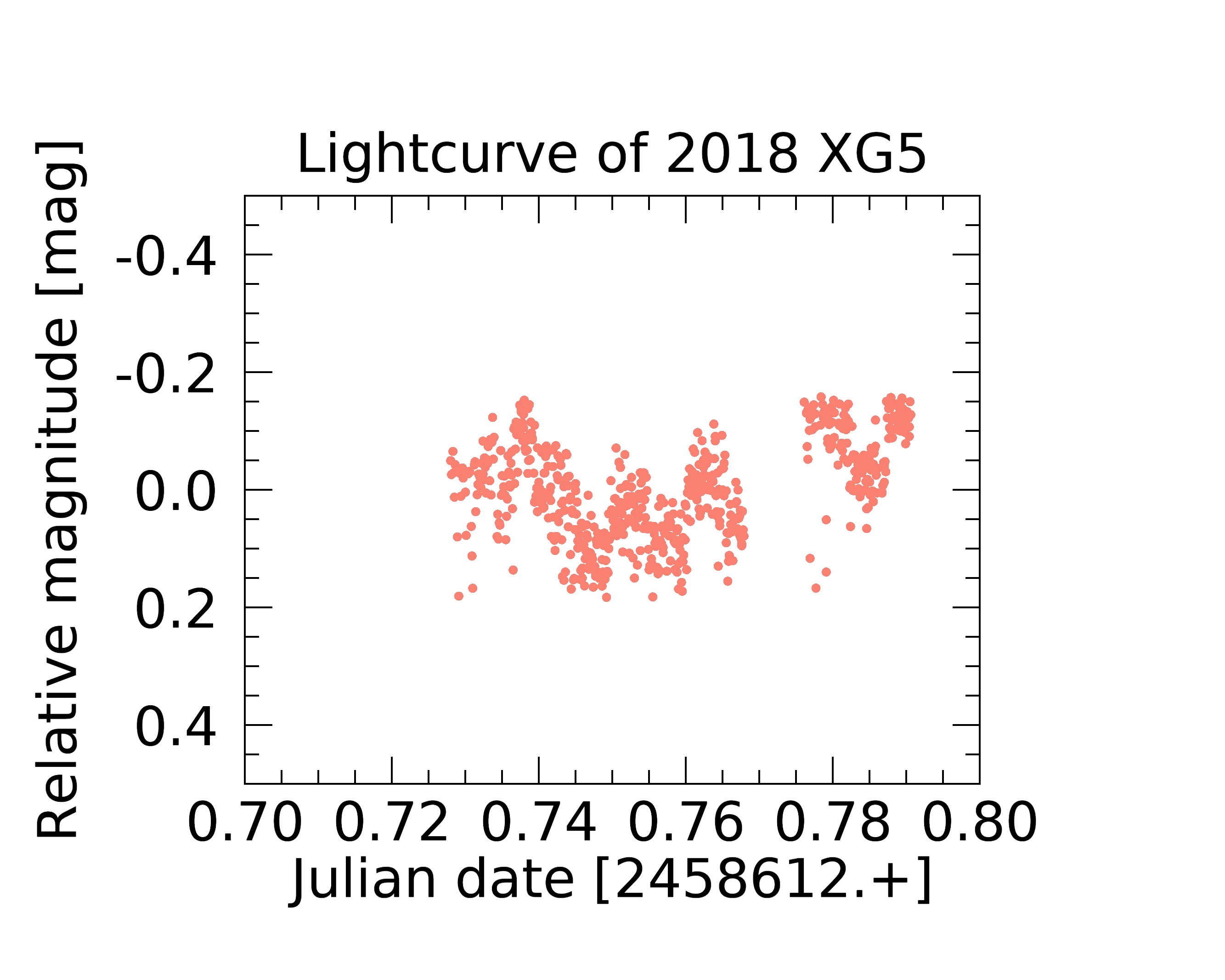}  
 \includegraphics[width=9cm,angle=0]{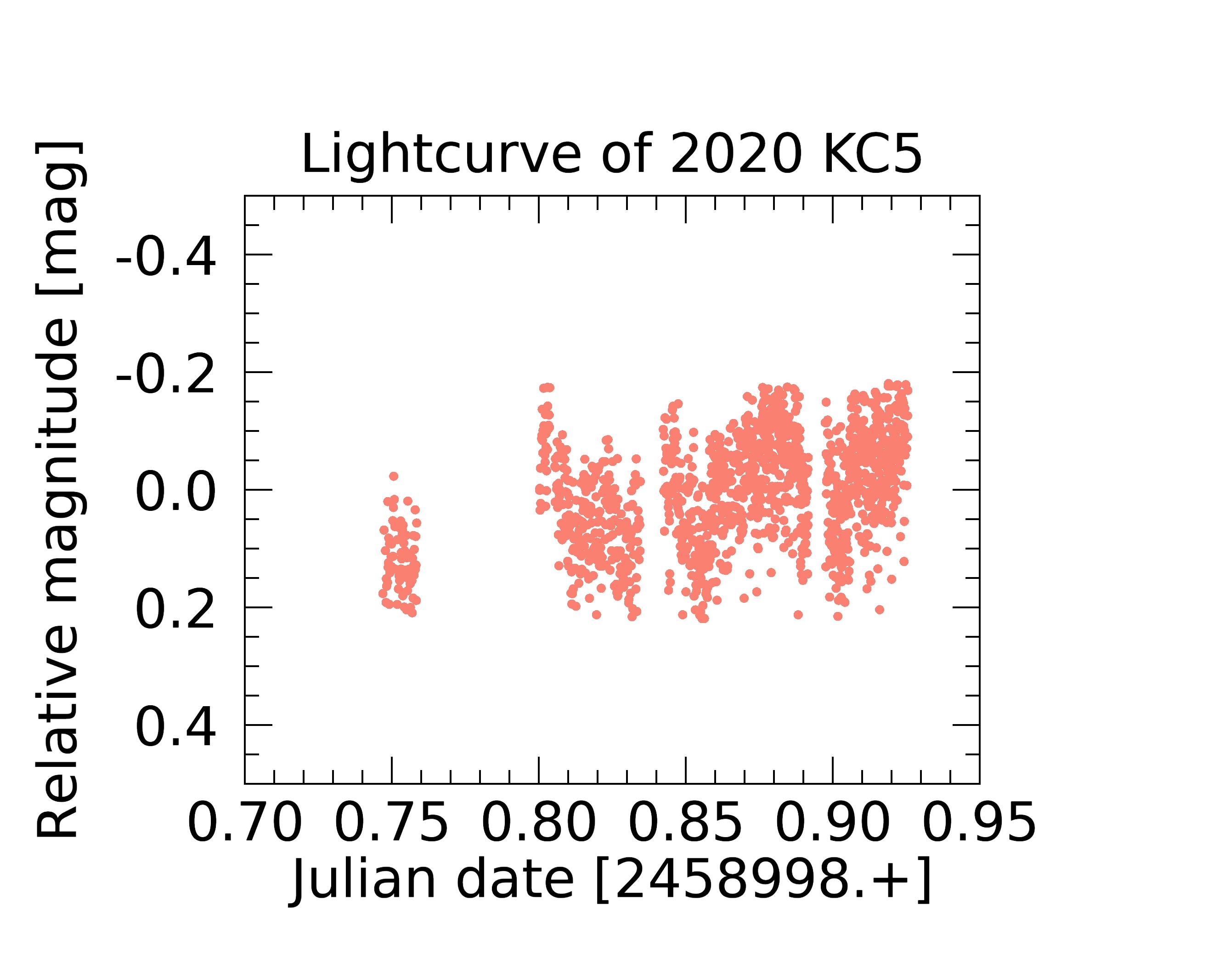}
 \includegraphics[width=9cm,angle=0]{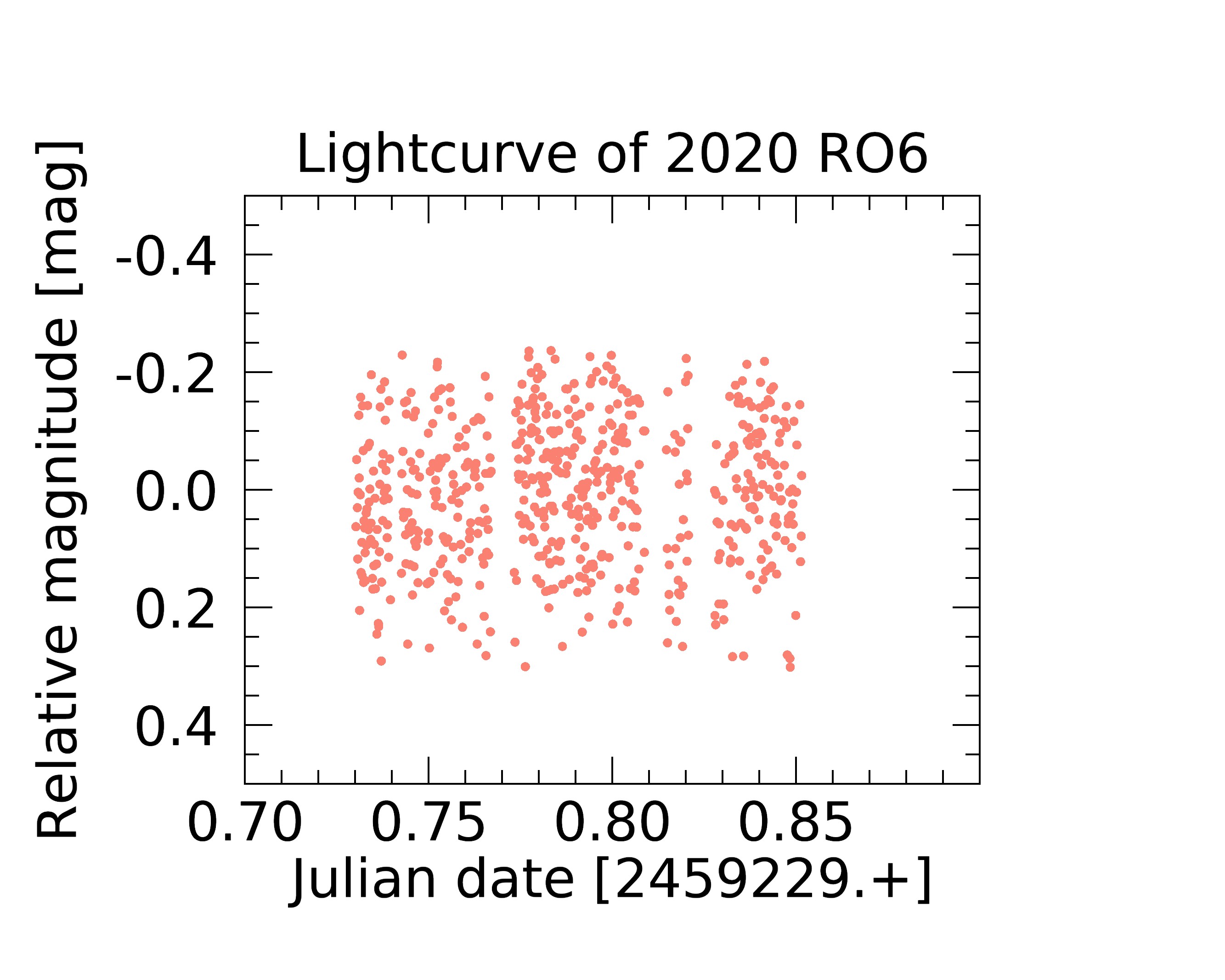}    
\caption{Flat lightcurves of NEOs included in the photometric study.}
\label{fig:Flat_lightcurves}

\end{figure*}

\clearpage

\textbf{Appendix B.4.}

\vspace{-5mm}

 \begin{figure*}[!h]
 \includegraphics[width=9cm,angle=0]{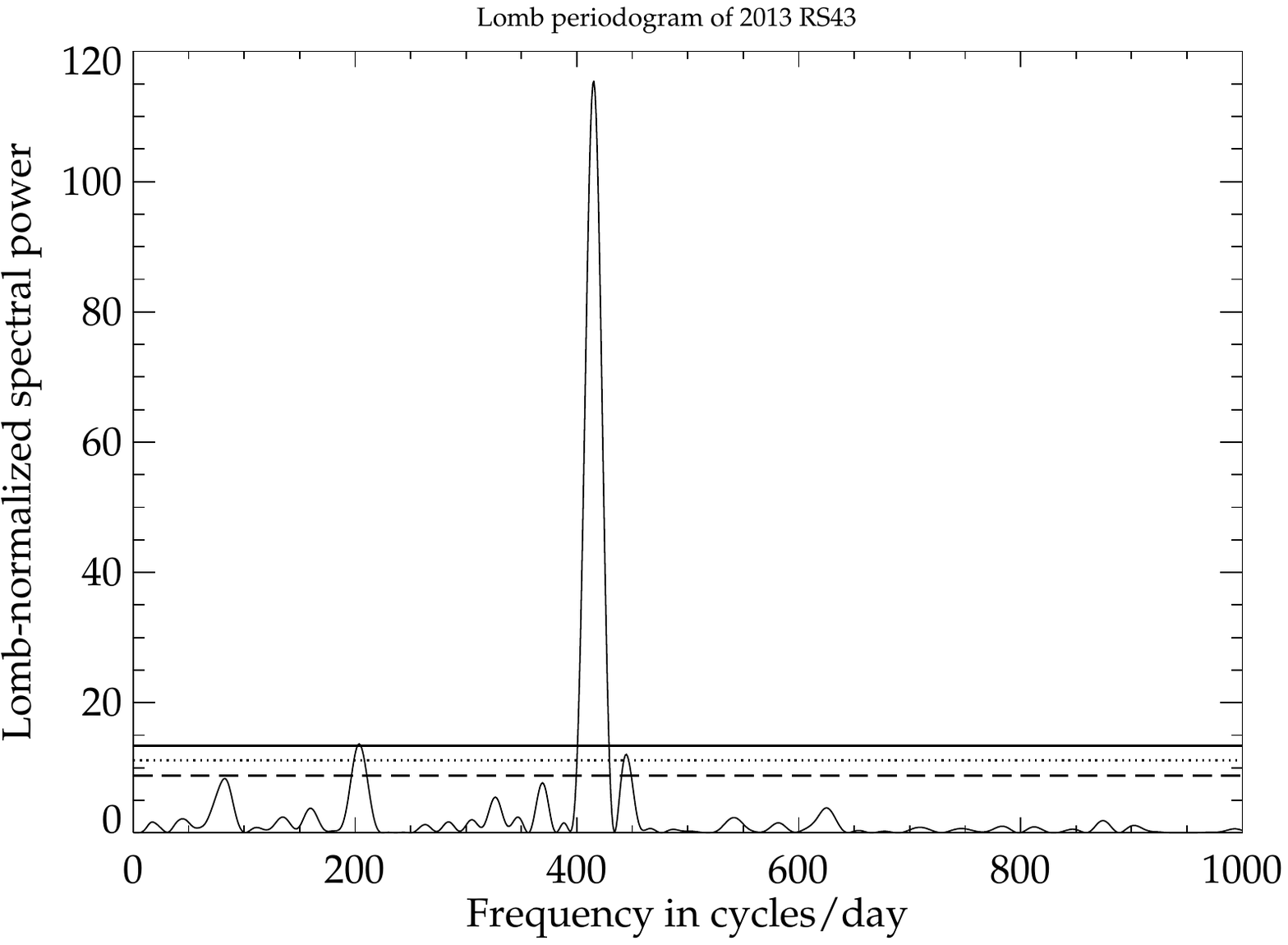}  
 \includegraphics[width=9cm,angle=0]{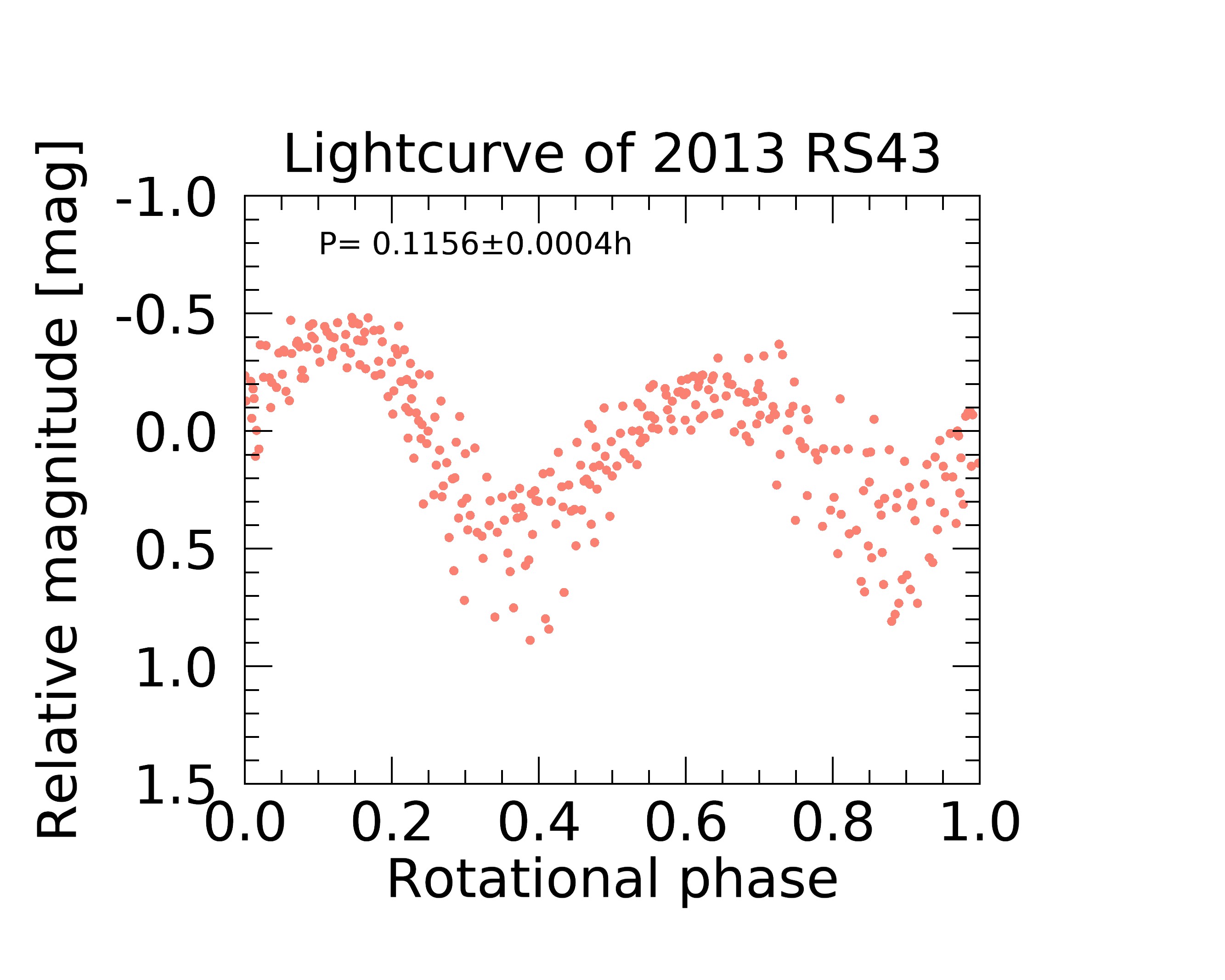}
 \includegraphics[width=9cm,angle=0]{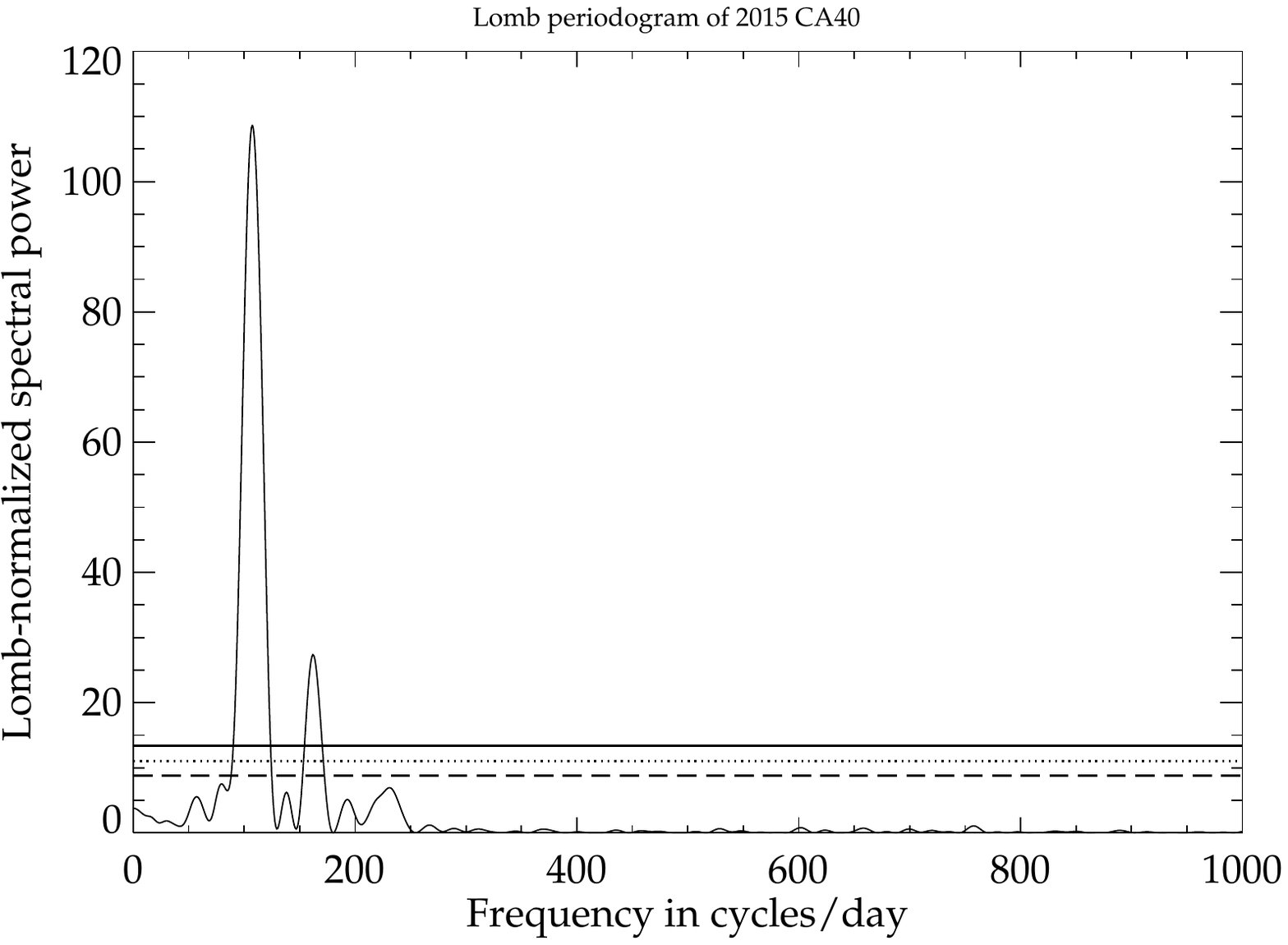}  
 \includegraphics[width=9cm,angle=0]{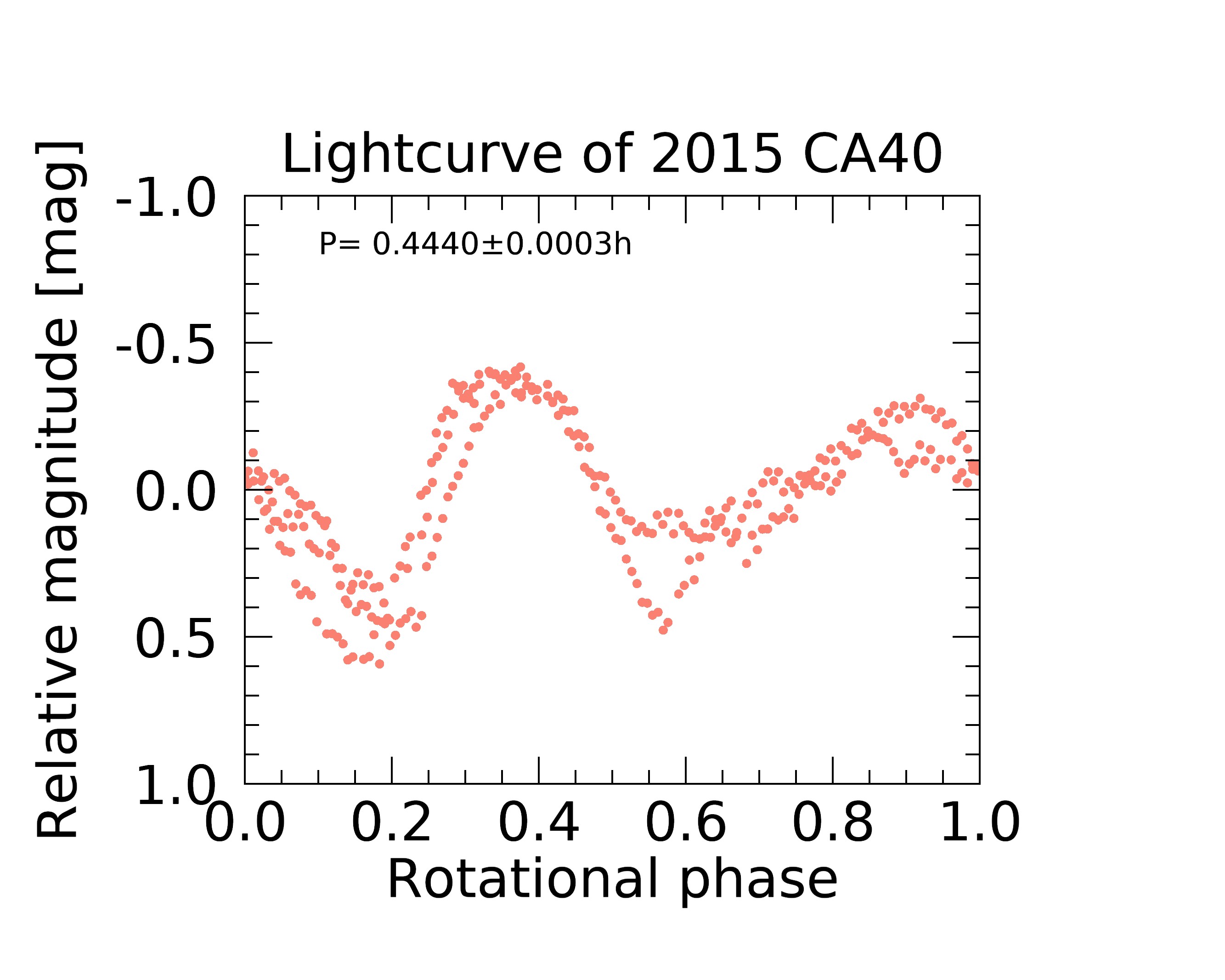}
 \includegraphics[width=9cm,angle=0]{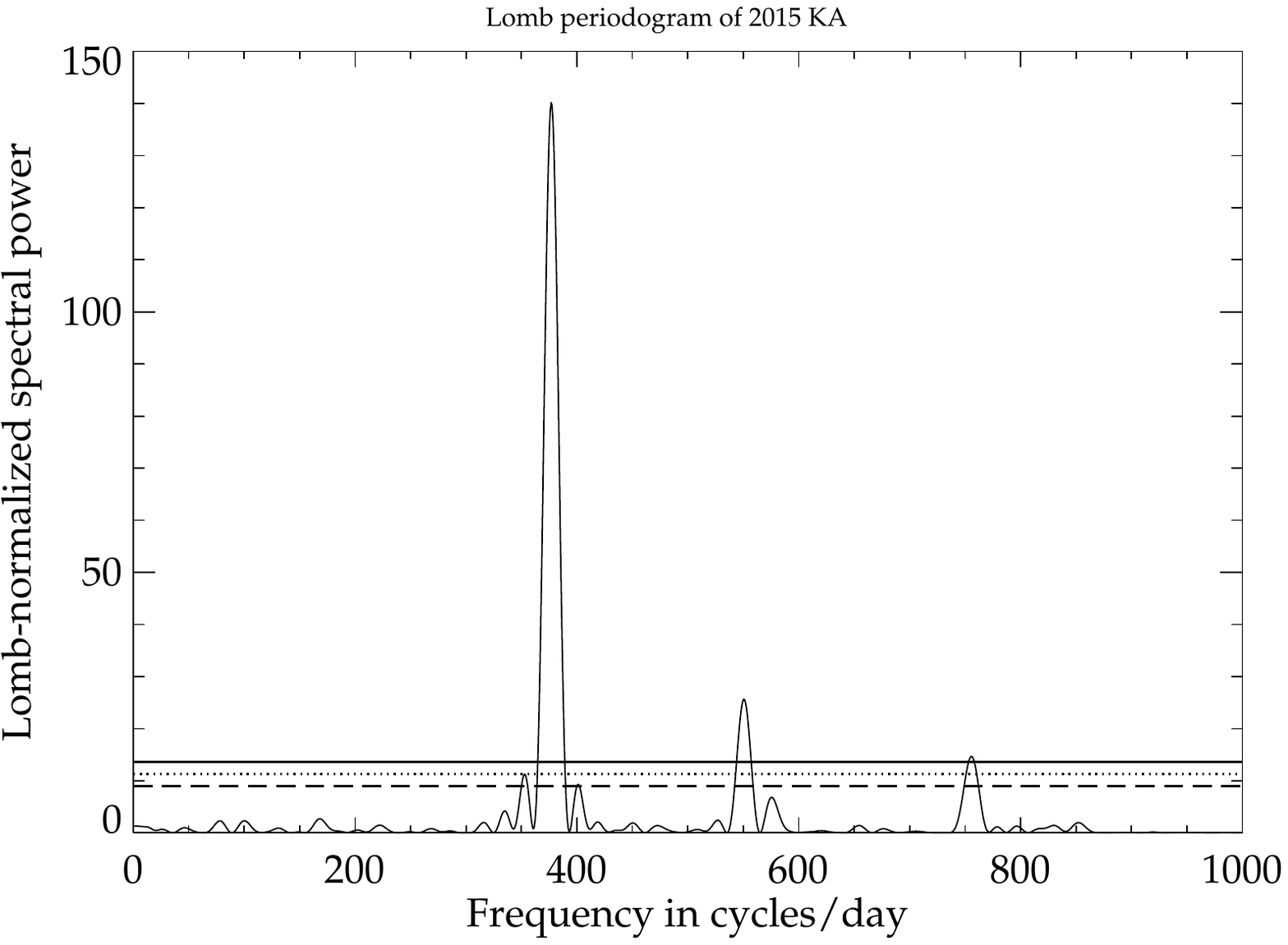}  
 \includegraphics[width=9cm,angle=0]{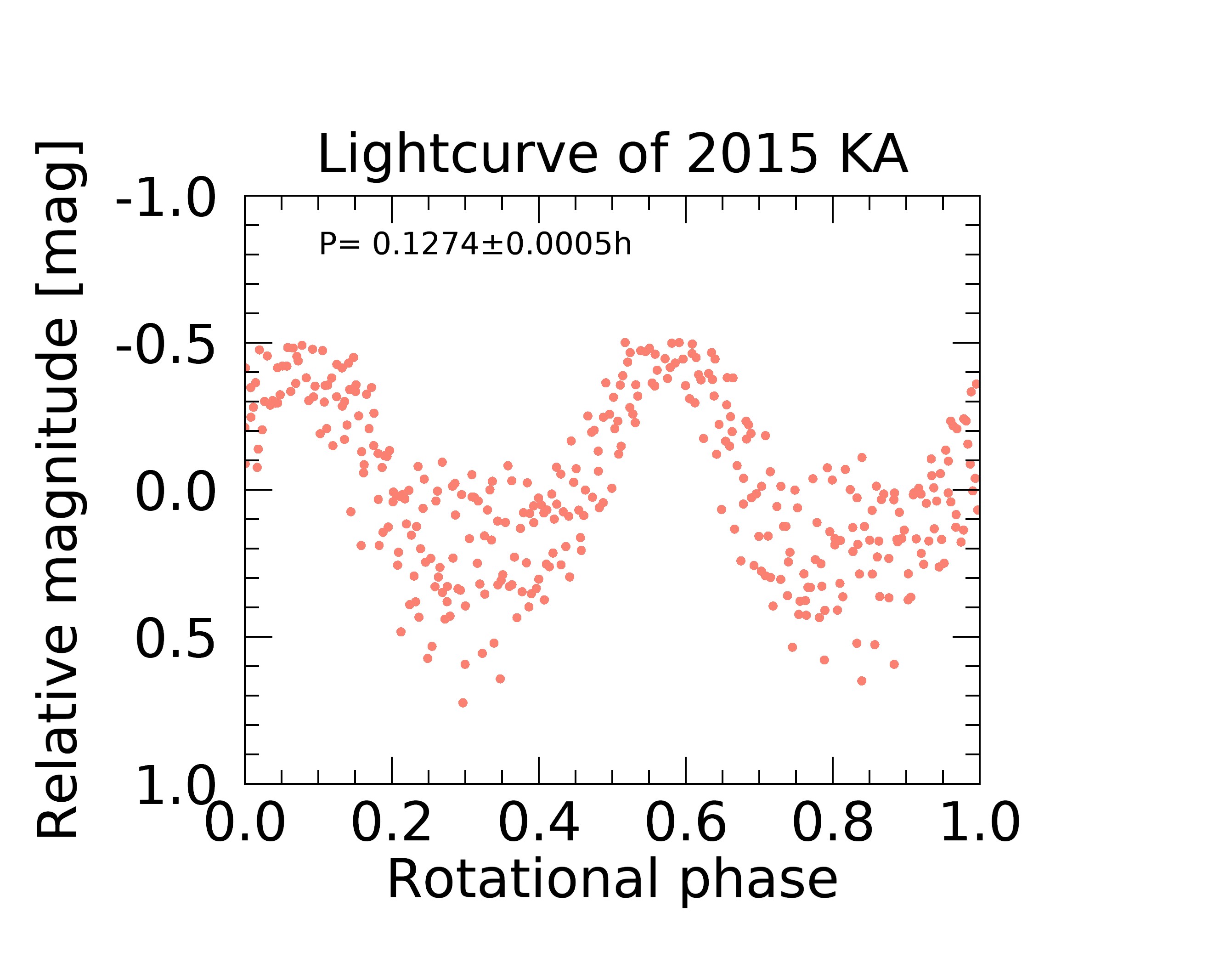}    
 
 \caption{Tumblers' lightcurves of NEOs included in the photometric study.}
\label{fig:Tumblers_lightcurves}

\end{figure*}

 \begin{figure*}
 \includegraphics[width=9cm,angle=0]{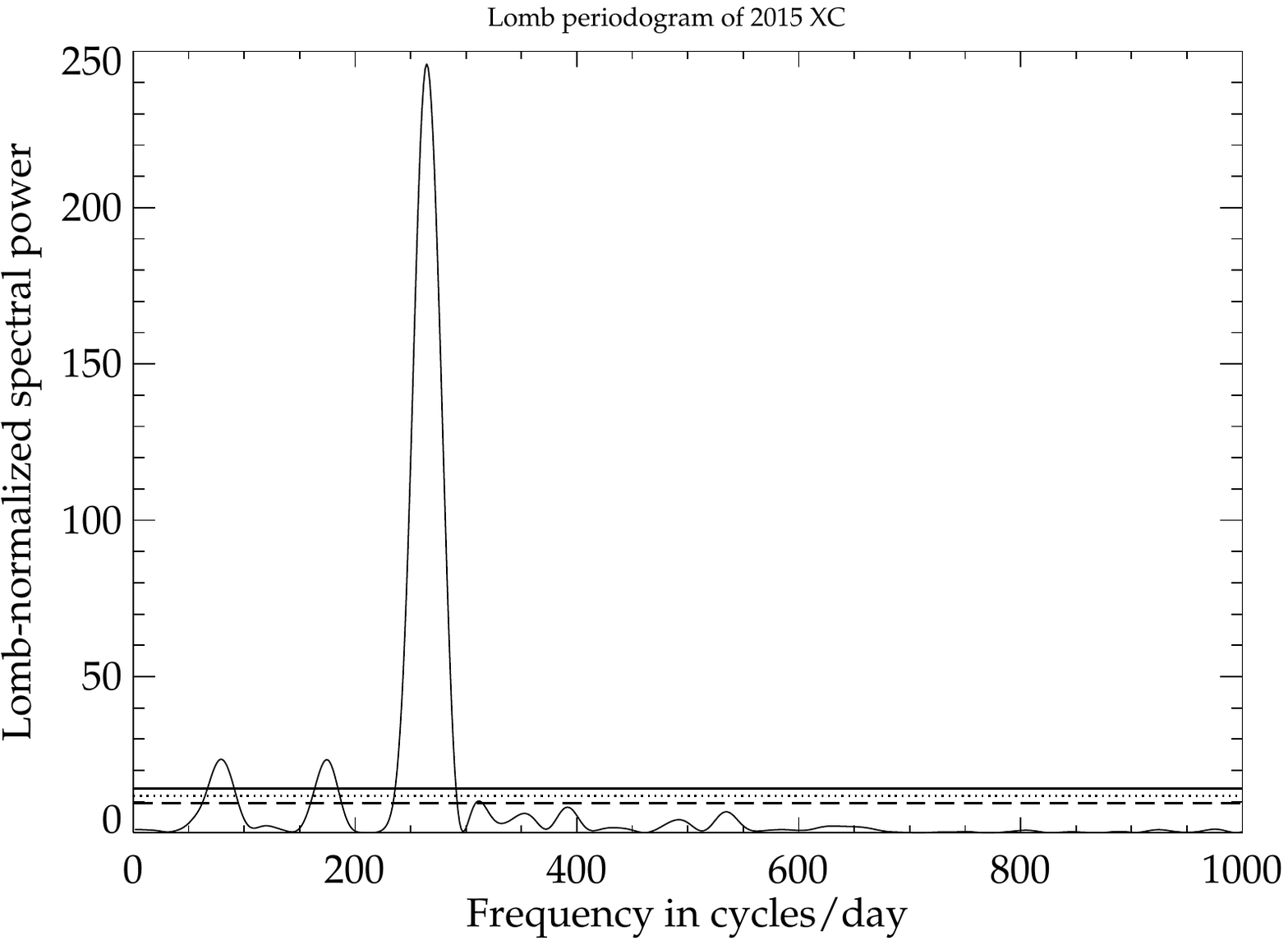}  
 \includegraphics[width=9cm,angle=0]{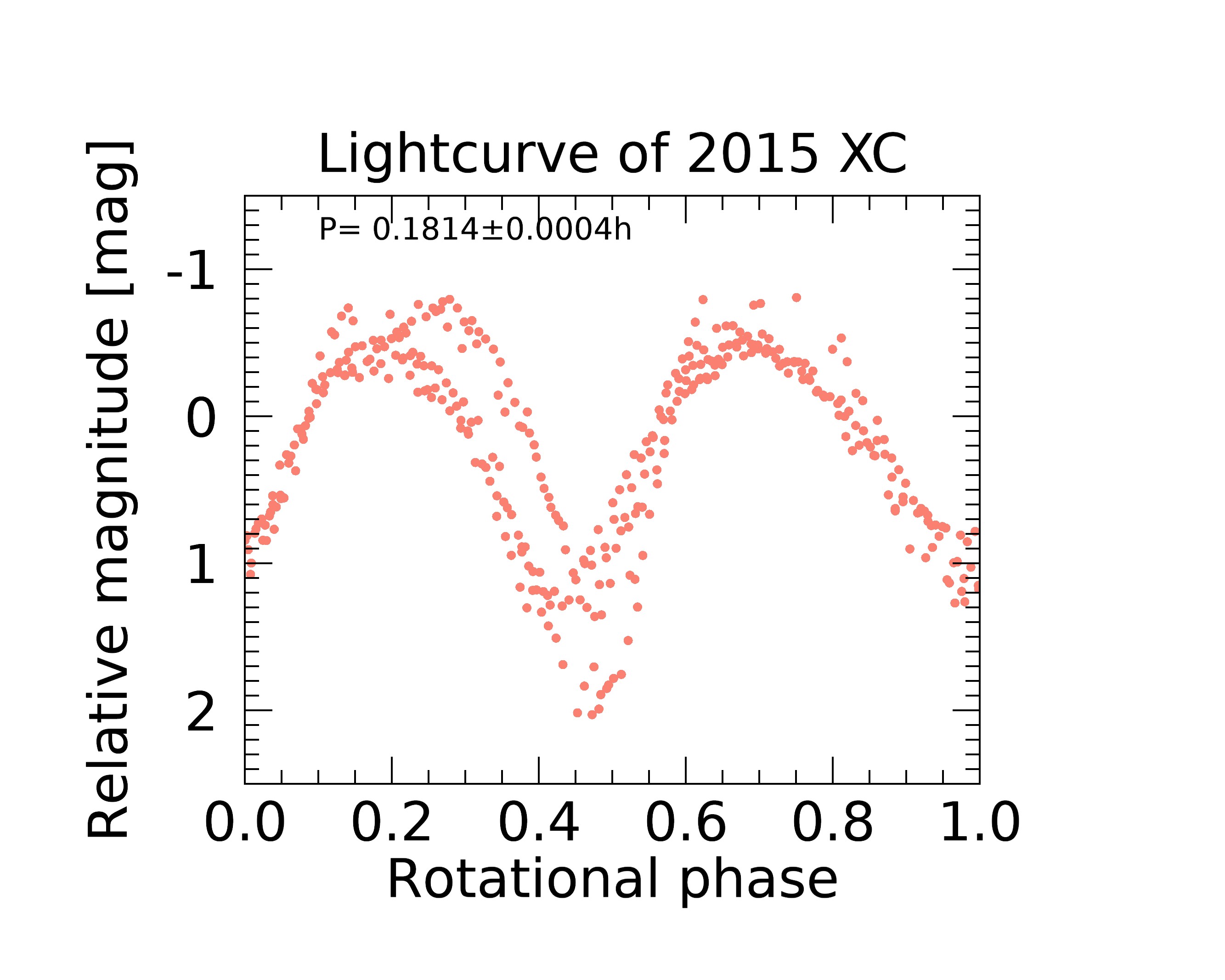}  
 \includegraphics[width=9cm,angle=0]{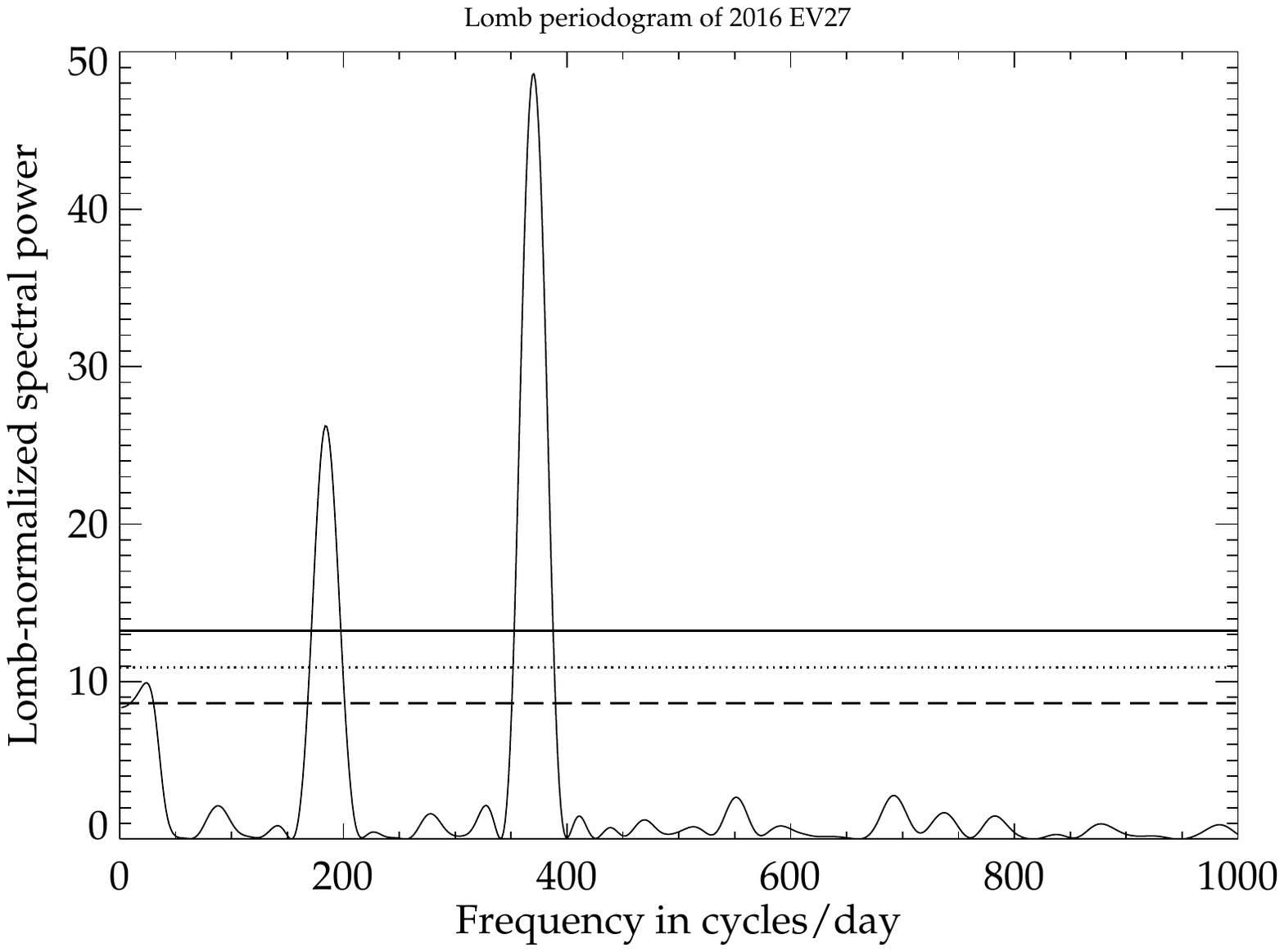}  
 \includegraphics[width=9cm,angle=0]{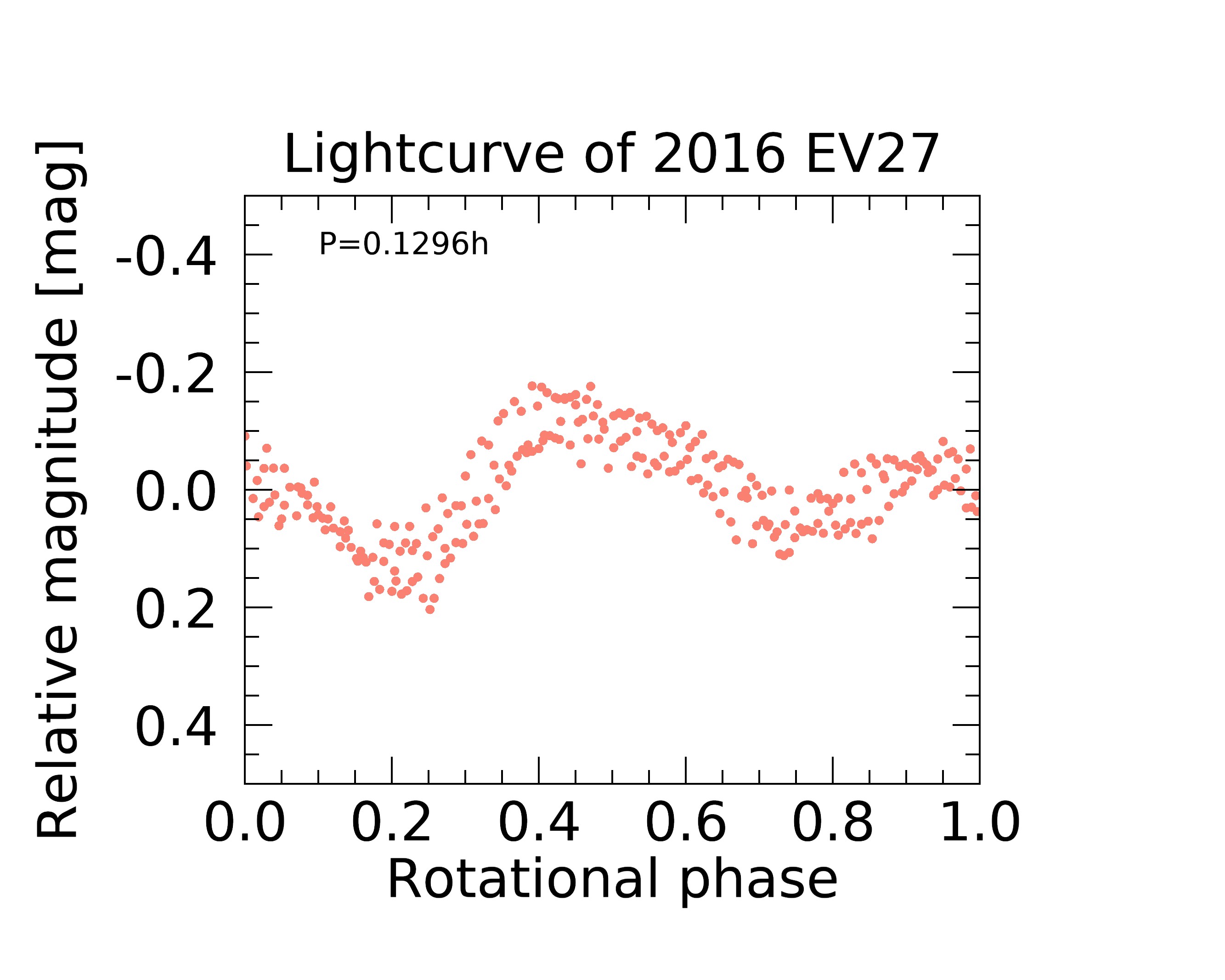}  
  \caption{Tumblers' lightcurves of NEOs included in the photometric study.}
\label{fig:Tumblers_lightcurves}

\end{figure*}

 \begin{figure*}
 \includegraphics[width=9cm,angle=0]{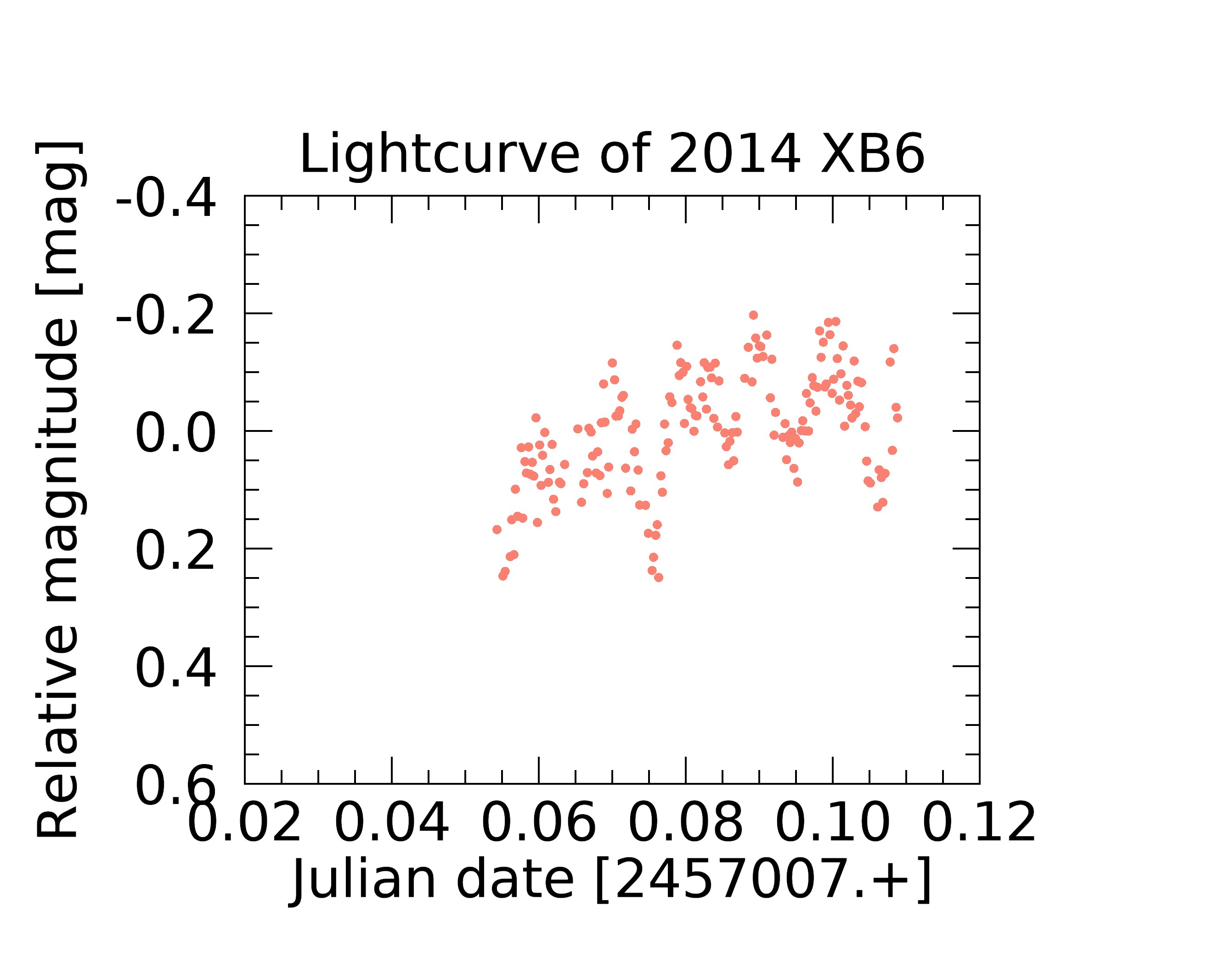}
 \includegraphics[width=9cm,angle=0]{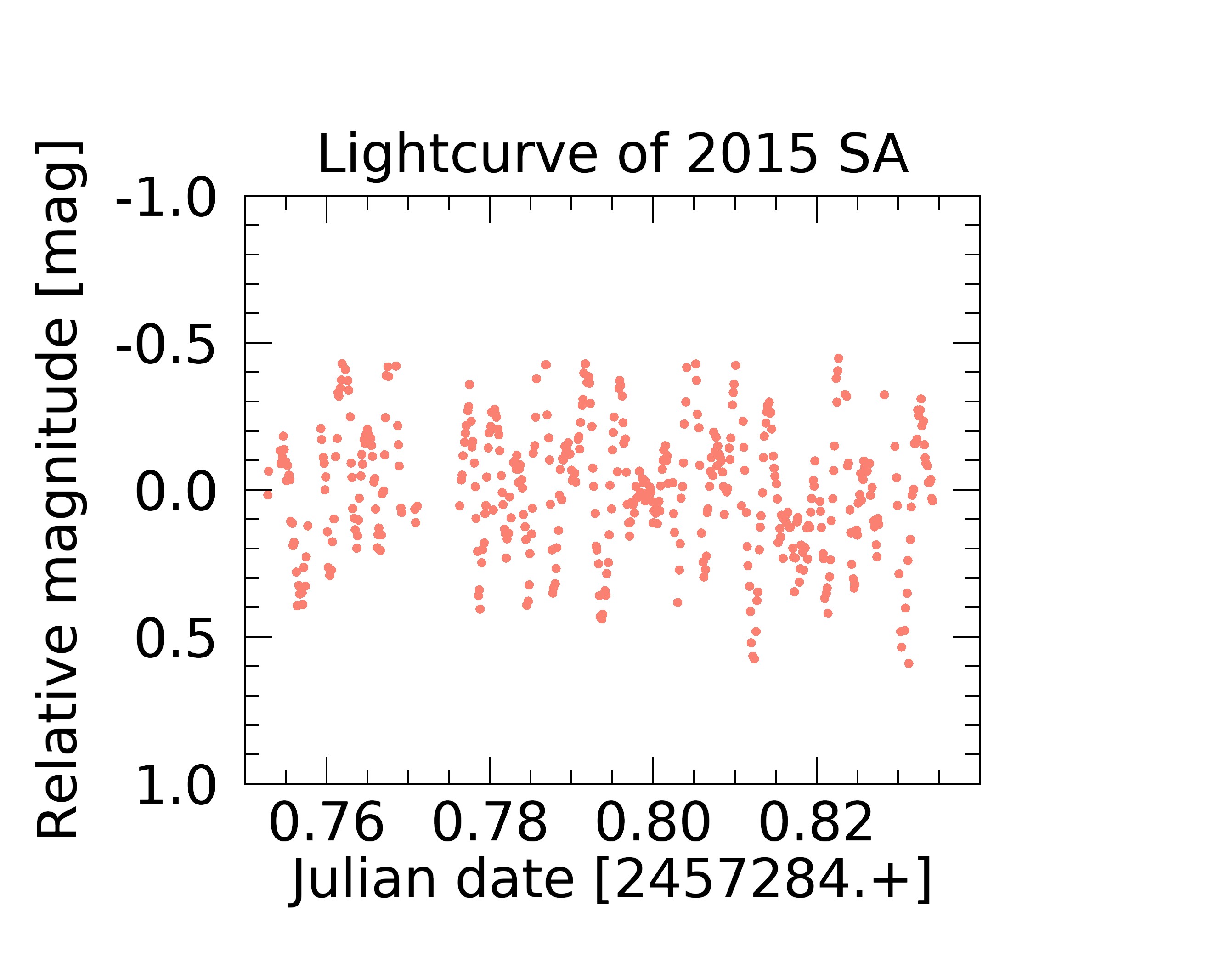}
 \includegraphics[width=9cm,angle=0]{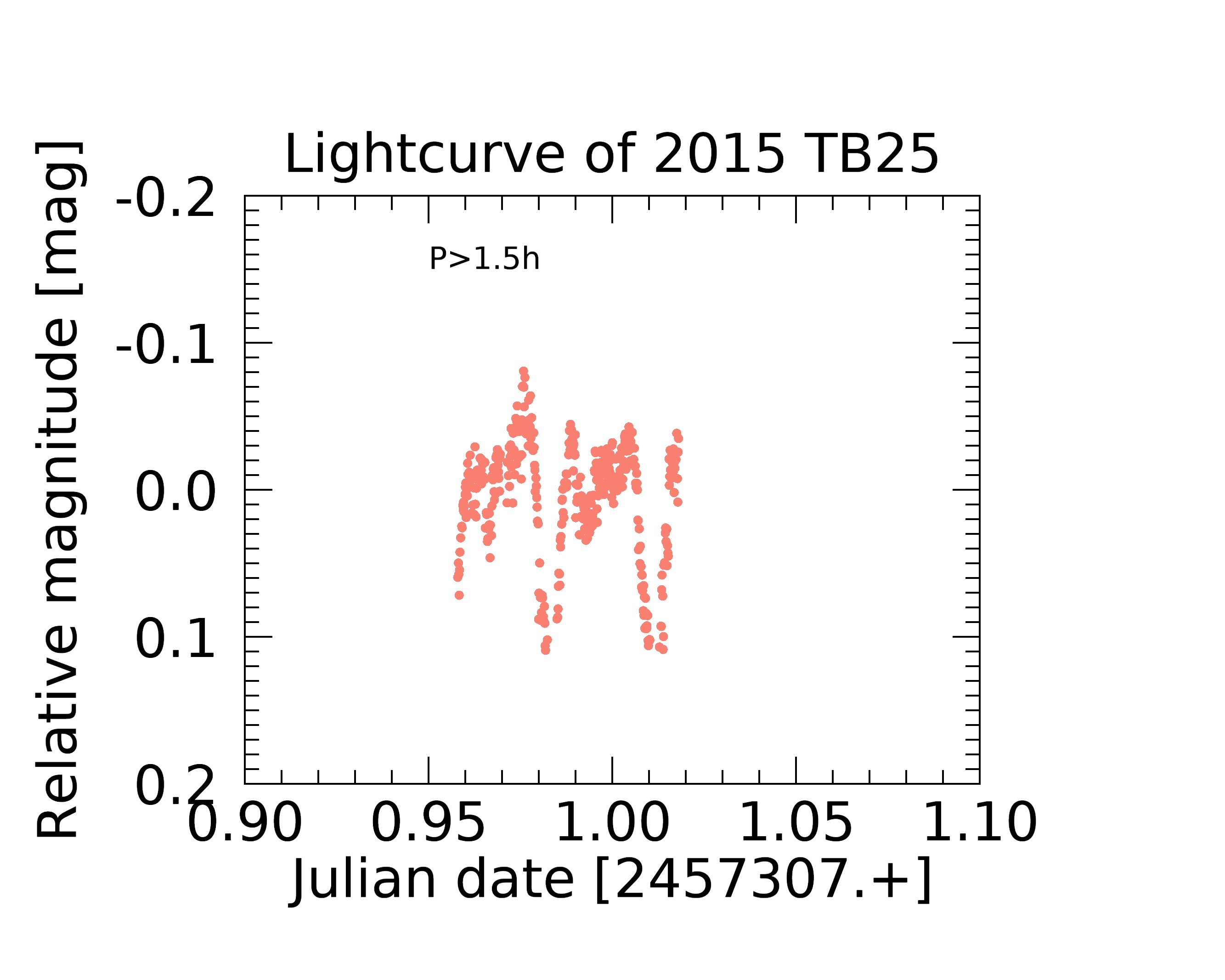}
 \includegraphics[width=9cm,angle=0]{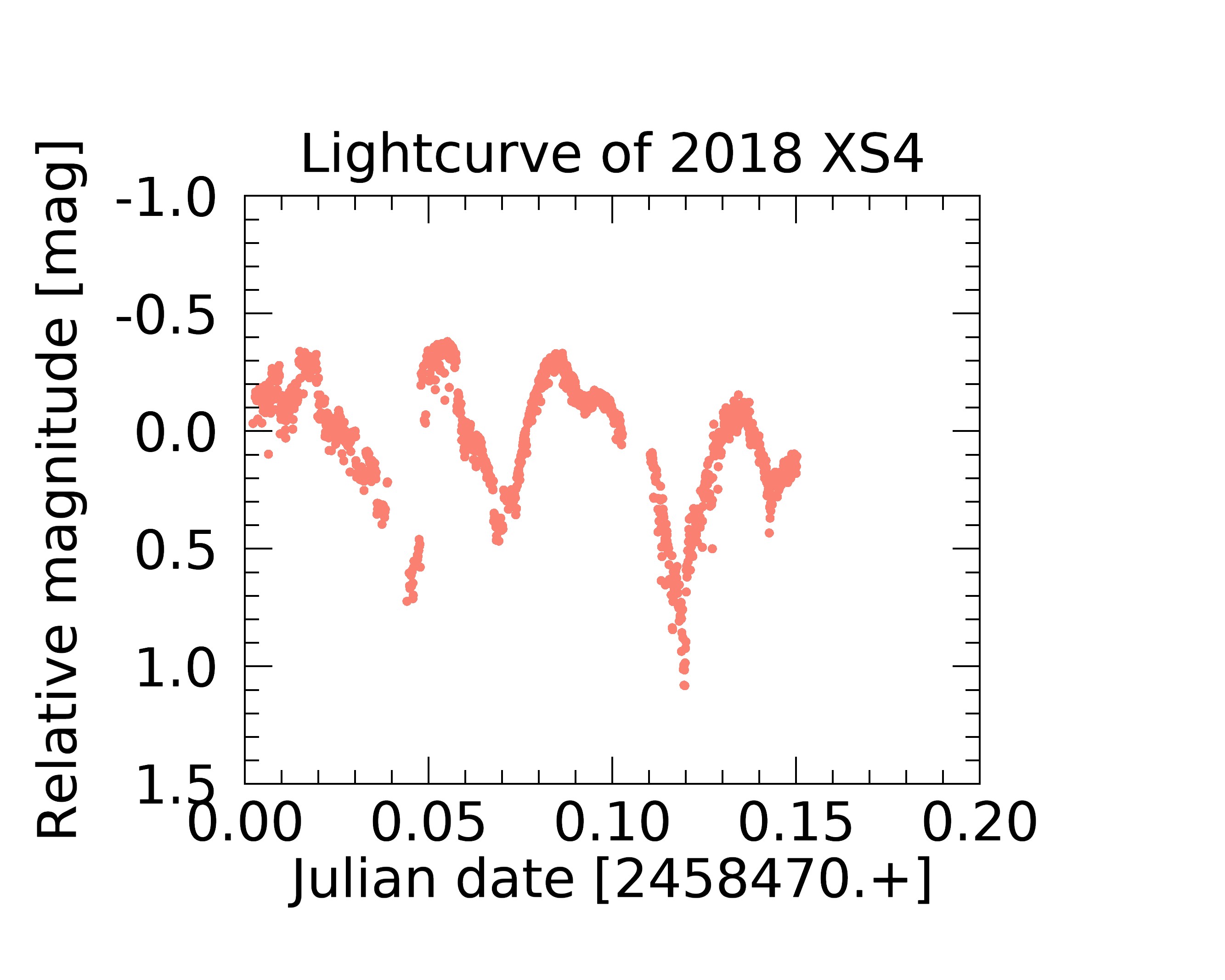}
  \caption{Tumblers' lightcurves of NEOs included in the photometric study.}
\label{fig:Tumblers_lightcurves}

\end{figure*}

\end{appendix}

\end{document}